\documentclass[pra,showpacs]{revtex4}

\usepackage[english]{babel}

\usepackage[pdfborder={0 0 0},colorlinks=true,linkcolor=blue,citecolor=blue]{hyperref}

\usepackage{epsfig}
\usepackage{amsmath}
\usepackage{amstext}
\usepackage{amssymb}
\usepackage{amsfonts}
\usepackage{graphicx}
\usepackage{color}

\def\beq{\begin{equation}}
\def\eeq{\end{equation}}
\def\bea{\begin{eqnarray}}
\def\eea{\end{eqnarray}}

\def\ra{\rangle}
\def\la{\langle}

\def\q{\mathbf{q}}
\def\k{\mathbf{k}}

\def\Q{\mathbf{Q}}

\newcommand{\su}{\uparrow}
\newcommand{\sd}{\downarrow}
\newcommand{\sgn}[1] {\mathrm{sgn}\left({#1}\right)}
\newcommand{\sgnnobr}[1] {\mathrm{sgn}{#1}}
\newcommand{\abs}[1] {\left|{#1}\right|}
\newcommand{\ii}{{\mathrm{i}}}

\newcommand{\nn}{\nonumber}
\newcommand{\vect}[1] {\mathbf{#1}}

\newcommand{\Tmat}{\mathcal{T}}
\newcommand{\Uimp}{\mathcal{U}}
\newcommand{\Vimp}{\mathcal{V}}
\newcommand{\iM}{\mathcal{I}}
\newcommand{\jM}{\mathcal{J}}
\newcommand{\imp}{\mathrm{imp}}

\newcommand{\GammaN}{\Gamma}
\newcommand{\sigmaN}{\sigma}
\newcommand{\etaN}{\eta}

\begin{document}

\pacs{74.20.Rp,74.25.-q,74.62.Dh}

\title{Impurities in multiband superconductors
\footnote{Published in\\ \href{https://doi.org/10.3367/UFNe.2016.07.037863}{\textit{Physics-Uspekhi} \textbf{59} (12), 1211-1240 (2016)}, DOI: 10.3367/UFNe.2016.07.037863 (in English) \\
\href{https://doi.org/10.3367/UFNr.2016.07.037863}{\textit{Usp. Fiz. Nauk} \textbf{186}, 1315-1347 (2016)}, DOI: 10.3367/UFNr.2016.07.037863 (in Russian)
}
}

\author{M.M. Korshunov}
\affiliation{Kirensky Institute of Physics, Federal Research Center KSC SB RAS, 660036, Krasnoyarsk, Russia\\
Siberian Federal University, 660041, Krasnoyarsk, Russia\\
e-mail: mkor@iph.krasn.ru}

\author{Yu.N. Togushova}
\affiliation{Siberian Federal University, 660041, Krasnoyarsk, Russia\\
e-mail: togushova@bk.ru}

\author{O.V. Dolgov}
\affiliation{Max-Planck-Institut f\"{u}r Festk\"{o}rperforschung, D-70569, Stuttgart, Germany\\
P.N. Lebedev Physical Institute RAS, 119991, Moscow, Russia\\
e-mail: o.dolgov@fkf.mpg.de}

\date{\today}

\begin{abstract}
Disorder - impurities and defects violating an ideal order - is always present in solids. It can result in interesting and sometimes unexpected effects in multiband superconductors. Especially if the superconductivity is unconventional thus having other than the usual $s$-wave symmetry. This paper uses the examples of iron-based pnictides and chalcogenides to examine how both nonmagnetic and magnetic impurities affect superconducting states with $s_\pm$ and $s_{++}$ order parameters. We show that disorder causes the transitions between $s_\pm$ and $s_{++}$ states and examine observable effects these transitions can produce.
\end{abstract}

\maketitle

\begin{flushright}
\scriptsize
\item[] \textit{
``For two dangers never cease threatening the world: order and disorder.''\\
Paul Valery, ``Crisis of the Mind'' (1919)
}
\end{flushright}

\begin{flushleft}
\textit{Received 18 May 2016\\
after revision 8 July 2016}
\end{flushleft}

\textbf{Keywords:} unconventional superconductors, iron pnictides, iron chalcogenides, impurity scattering

\tableofcontents


\section{Introduction}

Phenomena of superconductivity always attracted much attention of the scientific community. For the first time it was observed in Kamerlingh Onnes laboratory in 1911. It took half a century to develop the microscopic theory -- only in 1957 Bardeen, Cooper, and Schrieffer published an article~\cite{bcs}, where the superconductivity phenomena was explained by the formation of the condensate of Cooper pairs of electrons having opposite momenta and spins due to the electron-phonon interaction being attractive at small frequencies. Theory of superconductivity developed by Gor'kov, Abrikosov, and Dzyaloshinskii on the basis of Green's functions method~\cite{AGD1962eng} allowed to formalize the approach to the phenomena and describe many of its interesting features. From the second order phase transitions point of view, the superconductivity is a transition to the state with the broken gauge invariance. Phenomenological Ginzburg-Landau theory of superconductivity based on the free energy functional expansion depends on the complex superconducting order parameter $\Delta$. According to a theoretical-group classification, the order parameter of an ordinary superconductor in crystals obeying the tetragonal symmetry belongs to the simplest representation, $A_{1g}$ representation, and is isotropic in momentum space, i.e., has an $s$-wave symmetry~\cite{Volovik1984eng,Volovik1985}. In the simplest case, the gap in the spectrum of Fermi quasiparticles determined by the absolute value of the order parameter, $|\Delta|$. Superconductivity being the fundamental ground state occurs in almost all metals and doped semiconductors, which are not magnetic at low temperatures.

The superconductivity theory got an interesting development during investigations of superconducting states in $^3$He, heavy-fermion materials, and magnetic superconductors (see, e.g. Refs.~\cite{MineevSamokhin1998eng,Riseborough1998,VonsovskiiIzyumovKurmaev1982}). Important feature was the ``unconventional'', non-$s$-wave, symmetry of the order parameter. For example, in $^3$He order parameter has a $p$-wave symmetry.

The next important milestone is the discovery of the so-called high-temperature superconductivity (HTSC) in copper oxides, or, simply cuprates, in 1986~\cite{bednorz-muller}. One of their features is that the critical temperature of the transition to the superconducting state, $T_c$, was exceeding critical temperatures of superconductors known at that time by four to five times. Also, the cuprates have an unconventional symmetry of the superconducting gap, that is, most materials bear a $d$-wave order parameter belonging to the $B_{2g}$ representation of the tetragonal symmetry group~\cite{MineevSamokhin1998eng,SigristUeda}.

In 1990-th and 2000-th, the superconductivity in fullerides~\cite{Hebard1991} and magnesium diboride (MgB$_2$)~\cite{Nagamatsu2001} was discovered, which was explained in the framework of the electron-phonon interaction contrary to the cuprates. Characteristic feature of these materials was their significantly multiband nature, i.e., several bands originating from the mixture of different orbitals cross the Fermi level and form the multiply connected Fermi surface consisting of several sheets. Thus for the description of such systems, it is necessary to use multiband approach. On the contrary, in the cuprates, a single-band approach works well despite their multiband nature.

Discovery of a new class of superconductors in 2008 -- iron-based materials -- started a new phase of unconventional superconductivity studies~\cite{y_kamihara_08}. While the Fe-based systems have not lead to the technological breakthrough yet (these days, $T_c$ in bulk materials is only 15~K higher than that of MgB$_2$, besides, just as the cuprates, they are expensive to make and difficult to work on), conceptual importance of of their discovery is hard to overestimate. Indeed, as the cuprates, fullerides and magnesium diboride reveal many unusual features, however, Cooper pairing in them arise due to the electron-phonon interaction, while in the cuprates the mechanism of superconductivity has probably a non-phononic origin. Not surprisingly, there had been a growing feeling among physicists that phonon superconductivity will probably never grow past 50-60~K, while true high-temperature superconductivity is probably due to a strong-correlations and limited to the unique family of layered cuprates. What the discovery of the iron-based systems brought onto the table was the understanding that however unique cuprates may be, these features are not prerequisites for non-phonon, high-temperature superconductivity. And, if that is true, there are likely many other crystallochemical families to be discovered, some of which may have higher critical temperatures or be better suited for applications than cuprates and iron-based superconductors. For example, the discovery of superconductivity in sulfur hydrates with the record $T_c \approx 200$~K was claimed recently~\cite{Drozdov2014,Drozdov2015}.

Superconducting Fe-based materials can be divided into two subclasses, pnictides and chalcogenides. The square lattice of Fe is the basic element. Iron is surrounded by As or P situated in the tetrahedral positions within the first subclass and by Se, Te, or S within the second subclass. Fe $d$-orbitals are significantly overlapped and, apart from that, out-of-plane pnictogen or chalcogen are well hybridized with the $t_{2g}$-subset of the iron $d$-orbitals, and all of them contribute to the Fermi surface. Minimal model is than a significantly multiband model. In this regard, iron-based materials have more similarity to ruthenates and magnesium diboride than to cuprates. Multiband electronic structure of the cuprates can be described basically within an effective low-energy single-band model due to the dominating contribution of the in-plane $d_{x^2-y^2}$ copper orbital.

Different presently discussed mechanisms of Cooper pairs formation result in the distinct superconducting gap symmetry and structure in iron-based materials~\cite{HirschfeldKorshunov2011}. In particular, the random-phase approximation spin fluctuation (RPA-SF) approach in the clean limit gives the extended $s$-wave gap that changes sign between hole and electron Fermi surface sheets (the so-called $s_{\pm}$ state) as the main instability for the wide range of doping concentrations~\cite{mazin_08,Graser2009,k_kuroki_08,MaitiKorshunovPRB2011,Korshunov2014eng}. On the other hand, orbital fluctuations enhanced by the electron-phonon interaction promote the order parameter to have the sign-preserving order parameter, the so-called $s_{++}$ state~\cite{Kontani}. Electron-phonon interaction by itself (without Coulomb repulsion) also leads to the $s_{++}$ gap~\cite{Boeri_08,Kulic2009}. Thus, probing the gap structure is the fundamental problem that can help in elucidating the underlying mechanism of superconductivity.

Varying amounts of disorder are present in all actually existing materials. Moreover, cuprates and iron-based materials in most cases become superconducting when doped, i.e., some atoms are replaced by others and, consequently, potential is changed at sites where the replacement was made. In this regard, disorder is the inherent part of the observed picture of superconductivity and one has to bear a clear-eyed understanding of its role and impact on the features of studied systems.

\subsection{Comparison of iron pnictides and chalcogenides with cuprates}

High-$T_c$ cuprates are known for their high critical temperature, unconventional superconducting state, and unusual normal state properties. The Fe-based superconductors, with $T_c$ up to 58~K in bulk materials~\cite{SmFeAsOTc}
and probably up to 110~K in monolayer FeSe at the SrTiO$_3$ substrate~\cite{FeSeTc,FeSeARPES,HeFeSeAnneal,TanFeSeARPES,GeFeSe100K},
stand in second place after cuprates.
When superconductivity in the iron-based materials was discovered, the question immediately arose -- how similar are they to cuprates? Let us compare some of their properties.

At first glance, the phase diagrams of cuprates and many Fe-based superconductors are similar. In both cases the undoped materials exhibit antiferromagnetism, which vanishes with doping; superconductivity occurs at some nonzero doping and then disappears, such that $T_c$ forms a ``dome''. While in cuprates the long range ordered N\'eel phase vanishes before superconductivity occurs, in iron-based materials the competition between these orders can take several forms. In LaFeAsO, for example, there appears to be a transition between the magnetic and superconducting states at a critical doping value, whereas in the 122 systems (BaFe$_2$As$_2$ and alike) the superconducting phase coexists with magnetism over a finite range and then persists to higher doping. It is tempting to  conclude that the two classes of superconducting materials show generally very similar behavior, but there are profound differences as well. The first striking difference is that the undoped cuprates are Mott insulators, but iron-based materials are metals. This suggests that the Mott-Hubbard physics of a half-filled Hubbard model is not a good starting point for pnictides, although some authors have pursued strong-coupling approaches. It does not of course exclude effects of correlations in iron-based materials, but they may be moderate or small. In any case, density functional theory-based approaches describe the observed Fermi surface and band structure reasonably well for the whole phase diagram, contrary to the situation in cuprates, especially, in undoped and underdoped regimes.

The second important difference pertains to normal state properties. Underdoped cuprates reveal the pseudogap behavior in both one-particle and two-particle charge and/or spin excitations, while the similar robust behavior is absent in iron-based materials. Generally speaking, the term ``pseudogap'' imply the dip in the density of states near the Fermi level. There are, however, a wide variety of unusual features of pseudogap state in cuprates. For example, a strange metal phase near optimal doping in hole-doped cuprates is characterized by linear-$T$ resistivity over a wide range of temperatures. In iron-based materials, different temperature power laws for the resistivity, including linear $T$-dependence of the resistivity for some materials, have been observed near optimal doping and interpreted as being due to multiband physics and interband scattering~\cite{Golubov_10eng}. There are, however, indications of a pseudogap formation in densities of states of some pnictides band, see, e.g. Refs.~\cite{KordyukPseudogapReview,Kuchinskii2008eng}.

The mechanism of doping deserves additional discussion. Doping in cuprates is accomplished by replacing one of the spacer ions with another one with different valence like in La$_{2-x}$Sr$_x$CuO$_2$ and Nd$_{2-x}$Ce$_x$CuO$_2$ or adding extra out-of-plane oxygen like in YBa$_2$Cu$_3$O$_{6+\delta}$. The additional electron or hole is then assumed to dope the plane in an itinerant state. In iron-based materials, the nature of doping is not completely understood -- similar phase diagrams are obtained by replacing the spacer ion or by in-plane substitution of Fe with Co or Ni. For example, LaFeAsO$_{1-x}$F$_x$, Ba$_{1-x}$K$_x$Fe$_2$As$_2$ and Sr$_{1-x}$K$_x$Fe$_2$As$_2$ belong to the first case, while Ba(Fe$_{1-x}$Co$_x$)$_2$As$_2$ and Ba(Fe$_{1-x}$Ni$_x$)$_2$As$_2$ belong to the second one. Whether these heterovalent substitutions dope the FeAs or FeP plane as in the cuprates was not initially clear~\cite{Sawatzky2009}, but now it is well established that they affect the Fermi surface consistent with the formal electron count doping~\cite{Nakamura2011,Brouet2009}. Another mechanism to vary electronic and magnetic properties is via the possibility of isovalent doping with phosphorous in BaFe$_2$(As$_{1-x}$P$_x$)$_2$ or ruthenium in BaFe$_2$(As$_{1-x}$Ru$_x$)$_2$. ``Dopants'' can act as potential scatterers and change the electronic structure because of difference in ionic sizes or simply by diluting the magnetic ions with nonmagnetic ones. In iron-based materials, therefore, some of the doping mechanisms connected with the changes in the transition metal layer. But crudely the phase diagrams of all Fe-based materials are quite similar, challenging workers in the field to seek a systematic structural observable which correlates with the variation of $T_c$. Among several proposals, the height of the pnictogen or chalcogen above the Fe plane has frequently been noted as playing some role in the overall doping dependence~\cite{k_kuroki_09,Mizuguchi2010,Kuchinskii2010eng}.

It is well established that the superconducting state in the cuprates is universally $d$-wave. By contrast, the gap symmetry and structure of the iron-based materials can be quite different from material to material. Nevertheless, it seems quite possible that the ultimate source of the pairing interaction in both systems is fundamentally similar, although essential details such as pairing symmetry and the gap structure in the iron-based materials depend on the Fermi surface geometry, orbital character, and degree of correlations~\cite{HirschfeldKorshunov2011,Hirschfeld2016}.

\subsection{Role of disorder in cuprates}

Conventional superconductors act differently depending on the type of introduced impurities. So nonmagnetic impurities do not suppress superconducting critical temperature $T_c$ according to the Anderson's theorem~\cite{Anderson1959}, while on the contrary, magnetic impurities cause the $T_c$ suppression with the rate following the Abrikosov-Gor'kov theory~\cite{AGeng}.

Cuprate superconductors reveal more complicated picture. Phase diagram asymmetry for hole and electron doped cuprates is tightly related to the impact of nonmagnetic and magnetic impurities replacing copper sites on superconducting properties. In electron doped systems ($n$-type), the situation is analogous to the conventional superconductors -- nonmagnetic impurities weakly suppress $T_c$, while magnetic ones cause the collapse of superconductivity for the impurity concentration about one percent that is quite in agreement with the Abrikosov-Gor'kov theory. These results follows from studies of both polycrystalline samples~\cite{Tarascon1990,Jayaram1995} and Pr$_{2-x}$Ce$_x$Cu$_{1-y}M_y$O$_{4+z}$ monocrystals with $M$ = Ni, Co~\cite{Brinkmann1996}.

Contrary to the $n$-type cuprates, hole doped counterparts show different behavior. Early studies on YBa$_2$Cu$_3$O$_7$ (Y-123)~\cite{Markert1989} revealed the suppression of superconductivity via replacement of copper not only by magnetic (Fe, Co, Ni), but also by nonmagnetic (Zn, Al, Ga) ions. Note, however, that to compare effect of different types of impurities on $T_c$, it is more convenient to study the lanthanum-based system La$_{2-x}$Sr$_x$Cu$_{1-y}M_y$O$_4$ (with $y$ being the amount of $M$ = Fe, Co, Ni, Zn, Al, Ga), where all impurities are located in the CuO$_2$ layer, as opposed to Y-123 system. In the latter system, copper in-plane sites are replaced by divalent ions, while trivalent ions generally occupy Cu-O chains that reduces their effect on $T_c$ and thus complicates interpretation of results.

The problem described is absent in La$_{1.85}$Sr$_{0.15}$Cu$_{1-y}M_y$O$_4$, for which the following results were obtained in Ref.~\cite{Xiao1990}: both magnetic impurity, Co, and nonmagnetic impurities, Zn, Al, Ga, result in almost the same $T_c(y)$ dependence.
At the same time, Fe cause the most rapid suppression of $T_c$, while Ni gives the slowest decrease of it, though both should be magnetic due to their atomic structure. To clarify the relation between superconductivity and magnetic nature of impurities, the static susceptibility measurements were done~\cite{Xiao1990,Ting1992}. They revealed the presence of an effective magnetic moment at impurity site in all systems studied. Moreover, it become clear that the rate of $T_c$ suppression have a weak correlation with the impurity valence. Moreover, the magnitude of the moment correlates significantly with the critical impurity concentration at which $T_c$ vanishes. This argues in favor of the magnetic mechanism of pairbreaking and against pairbreaking originating from the change in the hole doping.

Authors of Ref.~\cite{Xiao1990} suggested a qualitative explanation for the $T_c(y)$ concentration dependence and for the magnetic properties of impurities. It is based on indications that all impurities with zero spin (nonmagnetic Zn, Al, Ga, as well as Co$^{3+}$ being in the low-spin state) induce effective magnetic moment that is close to Cu$^{2+}$ moment. That is, one has to consider the copper spin removed by the impurity. For the impurity with the open $d$-shell (Fe, Co, Ni), it is necessary to consider not only the removed copper spin, but also the own impurity moment. Experimental value of the Fe$^{3+}$ ion effective moment suggests that it is in a high-spin state with $S=5/2$ in the lanthanum system and it generates the effective moment significantly larger than the Cu$^{2+}$ moment. On the other hand, anomalously small experimental value of Ni$^{2+}$ effective moment suggests that the spin should be no more than $0.32$ instead of the expected $S=1$. The authors of Ref.~\cite{Xiao1990} explain this by the significant delocalization of the Ni spin state, in contrast to the strong localization of Fe state.

Besides the qualitative explanation of the anomalous result of Cu with Ni replacement, proper treatment of the multielectron effects in correlated band structure leads to the quantitative description of the $T_c(y)$ dependence~\cite{Ovchinnikov1995eng}. With the diamagnetic replacement of copper with zinc, the fraction of ions in configuration $d^{10}$ (Zn$^{2+}$) is equal to $y$. Model for such systems is the antiferromagnetic lattice of $S=1/2$ spins with one empty site that behaves as one paramagnetic center due to the uncompensation of sublattices. As for the copper replacement with nickel, the nickel ion Ni$^{2+}$ which formally should be in the $d^8$ state with the spin $S=1$, due to the strong intraatomic Coulomb repulsion have an intermediate valence. The probability of it being in the nonmagnetic $d^{10}$ state with the spin $S=0$ is equal to $u_0^2$ and the probability of $d^9$ state with $S=1/2$ is $v_0^2 = 1 - u_0^2$. As follows from the summary of optical, photoemission, and magnetic data on La$_2$CuO$_4$, weights of these states, $u_0^2$ and $v_0^2$, are expressed via such parameters of the multiband $p-d$ model of copper oxides~\cite{Gaididei1988} as energies of $p$ and $d$ holes in the crystal field and matrix elements of Coulomb interaction. This way, instead of nominal Ni$^{2+}$ state with $S=1$, nickel ion should have the effective spin $S=v_0^2 \cdot 1/2$. Calculated values of $v_0^2=0.72$ and $S=0.36$~\cite{Ovchinnikov1995eng} are in a good agreement with the experimental data. Therefore, with the substitution of nickel for copper, the amount of impurity ions in the $d^{10}$ state (the same state as zinc) is equal to $u_0^2 y$. Probability of ions to be in the $d^9$ state is $v_0^2 y$, and since their magnetic and charge characteristics are almost the same as of copper, such ions should not suppress superconductivity. The resulting ratio of $T_c(y)$ slopes for nickel and zinc impurities is $u_0^2=0.28$ that is close to the experimental value of $0.38$ for La$_{1.85}$Sr$_{0.15}$Cu$_{1-y}M_y$O$_4$~\cite{Xiao1990}.

The change of impurities effect with doping can be summarized as follows. Suppression of $T_c$ by impurities in overdoped systems does not depends on the doping $p$, while in the underdoped samples, it is strongly doping-dependent~\cite{Kluge1995}. Given that the pseudogap state occurs exactly at low doping, the observed $p$-dependence emphasize the importance of the ground state in the response of the system to the disorder.

Note, the alternative to the chemical introduction of impurities is the creation of defects via a fast neutron irradiation. Such a method benefits from avoiding some of the difficulties related to the replacement of some atoms with others. Suppression of $T_c$ in this case, as well as other physical characteristics of cuprates being irradiated by neutrons, are extensively described in Refs.~\cite{Davydov1988eng,Aleksashin1989,Ananyev1998eng}.

Summarizing, strong electronic correlations causing the formation of local moments due to the presence of formally nonmagnetic impurities complicates significantly the interpretation of effects of disorder on the $T_c$ suppression. Among other factors preventing the formulation of a consistent theory for the role of defects in superconductivity of cuprates are the absence of the theory for the correlated ground state, difficulties with controlling the defects parameters, and the presence of the anisotropy in impurity scattering. Since the detailed discussion of the cuprates physics with an important role of strong correlations is not the goal of the present review, here we mentioned only a few important points of impurity scattering. We direct the curious reader to other reviews like~\cite{Ovchinnikov1997eng,Kulic2000,Hussey2002,Hirschfeld2002,Balatsky2006,Alloul2009,Pogorelov2011}.

Also, we are not going further into the details of the $d$-wave superconductivity and related problems in the cuprates. This topic is extensively reviewed in many papers, concerning both theories of impurity scattering in a $d$-wave superconductor~\cite{Radtke1993,Preosti1994,Fehrenbacher1994,Balatsky1995,Haran1996,Franz1997,Kulic1997,Kulic1999,Chen2002,Fujita2005} and the effect of impurities on observable features of cuprates~\cite{Hirschfeld1989,Hirschfeld1993,Hirschfeld1994,Quinlan1996,Hirschfeld1997,Duffy2001,Atkinson2002,Tsai2002,Zhu2003}.
Let us just mention that the single-band $d$-wave superconductor can be approximately treated as the two-band superconductor with the opposite signs of gaps in different bands -- the analogy of the $s_\pm$ state~\cite{Muzikar1996}. In other words, parts of the Fermi surface with different signs of the order parameter are considered as originating from different bands. Though this is a rough approximation, it can give some qualitative results.

\subsection{Specific features of iron-based superconductors}

Iron under normal conditions is ferromagnetic. Under the pressure, however, once the Fe atoms form an hcp lattice, iron becomes nonmagnetic and even superconducting at $T<2$К~\cite{Shimizu2001} most probably due to the electron-phonon interaction~\cite{Bose2003}. On the other hand, iron-based superconductors are the quasi-two-dimensional materials with the conducting square lattice of Fe ions. Fermi level is occupied by the $3d^6$ states of Fe$^{2+}$. It was established in the early DFT (Density Functional Theory) calculations~\cite{s_lebegue_07,d_singh_08,mazin_08}, which are in a quite good agreement with the results of quantum oscillations and ARPES (Angle-Resolved Photoemission Spectroscopy). All five orbitals, $d_{x^2-y^2}$, $d_{3z^2-r^2}$, $d_{xy}$, $d_{xz}$, and $d_{yz}$, are near or at the Fermi level. This results in the significantly ``multiorbital'' and multiband low-energy electronic structure, which could not be described within the single-band model. For example, within the five-orbital model~\cite{Graser2009} correctly reproducing the DFT band structure~\cite{c_cao_08}, the Fermi surface comprised of four sheets: two hole pockets around the $(0,0)$ point and two electron pockets around $(\pi,0)$ and $(0,\pi)$ points. Such $\k$-space geometry results in the possibility of the spin-density wave (SDW) instability due to the nesting between hole and electron Fermi surface sheets at the wave vector $\Q=(\pi,0)$ or $(0,\pi)$. Upon doping $x$ the long-range SDW order is destroyed. If electrons are doped, then for the large $x$ hole pockets disappear leaving only electron Fermi surface sheets that is observed in K$_x$Fe$_{2-y}$Se$_2$ and in FeSe monolayers~\cite{FeSeARPES}. Upon increase of the hole doping, first, a new hole pocket appears around $(\pi,\pi)$ point and then electron sheets vanish. KFe$_2$As$_2$ corresponds to the latter case. ARPES confirms that the maximal contribution to the bands at the Fermi level comes from the $d_{xz,yz}$ and $d_{xy}$ orbitals~\cite{Kordyuk,Brouet}. At the same time, as will be pointed out later, the presence of a few pockets and the multiorbital band character significantly affect the superconducting pairing.

Soon after high quality samples of cuprates were prepared, the $d_{x^2-y^2}$ symmetry of the gap, with $\cos k_x-\cos k_y$ structure, was empirically established by penetration depth, ARPES, NMR and phase sensitive Josephson tunneling experiments. No similar consensus on any universal gap structure has been reached even after several years of intensive research on the high-quality monocrystals of iron-based superconductors. There is strong evidence that small differences in electronic structure can lead to a strong diversity in superconducting gap structures, including nodal states and states with a full gap at the Fermi surface. The actual symmetry class of most of the materials may be of generalized $A_{1g}$ (extended $s$-wave symmetry) type, probably involving a sign change of the order parameter between Fermi surface sheets or its parts~\cite{HirschfeldKorshunov2011}. Understanding the symmetry character of the superconducting ground states as well as the detailed structure of the order parameter should provide clues to the microscopic pairing mechanism in the iron-based materials and thereby lead to a deeper understanding of the phenomenon of high-temperature superconductivity.

The group theoretical classification of gap structures in unconventional superconductors is rather complicated and has been reviewed in, e.g. Ref.~\cite{SigristUeda}. In the absence of spin-orbit coupling, the total spin of the Cooper pair is well-defined and can be either $S=1$ or $S=0$. The easiest and the most accurate way to probe whether the pair is spin-triplet is via the Knight shift measurements. These experiments have been performed on several iron-based materials including Ba(Fe$_{1-x}$Co$_x$)$_2$As$_2$~\cite{f_ning_08}, LaFeAsO$_{1-x}$F$_x$~\cite{h_grafe_08}, PrFeAsO$_{0.89}$F$_{0.11}$~\cite{k_matano_08}, Ba$_{1-x}$K$_x$Fe$_2$As$_2$~\cite{MatanoBKFA,m_yashima_09}, LiFeAs~\cite{Jeglic111,Li111}, and BaFe$_2$(As$_{0.67}$P$_{0.33}$)$_2$~\cite{Nakai2010}. It was found that the Knight shift decreases in all crystallographic directions. This effectively excluded triplet symmetries such as $p$-wave or $f$-wave.

Having excluded the spin-triplet states, we focus first on simple tetragonal point group symmetry. In a three-dimensional tetragonal system, group theory allows only for five one-dimensional irreducible representations according to how the order parameter transforms under rotations by $90^\circ$ and other operations of the tetragonal group: $A_{1g}$ (``$s$-wave''), $B_{1g}$ (``$d$-wave'', $x^2-y^2$), $B_{2g}$ (``$d$-wave'', $xy$), $A_{2g}$ (``$g$-wave'', $xy(x^2-y^2)$), and $E_g$ (``$d$-wave'', $xz,yz$). Note that the $s_{++}$ and $s_{\pm}$ states all have the same \textit{symmetry}, i.e., neither changes sign if the crystal axes are rotated by $90^\circ$. By contrast, the $d$-wave state changes sign under a such rotation. Note further that the mere existence of the hole and electron pocket lead to new ambiguities in the sign structure of the various states. In addition to a global change of sign, which is equivalent to a gauge transformation, one can have individual rotations on single pockets and still preserve symmetry. For example, if for the $d$-wave case one rotates the gap on the hole pocket by a $90^\circ$ but keeps the electron pocket signs fixed, it still represents a $B_{1g}$ state. $B_{2g}$ states are also possible by symmetry and would have nodes on the electron pockets. Further, more complicated, gap functions with differing relative phases become possible when more pockets are present and when three-dimensional effects are included.

It is important to note that, while $d$-wave does not necessarily imply the existence of gap nodes, in combination with a quasi-two-dimensional Fermi surface at the center of the Brillouin zone such nodes are unavoidable: either vertical for the $B_{1g}$, $B_{2g}$, and $A_{2g}$ symmetries, or horizontal, for the $E_g$ symmetry. Since such a Fermi surface exists in pnictides, experimentally proved absence of nodes on it would evidence against the $d$-wave symmetry. As for experiments, the surface probe such as ARPES show full gaps at the central Fermi surface sheet. Moreover, the full gap at the whole Fermi surface observed in tunneling and bulk probes in hole-doped systems as well as in materials with a small electron doping.

There are also direct experiments that provide evidences against $d$-wave. The Josephson current in the $c$-direction when the studied superconductors is coupled to a known $s$-wave superconductor would confirm the $s$-type of the former. Exactly such current was observed in the 122 single crystals~\cite{Greene_tunneling}.

Another piece of evidence comes from the absence of the so-called anomalous Meissner effect (or Wohlleben effect)~\cite{KAM}. This effect appears in polycrystalline samples with random orientation of grains. It was predicted in the beginning of the cuprates era~\cite{Geshkenbein1986eng} and since then it has been routinely observed only in $d$-wave superconductors. The Wohlleben effect appear due to the fact that the response to a weak external magnetic field is paramagnetic, i.e., opposite to the standard diamagnetic response of an $s$-wave superconductor. This happens because half of weak links have a zero phase shift, and other half have the $\pi$ phase shift.

The described separate pieces of evidence strongly suggest that the pairing symmetry is $s$-wave, and not $d$-wave. However, we want to stress that direct testing similar to that performed in cuprates, namely a single-crystal experiment with a $90^\circ$ Josephson junction forming a closed loop, is still missing, and it is highly desirable to make an ultimate conclusion.

Also it should be borne in mind that nothing forbids different iron-based materials from having different order parameter symmetries, although our previous experience with other superconductors tends to argue against this. Indeed, there are several theories claiming that, while most iron-based systems have the $s$-wave gap symmetry, those with unusual Fermi surfaces with either electron or hole pockets can have the $d$-wave symmetry of the order parameter~\cite{Maier2011,Wang2011,Das2011,MaitiKorshunovPRL2011,Mazin2011,Korshunov2014eng}.

\begin{figure}
\centering
\includegraphics[width=0.5\textwidth]{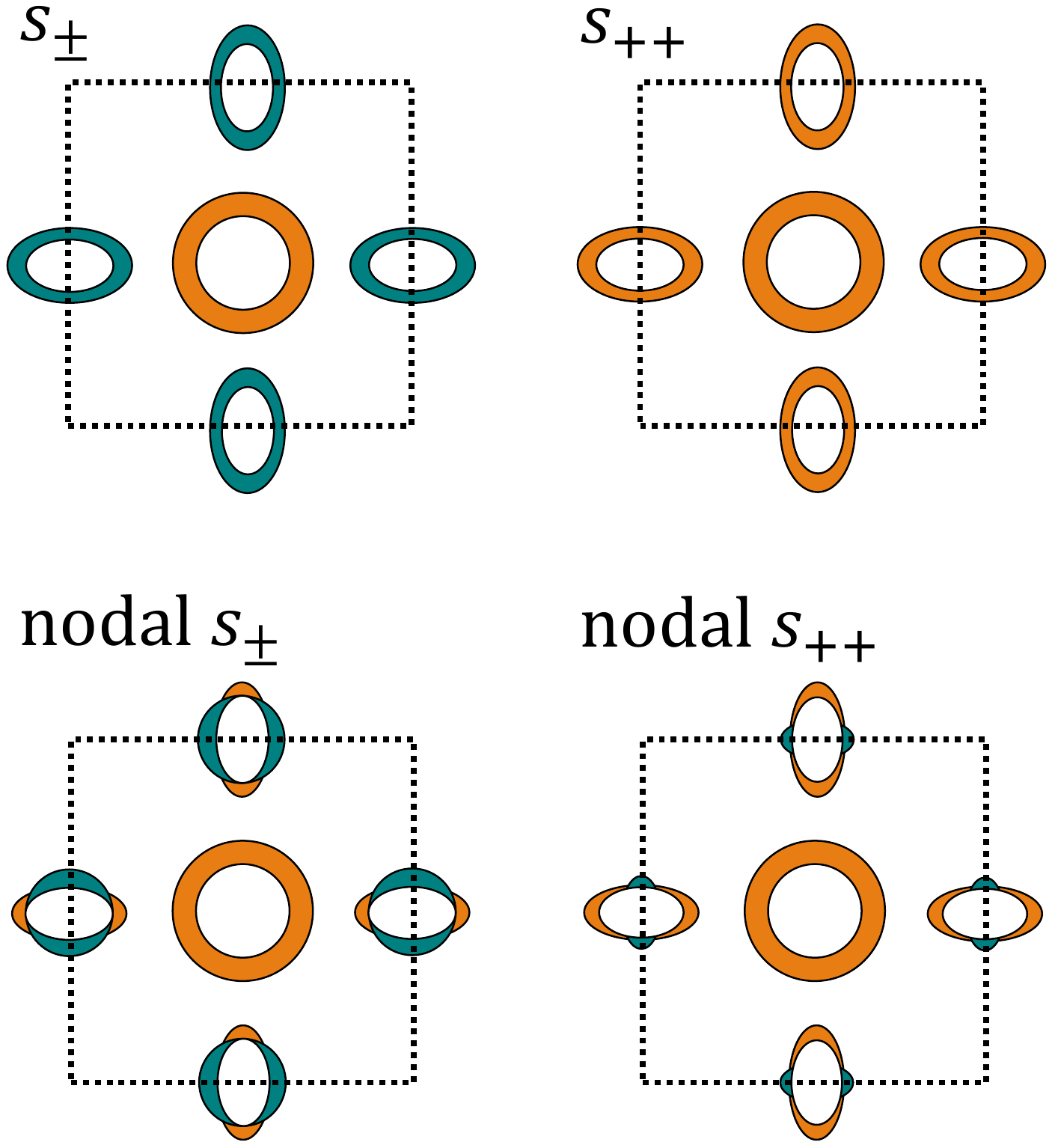}
\caption{(Color online.) Cartoon of four types of order parameter \textit{structures} having the $s$-wave symmetry in the two-dimensional Brillouin zone (dashed square) corresponding to one iron per unit cell. Different colors stands for different signs of the gap.
\label{fig:gaps}}
\end{figure}

Note that the term \textit{symmetry} should be distinguished from the term \textit{structure} of the gap. Latter we use to designate the $\k$-dependent variation of an order parameter within a given symmetry class. Gaps with the same symmetry may have very different structures. Let us illustrate this for the $s$-wave symmetry (Figure~\ref{fig:gaps}). Fully gapped $s$-states without nodes at the Fermi surface differ only by a relative gap sign between the hole and electron pockets, which is positive in the $s_{++}$ state and negative in the $s_\pm$ state. On the other hand, in the nodal $s$-states, the gap vanishes at certain points on the electron pockets. These states are called ``nodal $s_\pm$'' (``nodal $s_{++}$'') and are characterized by the opposite (same) averaged signs of the order parameter on the hole and electron pockets. Nodes of this type are sometimes described as ``accidental'', since their existence is not dictated by symmetry in contrast to the symmetry nodes of the $d$-wave gap. Therefore, they can be removed continuously, resulting in either an $s_\pm$ or an $s_{++}$ state~\cite{v_mishra_09,Mizukami2014}.

Superconducting states with different symmetries and structures of order parameters act differently being subject to the disorder. As we have mentioned earlier, in the single-band $s$-wave superconductors, nonmagnetic impurities do not suppress $T_c$, while magnetic impurities do it in accordance with the Abrikosov-Gor'kov theory~\cite{AGeng}. In the unconventional superconductors, suppression of the critical temperature as a function of a parameter $\Gamma$ characterizing impurity scattering may follow a complicated though a particular law. That is why the desire of many authors to attribute the observed $T_c(\Gamma)$ dependence as pointing towards a particular gap structure is not surprising.

Several experiments on iron-based systems show that the $T_c$ suppression is much weaker than expected in the framework of the Abrikosov-Gor'kov theory for both nonmagnetic~\cite{Karkin2009,Cheng2010,Li2010,Nakajima2010,Tropeano2010,Kim2014,Prozorov2014} and magnetic disorder~\cite{Cheng2010,Tarantini2010,Tan2011,Grinenko2011,Li2012}. It is worth advising the reader to interpret $T_c$ suppression results with caution, for several reasons. First, in some cases not all the nominal concentration of impurity substitutes in the crystal. Second, ``slow" and ``fast" $T_c$ suppression cannot be determined by plotting $T_c$ vs impurity concentration, but only vs a scattering rate directly comparable to the theoretical scattering rate, which is generally difficult to determine from experiments. The alternative is to plot $T_c$ vs residual resistivity change $\Delta \rho$, but, first, this is only possible if the $\rho(T)$ curve shifts rigidly with disorder, and, second, if comparisons with theory include a proper treatment of the transport rather than the quasiparticle lifetime. Finally, the effect of a chemical substitution in a iron-based superconductor is quite clearly not describable solely in terms of a potential scatterer, but the impurity may dope the system or cause other electronic structure changes which influence the pairing interaction. Most promising alternative are the irradiation experiments since the disorder is introduced without altering the chemical composition of the studied material. Experiments of this kind include irradiation by protons~\cite{Nakajima2010,Schilling2016,Smylie2016}, neutrons~\cite{Karkin2009}, electrons~\cite{Strehlow2014,Cho2014,Prozorov2014,Mizukami2014}, and heavy ions~\cite{Kim2010,Murphy2013,Salovich2013}. There are some specific complications though. For example, consider the works on neutron irradiation by Karkin et al.~\cite{Karkin2009}. As seen from the other work of the same group~\cite{Gerashenko2009}, the structure of the studied material is changing after the neutron irradiation. The doping is also accompanied by the changes of structural parameters that correlates with the changes of $T_c$~\cite{Zhao2008,k_kuroki_09,Mizuguchi2010,Kuchinskii2010eng}. And it seems that the problem of separating the role of defects and changes of structural parameters with neutron irradiation is not a simple one. Given the many uncertainties present in the basic modeling of a single impurity, as well as the multiband nature of the iron-based materials, it is reasonable to assume that systematic disorder experiments may not play a decisive role in determining the order parameter symmetry and structure. Nevertheless, one can extract useful information from the qualitative effects appearing on the level of simple multiband models of disorder~\cite{EfremovKorshunov2011,KorshunovMagn2014}.

In this review, we demonstrate the basics of impurity effects on the multiband superconductivity using the simple model for iron-based materials as an example. In particular, within the $\Tmat$-matrix approximation, we discuss the role of the scattering on nonmagnetic and magnetic impurities for the $s_\pm$ and $s_{++}$ states in a two-band model. We show that for the finite nonmagnetic impurity scattering rate, the transition from $s_\pm$ to $s_{++}$ occurs, i.e., one of two gaps changes the sign going through zero. The transition happens for the positive sign of the averaged over the bands superconducting coupling constant. At the same time, $T_c$ stays finite and almost independent of the impurity scattering rate that is proportional to the impurity concentration and magnitude of the scattering potential. There are two cases for scattering on magnetic impurities, when the transition temperature $T_c$ is not fully suppressed in contrast to the Abrikosov-Gor'kov theory, but a saturation of it appears in the regime of the large scattering rate. The first case is characterized by purely interband impurity scattering. At the same time, the $s_\pm$ gap is preserved, while the $s_{++}$ state transforms into the $s_\pm$ state with increasing magnetic disorder. The second case corresponds to the unitary limit with the gap structure remaining intact. The reason for the $s_\pm \leftrightarrow s_{++}$ transitions is the following -- if one of the two competing superconducting interactions leads to the state robust against impurity scattering, then although it was subdominating in the clean limit, it should become dominating while the other state is destroyed by impurity scattering. Since the transitions between the $s_\pm$ and $s_{++}$ states go through the gapless regime, they should reveal themselves in thermodynamic and transport properties of the system and thus be observable in optical and tunneling experiments, as well as in the photoemission spectroscopy.
Because one of the gaps vanishes near the transition, ARPES should reveal the gapless spectra and in the optical conductivity, the transition should results in the ``restoring'' of the Drude frequency dependence of $\mathrm{Re}\sigma(\omega)$. We left behind the complicated question of nonmagnetic and magnetic scattering channels coexistence due to its poor development in the multiband case at the time of writing.

The structure of the review is the following. In Section~\ref{sec:Tmat} we present the Eliashberg formalism for the multiband superconductor and the $\Tmat$-matrix approximation for the impurity self-energy. Than the approximation is applied to the simple two-band model, in which either $s_\pm$ or $s_{++}$ state occurs depending on the parameters. Section~\ref{sec:Born} contains the discussion of the qualitative impurity scattering effects in the Born limit. In Sections~\ref{sec:nonmag} and~\ref{sec:magn} the role of nonmagnetic and magnetic impurities is described correspondingly. Section~\ref{sec:experimentsreview} is devoted to the short review of the experimental findings on the impact of impurities on the superconducting state of pnictides and chalcogenides. In Section~\ref{sec:experiments} we discuss the effect of disorder on such experimentally observable dynamical characteristics as a density of states, the spectral function, an optical conductivity, and the magnetic field penetration depth. Conclusions are contained in the final Section~\ref{sec:conclusion}.

\section{Strong coupling formalism and the $\Tmat$-matrix approximation \label{sec:Tmat}}

For the sake of simplicity, we consider a two-band model with the interaction leading to the superconductivity with the spin-singlet order parameter that is isotropic in each band. Results can be easily generalized for the larger number of bands, as will follow from equations below. Isotropy of the order parameter allows to obtain some results analytically, though it is a heavy restriction of the theory. On the other hand, with it one can pursue the approximate treatment of the superconductors with the sign-changing gap, like the $d$-wave cuprates, where parts of the Fermi surface with different signs of the gap can be roughly considered as contributions from different bands~\cite{Muzikar1996}.

For the considered task of impurity scattering, Hamiltonian can be written in the following form:
\begin{equation}
H = \sum\limits_{\k,\alpha, \sigma} \xi_{\k \alpha} c_{\k \alpha \sigma}^{\dag} c_{\k \alpha \sigma} + H_{sc} + H_{\imp},
\label{eq.H}
\end{equation}
where $c_{\k \alpha \sigma}$ is the annihilation operator of the electron with a momentum $\k$, spin $\sigma$, and a band index $\alpha$ that equals to $a$ (first band) or $b$ (the second one), $\xi_{\k \alpha}$ is the electron dispersion that, for simplicity, we treat as linearized near the Fermi level, $\xi_{\k \alpha} = \vect{v}_{F \alpha} (\k - \k_{F \alpha})$, with $\vect{v}_{F\alpha}$ and $\k_{F \alpha}$ being the Fermi velocity and the Fermi momentum of the band $\alpha$, respectively.

Superconductivity occurs in our system due to the interaction $H_{sc}$. It has different form for different mechanisms of pairing. That is, when the superconductivity is mediated by the spin and/or orbital fluctuations, it is the on-site Coulomb (Hubbard) electron-electron interaction~\cite{Castallani1978,Oles1983,k_kuroki_08,Graser2009},
\bea
 H_{sc}^{sf} &=& U \sum_{f, l} n_{f l \su} n_{f l \sd} + U' \sum_{f, l < l'} n_{f l} n_{f l'} + \nn\\
 &+& J \sum_{f, l < l'} \sum_{\sigma,\sigma'} c_{f l \sigma}^\dag c_{f l' \sigma'}^\dag c_{f l \sigma'} c_{f l' \sigma}
 + J' \sum_{f, l \neq l'} c_{f l \su}^\dag c_{f l \sd}^\dag c_{f l' \sd} c_{f l' \su},
\label{eq.Hsf}
\eea
where $n_{f l} = n_{f l \su} + n_{f l \sd}$ is the number of particles operator, $f$ is the site index, $l$ and $l'$ are orbital indices, $U$ and $U'$ are intra- and interorbital Hubbard repulsions, $J$ is the Hund's exchange, and $J'$ is the pair-hopping. Usually, parameters obey the spin-rotational invariance, that leads to relations $U' = U - 2J$ and $J' = J$ thus reducing the number of free parameters in the theory.

In the case of electron-phonon interaction inducing the superconductivity, one of the examples of Hamiltonian is
\begin{equation}
H_{sc}^{e-ph} = \sum\limits_{\q, \lambda} \omega_{\q \lambda} \left( b_{\q \lambda}^\dag b_{\q \lambda} + \frac{1}{2} \right) + \frac{1}{\sqrt{N}} \sum\limits_{\k,\q,\lambda, \alpha, \sigma} g_\lambda(\k,\q) \left( b_{\q \lambda} + b_{-\q \lambda}^\dag \right) c_{\k+\q \alpha \sigma}^\dag c_{\k \alpha \sigma}.
\label{eq.Heph}
\end{equation}
Here, $b_{\q \lambda}$ is the annihilation operator of the phonon with momentum $\q$, polarization $\lambda$, and frequency $\omega_{\q \lambda}$, $g_\lambda(\k,\q)$ is the electron-phonon interaction matrix element.

Hereafter we assume that the problem of finding the effective dynamical superconducting interaction is already solved and both coupling constants and the bosonic spectral function are obtained. Latter describes the effective electron-electron interaction via an intermediate boson. In the case of Hubbard interaction~(\ref{eq.Hsf}), intermediate excitations are spin or charge fluctuations, while in the case of electron-phonon interaction~(\ref{eq.Heph}) those are phonons. Moreover, if in the case of phonons the retarded nature of the interaction in obvious from the beginning, for the Hubbard Hamiltonian it reveals only after the summation of particular diagram series~\cite{BerkSchrieffer}. Nature of the effective dynamical interaction is not important for the following analysis of the role played by the disorder in a superconducting state. Rather important is the fact, that the corresponding bosonic spectral function is maximal at small frequencies and drops down with further increase of frequency. For example, inelastic neutron scattering experiments confirm such a behavior for spin fluctuations.

Note, though the dynamical interaction have a complicated structure and it is hard to write it in a unified form, everything becomes simplified in a mean field approximation and can be cast in the following Hamiltonian,
\begin{equation}
H_{sc}^{MF} = \sum\limits_{\k, \alpha} \left( \Delta_{\alpha} c_{\k \alpha \uparrow}^{\dag} c_{-\k \alpha \downarrow}^{\dag} + h.c. \right),
\label{eq.HscMF}
\end{equation}
where $\Delta_{\alpha}$ is a mean field spin-singlet order parameter. For example, $\sgnnobr{\Delta_{a}} = \sgnnobr{\Delta_{b}}$ for the two-band superconductor in the $s_{++}$ state, while for the $s_{\pm}$ state it is $\sgnnobr{\Delta_{a}} = -\sgnnobr{\Delta_{b}}$.

Impurity scattering is described by the $H_{\imp}$ term containing nonmagnetic ($\Uimp$) and magnetic ($\Vimp$) impurity scattering potentials:
\begin{equation}
H_{\imp} = \sum\limits_{\mathbf{R}_{i}, \sigma, \sigma', \alpha, \beta} \left( \Uimp_{\mathbf{R}_{i}}^{\alpha \beta} \delta_{\sigma \sigma'} + \Vimp_{\mathbf{R}_{i}}^{\alpha \beta} \hat{S}_{\mathbf{R}_{i}} \cdot \hat{\sigma}_{\sigma \sigma'} \right) c_{\mathbf{R}_{i} \alpha \sigma}^{\dag} c_{\mathbf{R}_{i} \beta \sigma'}, \label{eq.Himp}
\end{equation}
where $\hat{S}_{\mathbf{R}_{i}}$ is the operator of an impurity spin at site $\mathbf{R}_{i}$ with the spin quantum number $S_{\mathbf{R}_{i}}$, and $\hat{\mathbf{\sigma}}$ are the Pauli spin matrices\footnotemark[1].

\footnotetext[1]{$\hat{\sigma}_{0} = \left( \begin{array}{cc}
 1 & 0 \\
 0 & 1 \\
\end{array} \right)$, $\hat{\sigma}_{1} = \left( \begin{array}{cc}
 0 & 1 \\
 1 & 0 \\
\end{array} \right)$, $\hat{\sigma}_{2} = \left( \begin{array}{cc}
 0 & -\ii \\
 \ii & 0 \\
\end{array} \right)$, $\hat{\sigma}_{3} = \left( \begin{array}{cc}
 1 & 0 \\
 0 & -1 \\
\end{array} \right)$}

In the following, we use Eliashberg approach generalized for the multiband superconductors~\cite{allen}. To describe thermodynamics of the superconducting state, we are interested in Green's function $\hat{\mathbf{G}}(\k,\omega_n)$ of the quasiparticle with momentum $\k$ and Matsubara frequency $\omega_n = (2 n + 1) \pi T$. Green's function is a matrix in the band space and combined Nambu and spin spaces (we indicate quantities in the band space by the bold face and quantities in the combined Nambu and spin spaces by the hat). For the definiteness, we assume that the index $\alpha = a, b$ denotes the band space, Pauli matrices $\hat{\tau}_{i}$ and $\hat{\sigma}_{i}$ denote the Nambu ($\hat{\tau}_{i}$) and spin ($\hat{\sigma}_{i}$) spaces. As a result of the direct product (operation $\otimes$) of all matrices, for the two-band model we have Green's function with the dimension $8 \times 8$.

Dyson equation
\beq
\hat{\mathbf{G}}(\k,\omega_n) = \left[\hat{\mathbf{G}}_0^{-1}(\k,\omega_n) - \hat{\mathbf{\Sigma}}(\k,\omega_n)\right]^{-1}
\label{eq.Gfull}
\eeq
establish connection between the full Green's function, the ``bare'' Green's function (without interelectron interactions and impurities),
\beq
\hat{G}_0^{\alpha \beta}(\k,\omega_n) = \left[ \ii \omega_n \hat{\tau}_{0} \otimes \hat{\sigma}_{0} - \xi_{\k \alpha} \hat{\tau}_{3} \otimes \hat{\sigma}_{0} \right]^{-1} \delta_{\alpha \beta}
\eeq
and the self-energy matrix $\hat{\mathbf{\Sigma}}(\k,\omega_n)$. Further we assume that the latter does not depends on the wave vector $\k$ but keep the frequency and band indices dependencies,
\beq
\hat{\mathbf{\Sigma}}(\omega_n) = \sum_{i=0}^{3} \Sigma_{(i) \alpha \beta}(\omega_n) \hat{\tau}_i.
\eeq
In this case, the problem can be simplified by averaging over $\k$. Thus, all equations will be written in terms of quasiclassical $\xi$-integrated Green's functions,
\beq
\hat{\mathbf{g}}(\omega_n) = \int d \xi \hat{\mathbf{G}}(\k, \omega_n) =
\left(
\begin{array}{cc}
\hat{g}_{an} & 0 \\
0 & \hat{g}_{bn}
\end{array}
\right),
\label{eq.g}
\eeq
where
\beq
\hat{g}_{\alpha n} = g_{0\alpha n} \hat{\tau}_{0} \otimes \hat{\sigma}_{0} + g_{2\alpha n} \hat{\tau}_{2}\otimes \hat{\sigma}_{2}.
\label{eq.g.alpha}
\eeq
Here, $g_{0\alpha n}$ and $g_{2\alpha n}$ are the normal and anomalous (Gor'kov) $\xi$-integrated Green's functions in the Nambu representation,
\beq
g_{0\alpha n} = -\frac{\ii \pi N_{\alpha} \tilde{\omega}_{\alpha n}}{\sqrt{\tilde{\omega}_{\alpha n}^{2}+\tilde{\phi}_{\alpha n}^{2}}}, \;\;\; g_{2\alpha n} = -\frac{\pi N_{\alpha} \tilde{\phi}_{\alpha n}}{\sqrt{\tilde{\omega}_{\alpha n}^{2}+\tilde{\phi}_{\alpha n}^{2}}}.
\label{eq.g02}
\eeq
They depend on the density of states per spin at the Fermi level in the corresponding band ($N_{a,b}$), and on the renormalized by the self-energy order parameter $\tilde{\phi}_{\alpha n}$ and frequency $\tilde{\omega}_{\alpha n}$,
\begin{eqnarray}
\ii \tilde\omega_{\alpha n} &=& \ii \omega_n - \Sigma_{0\alpha}(\omega_n) - \Sigma_{0\alpha}^{\imp}(\omega_n), \label{eq.omega.tilde} \\
\tilde\phi_{\alpha n} &=& \Sigma_{2\alpha}(\omega_n) + \Sigma_{2\alpha}^{\imp}(\omega_n). \label{eq.phi.tilde}
\end{eqnarray}
Often, it is convenient to introduce the renormalization factor $Z_{\alpha n} = \tilde{\omega}_{\alpha n} / \omega_n$ that enters the gap function $\Delta_{\alpha n} = \tilde{\phi}_{\alpha n} / Z_{\alpha n}$. It is the gap function that generates peculiarities in the density of states.

\begin{figure}
\centering
\includegraphics[width=0.5\textwidth]{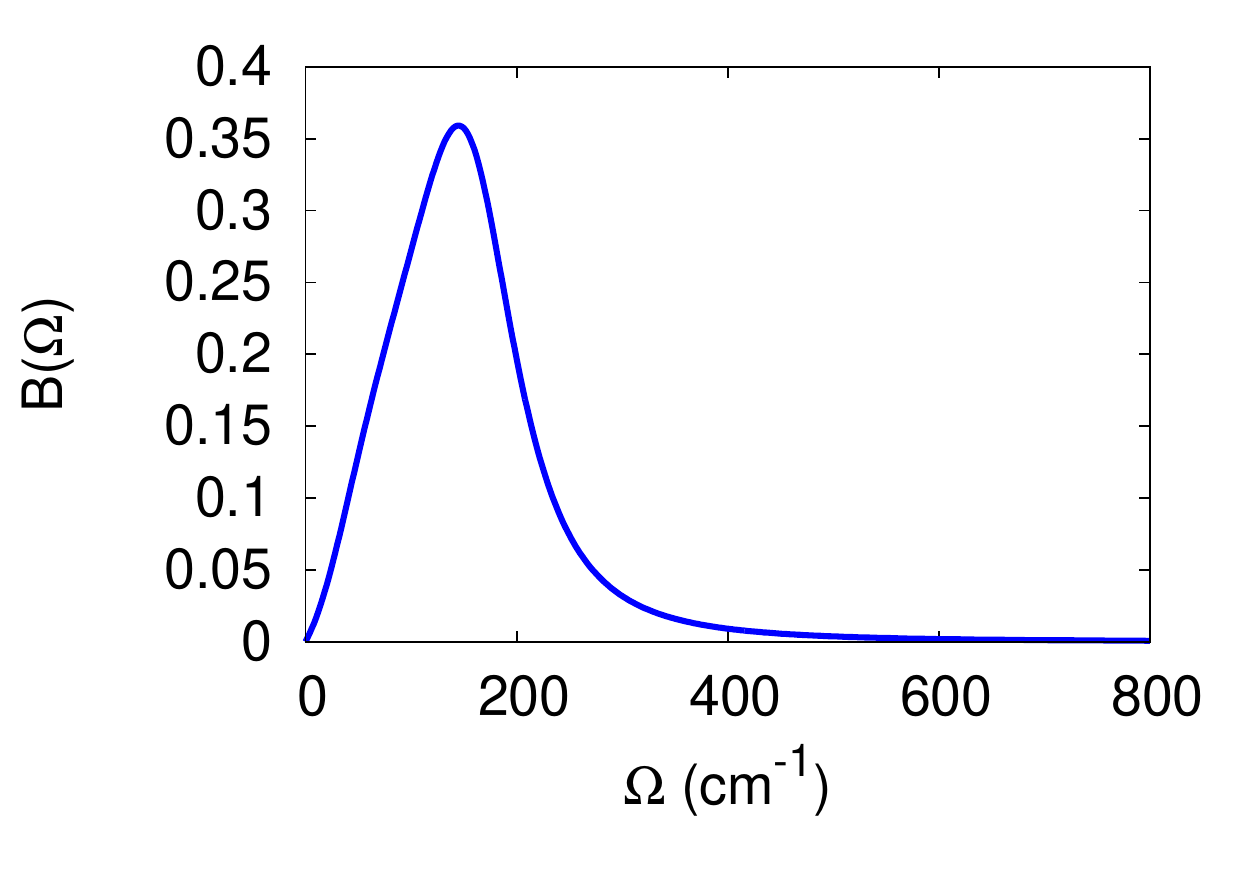}
\caption{Spectral function $B(\Omega)$ reproducing the frequency dependence of spin fluctuations~\cite{ParkerKorshunov2008,Popovich2010,Charnukha2011}.
\label{fig:sfspektr}}
\end{figure}

A part of the self-energy due to spin fluctuations or any other retarded interaction (electron-phonon, retarded Coulomb interaction) can be written in the following way:
\bea
\Sigma_{0\alpha}(\omega_n) &=& T \sum\limits_{\omega_n',\beta} \lambda^{Z}_{\alpha\beta}(n-n') \frac{g_{0\beta n'}}{N_\beta}, \label{eq:SigmaSF0} \\
\Sigma_{2\alpha}(\omega_n) &=& -T \sum\limits_{\omega_n',\beta} \lambda^{\phi}_{\alpha\beta}(n-n') \frac{g_{2\beta n'}}{N_\beta},
\label{eq:SigmaSF2}
\eea
Coupling functions,
\beq
 \lambda^{\phi,Z}_{\alpha\beta}(n-n') = 2 \lambda^{\phi,Z}_{\alpha\beta} \int^{\infty}_{0} d\Omega \frac{\Omega B(\Omega)}{(\omega_n-\omega_{n'})^{2} + \Omega^{2}}, \nn
\eeq
depend on coupling constants $\lambda^{\phi,Z}_{\alpha \beta}$, which include density of states $N_{\beta}$ in themselves, and on the normalized bosonic spectral function $B(\Omega)$, shown in Figure~\ref{fig:sfspektr}. The matrix elements $\lambda^\phi_{\alpha \beta}$ can be positive (attractive) as well as negative (repulsive) due to the interplay between spin fluctuations and electron-phonon coupling~\cite{BerkSchrieffer,ParkerKorshunov2008}, while the matrix elements $\lambda^Z_{\alpha \beta}$ are always positive. For the simplicity we set $\lambda^Z_{\alpha \beta} = |\lambda^\phi_{\alpha \beta}| \equiv |\lambda_{\alpha \beta}|$ and neglect possible anisotropy in each order parameter $\tilde\phi_{\alpha n}$. Effects due to the anisotropy in the $s_\pm$ state have been examined in, e.g., Ref.~\cite{v_mishra_09}.

\begin{figure}
\centering
\includegraphics[width=0.6\textwidth]{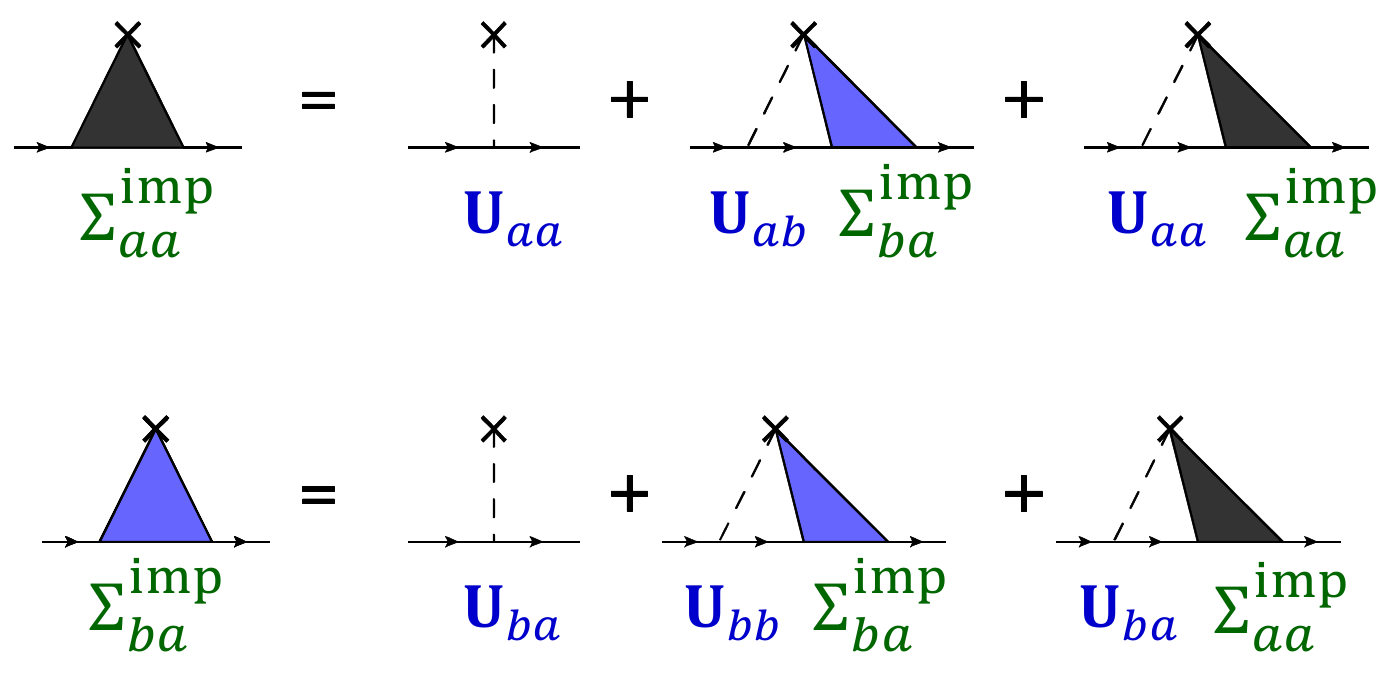}
\caption{System of equations for the intra- and interband parts of the impurity self-energy $\hat{\mathbf{\Sigma}}^{\imp}$ in the $\Tmat$-matrix self-consistent approximation~\cite{Ohashi2004}. Here, $\mathbf{U}_{aa(bb)}$ and $\mathbf{U}_{ab(ba)}$ are the intraband and interband components of the impurity potential, respectively.
\label{fig:TmatrixEq}}
\end{figure}

We use a noncrossing approximation (graphically shown in Figure~\ref{fig:TmatrixEq}) to calculate the impurity self-energy $\hat{\mathbf{\Sigma}}^{\imp}$:
\begin{equation}
\hat{\mathbf{\Sigma}}^{\imp}(\omega_n) = n_{\imp} \hat{\mathbf{U}} + \hat{\mathbf{U}} \hat{\mathbf{g}}(\omega_n) \hat{\mathbf{\Sigma}}^{\imp}(\omega_n),
\label{eq.tmatrix}
\end{equation}
where $\hat{\mathbf{U}}$ is the matrix of the impurity potential, and $n_{\imp}$ is the concentration of impurities. Equation~(\ref{eq.tmatrix}) represents the $\Tmat$-matrix approximation.

The impurity scattering matrix $\mathbf{\hat{U}}$ is derived from the Hamiltonian~(\ref{eq.Himp}). The procedure of further calculations is the following: i) solve equation~(\ref{eq.tmatrix}), ii) calculate renormalizations of frequency~(\ref{eq.omega.tilde}) and order parameter~(\ref{eq.phi.tilde}) self-consistently, iii) use them to obtain Green's functions~(\ref{eq.g02}) and~(\ref{eq.g}).

Solution of the equation~(\ref{eq.tmatrix}) depends on the explicit form of the impurity potential. Further we consider the two cases separately: nonmagnetic ($\Uimp_{\mathbf{R}_{i}}^{\alpha \beta} \neq 0$, $\Vimp_{\mathbf{R}_{i}}^{\alpha \beta} = 0$) and magnetic ($\Uimp_{\mathbf{R}_{i}}^{\alpha \beta} = 0$, $\Vimp_{\mathbf{R}_{i}}^{\alpha \beta} \neq 0$) impurities.

At the end of the present Section, we write expressions for some of observables, which affected by the detailed structure of impurity scattering. In the first place, this is the density of states that can be measured in tunneling experiments and in ARPES,
\beq
N(\omega) = \sum_{\alpha} N_{\alpha}(\omega) = \sum_{\k, \alpha} A_{\alpha}(\k,\omega) = - \frac{1}{\pi} \sum_{\alpha} \mathrm{Im} g_{0\alpha}(\omega),
\label{eq.N}
\eeq
where $g_{0\alpha}(\ii \omega_n \to \omega + \ii \delta)$ is the retarded Green's function that is Matsubara Green's function~(\ref{eq.g02}) analytically continued to the real frequency axis, $N_{\alpha}(\omega)$ is the partial density of states for band $\alpha$, and $A_{\alpha}(\k,\omega)$ is the quasiparticle spectral function, $\omega$ is the real frequency, and $\delta \to 0+$.

Another important characteristic of the superconductor is the temperature dependence of the London magnetic field penetration depth $\lambda_{L}$. In the local limit, it is related to the imaginary part of the optical conductivity,
\begin{equation}
 \frac{1}{\lambda_{L, x x'}^2} = \lim_{\omega \to 0} \frac{4 \pi \omega}{c^2} \mathrm{Im} \sigma^{x x'}(\omega, \q = 0),
 \label{eq:pen-london}
\end{equation}
where $x$ and $x'$ are axes directions of the Cartesian coordinates, $c$ is the velocity of light, and $\sigma^{x x'}(\omega, \q = 0)$ is the optical conductivity at zero momentum $\q$ (in the local, i.e., London, limit). If we neglect the effects of strong coupling and, in general, Fermi-liquid effects, then for the clean uniform superconductor at zero temperature we have $1 / \lambda_{L, x x'} = \omega_{P\alpha}^{x x'} / c$, where $\omega_{P\alpha}^{x x'} = \sqrt{ 8 \pi e^2 N_\alpha(0) \la v_{F\alpha}^{x} v_{F\alpha}^{x'} \ra}$ is the electron plasma frequency. For the impurity scattering, vertex corrections from noncrossing diagrams vanish due to the $\q = 0$ condition. Thus, penetration depth for the multiband system can be calculated via the following expression,
\beq
\frac{1}{\lambda_{L, x x'}^2} = \sum\limits_{\alpha} \left( \frac{\omega_{P\alpha}^{x x'}}{c} \right)^2 T \sum\limits_{n} \frac{ g_{2\alpha n}^2}{\pi N_{\alpha}^2 \sqrt{\tilde{\omega}_{\alpha n}^2 + \tilde{\phi}_{\alpha n}^2} }.
\label{eq.lambda}
\eeq
One can introduce a so-called ``superfluid plasma frequency'' $\omega_{SF}^{x x'} = c / \lambda_{L, x x'}$. It is often mentioned, that this function corresponds to the charge density of the superfluid condensate. Note, this is true only in the noninteracting clean system at zero temperature.

We consider hereafter the square lattice with $x$ and $x'$ in the $ab$-plane. In this case, we denote the penetration depth and corresponding plasma frequency by $\lambda_{L}$ and $\omega_{P\alpha}$, respectively.

Optical conductivity is the third important observable characteristic. In the local (London) limit with $\q = 0$ in the $ab$-plane, it is equal to
\beq
 \sigma(\omega) = \sum\limits_{\alpha} \sigma_\alpha(\omega) = \ii \sum\limits_{\alpha} \frac{\Pi_\alpha^{xx}(\ii \omega_m \to \omega + \ii \delta)}{\omega},
 \label{eq:sigma}
\eeq
where a polarization operator is
\beq
 \Pi_\alpha^{x x'}(\omega_m) = \frac{T}{N_{\k}} \sum\limits_{\k, \omega_n} \mathrm{Tr} \, e v_{\alpha}^{x} \hat\tau_0 \otimes \hat\sigma_0 \hat{G}^{\alpha\alpha}(\k, \omega_n + \omega_m) \hat{G}^{\alpha\alpha}(\k, \omega_n) \hat\gamma_{\alpha}^{x'}.
\eeq
Here, $N_{\k}$ is the normalization coefficient of the sum over momenta, $\hat\gamma_{\alpha}^{x'}$ is a vertex function, and the trace is taken over the Nambu and spin spaces. As was mentioned before, vertex corrections from noncrossing diagrams for the impurity scattering at $\q = 0$ vanish due to the vector nature of the optical conductivity vertex and the scalar character of the impurity scattering. Thus the zeroth order is a good approximation for the vertex~\cite{Bickers1990}, in which the vertex is equal to $e v_{\alpha}^{x'} \hat\tau_0 \otimes \hat\sigma_0$. It is also convenient to transform from the summation over momenta to the integration over energy and averaging over the Fermi surface. Latter results in $2 e^2 N_\alpha(0) \left< v_{F\alpha}^{x} v_{F\alpha}^{x'} \right> \approx \left(\omega_{P\alpha}^{x x'}\right)^2 / 4\pi $. After the transformation, the polarization operator becomes equal to~\cite{Bickers1990}:
\beq
 \Pi_\alpha^{x x'}(\omega_m) = \frac{\left(\omega_{P\alpha}^{x x'}\right)^2}{4\pi} \pi T \sum\limits_{\omega_n} S_{\alpha}^{nm},
 \label{eq:Pi_iomega_m}
\eeq
where $S_{\alpha}^{nm} = \tilde{\phi}_{\alpha n}^2 / Q_{\alpha n}^3$ for $m = 0$, $S_{\alpha}^{nm} = 1 / Q_{\alpha n}$ for $m = -2n - 1$, and
\beq
 S_{\alpha}^{nm} = \frac{\tilde{\omega}_{\alpha n} \left( \tilde{\omega}_{\alpha n} + \tilde{\omega}_{\alpha n+m} \right) + \tilde{\phi}_{\alpha n} (\tilde{\phi}_{\alpha n} - \tilde{\phi}_{\alpha n+m})}{Q_{\alpha n} P_{\alpha nm}}
 - \frac{\tilde{\omega}_{\alpha n+m} \left( \tilde{\omega}_{\alpha n+m} + \tilde{\omega}_{\alpha n} \right) + \tilde{\phi}_{\alpha n+m} (\tilde{\phi}_{\alpha n+m} - \tilde{\phi}_{\alpha n})}{Q_{\alpha n+m} P_{\alpha nm}} \nn
\eeq
in all other cases. Here, $Q_{\alpha n} = \sqrt{\tilde{\omega}_{\alpha n}^2 + \tilde{\phi}_{\alpha n}^2}$ and $P_{\alpha nm} = \tilde{\omega}_{\alpha n}^2 - \tilde{\omega}_{\alpha n+m}^2 + \tilde{\phi}_{\alpha n}^2 - \tilde{\phi}_{\alpha n+m}^2$.

To obtain the optical conductivity, one has to perform an analytical continuation of the polarization operator given above to real frequencies ($\ii\omega_m \to \omega + \ii \delta$). Another approach is to make the analytical continuation together with the integration~\cite{Nam1967_I,Nam1967_II,Lee1989,Dolgov1990,Marsiglio1991,Akis1991}. This leads to the polarization operator in the following form:
\bea
 \Pi_\alpha^{x x}(\omega) &=& \frac{\left(\omega_{P\alpha}^{x x}\right)^2}{4\pi}
 \int d\omega' \left[ \frac{\tanh\left[\omega_{-}/(2T)\right]}{Q^{R}_{+} + Q^{R}_{-}} \right. \left( 1 - \frac{\tilde{\omega}_{-}^{R} \tilde{\omega}_{+}^{R} + \tilde{\phi}_{-}^{R} \tilde{\phi}_{+}^{R}}{Q_{-}^{R} Q_{+}^{R}} \right) \nn\\
 &-& \frac{\tanh\left[\omega_{+}/(2T)\right]}{Q^{A}_{+} + Q^{A}_{-}} \left( 1 - \frac{\tilde{\omega}_{-}^{A} \tilde{\omega}_{+}^{A} + \tilde{\phi}_{-}^{A} \tilde{\phi}_{+}^{A}}{Q_{-}^{A} Q_{+}^{A}} \right) \nn\\
 &-& \frac{\tanh\left[\omega_{+}/(2T)\right] - \tanh\left[\omega_{-}/(2T)\right]}{Q^{R}_{+} - Q^{A}_{-}} \left. \left(1 - \frac{\tilde{\omega}_{-}^{A} \tilde{\omega}_{+}^{R} + \tilde{\phi}_{-}^{A} \tilde{\phi}_{+}^{R}}{Q_{-}^{A} Q_{+}^{R}} \right) \right],
 \label{eq:Pi_omega}
\eea
where $Q^{R,A}_{\pm} = \sqrt{ \left(\tilde{\omega}_{\pm}^{R,A}\right)^2 - \left(\tilde\phi_{\pm}^{R,A}\right)^2}$, indices $\pm$ set the frequency $\omega_{\pm} = \omega' \pm \omega/2$ which enters the corresponding function, the band index $\alpha$ is omitted in integrand, and indices $R$ and $A$ refers to retarded and advanced branches of a complex function $F$, i.e., $F^{R(A)} = \mathrm{Re} F \pm \ii \mathrm{Im} F$.

Note that the optical conductivity in the normal state is
\beq
 \sigma^N(\omega) = \sum\limits_{\alpha} \frac{\left(\omega_{P\alpha}^{x x}\right)^2}{8 \ii \pi \omega} \int\limits_{-\infty}^{+\infty} dz \frac{\tanh\left[(z+\omega)/(2T)\right] - \tanh\left[z/(2T)\right]}{\tilde\omega_\alpha(z+\omega) - \tilde\omega_\alpha(z)}.
\label{eq:sigmaN}
\eeq

In the following, we will use expressions from this Section to describe properties of different systems and their observable characteristics.

\section{Born approximation for one- and two-band superconductors \label{sec:Born}}

\subsection{Qualitative analysis}

\begin{figure}
\centering
\includegraphics[width=0.9\textwidth]{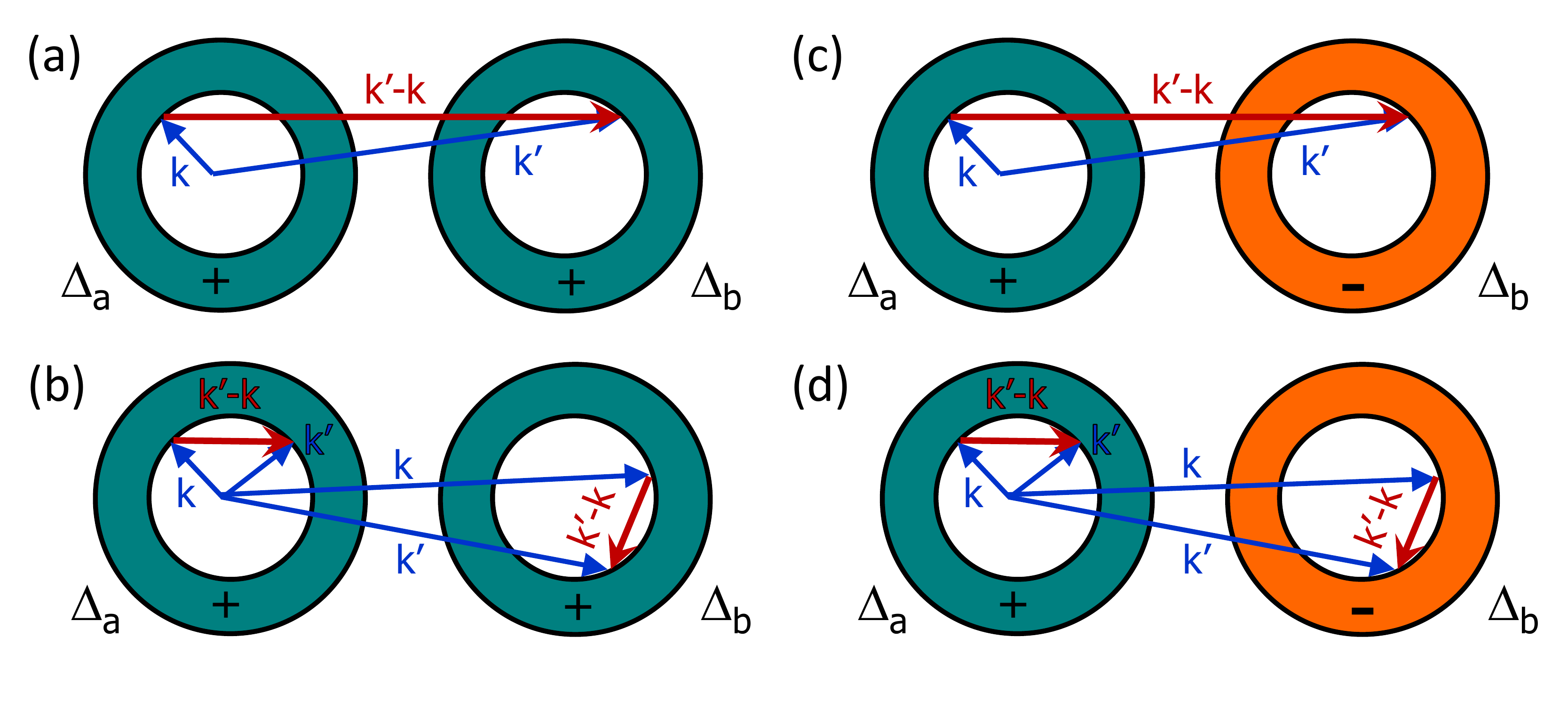}
\caption{Cartoon of two Fermi surfaces with the superconducting gaps $\Delta_a$ and $\Delta_b$ having the same signs (a,b) and having opposite signs (c,d). Interband impurity scattering (panels~a and c) mixes states with $\Delta_a$ and $\Delta_b$, while the intraband scattering (panels~b and d) involves states within each Fermi surface.}
\label{fig:impurities_2band}
\end{figure}

Nonmagnetic impurities in a conventional two-band superconductor with two isotropic gaps lead to scattering of quasiparticles between bands or within each band. Interband processes shown in Figure~\ref{fig:impurities_2band} result in averaging of gaps and, therefore, to the initial suppression of $T_c$, after which $T_c$ saturates and stays constant until localization effects become important~\cite{Sadovskii1997,SadovskiiBook2000}. Interband scattering in a two-band system with the sign-changing order parameter leads to a much more complicated behavior~\cite{Muzikar1996,Golubov1997,Kulic1999,Ohashi2004}. In this case, nonmagnetic impurities with the interband component of the scattering potential destroy the superconductivity even for equal magnitudes of gaps and densities of states (the so-called symmetric model). The reason is quite simple -- interband scattering results in the averaging of gaps in two differen bands and since $\Delta_a$ and $\Delta_b$ have opposite signs in the $s_\pm$ state, their average goes to zero. Then $T_c$ would vanish for the finite critical impurity concentration similar to the theory of magnetic impurity scattering in a single-band $s$-wave superconductor~\cite{AGeng}. Such characteristic feature of iron-based superconductors was quickly noticed by different groups of researchers~\cite{mazin_08,ParkerKorshunov2008,Chubukov2008,Senga2008,Bang2009}.

As for the effect of the nonmagnetic and magnetic disorder on a multiband anisotropic superconductor, the simple (and naive!) qualitative rule of thumb is as follows: when a nonmagnetic impurity scatters a pair from one point on the Fermi surface into another point, such that the order parameter does not change sign, scattering is not pair breaking; if the order parameter flips its sign, it is pair breaking. For a magnetic impurity, the opposite is true: scattering with an order parameter sign change is not pairbreaking, otherwise it is. However, as follows from calculations for some particular cases, such a naive qualitative rule collapses and quite unexpected results appear, which we discuss in Sections~\ref{sec:nonmag}--\ref{sec:experiments}. But before going further, we use a simplest, Born, limit in Section~\ref{subsec:Born} to demonstrate the general results in single- and two-band models.

\subsection{Clean and Born limits \label{subsec:Born}}

Here we consider a \textit{weak coupling} example, i.e., the case of $\lambda_{\alpha \beta} \ll 1$, and describe the effect of disorder on the macroscopic characteristic -- critical temperature of the superconducting transition $T_c$. Scattering on static impurities would results only in decrease of $T_c$ from its clean limit value $T_{c0}$.

Order parameter $\Delta_{\alpha n}$ in a multiband superconductor in the clean limit is a solution of an equation that follows from expressions~(\ref{eq.phi.tilde}) and~(\ref{eq:SigmaSF2}). Neglecting the frequency dependence of the coupling functions (that is the essence of the weak coupling), we have
\beq
\Delta_{\alpha} = -T \sum\limits_{\omega_n, \beta} \lambda_{\alpha \beta} \frac{g_{2\beta n}}{N_{\beta}}, \label{eq.Delta.clean}
\eeq
Summation over the Matsubara frequencies $\omega_n$ goes until a cut-off frequency $\omega_c$. In the limit $T \to T_{c0}$, we have $\Delta_{\alpha} \to 0$ and $g_{2\alpha n} \to -\pi N_{\alpha} \Delta_{\alpha} / \abs{\omega_n}$. Then the expression~(\ref{eq.Delta.clean}) becomes equation for the critical temperature in the clean limit, i.e., $T_{c0}$,
\begin{equation}
1 = \pi T_{c0} \sum\limits_{\beta} \lambda_{\alpha \beta} \sum\limits_{\omega_n} \frac{1}{\Delta_{\alpha}} \frac{\Delta_{\beta}}{\abs{\omega_n}} = 2 \pi T_{c0} \sum\limits_{\beta} \lambda_{\alpha \beta} \sum\limits_{\omega_n \geq 0} \frac{1}{\Delta_{\alpha}} \frac{\Delta_{\beta}}{\omega_n}. \label{eq.Tc.clean}
\end{equation}

Single-band case is realized for $\lambda_{\alpha \beta} = \lambda \delta_{\alpha, \beta}$:
\bea
1 &=& 2 \pi T_{c0} \lambda \sum\limits_{\omega_n \geq 0}^{\omega_c} \frac{1}{\omega_n} = \lambda \sum\limits_{n \geq 0}^{N_c} \frac{1}{n+1/2} \nn\\
&=& \lambda \left[ \Psi\left(N_c + 3/2\right) - \Psi\left(1/2\right) \right] \nn\\
&=& \lambda \left[ \Psi\left(\frac{\omega_c}{2 \pi T_{c0}} + 1\right) + C \right]
\to \lambda \left[ \ln\frac{\omega_c}{2 \pi T_{c0}} + C \right], \nn
\eea
where we have taken into account that $\omega_n = (2 n + 1) \pi T_{c0}$, $(2 N_c + 1) \pi T_{c0} = \omega_c$, $\Psi(1/2) = -\gamma - 2 \ln{2} \equiv -C$, and $\Psi(z+1) \to \ln{z}$ for $z \to \infty$. Here, $\Psi$ is the digamma function and $\gamma$ is the Euler constant. Since $\exp(C)/2\pi \approx 1.13$, the solution of the equation above gives the well-known expression for the superconducting critical temperature $T_{c0} = 1.13 \omega_c e^{-1/\lambda}$. Note again that the coupling constant includes density of states, $\lambda \propto N_a$.

Expression~(\ref{eq.Tc.clean}) in a two-band case is the system of two equations on gaps $\Delta_a$ and $\Delta_b$. From the consistency condition of the system (determinant of the corresponding matrix should be equal to zero), one can derive the expression for $T_{c0}$,
\begin{equation}
\ln{\frac{1.13 \omega_c}{T_{c0}}} = \max{ \left[ \frac{\lambda_{aa} + \lambda_{bb} \pm \sqrt{\left( \lambda_{aa} - \lambda_{bb} \right)^2 + 4 \lambda_{ab} \lambda_{ba}}}{2 \left( \lambda_{aa} \lambda_{bb} - \lambda_{ab} \lambda_{ba} \right) }\right] }. \label{eq.Tc.clean.2band}
\end{equation}

One can consider the simplest case of treating the impurity scattering by replacing the ``bare'' Green's function $g_{2\beta n}$ in equation~(\ref{eq.Delta.clean}) by the full one~(\ref{eq.g02}) containing the renormalized order parameter $\tilde{\phi}_{\alpha n}$ and frequency $\tilde{\omega}_{\alpha n}$,
\beq
\Delta_{\alpha} = \pi T \sum\limits_{\omega_n, \beta} \lambda_{\alpha \beta} \frac{\tilde\phi_{\beta n}}{Q_{\beta n}}, \label{eq.Delta.notclean}
\eeq
where $Q_{\beta n} = \sqrt{\tilde{\omega}_{\beta n}^2 + \tilde{\phi}_{\beta n}^2}$. For $T \to T_c$, we have $g_{0\alpha n} \to -\ii \pi N_{\alpha} \tilde\omega_{\alpha n} / \abs{\tilde\omega_{\alpha n}} = -\ii \pi N_{\alpha} \sgn{\omega_{n}}$ and $g_{2\alpha n} \to -\pi N_{\alpha} \tilde\phi_{\alpha n} / \abs{\omega_{\alpha n}}$. Equation for the critical temperature becomes
\begin{equation}
1 = 2 \pi T_c \sum\limits_{\beta} \lambda_{\alpha \beta} \sum\limits_{\omega_n \geq 0} \left. \left[ \frac{1}{\Delta_{\alpha}} \frac{\tilde{\phi}_{\beta n}}{\tilde{\omega}_{\beta n}} \right] \right\vert_{T \to T_c}.
\label{eq.Tc.notclean}
\end{equation}
As follows from a comparison with equation~(\ref{eq.Tc.clean}), $T_c$ does not depends on impurity scattering if the following condition is satisfied:
\beq
\frac{\tilde{\phi}_{\beta n}}{\tilde{\omega}_{\beta n}} = \frac{\Delta_{\beta}}{\omega_n}.
\label{eq:condTc}
\eeq

In the Born approximation, only contribution of double scattering at the same impurity is allowed, $\hat{\mathbf{\Sigma}}^{\imp}(\omega_n) \approx n_{\imp} \hat{\mathbf{U}} + n_{\imp} \hat{\mathbf{U}} \hat{\mathbf{g}}(\omega_n) \hat{\mathbf{U}}$. One can derive now expressions for frequency and order parameters,
\bea
\tilde{\omega}_{a n} &=& \omega_n + \gamma_{aa} \frac{\tilde\omega_{a n}}{Q_{a n}} + \gamma_{ab} \frac{\tilde\omega_{b n}}{Q_{b n}}, \label{eq.omegaBorn} \\
\tilde{\phi}_{a n} &=& \Delta_a \pm \gamma_{aa} \frac{\tilde\phi_{a n}}{Q_{a n}} \pm \gamma_{ab} \frac{\tilde\phi_{b n}}{Q_{b n}}, \label{eq.phiBorn}
\eea
where $\gamma_{\alpha \beta} \propto n_{\imp} (\mathbf{U})_{\alpha \beta}^2$ is the scattering rate parameter; the sign $+$ ($-$) corresponds to nonmagnetic (magnetic) impurities. Difference in sign for magnetic disorder occurs due to the spin operators accompanying the impurity potential $\Vimp_{\mathbf{R}_{i}}^{\alpha \beta}$ in $H_{\imp}$.

Firstly, we consider the single-band case. Then $\gamma_{ab} = 0$ and $\tilde{\omega}_{a n} = \omega_n + \gamma_{aa} \frac{\tilde\omega_{a n}}{Q_{a n}}$, $\tilde{\phi}_{a n} = \Delta_a \pm \gamma_{aa} \frac{\tilde\phi_{a n}}{Q_{a n}}$, $\gamma_{aa} = 2\pi N_a n_{\imp} \Uimp^2$. Equations for nonmagnetic impurities are $\tilde{\omega}_{a n} \left( 1 - \gamma_{aa} / Q_{a n} \right) = \omega_n$ and $\tilde{\phi}_{a n} \left( 1 - \gamma_{aa} / Q_{a n} \right) = \Delta_a$, which immediately lead to the relation~(\ref{eq:condTc}). Therefore, $T_c$ is independent of impurity concentration. This is the essence of Anderson's theorem.

For magnetic impurities, $\tilde{\omega}_{a n} \left( 1 - \gamma_{aa} / Q_{a n} \right) = \omega_n$ and $\tilde{\phi}_{a n} \left( 1 + \gamma_{aa} / Q_{a n} \right) = \Delta_a$. Thus, the condition~(\ref{eq:condTc}) is violated. Instead of it we have
\bea
\left. \left[ \frac{\tilde{\phi}_{a n}}{\Delta_a} \frac{1}{\tilde{\omega}_{a n}} \right] \right\vert_{T \to T_c} &=& \left. \left[ \frac{1}{1 + \gamma_{aa} / Q_{a n}} \frac{1}{\tilde{\omega}_{a n}} \right] \right\vert_{T \to T_c} \nn\\
&=& \frac{1}{\left. \tilde{\omega}_{a n} \right\vert_{T \to T_c} + \gamma_{aa}} = \frac{1}{\omega_n + \gamma_{aa} \tilde{\omega}_{a n}/\abs{\tilde{\omega}_{a n}} + \gamma_{aa}} \nn\\
&=& \frac{1}{\omega_n + 2\gamma_{aa}}, \nn
\eea
since for the $T_c$ equation we are interested in the case of $\omega_n \geq 0$. In the equation for the critical temperature, additional factor of $2\gamma_{aa}$ appears in a denominator,
\bea
1 &=& 2 \pi T_c \lambda \sum\limits_{\omega_n \geq 0}^{\omega_c} \frac{1}{\omega_n + 2\gamma_{aa}} \nn\\
&=& \lambda \left[ \Psi\left(\frac{\omega_c}{2 \pi T_c} + \frac{\gamma_{aa}}{\pi T_c} + 1\right) - \Psi\left(\frac{\gamma_{aa}}{\pi T_c} + \frac{1}{2}\right) \right]. \nn
\eea
In the $\omega_c \to \infty$ limit, the equation takes the form
\beq
1 = \lambda \left[ \ln{\frac{\omega_c}{2 \pi T_c}} - \Psi\left(\frac{\gamma_{aa}}{\pi T_c} + \frac{1}{2}\right) \right]. \nn
\eeq
Combining this equation with the corresponding expression in the clean limit, $1 = \lambda \left[ \ln{\omega_c/\left(2 \pi T_{c0}\right)} - \Psi(1/2) \right]$, we finally have
\beq
\ln{\frac{T_{c0}}{T_c}} = \Psi\left( \frac{\gamma_{aa}}{\pi T_c} + \frac{1}{2} \right) - \Psi\left( \frac{1}{2} \right),
\label{eq.AG}
\eeq
that is the formula for $T_c$ suppression according to the Abrikosov-Gor'kov theory~\cite{AGeng}.

Let us consider now the two-band case. For impurity scattering within only one band (``intraband impurities''), $\gamma_{ab} = 0$, equations~(\ref{eq.omegaBorn}) and~(\ref{eq.phiBorn}) for different bands are not coupled. Therefore, all conclusions made for the single-band case above are also true here for each band in the presence of nonmagnetic as well as magnetic impurities.

When both intra- and interband nonmagnetic impurity scattering channels are present, from equations~(\ref{eq.omegaBorn}) and~(\ref{eq.phiBorn}) we have:
\bea
\tilde{\omega}_{a n} \left( 1 - \frac{\gamma_{aa}}{Q_{a n}} - \frac{\gamma_{ab}^2}{Q_{a n}} \frac{1}{Q_{b n} - \gamma_{bb}} \right) &=& \omega_n \left( 1 + \frac{\gamma_{ab}}{Q_{b n} - \gamma_{bb}} \right), \nn\\
\tilde{\phi}_{a n} \left( 1 - \frac{\gamma_{aa}}{Q_{a n}} - \frac{\gamma_{ab}^2}{Q_{a n}} \frac{1}{Q_{b n} - \gamma_{bb}} \right) &=& \Delta_a + \Delta_b \frac{\gamma_{ab}}{Q_{b n} - \gamma_{bb}}. \nn
\eea
Evidently, if $\Delta_a = \Delta_b$, then condition~(\ref{eq:condTc}) is held and thus $T_c$ is independent of the disorder. Therefore there is no impurity effect on the multiband isotropic $s$-wave superconducting state. If, however, $\Delta_a \neq \Delta_b$, then condition~(\ref{eq:condTc}) is violated and $T_c$ will be suppressed by impurity scattering.

For the magnetic impurity, equations~(\ref{eq.omegaBorn}) and~(\ref{eq.phiBorn}) lead to
\bea
\tilde{\omega}_{a n} \left( 1 - \frac{\gamma_{aa}}{Q_{a n}} - \frac{\gamma_{ab}^2}{Q_{a n}} \frac{1}{Q_{b n} - \gamma_{bb}} \right) &=& \omega_n \left( 1 + \frac{\gamma_{ab}}{Q_{b n} - \gamma_{bb}} \right), \nn\\
\tilde{\phi}_{a n} \left( 1 + \frac{\gamma_{aa}}{Q_{a n}} - \frac{\gamma_{ab}^2}{Q_{a n}} \frac{1}{Q_{b n} + \gamma_{bb}} \right) &=& \Delta_a - \Delta_b \frac{\gamma_{ab}}{Q_{b n} + \gamma_{bb}}. \nn
\eea
Obviously, the condition~(\ref{eq:condTc}) is held only for $\Delta_b = - \Delta_a$ and $\gamma_{aa} = \gamma_{bb} = 0$. That is, the $s_\pm$ state with the equal absolute values of gaps is not susceptible to the scattering by magnetic impurities having  interband scattering channel only (``interband impurities''). In all other cases, $T_c$ would decrease with increasing concentration and potential of magnetic impurities.

\section{Nonmagnetic impurities in two-band superconductors \label{sec:nonmag}}

Since now we know what happens in the simplest cases, we move on to solving Eliashberg equations in the $\Tmat$-matrix approximation~(\ref{eq.tmatrix}). Here we consider nonmagnetic impurities.

As was mentioned in Section~\ref{sec:Born}, in the $s_\pm$ state, any nonmagnetic impurity scattering \textit{only} between the bands with different signs of the gaps leads to suppression of the critical temperature $T_c$ similar to magnetic impurity scattering in a single-band BCS superconductor~\cite{Golubov1995,Golubov1997}. Then $T_c$ is determined from the Abrikosov-Gor'kov formula~(\ref{eq.AG}). Critical impurity scattering rate $\Gamma$ determined by the equation $T_c(\Gamma^\mathrm{crit}) = 0$ satisfies the relation $\Gamma^\mathrm{crit}/T_{c0} \approx 1.12$ in the Abrikosov-Gor'kov theory. On the other hand, several experiments on iron-based superconductors, for example, introduction of zinc or a proton irradiation~\cite{Cheng2010,Li2010,Nakajima2010,Tropeano2010}, show that $T_c$ suppression is much weaker than expected in the framework of the Abrikosov-Gor'kov theory. Therefore, it was even suggested that the $s_\pm$ state is not realized in these systems and the order parameter should be of the $s_{++}$-type~\cite{Kontani,Bang2009}.

The problem of disorder in iron-based superconductors is much more intricate than the simple arguments suggest. Even assuming isotropic gaps on two different Fermi surface sheets and nonmagnetic scattering, one finds the suppression of superconductivity for a system with mainly intraband scattering to be slower than expected. The Anderson's theorem is applicable in the limit of pure intraband scattering, the system is ``insensitive'' to signs of gaps, and $T_c$ is not suppressed.

Therefore, the $T_c$ suppression rate depends on the ratio of intra- and interband scattering rates, and making conclusions about the superconducting state on the basis of systematic disorder studies is harder than in the single-band case. One approach to the problem is to try to determine intra- and interband impurity potentials from first principles methods for different materials and types of impurities~\cite{a_kemper_09,Nakamura2011,mazin_08}, however, quantitative applicability of band structure calculations here is questionable.

\subsection{Solution of Eliashberg equations in the $\Tmat$-matrix approximation}

In the case of a nonmagnetic disorder, we can simplify the problem by reducing the dimension of matrices due to the spin degeneracy. Thus instead of expressions~(\ref{eq.g}) and~(\ref{eq.g.alpha}) we have $4 \times 4$ quasiclassical matrix Green's function in Nambu and band spaces,
%
\begin{equation}
 \hat{\mathbf{g}}(\omega_n)=\left(
 \begin{array}{cc}
  g_{0a n} & 0 \\
  0 & g_{0b n}
 \end{array}
 \right) \otimes \hat\tau_0 + \left(
 \begin{array}{cc}
  g_{2a n} & 0 \\
  0 & g_{2b n}
 \end{array}
 \right)\otimes \hat\tau_2, \label{eq.g.nonmagn}
\end{equation}
where $\tau_i$ are Pauli matrices corresponding to the Nambu space.

The impurity potential matrix entering the $\Tmat$-matrix equation~(\ref{eq.tmatrix}) is $\hat{\mathbf{U}} = \mathbf{U} \otimes \hat\tau_3$, where $(\mathbf{U})_{\alpha \beta} = \Uimp_{\mathbf{R}_{i}}^{\alpha \beta}$. Without loss of generality we set $\mathbf{R}_{i} = 0$ for the single impurity problem studied here. For simplicity intraband and interband parts of the impurity potential are set equal to $v$ and $u$, respectively, such that $(\mathbf{U})_{\alpha \beta} = (v-u) \delta_{\alpha \beta} + u$.

From equations~(\ref{eq.tmatrix}) and~(\ref{eq.g.nonmagn}) we then have
\bea
\hat{\Sigma}_{aa}^{\imp} & = & n_{\imp} v\hat{\tau}_{3} + v\hat{\tau}_{3}(g_{0an}\hat{\tau}_{0} + g_{2an}\hat{\tau}_{2})\hat{\Sigma}_{aa}^{\imp} + u\hat{\tau}_{3}(g_{0bn}\hat{\tau}_{0} + g_{2bn}\hat{\tau}_{2})\hat{\Sigma}_{ba}^{\imp}, \\
\hat{\Sigma}_{ba}^{\imp} & = & n_{\imp} u\hat{\tau}_{3} + u\hat{\tau}_{3}(g_{0an}\hat{\tau}_{0} + g_{2an}\hat{\tau}_{2})\hat{\Sigma}_{aa}^{\imp} + v\hat{\tau}_{3}(g_{0bn}\hat{\tau}_{0} + g_{2bn}\hat{\tau}_{2})\hat{\Sigma}_{ba}^{\imp}. 
\eea
Renormalizations of frequencies and gaps come from $\Sigma^{\imp}_{0a} = \frac{1}{2} \mathrm{Tr}\left[\hat{\Sigma}_{aa}^{\imp} \cdot \hat{\tau}_0 \right]$ and $\Sigma^{\imp}_{2a} = \frac{1}{2} \mathrm{Tr}\left[\hat{\Sigma}_{aa}^{\imp} \cdot \hat{\tau}_2 \right]$, respectively. Equations for $\Sigma^{\imp}_{0b}$ and $\Sigma^{\imp}_{2b}$ are derived via replacement $a \leftrightarrow b$ in the equations above. Considering the relation $g_{0\alpha n}^2 - g_{2\alpha n}^2 = -\pi^2 N_{\alpha}^2$, we derive the following solution for $\Sigma_{0a}^{\imp}$ and $\Sigma_{1a}^{\imp}$:
\begin{eqnarray}
\Sigma_{0a}^{\imp} &=& \frac{n_{\imp}}{D} \left[ g_{0bn} u^2 + g_{0an} v^2 + g_{0an} \left( u^2 - v^2 \right)^2 \pi^2 N_b^2 \right], \label{eq.omegaGen0} \\
\Sigma_{2a}^{\imp} &=& -\frac{n_{\imp}}{D} \left[ g_{2bn} u^2 + g_{2an} v^2 + g_{2an} \left( u^2 - v^2 \right)^2 \pi^2 N_b^2 \right],
\label{eq.SigmaGen2}
\end{eqnarray}
where
\beq
 D = 1 + \pi^2 N_a^2 v^2 + \pi^4 N_a^2 N_b^2 \left( u^2 - v^2 \right)^2 + \pi^2 N_b^2 v^2 - 2 u^2 \left( g_{0an} g_{0bn} - g_{2an} g_{2bn} \right). \nn
\eeq

In the following, apart from the general case we also consider two important limits: the Born, weak scattering, limit with $\pi u N_{a,b} \ll 1$, and the opposite limit of a very strong scattering with $\pi u N_{a,b} \gg 1$, called the unitary limit.

It is convenient to introduce the generalized cross-section parameter
\beq
\sigmaN = \frac{\pi^2 N_a N_b u^2}{1 + \pi^2 N_a N_b u^2} \to \left\{
\begin{array}{l}
0, \text{Born limit} \\
1, \text{unitary limit}
\end{array}
\right.
\eeq
and the impurity scattering rate
\beq
\GammaN_{a(b)} = 2 n_{\imp} \pi N_{b(a)} u^2 (1 - \sigmaN) = \frac{2 n_{\imp} \sigmaN}{\pi N_{a,b}} \to \left\{
\begin{array}{l}
2 n_{\imp}\pi N_{b,a} u^2, \text{Born limit} \\
2 n_{\imp}/\left( \pi N_{a,b} \right), \text{unitary limit}
\end{array}
\right.
\eeq
Parameter $\etaN$ is controlling the ratio of intra- and interband scattering potentials,
\beq
 v = \etaN u.
\eeq

Using the introduced notations, we rewrite equations for the frequency~(\ref{eq.omega.tilde}) and the order parameter~(\ref{eq.phi.tilde}) taking the impurity self-energy~(\ref{eq.omegaGen0})-(\ref{eq.SigmaGen2}) into account:
\begin{eqnarray}
\tilde{\omega}_{an} &=& \omega_n + \ii \Sigma_{0a}(\omega_n) + \frac{\GammaN_a}{2 D} \left[ \sigmaN \frac{\tilde{\omega}_{an}}{Q_{an}} (1 - \etaN^2)^2 + (1 - \sigmaN) \left( \frac{N_a \tilde{\omega}_{an}}{N_b Q_{an}} \etaN^2 + \frac{\tilde{\omega}_{bn}}{Q_{bn}} \right) \right],
\label{eq.omega.nonmagn}
\\
\tilde{\phi}_{an} &=& \Sigma_{2a}(\omega_n) + \frac{\GammaN_a}{2 D} \left[ \sigmaN \frac{\tilde{\phi}_{an}}{Q_{an}} (1 - \etaN^2)^2 + (1-\sigmaN) \left( \frac{N_a \tilde{\phi}_{an}}{N_b Q_{an}} \etaN^2 + \frac{\tilde{\phi}_{bn}}{Q_{bn}} \right) \right],
\label{eq.phi.nonmagn}
\end{eqnarray}
where
\beq
 D = (1-\sigmaN)^2 + \sigmaN (1-\sigmaN) \left( 2 \frac{\tilde{\omega}_{an} \tilde{\omega}_{bn} + \tilde{\phi}_{an} \tilde{\phi}_{bn}}{Q_{an} Q_{bn}} + \frac{N_a^2 + N_b^2}{N_a N_b} \etaN^2 \right) + \sigmaN^2 (1 - \etaN^2)^2. \nn
\eeq

Let's examine important limiting cases. In the Born limit, we have $\sigmaN \to 0$ (weak scattering, $\pi u N_{a,b} \ll 1$) thus $D = 1$, $\GammaN_a = 2 n_{\imp} \pi N_b u^2$, and
\begin{eqnarray}
\tilde{\omega}_{an} &=& \omega_{n} + \ii \Sigma_{0a}(\omega_n) + \frac{\gamma_{aa}}{2} \frac{\tilde{\omega}_{an}}{Q_{an}} + \frac{\gamma_{ab}}{2} \frac{\tilde{\omega}_{bn}}{Q_{bn}},
\label{eq.omega.interBorn}
\\
\tilde{\phi}_{an} &=& \Sigma_{2a}(\omega_n) + \frac{\gamma_{aa}}{2} \frac{\tilde{\phi}_{an}}{Q_{an}} + \frac{\gamma_{ab}}{2} \frac{\tilde{\phi}_{bn}}{Q_{bn}},
\label{eq.phi.interBorn}
\end{eqnarray}%
where $\gamma_{aa} = 2 \pi n_{\imp} N_a u^2 \etaN^2$ and $\gamma_{ab} = 2 \pi n_{\imp} N_b u^2$. Evidently, for the finite interband scattering $\gamma_{ab}$, i.e., finite $\etaN$, different bands are mixed in equations. This leads to the suppression of $T_c$ similar to the one following from the Abrikosov-Gor'kov expression~(\ref{eq.AG}).

In the unitary limit we have $\sigmaN \to 1$ (strong scattering, $\pi u N_{a,b} \gg 1$), $\GammaN_a = 2 n_{\imp} / (\pi N_a)$, and we have to consider two cases:

1) Uniform impurity potential with $\etaN = 1$. Than we have
\begin{eqnarray}
\tilde{\omega}_{an} &=& \omega_n + \ii \Sigma_{0a}(\omega_n) + \frac{n_{\imp}}{\pi N_a N_b D_{uni}} \left[ N_a \frac{\tilde{\omega}_{an}}{Q_{an}} + N_b \frac{\tilde{\omega}_{bn}}{Q_{bn}} \right], \label{eq.omega.uni.eta1.nonmag}
\\
\tilde{\phi}_{a n} &=& \Sigma_{2a}(\omega_n) + \frac{n_{\imp}}{\pi N_a N_b D_{uni}} \left[ N_a \frac{\tilde{\phi}_{an}}{Q_{an}} + N_b \frac{\tilde{\phi}_{bn}}{Q_{bn}} \right],
\label{eq.phi.uni.eta1.nonmag}
\end{eqnarray}
where
\beq
 D_{uni} = 2 \frac{\tilde{\omega}_{an} \tilde{\omega}_{bn} + \tilde{\phi}_{an} \tilde{\phi}_{bn}}{Q_{an} Q_{bn}} + \frac{N_a^2 + N_b^2}{N_a N_b}. \nn
\eeq
Obviously, different bands are mixed in equations for renormalized frequency and order parameter, so we have a suppression of $T_c$.

2) All other cases with $\etaN \neq 1$. We have
\begin{eqnarray}
\tilde{\omega}_{an} &=& \omega_n + \ii \Sigma_{0a}(\omega_n) + \frac{n_{\imp}}{\pi N_a} \frac{\tilde{\omega}_{an}}{Q_{an}}, \label{eq.omega.uni.nonmag} \\
\tilde{\phi}_{a n} &=& \Sigma_{2a}(\omega_n) + \frac{n_{\imp}}{\pi N_a} \frac{\tilde{\phi}_{an}}{Q_{an}}. \label{eq.phi.uni.nonmag}
\end{eqnarray}
We get the same result, as for the intraband impurities since the other band ($b$) does not contribute to the equations. Surprisingly, but here the Anderson's theorem works independent of the gap signs in different bands. Thus, $T_c$ should be finite for any impurity concentration.

Therefore, there is a special case of $T_c$ suppression in the unitary limit for the uniform impurity potential $\etaN = 1$. Such situation arise due to the structure of the denominator $D$ in equations~(\ref{eq.omega.nonmagn})-(\ref{eq.phi.nonmagn}). It vanishes for $\etaN = \sigmaN = 1$ and one has to accurately take the limit $\etaN \to 1$ first, and only then put $\sigmaN \to 1$. It is the $\etaN = 1$ case, that was considered in Ref.~\cite{Bang2009}. For all other values of $\etaN$ (even for a slight difference between intra- and interband potentials) impurities are not going to affect the critical temperature. Of course, from the physical point of view former situation is improbable since it is hard to imagine an impurity in a multiorbital system that has equal strength of intra- and interband scattering.

\subsection{Critical temperature of the superconducting transition}

At $T \to T_c$, equations becomes significantly simplified because the order parameter vanishes and $Q_{\alpha n} = \sqrt{\tilde{\omega}_{\alpha n}^2 + \tilde{\phi}_{\alpha n}^2} \to \left| \tilde{\omega}_{\alpha n} \right|$. Thus the linearized Eliashberg equations~(\ref{eq.omega.tilde})-(\ref{eq.phi.tilde}) for the renormalization factors $Z_{\alpha n} = \tilde{\omega}_{\alpha n}/\omega _{n}$ and gap functions $\Delta_{\alpha n} = \tilde{\phi}_{an} / Z_{\alpha n}$~\cite{allen} considering expressions~(\ref{eq.omega.nonmagn})-(\ref{eq.phi.nonmagn}) are rewritten as follows:
\begin{eqnarray}
 Z_{\alpha n} &=& 1 + \sum_{\beta} \frac{\tilde{\Gamma}_{\alpha \beta}}{|\omega_{n}|} + \pi T_c \sum\limits_{\omega_{n'}, \beta} |\lambda_{\alpha \beta}(n-n')| \frac{\sgn{\omega_{n'}}}{\omega_{n}}, \label{eq:Elias1} \\
 Z_{\alpha n} \Delta_{\alpha n} &=& \sum_{\beta} \frac{\tilde{\Gamma}_{\alpha \beta} \Delta_{\beta n}}{|\omega_{n}|} + \pi T_c \sum\limits_{\omega_{n'}, \beta} \lambda _{\alpha \beta}(n-n') \frac{\Delta_{\beta n'}}{|\omega_{n'}|}, \label{eq:Elias2}
\end{eqnarray}
where we have introduced renormalized impurity scattering rates $\tilde\Gamma_{\alpha \beta}$~\cite{EfremovKorshunov2011}:
\bea
 \tilde\Gamma_{ab(ba)} &=& \GammaN_{a(b)} \frac{(1 - \sigmaN)}{\sigmaN (1 - \sigmaN) \etaN^2 \frac{(N_a + N_b)^2}{N_a N_b} + (\sigmaN \etaN^2 - 1)^2}, \label{eq.Gamma.ab} \\
 \tilde\Gamma_{aa} &=& \GammaN_{a} \frac{\sigmaN (1 - \etaN^2)^2 + (1 - \tilde{\sigma}) \etaN^2 \frac{N_a}{N_b}}{\sigmaN (1 - \sigmaN) \etaN^2 \frac{(N_a + N_b)^2}{N_a N_b} + (\sigmaN \etaN^2 - 1)^2}, \label{eq.Gamma.aa}
\eea
After substitution of $Z_{\alpha n}$ from~(\ref{eq:Elias1}) to~(\ref{eq:Elias2}) we obtain equation for the critical temperature $T_c$:
\beq
\Delta_{\alpha n} + \pi T_c \sum_{n', \beta} \left[ \left| \lambda_{\alpha\beta}(n - n') \right| \sgn{\omega_{n'}} \frac{\Delta_{\alpha n}}{\omega_n} - \lambda_{\alpha\beta}(n - n') \frac{\Delta_{\beta n'}}{|\omega_{n'}|} \right] + \sum_{\beta} \tilde\Gamma_{\alpha\beta} \frac{\Delta_{\alpha n} - \Delta_{\beta n}}{|\omega_n|} = 0.
\label{eq.Tc.nonmagn}
\eeq
Last term is finite only for $\alpha \neq \beta$. Therefore, intraband terms $\propto \tilde\Gamma_{aa}$ and $\tilde\Gamma_{bb}$ are cancelled and do not contribute to $T_c$ in agreement with the Anderson's theorem. From the expression for scattering rates~(\ref{eq.Gamma.ab}), we recover explicitly the well-known but counterintuitive result that in the unitary limit $\tilde\Gamma_{ab} = 0$, that is, nonmagnetic impurities do not affect $T_{c}$ in the $s_{\pm }$ state~\cite{Kulic1999,Ohashi2004}.

Since $T_c$ depends only on parameter $\tilde\Gamma_{ab}$, we call it the effective impurity scattering rate.

\subsection{Results of the numerical solution}

\begin{figure}
\centering
\includegraphics[width=0.7\textwidth]{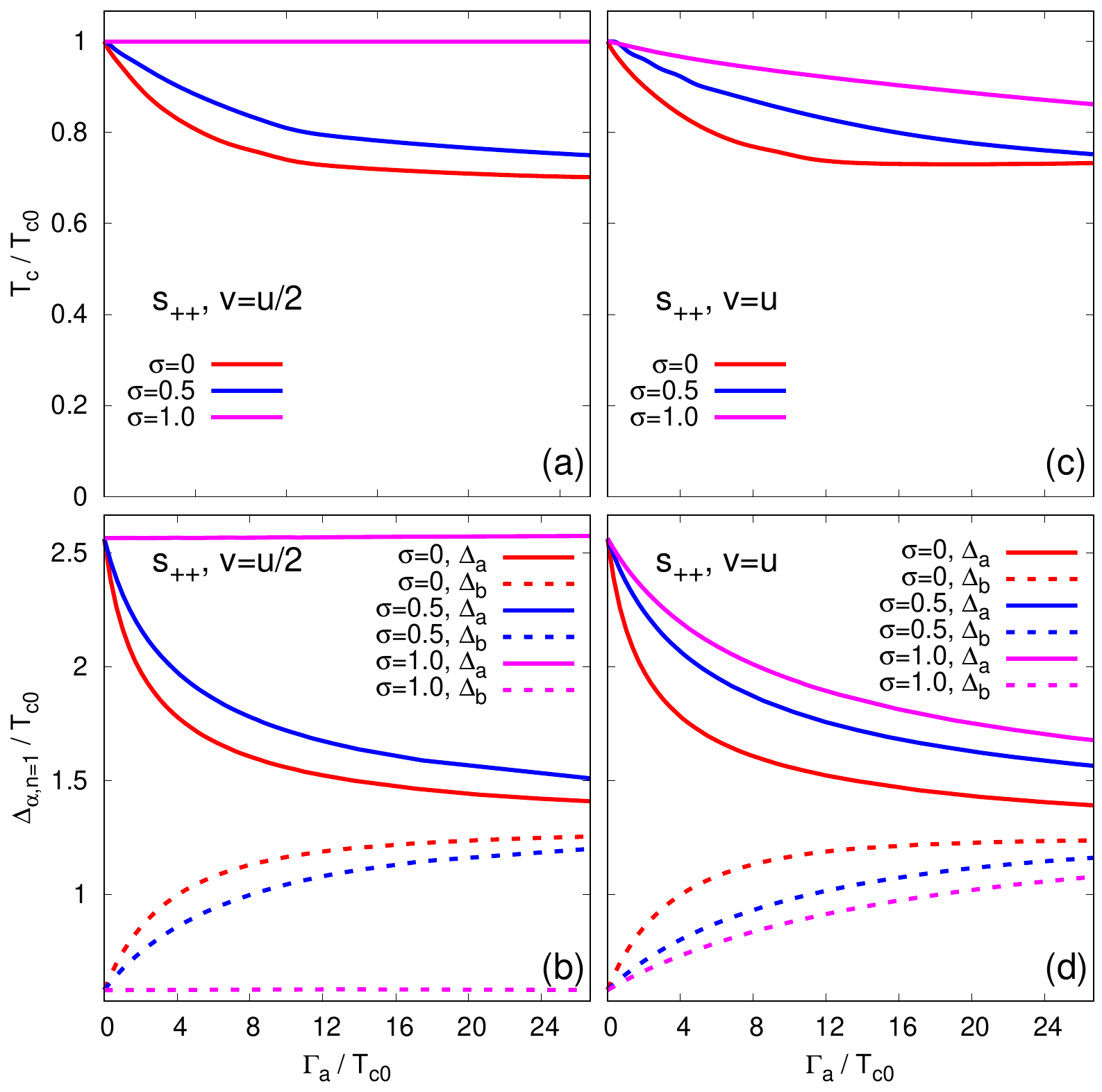}
\caption{(Color online.) Dependencies of $T_c$ (a,c) and the order parameter $\Delta_{\alpha n=1}$ at $T = 0.016 T_{c0}$
(b,d) on the impurity scattering rate $\GammaN_a$ for the $s_{++}$ state with $v = u/2$ (panels~a and~b) and $v = u$ (panels~c and~d). For $v = u$, the suppression of superconductivity occurs even in the unitary limit. Here coupling constant are $(\lambda_{aa},\lambda_{ab},\lambda_{ba},\lambda_{bb}) = (3,0.2,0.1,0.5)$, that gives $T_{c0}=43.1$~K.}
\label{fig:sppTcDeltaNonmag_vu_abcd}
\end{figure}

To determine $T_c$, we solve numerically either equation~(\ref{eq.Tc.nonmagn}) or Eliashberg equations~(\ref{eq:Elias1})-(\ref{eq:Elias2}) and vary $T$ to find highest temperature at which nontrivial solution exists~\cite{EfremovKorshunov2011}. For definiteness we choose $N_b / N_a = 2$. Resulting $T_c$ and gap $\Delta_{\alpha n=1}$ as functions of $\GammaN_a$ in $s_{++}$ state are shown in Figure~\ref{fig:sppTcDeltaNonmag_vu_abcd}. Generally, the superconductivity is not suppressed completely though there is an initial drop of $T_c$ due to the scattering between bands with initially unequal gaps. Note, the system in the unitary limit seems to don't care about disorder -- neither critical temperature nor gaps depend on $\GammaN_a$. As seen from equations~(\ref{eq.omega.uni.eta1.nonmag})-(\ref{eq.phi.uni.eta1.nonmag}), there is, however, an isolated point, $\etaN = 1$, corresponding to the vanishing of determinant $D$. That is, superconductivity is suppressed for the uniform impurity potential, $v = u$, see Figure~\ref{fig:sppTcDeltaNonmag_vu_abcd}.

\begin{figure}
\centering
\includegraphics[width=0.6\textwidth]{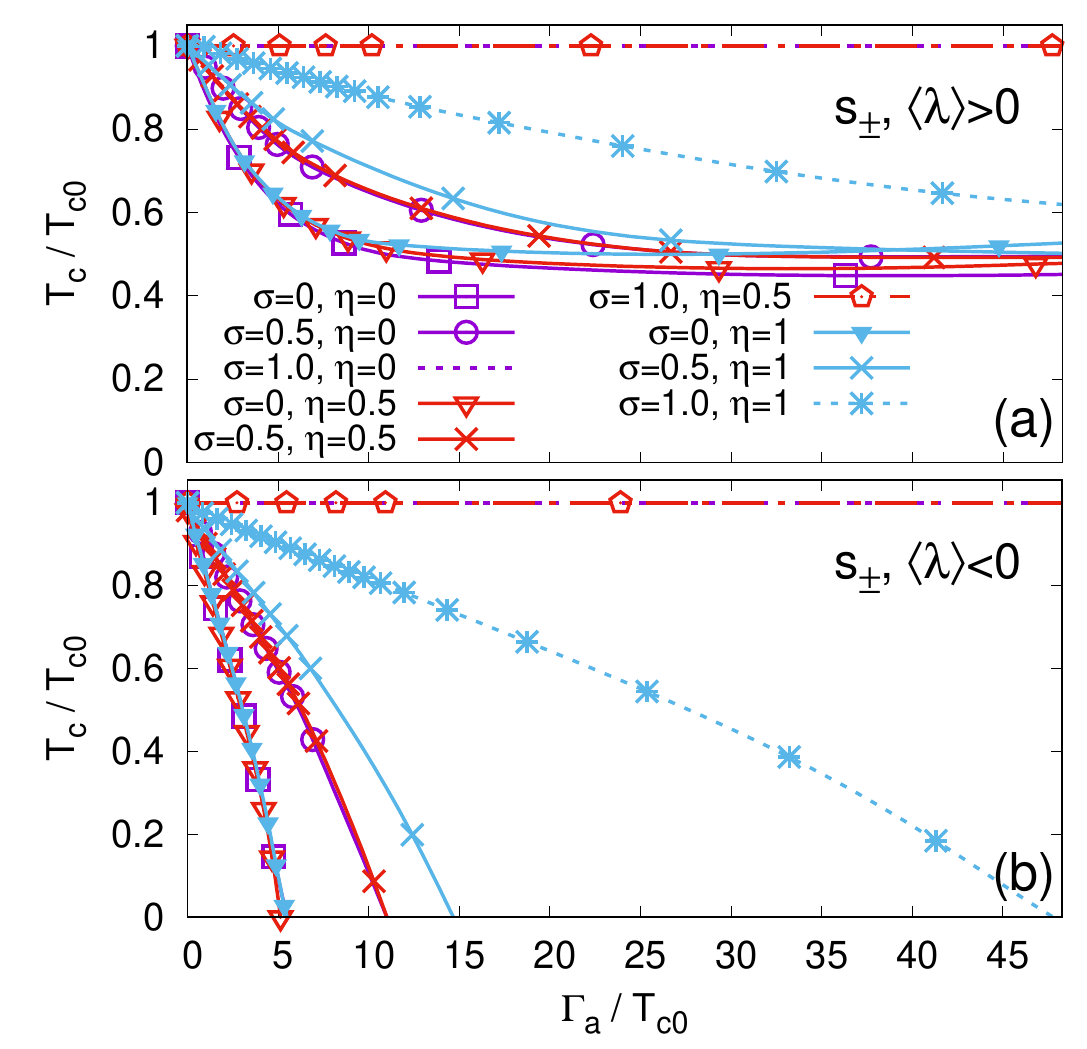}
\caption{Critical temperature for various $\sigmaN$ and $\etaN$ as a function of the impurity scattering rate $\GammaN_a$ for different signs of average coupling constant $\la \lambda \ra$.}
\label{fig:TcGamma}
\end{figure}

\begin{figure}
\centering
\includegraphics[width=0.6\textwidth]{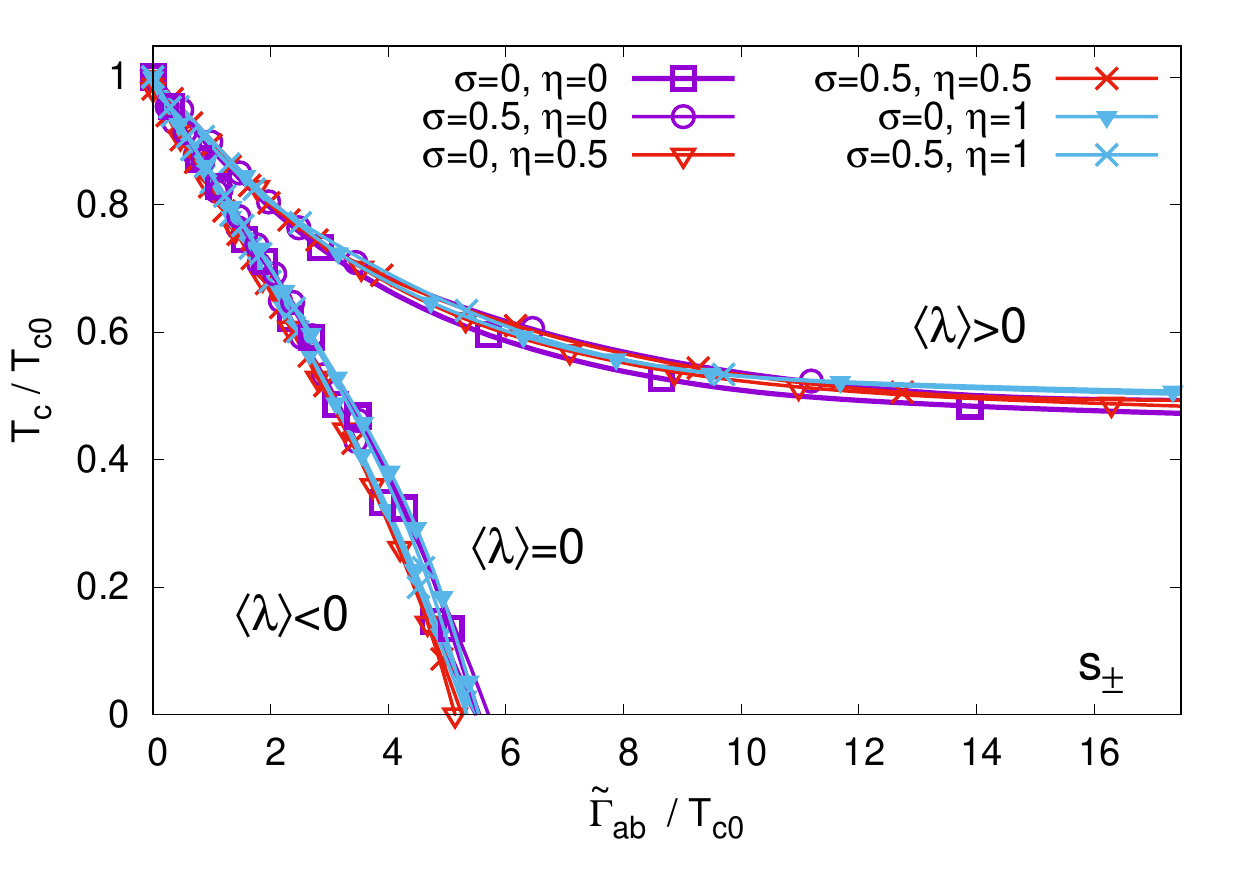}
\caption{(Color online.) $T_c$ for various $\sigmaN$ and $\etaN$ as a function of the effective interband scattering rate $\tilde\Gamma_{ab}$. Note that curves for different sets of $\sigmaN$ and $\etaN$ overlap and fall onto one of the three universal curves depending on the sign of $\la \lambda \ra$. Visible deviations originates from the numerical calculating errors. Curve for the case of $\la \lambda \ra = 0$ is situated slightly above the curve for $\la \lambda \ra < 0$ and they almost overlap.}
\label{fig:TcGammaUni}
\end{figure}

Figure~\ref{fig:TcGamma} shows $T_c$ as a function of $\GammaN_a$ for the $s_\pm$ state. As follows from calculations, $T_c$ behavior is qualitatively different for different signs of the coupling constant averaged over the Fermi surface~\cite{EfremovKorshunov2011},
\beq
\la \lambda \ra \equiv (\lambda_{aa} + \lambda_{ab}) N_a / N + (\lambda_{ba} + \lambda_{bb}) N_b / N,
\eeq
where $N = N_a + N_b$ is the total density of states in the normal phase. We choose the following coupling constants for illustrative purpose: $(\lambda_{aa},\lambda_{ab},\lambda_{ba},\lambda_{bb}) = (3,-0.2,-0.1,0.5)$ for $\la \lambda \ra > 0$~\cite{Popovich2010,Charnukha2011}, $(1,-2,-1,1)$ for $\la \lambda \ra < 0$, and $(2,-2,-1,1)$ for $\la \lambda \ra = 0$. For the first set in the clean limit, critical temperature is $T_{c0}=30$~cm$^{-1}$, for the second set it is $T_{c0}=27.96$~cm$^{-1}$, and for the third set it is $T_{c0}=31.47$~cm$^{-1}$, which correspond to 43.1~K, 40.2~K, and 45.2~K. Note, the strongest $T_c$ suppression occurs in the Born limit for the pure interband potential, i.e., $\etaN = 0$. In the opposite limit of pure intraband scattering with $u = 0$ ($\etaN \to \infty$), pairbreaking is absent because $\tilde\Gamma_{ab} \to 0$. Such situation appears in the unitary limit. As for the dependence of $T_c$ on  $\tilde\Gamma_{ab}$~(\ref{eq.Gamma.ab}) that is shown in Figure~\ref{fig:TcGammaUni}, all cases with different sets of $\sigmaN$ and $\etaN$ fall onto one of the universal $T_c$ curves depending on the sign of the coupling constant averaged over the Fermi surface, $\la \lambda \ra$. It is clearly seen from Figure~\ref{fig:TcGammaUni} that depending on the sign of $\la \lambda \ra$, one gets two types of $T_c$ behavior for the $s_\pm$ state: (1) the critical temperature vanishes at a finite impurity scattering rate $\tilde\Gamma_{ab}^\mathrm{crit}$ for $\la \lambda \ra < 0$, and (2) for $\la \lambda \ra > 0$, the critical temperature remains finite at $\tilde\Gamma_{ab} \to \infty$. In the marginal case of $\la \lambda \ra = 0$ we find that $\tilde\Gamma_{ab}^\mathrm{crit} \to \infty$ but with exponentially small $T_c$. Therefore, we have found a universal behavior of $T_c$ controlled by a single parameter $\la \lambda \ra$.


\begin{figure}
\centering
\includegraphics[width=0.7\textwidth]{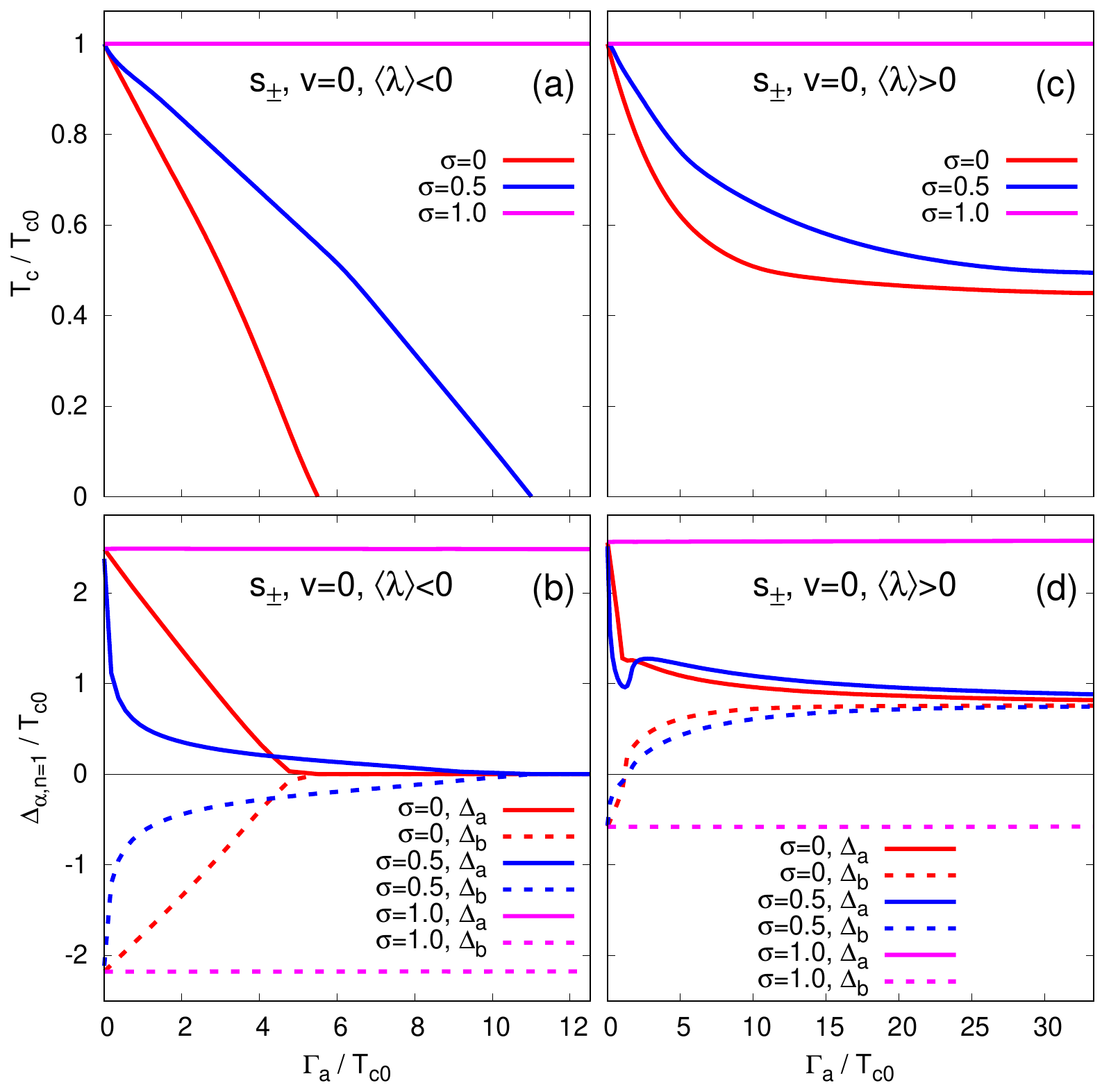}
\caption{(Color online.) Dependence of $T_c$ (a,c) and Matsubara gaps $\Delta_{\alpha n=1}$ (b,d) on the impurity scattering rate $\GammaN_a$ for the $s_\pm$ state with $v = 0$ for $\la \lambda \ra < 0$ (panels~a and~b) and for $\la \lambda \ra > 0$ (panels~c and~d). For $\la \lambda \ra < 0$, gaps in both bands vanish making $T_c$ drops to zero. For $\la \lambda \ra > 0$, smaller gap $\Delta_b$ crosses zero and its sign become the same as the sign of the larger gap $\Delta_a$, i.e., the system experience transition to the $s_{++}$ state. Unitary limit is always an exceptional case with constant $T_c$ and gaps. Gaps are shows for the Matsubara frequency $\omega_n = \pi T (2n+1)$ with $n = 1$ at $T = 0.016 T_{c0}$.
}
\label{fig:spmTcDeltaNonmag_v0_abcd}
\end{figure}

\begin{figure}
\centering
\includegraphics[width=0.7\textwidth]{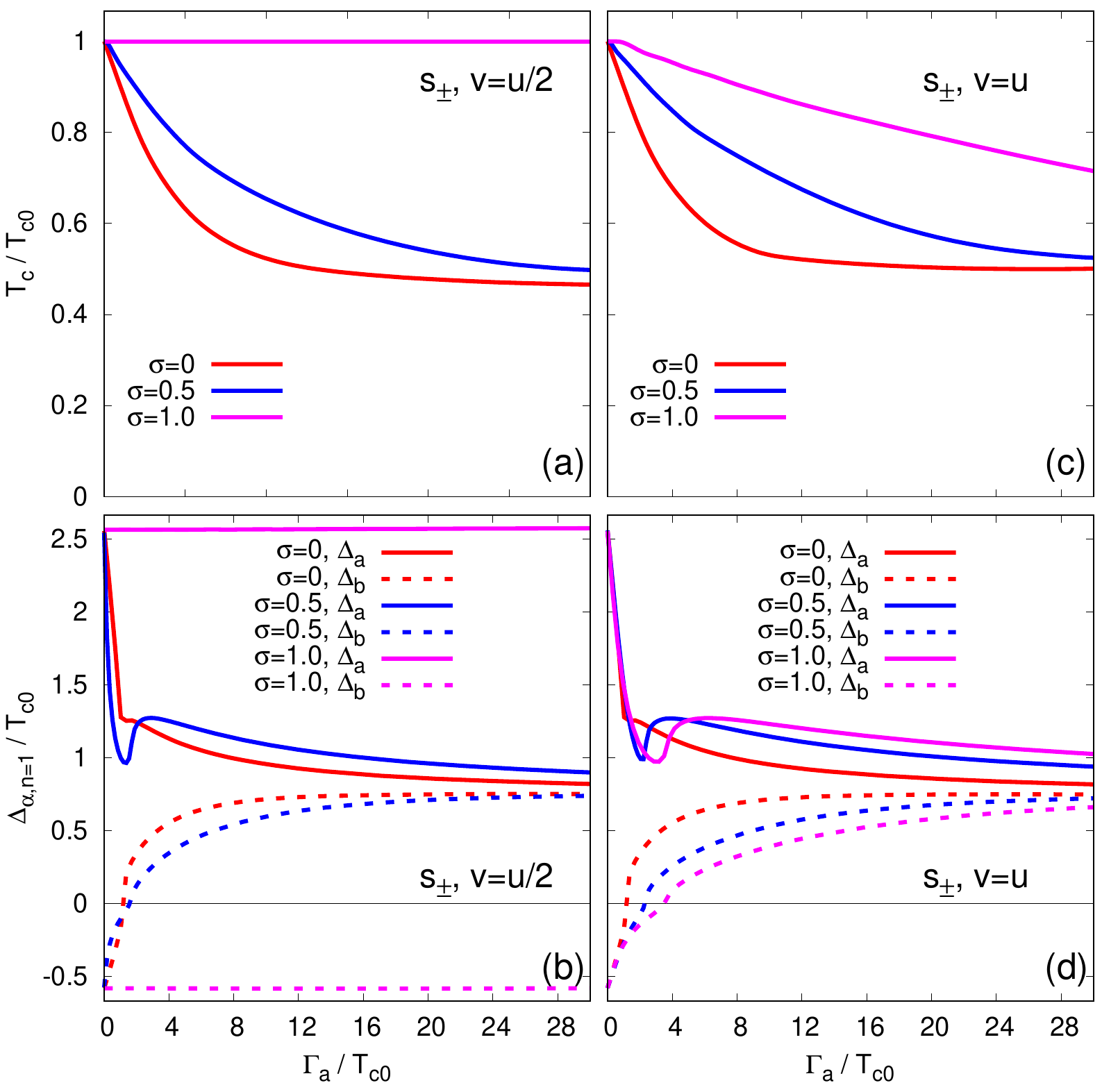}
\caption{(Color online.) Dependence of $T_c$ (a,c) and Matsubara gaps $\Delta_{\alpha n=1}$ (b,d) on the impurity scattering rate $\GammaN_a$ for the $s_\pm$ state with $\la \lambda \ra > 0$ with $v = u/2$ (panels~a and~b) and with $v = u$ (panels~c and~d). In both cases, the $s_\pm \to s_{++}$ transition occurs. For $v = u$, the superconductivity is suppressed even in the unitary limit.}
\label{fig:spmTcDeltaNonmag_vu_abcd}
\end{figure}

While the behavior of type-(1) systems is in agreement with the qualitative statement that the $s_\pm$ superconductivity is destroyed by the nonmagnetic interband impurities due to the ``mixing'' of gaps with different signs~\cite{Golubov1995,Golubov1997}, the behavior of type-(2) with $\la \lambda \ra > 0$ is surprising. To understand what happens in this case, we calculated the gap $\Delta_{\alpha n}$ for the first Matsubara frequency $n = 1$ at $T = 0.016 T_{c0}$.
Results are shown in Figures~\ref{fig:spmTcDeltaNonmag_v0_abcd} and~\ref{fig:spmTcDeltaNonmag_vu_abcd} for zero and finite intraband potential $v$, respectively. Coupling constants $\lambda_{\alpha \beta}$ are chosen to have $T_{c0} \approx 40$~K. It is seen that gaps on both bands, $\Delta_{a(b)n}$, converge to the same value, $\Delta_{\Gamma_{a(b)} \to \infty}$, while $T_c$ quickly saturates. The initially negative order parameter $\Delta_{bn}$ (corresponding to the smaller gap) increases and at some point crosses zero and becomes positive. After that since gaps signs for both bands are equal, we have the $s_{++}$ state. Due to the Anderson's theorem, this state is robust against impurity scattering thus having the finite $T_c$ up to $\Gamma_a \to \infty$. Therefore, $T_c$ stays finite in type-(2) systems due to the $s_\pm \to s_{++}$ transition.

\begin{figure}
\centering
\includegraphics[width=0.7\textwidth]{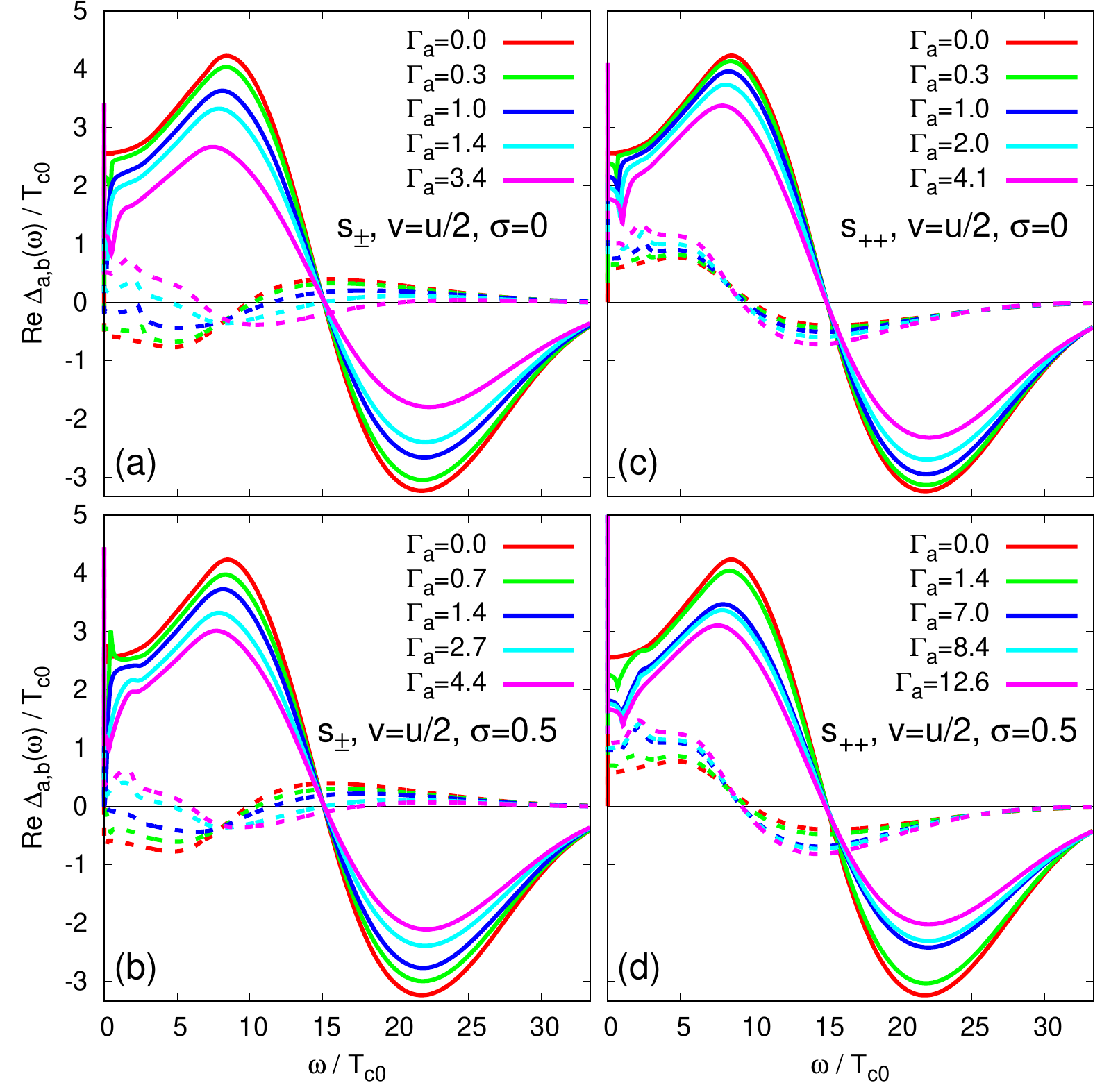}
\caption{(Color online.) Frequency dependence of a real part of the gap function $\mathrm{Re}\Delta_\alpha(\omega)$ for various $\GammaN_a$ for the $s_\pm$ (a,b) and the $s_{++}$ (c,d) superconductors with $v = u/2$. $\GammaN_a$ is given in units of $T_{c0}$. The gap in the band $\alpha = a$ ($\alpha = b$) is shown by the straight (dashed) curve. The Born limit, $\sigma = 0$, is shown in panels~a and~c, while the intermediate scattering limit with $\sigma = 0.5$ is shown in panels~b and~d.}
\label{fig:spmsppDeltaPadeNonmag}
\end{figure}

The transition is also seen in gap functions $\mathrm{Re}\Delta_\alpha(\omega)$ analytically continued to real frequencies, which are shown in Figure~\ref{fig:spmsppDeltaPadeNonmag}.

Similar to the $s_{++}$ state, there is no effect of disorder on the critical temperature and gaps in the unitary limit except for the case of $\etaN = 1$, where the $s_\pm \to s_{++}$ transition occurs (see Figure~\ref{fig:spmTcDeltaNonmag_vu_abcd}). Latter again makes the case of uniform scattering somehow unique~\cite{Kulic1999}.


There is a simple physical reason for the transition: with increasing interband scattering, order parameters on different Fermi surfaces ``mixes'' due to the scattering processes and converge to the same value. At the same time, larger gap ``attracts'' the smaller one that crosses zero and changes its sign. Similar effects were discussed in Refs.~\cite{Schopohl1977,Golubov1995,Golubov1997,Ummarino2007} for the two-band $s_{++}$ superconductor and in Ref.~\cite{v_mishra_09}, where node lifting in the extended $s_\pm$ state at the electron pocket was investigated. The discovered $s_\pm \to s_{++}$ transition allows to explain the much slower suppression of critical temperature than that following from the well-known Abrikosov-Gor'kov equation. Qualitatively this result was confirmed by agreement with the numerical solution of the Bogoliubov-de Gennes equations~\cite{Yao2012,Chen2013}.

\section{Magnetic disorder in multiband superconductors \label{sec:magn}}

Here we focus on magnetic impurities and their effect on the properties of $s_\pm$ and $s_{++}$ models. We show that there are few cases when the critical temperature $T_c$ saturates and stays finite in contrast to $T_c$ following from the prediction of the Abrikosov-Gor'kov theory for single-band superconductors~\cite{KorshunovMagn2014}.

\subsection{Eliashberg equations in the $\Tmat$-matrix approximation}

In the case of magnetic impurities, we have to consider Green's function matrix entering equation~(\ref{eq.g}) with the dimension $8 \times 8$. This considerably complicates the problem in comparison with the study of the nonmagnetic disorder. The impurity potential for the non-correlated impurities can be written as $\hat{\mathbf{U}} = \mathbf{V} \otimes \hat{S}$, where
\begin{equation}
\hat{S} = \left(
    \begin{array}{cc}
    \vec{\hat{\sigma}} \cdot \vec{S} & 0 \\
    0 & -(\vec{\hat{\sigma}} \cdot \vec{S})^{T}
    \end{array}
\right)
\end{equation}
is the $4 \times 4$ matrix with $(...)^{T}$ being the matrix transpose and $\vec{S} = \left( S_x, S_y, S_z \right)$ being the classic spin vector~\cite{ambeg}. The vector $\vec{\hat{\sigma}}$ is composed of Pauli $\tau$-matrices, $\vec{\hat{\sigma}} = \left( \hat{\tau}_1, \hat{\tau}_2, \hat{\tau}_3 \right)$. The potential strength is determined by $(\mathbf{V})_{\alpha \beta} = \Vimp_{\mathbf{R}_i = 0}^{\alpha \beta}$. For simplicity, intraband and interband parts of the potential are set equal to $\iM$ and $\jM$, respectively, such that $(\mathbf{V})_{\alpha \beta} = (\iM-\jM) \delta_{\alpha \beta} + \jM$. Than $\mathbf{V}$ is given by
\begin{equation}
\mathbf{V}=\left(
    \begin{array}{cc}
    \iM & \jM \\
    \jM & \iM
    \end{array}
\right).
\end{equation}
Components of the impurity potential matrix $\hat{\mathbf{U}}$ is then $\hat{U}_{aa,bb} = \iM \hat{S}$ and $\hat{U}_{ab,ba} = \jM \hat{S}$, and the matrix itself is given by
\beq
 \hat{\mathbf{U}} =
 \left(
    \begin{array}{cc}
    \iM \hat{S} & \jM \hat{S} \\
    \jM \hat{S} & \iM \hat{S}
    \end{array}
\right).
\label{eq.U}
\eeq

Coupled $\Tmat$-matrix equations~(\ref{eq.tmatrix}) for $aa$ and $ba$ components of the self-energy in the introduced notations become
\bea
\hat{\Sigma}_{aa}^{\imp} &=& n_{\imp} \hat{U}_{aa} + \hat{U}_{aa} \hat{g}_a \hat{\Sigma}_{aa}^{\imp} + \hat{U}_{ab} \hat{g}_b \hat{\Sigma}_{ba}^{\imp},
\label{eq.Sigma_aa} \\
\hat{\Sigma}_{ba}^{\imp} &=& n_{\imp} \hat{U}_{ba} + \hat{U}_{ba} \hat{g}_a \hat{\Sigma}_{aa}^{\imp} + \hat{U}_{bb} \hat{g}_b \hat{\Sigma}_{ba}^{\imp}.
\label{eq.Sigma_ba}
\eea
Solution of the system in the matrix form is
\bea
\hat{\Sigma}_{aa}^{\imp} &=& n_{\imp} \left[ \hat{1} - \hat{U}_{aa} \hat{g}_a - \hat{U}_{ab} \hat{g}_b \hat{\zeta} \hat{U}_{ba} \hat{g}_a \right]^{-1} \left( \hat{U}_{aa} + \hat{U}_{ab} \hat{g}_b \hat{\zeta} \hat{U}_{ba} \right), \\
\hat{\Sigma}_{ba}^{\imp} &=& \hat{\zeta} \hat{U}_{ba} \left( n_{\imp} + \hat{g}_a \hat{\Sigma}_{aa}^{\imp} \right),
\eea
where $\hat{\zeta} = \left[ \hat{1} - \hat{U}_{bb} \hat{g}_b \right]^{-1}$. Renormalizations of frequencies and gaps come from
\bea
\Sigma^{\imp}_{0a} &=& \frac{1}{4} \mathrm{Tr}\left[\hat{\Sigma}_{aa}^{\imp} \cdot \left( \hat{\tau}_0 \otimes \hat{\sigma}_0 \right) \right], \\
\Sigma^{\imp}_{2a} &=& \frac{1}{4} \mathrm{Tr}\left[\hat{\Sigma}_{aa}^{\imp} \cdot \left( \hat{\tau}_2 \otimes \hat{\sigma}_2 \right) \right].
\eea
Equations for $\Sigma^{\imp}_{0b}$ and $\Sigma^{\imp}_{2b}$ are derived from equations above via replacement $a \leftrightarrow b$.

We assume that spins are not polarized and $s^2 = \la S^2 \ra = S(S+1)$. Since $s$ enters everywhere together with the components of the impurity potential, $\iM$ and $\jM$ (see expression~(\ref{eq.U}) for $\hat{\mathbf{U}}$), without loss of generality, later we set $s = 1$ assuming that $\iM$ and $\jM$ are renormalized to include $s$ in themselves.

As follows from the calculations, similar to results presented in Section~\ref{sec:nonmag}, expressions for $\Sigma^{\imp}_{0\alpha}$ and $\Sigma^{\imp}_{2\alpha}$ are proportional to the impurity scattering rate $\Gamma_{a,b}$ and contain the generalized cross-section parameter $\sigma$ that helps to control the approximation for the ``strength'' of impurity scattering. Latter ranges from the Born limit (weak scattering, $\pi \jM N_{a,b} \ll 1$) to the unitary limit (strong scattering, $\pi \jM N_{a,b} \gg 1$):
\bea
\Gamma_{a,b} &=& 2 \pi n_{\imp} \jM^2 (1-\sigma) N_{b,a} = \frac{2 n_{\imp} \sigma}{\pi N_{a,b}} \to \left\{
    \begin{array}{l}
    2 \pi \jM^2 n_{\imp} N_{b,a}, \text{Born limit}\\
    \frac{2 n_{\imp}}{\pi N_{a,b}}, \text{unitary limit}
    \end{array}
  \right.
\\
\sigma &=& \frac{\pi^2 \jM^2 N_a N_b}{1 + \pi^2 \jM^2 N_a N_b}
\to \left\{
    \begin{array}{l}
    0, \text{Born limit}\\
    1, \text{unitary limit}
    \end{array}
  \right.
\eea
Also, we introduce the parameter $\eta$ to control the ratio of intra- and interband scattering potentials, $\iM = \jM \eta$.

Expressions for $\Sigma_{0(2)a}^{\imp}$ at arbitrary temperature and $\eta$ are too complicated and non-informative to write them here. It is much more convenient to consider limiting cases. We also consider the three special forms of the impurity potential: the uniform potential with $\eta = 1$ ($\iM = \jM$), the interband-only potential with $\eta = 0$ ($\iM = 0$, $\jM \neq 0$), and the intraband-only potential with $\jM = 0$, $\iM \neq 0$ (formally, $\eta = \infty$).

The Born limit corresponds to $\sigma = 0$. Eliashberg equations~(\ref{eq.omega.tilde})-(\ref{eq.phi.tilde}) are then written as follows:
\bea
\tilde\omega_{an} = \omega_n + \ii \Sigma_{0a}(\omega_n) + \pi \jM^2 n_{\imp} \left(\eta^2 N_a \frac{\tilde\omega_{an}}{Q_{an}} + N_b \frac{\tilde\omega_{bn}}{Q_{bn}} \right), \label{eq.omega_magn_Born} \\
\tilde\phi_{an} = \Sigma_{2a}(\omega_n) -\pi \jM^2 n_{\imp} \left(\eta^2 N_a \frac{\tilde\phi_{an}}{Q_{an}} + N_b \frac{\tilde\phi_{bn}}{Q_{bn}}\right).
\label{eq.phi_magn_Born}
\eea
One of the significant differences of this expression from analogous for the nonmagnetic impurities (see (\ref{eq.omega.interBorn}) and (\ref{eq.phi.interBorn})) is the minus sign before the term originating from impurity scattering in equation~(\ref{eq.phi_magn_Born}).
In the presence of interband-only scattering ($\eta = 0$), we derive here the remarkable result,
\bea
\tilde\omega_{an} = \omega_n + \ii \Sigma_{0a}(\omega_n) + \pi \jM^2 n_{\imp} N_b \frac{\tilde\omega_{bn}}{Q_{bn}}, \\
\tilde\phi_{an} = \Sigma_{2a}(\omega_n) -\pi \jM^2 n_{\imp} N_b \frac{\tilde\phi_{bn}}{Q_{bn}}.
\eea
Indeed, for the $s_{++}$ state we have $\sgnnobr{\tilde\phi_{bn}} = \sgnnobr{\tilde\phi_{an}}$ and equations written above correspond to the generalization of the Abrikosov-Gor'kov theory to the two-band case, therefore, impurities should suppress superconductivity. This, however, as we will see later, is not always true due to the complicated structure of equations and their self-consistent solution may lead to the unexpected results. On the other hand, for the $s_\pm$ state we have $\sgnnobr{\tilde\phi_{bn}} = - \sgnnobr{\tilde\phi_{an}}$ and the sign of the last term in~(\ref{eq.phi_magn_Born}), originating from impurity scattering, changes and equations become similar to expressions for the two-band superconductor with the \textit{nonmagnetic impurities}. That is, $T_c$ is not suppressed by disorder except for the $\eta = 1$ case.

For the uniform impurity potential we have $\eta = 1$, then $\tilde\omega_{an} = \omega_n + \ii \Sigma_{0a}(\omega_n) + \pi \jM^2 n_{\imp} \left(N_a \frac{\tilde\omega_{an}}{Q_{an}} + N_b \frac{\tilde\omega_{bn}}{Q_{bn}}\right)$ and $\tilde\phi_{an} = \Sigma_{2a}(\omega_n) - \pi \jM^2 n_{\imp} \left(N_a \frac{\tilde\phi_{an}}{Q_{an}} + N_b \frac{\tilde\phi_{bn}}{Q_{bn}}\right)$. Here contributions from both $a$ and $b$ bands are mixed so we expect a suppression of $T_c$ by the disorder~\cite{Korshunov2016}.

When the interband component is absent ($\eta = \infty$), equations for different bands are decoupled,
\bea
\tilde\omega_{an} &=& \omega_n + \ii \Sigma_{0a}(\omega_n) + \pi \iM^2 n_{\imp} N_a \frac{\tilde\omega_{an}}{Q_{an}}, \nn\\
\tilde\phi_{an} &=& \Sigma_{2a}(\omega_n) - \pi \iM^2 n_{\imp} N_a \frac{\tilde\phi_{an}}{Q_{an}}, \nn
\eea
and we have the suppression of superconductivity in each band following the Abrikosov-Gor'kov theory.

It is remarkable that equations in the unitary limit are exactly the same as in the unitary limit for the \textit{nonmagnetic impurities}~(\ref{eq.omega.uni.eta1.nonmag})--
(\ref{eq.phi.uni.nonmag}). Therefore, all conclusions about suppression of superconductivity for $\eta \neq 1$ and $\eta = 1$ are the same.

Now we write down Eliashberg equations for special forms of the impurity potential. For the intraband-only impurity potential ($\iM = 0$), terms in equations corresponding to bands $a$ and $b$ are separated,
\bea
\tilde\omega_{an} = \omega_n + \ii \Sigma_{0a}(\omega_n) + \frac{\Gamma_a}{2 D} \left[ \sigma \frac{\tilde\omega_{an}}{Q_a} + (1 - \sigma) \frac{\tilde\omega_{bn}}{Q_b} \right], \\
\tilde\phi_{an} = \Sigma_{2a}(\omega_n) + \frac{\Gamma_a}{2 D} \left[ \sigma \frac{\tilde\phi_{an}}{Q_a} - (1 - \sigma) \frac{\tilde\phi_{bn}}{Q_b} \right],
\eea
where
\beq
D = 1 - 2 (1-\sigma) \sigma \left( 1 - \frac{\tilde\omega_{an} \tilde\omega_{bn} - \tilde\phi_{an} \tilde\phi_{bn}}{Q_a Q_b} \right). \nn
\eeq

For the impurity potential scattering solely between different bands ($\jM = 0$), equations for different bands decouple:
\bea
\tilde\omega_{an} &=& \omega_n + \ii \Sigma_{0a}(\omega_n) + \Gamma_a \frac{N_a}{2 D}\frac{\tilde\omega_{an}}{Q_a} \left[ \sigma N_a + (1 - \sigma) N_b \right], \\
\tilde\phi_{an} &=& \Sigma_{2a}(\omega_n) + \Gamma_a \frac{N_a}{2 D} \frac{\tilde\phi_{an}}{Q_a} \left[ \sigma N_a - (1 - \sigma) N_b \right],
\eea
where
\beq
D = \sigma^2 N_a^2 + (1-\sigma)^2 N_b^2 + 2 \sigma (1-\sigma) N_a N_b \frac{ \tilde\omega_{an}^2 - \tilde\phi_{an}^2}{Q_a^2}. \nn
\eeq

\subsection{Results of calculations \label{subsec:magnimpresults}}

Following results were obtained by solving self-consistently frequency and gap equations~(\ref{eq.omega.tilde}) and~(\ref{eq.phi.tilde}) with the impurity self-energy from the solution of equations~(\ref{eq.Sigma_aa}),~(\ref{eq.Sigma_ba}) for both arbitrary finite temperature below $T_c$ and at $T_c$~\cite{KorshunovMagn2014}. Hereafter, for illustrative purpose we consider the case of $N_b / N_a = 2$ and choose coupling constants as $(\lambda_{aa},\lambda_{ab},\lambda_{ba},\lambda_{bb}) = (3,-0.2,-0.1,0.5)$ for the $s_\pm$ state with $\la \lambda \ra > 0$~\cite{Popovich2010,Charnukha2011}, and as $(3,0.2,0.1,0.5)$ for the $s_{++}$ state. Critical temperature in the clean limit for both sets is $T_{c0}=30$~cm$^{-1}$ that corresponds to 43.1~K.

\begin{figure}[ht]
\centering
\includegraphics[width=0.7\textwidth]{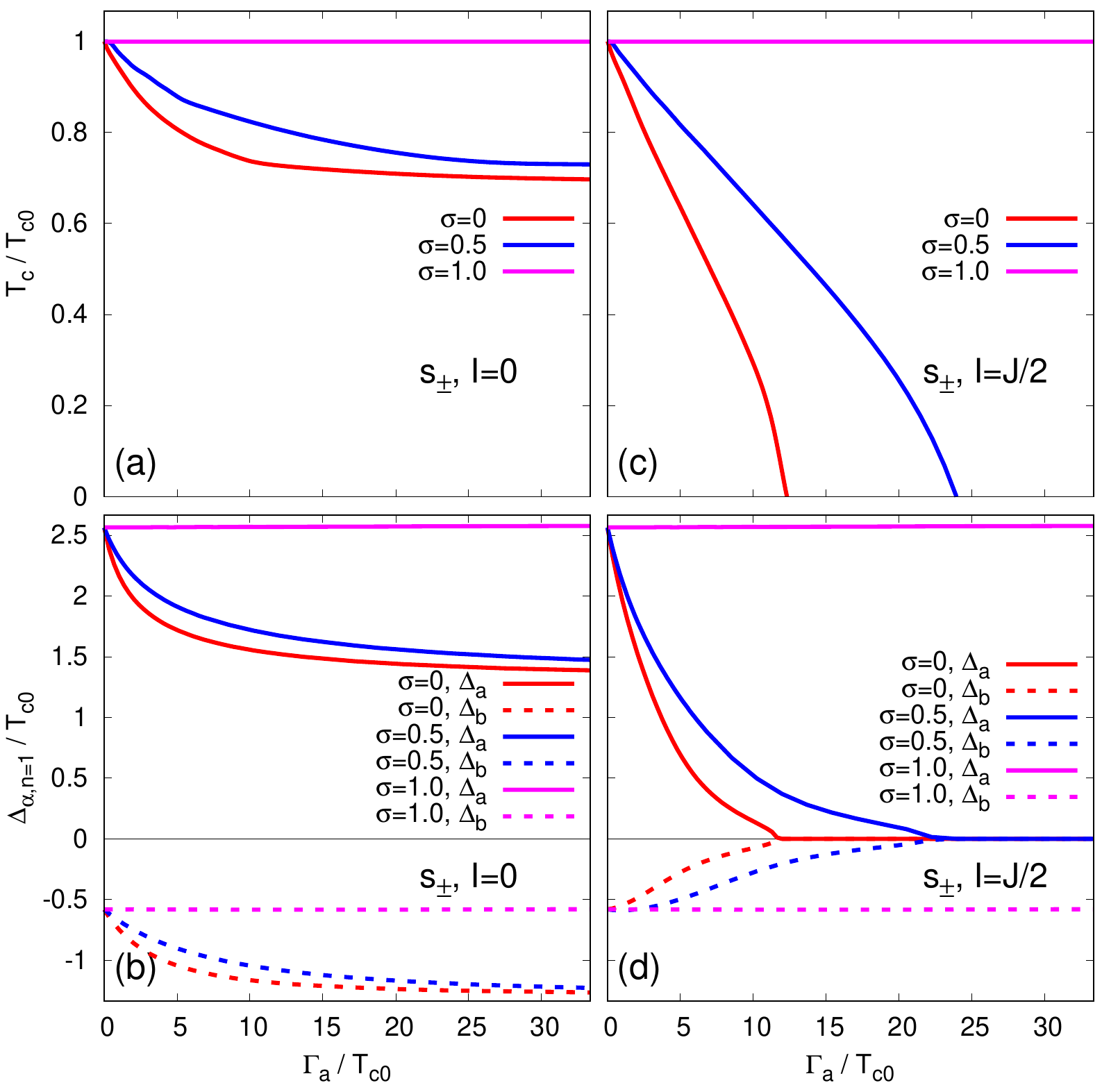}
\caption{(Color online.)  Dependence of $T_c$ (a,b) and Matsubara gaps $\Delta_{\alpha n=1}$ (c,d) on the impurity scattering rate $\Gamma_a$ for the $s_\pm$ superconductor with only interband scattering, $\iM=0$, in panels~a and~c, and with $\iM=\jM/2$ in panels~b and~d.}
\label{fig:spmTcDelta}
\end{figure}
\begin{figure}[ht]
\centering
\includegraphics[width=0.7\textwidth]{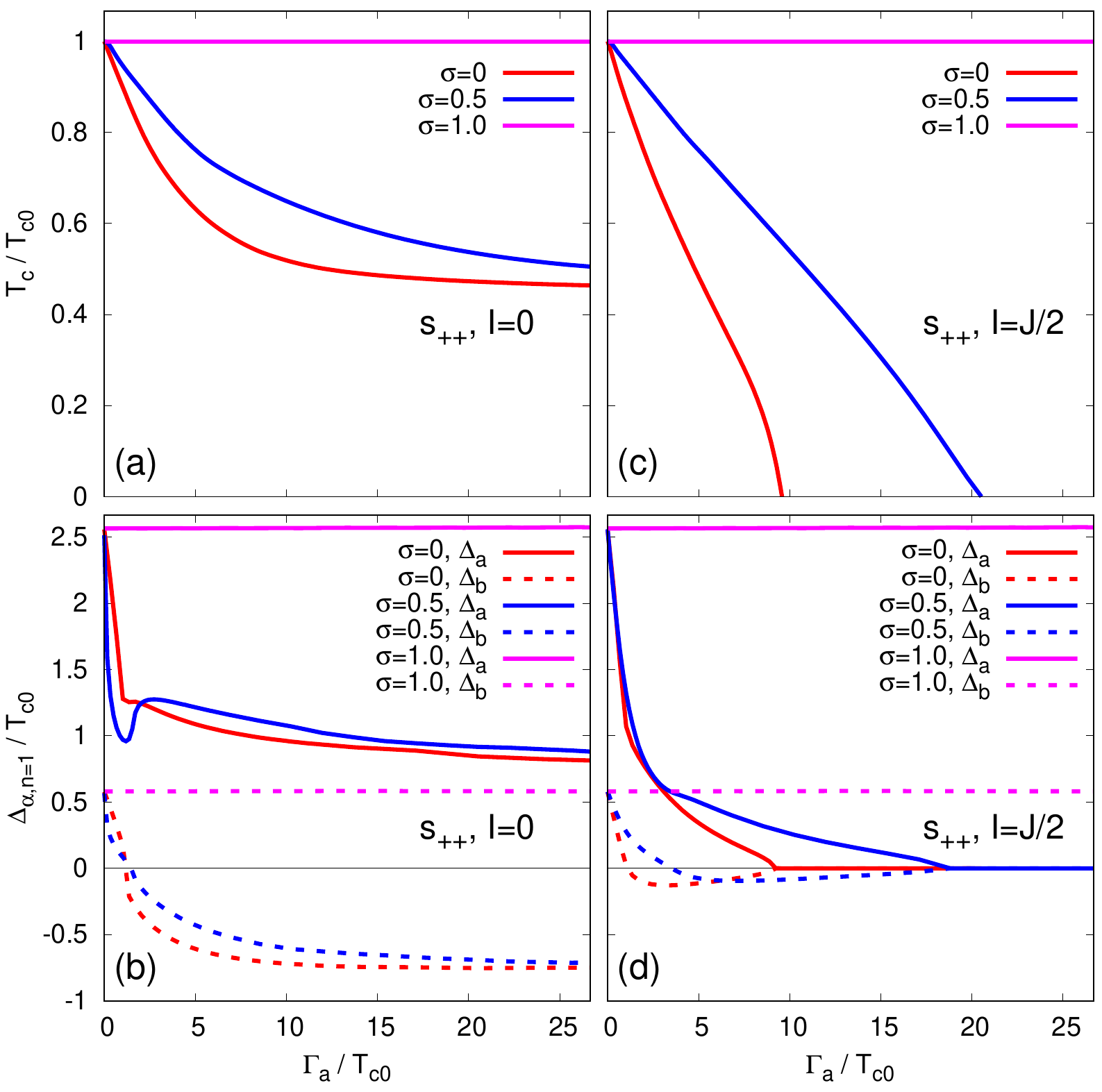}
\caption{(Color online.) Dependence of $T_c$ (a,b) and Matsubara gaps $\Delta_{\alpha n=1}$ (c,d) on the impurity scattering rate $\Gamma_a$ for the $s_{++}$ superconductor with only interband scattering, $\iM=0$, in panels~a and~c, and with $\iM=\jM/2$ in panels~b and~d.}
\label{fig:sppTcDelta}
\end{figure}
\begin{figure}[ht]
\centering
\includegraphics[width=0.7\textwidth]{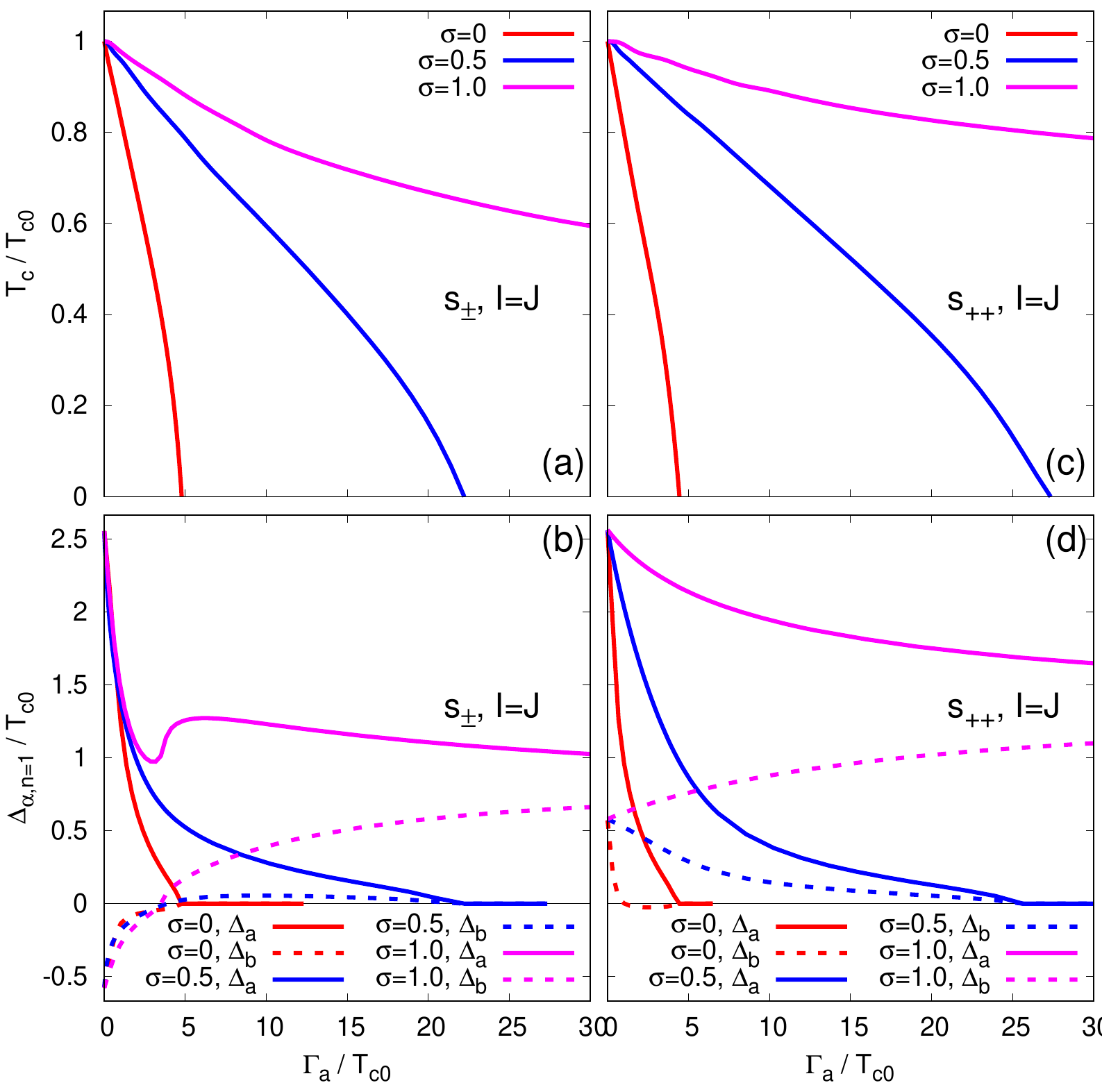}
\caption{(Color online.) The same as in Figures~\ref{fig:spmTcDelta} and~\ref{fig:sppTcDelta} but for the special case of $\iM = \jM$.}
\label{fig:spmsppTcDeltaIeqJ}
\end{figure}
\begin{figure}[ht]
\centering
\includegraphics[width=0.7\textwidth]{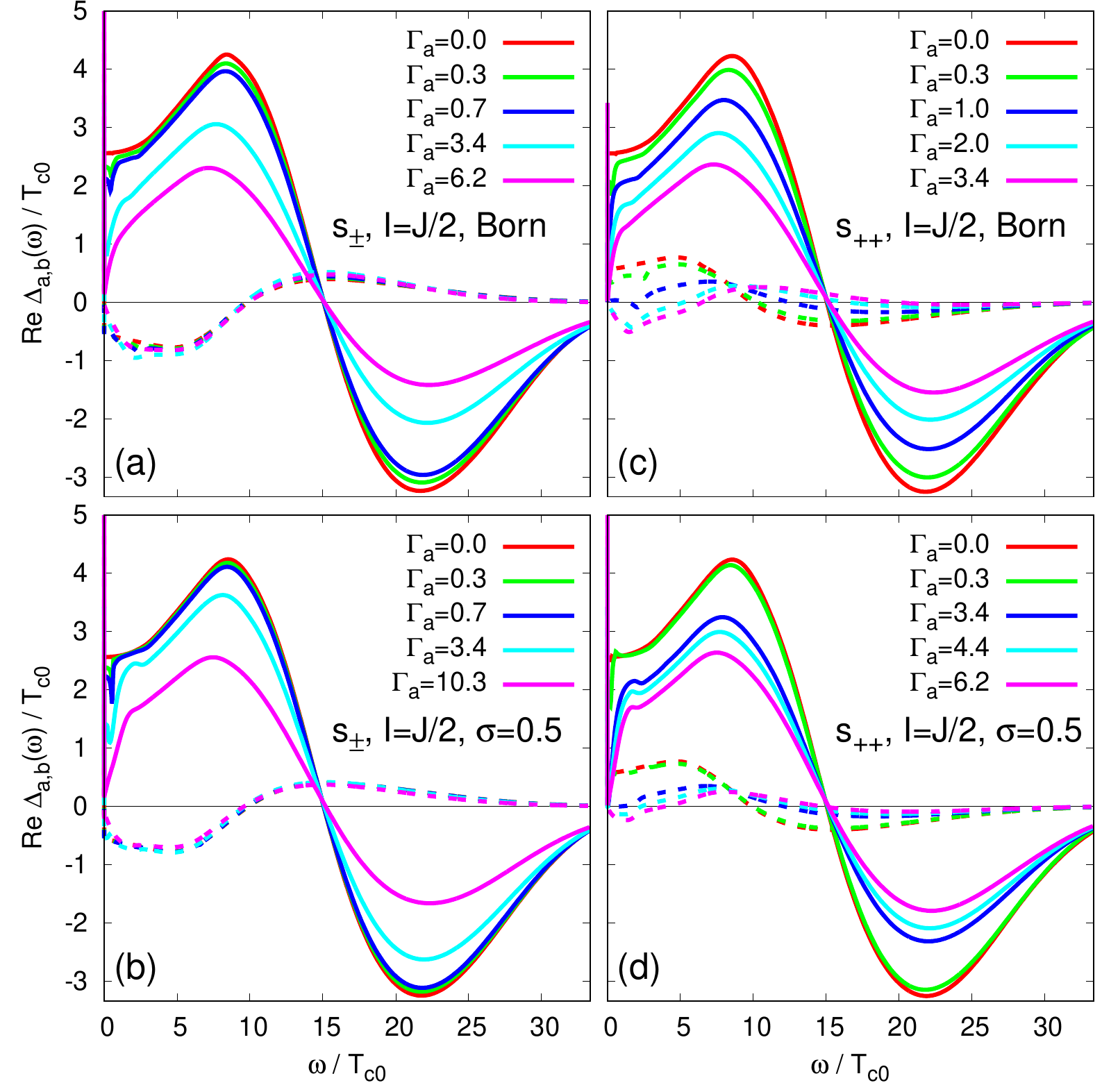}
\caption{(Color online.) Frequency dependence of the gap function $\mathrm{Re}\Delta_\alpha(\omega)$ for various values of $\Gamma_a$ for the $s_\pm$ (a,b) and the $s_{++}$ (c,d) superconductors with $\iM = \jM/2$. $\GammaN_a$ is given in units of $T_{c0}$. Gap in the band $\alpha = a$ ($\alpha = b$) is shown by the straight (dashed) curve. Born limit, $\sigma = 0$, is shown in panels~a and~c, while the intermediate scattering limit with $\sigma = 0.5$ is shown in panels~b and~d.}
\label{fig:spmsppDeltaPade}
\end{figure}

In Figures~\ref{fig:spmTcDelta}--\ref{fig:spmsppTcDeltaIeqJ} we plot $T_c$ and Matsubara gaps $\Delta_{\alpha n}$ for the first Matsubara frequency $\omega_{n=1} = 3 \pi T$ as functions of $\Gamma_a$ for various values of $\sigma$ for both $s_\pm$ and $s_{++}$ superconductors. The real part of the analytical continuation of $\Delta_{\alpha n}$ to real frequencies, the gap function $\mathrm{Re}\Delta_\alpha(\omega)$, is shown in Figure~\ref{fig:spmsppDeltaPade}.

First, we discuss the $s_\pm$ state. $T_c$ becomes insensitive to impurities for the pure interband scattering, $\iM = 0$. This partially confirms qualitative arguments that the $s_\pm$ state with magnetic impurities behaves like the $s_{++}$ state with the nonmagnetic disorder~\cite{Golubov1995,Golubov1997} and agrees with theoretical calculations in the Born limit~\cite{Li2009}. For the initially unequal gaps, $|\Delta_a| \neq |\Delta_b|$, there is an initial decrease of $T_c$ for small $\Gamma_a$ until the renormalized gaps become equal and then $T_c$ saturate since the analog of Anderson's theorem achieved. For the finite $\iM$, intraband scattering on the magnetic disorder averages gaps to zero thus suppress $T_c$. On the other hand, in the unitary limit ($\sigma=1$) at $T \to T_c$ we have
\bea
 \tilde\omega_{a n} &=& \omega_n + \ii \Sigma_{0a}(\omega_n) + \frac{\Gamma_a}{2} \sgn{\omega_n}, \nn \\
 \tilde\phi_{a n} &=& \Sigma_{2a}(\omega_n) + \frac{\Gamma_a}{2} \frac{\tilde\phi_{a n}}{\left|\tilde\omega_{a n}\right|} \nn
\eea
for arbitrary value of $\eta$, including the case of intraband-only impurities, $1/\eta = 0$. This form of equations is the same as for nonmagnetic impurities and thus there is no impurity contribution to the $T_c$ equation in analogy to the Anderson's theorem. The only exception here is the special case of uniform impurities, $\eta = 1$, when
\bea
 \tilde\omega_{a n} &=& \omega_n + \ii \Sigma_{0a}(\omega_n) + \frac{n_{\imp}}{\pi \left(N_a+N_b\right)} \sgn{\omega_n}, \nn \\
 \tilde\phi_{a n} &=& \Sigma_{2a}(\omega_n) + \frac{n_{\imp}}{\pi \left(N_a+N_b\right)^2} \left(N_a \frac{\tilde\phi_{a n}}{\left|\tilde\omega_{a n}\right|} + N_b \frac{\tilde\phi_{b n}}{\left|\tilde\omega_{b n}\right|} \right). \nn
\eea
Both gaps are mixed in equation for $\tilde\phi_{a n}$ thus they tend to zero with increasing amount of disorder. That's also true away from the unitary limit (see Figure~\ref{fig:spmsppTcDeltaIeqJ}) and that's the source of claim that the uniform potential with $\iM = \jM$ is a special case with the strongest $T_c$ suppression.

In general, multiband $s_{++}$ state should always be fragile against paramagnetic disorder since magnetic scattering between bands having the gaps of the same sign equivalent to the pairbreaking scattering within the single (quasi)isotropic band. Surprisingly, we find a regime with the saturation of $T_c$ for the finite amount of disorder right after the initial downfall (similar to the one following from the Abrikosov-Gor'kov theory) (Figure~\ref{fig:sppTcDelta}b). The saturation of $T_c$ is observed for the interband-only impurities, while the presence of the intraband magnetic disorder finally suppress $T_c$ to zero. However, depending on the ``strength'' of scattering $\sigma$, a decrease of $T_c$ may be quite slow compared to the one predicted by the Abrikosov-Gor'kov theory.

To understand the origin of the $T_c$ saturation we analyzed the gap function dependence on the scattering rate $\Gamma_a$ (see Figure~\ref{fig:spmTcDelta}). For the $s_{++}$ state after the certain value of the scattering rate, the smaller gap, $\Delta_b$, becomes negative. What we see is the $s_{++} \to s_\pm$ transition. As soon as system becomes effectively $s_\pm$, scattering on magnetic impurities cancels out in the $T_c$ equation, similar to the Anderson's theorem, and $T_c$ saturates. Before the saturation, the initial downfall akin to the one following from the Abrikosov-Gor'kov theory occurs. The transition is also seen in the frequency dependence of the gap function on a real frequency axis (see Figure~\ref{fig:spmsppDeltaPade}).

Similar to the $s_\pm \to s_{++}$ transition for the nonmagnetic disorder, there is a simple physical argument behind the $s_{++} \to s_\pm$ transition here. Namely, with increasing interband magnetic disorder, the gap functions on the different Fermi surfaces tend to the same value and if one of the gaps is smaller than another, it cross zero and change sing. A similar effect has been mentioned in Refs.~\cite{Schopohl1977,Golubov1995,Golubov1997} for a two-band systems with $s_{++}$ symmetry in the Born limit.

Note that here we do not consider a time-reversal symmetry broken $s_\pm + \ii s_{++}$ state. It may appear at $T \lesssim T_c$ in cases when translational symmetry is violated~\cite{Stanev2012}.

\section{Experimental situation with the disorder-induced superconductivity suppression in iron-based materials \label{sec:experimentsreview}}

Presently, there are not so many experimental studies of impurity effects on the superconducting state of iron pnictides and chalcogenides. Moreover, it is hard to determine exactly whether the impurity is nonmagnetic or magnetic because of the possible magnetic moment induced by nonmagnetic ions or irradiated particles, e.g. neutrons~\cite{Karkin2009}. Other concomitant difficulties in the results interpretation include changes of the crystal structure with the replacement of one ion with another and possible effective doping also affecting the superconducting critical temperature. That is why further we describe effect of various kinds of disorder on the critical temperature $T_c$ without going into the details of the nature of the disorder.

Let us systemize the data in the following way: first, we discuss the subset of works on introducing impurities via replacing one ion with another, and second, we make a short review of irradiation studies.

The chemical substitution of iron with copper or nickel in the 122 system, Ba$_{0.6}$K$_{0.4}$(Fe$_{1-x}M_x$)$_2$As$_2$ with $M$ = Cu or Ni, resulted in the full suppression of $T_c$ for $x \sim 0.1$ with the rates of $-3.5$~K per 1\% of Cu and $-2.9$~K per 1\% of Ni~\cite{Cheng2013}.
With the chemical substitution in Ba$_{0.5}$K$_{0.5}$(Fe$_{1-x}M_x$)$_2$As$_2$ of iron with zinc ($M$ = Zn), the effect on $T_c$ is practically absent, while with a change for manganese ($M$ = Mn), $T_c$ is completely suppressed for $x \sim 0.08$~\cite{Cheng2010}.
Another study of Ba$_{0.5}$K$_{0.5}$Fe$_{2-2x}M_{2x}$As$_2$ system with $M$ = Fe, Mn, Ru, Co, Ni, Cu, Zn, revealed that all types of chemical substitution result in full $T_c$ suppression except for $M$ = Ru, when $T_c$ changes quite weakly~\cite{Li2012}. Rates of suppression for Mn, Co, Ni, Cu, and Zn are equal to $6.98$, $1.73$, $2.21$, $2.68$, and $2.22$~K per 1\% of Fe replacement with these atoms, respectively. Difference in $T_c$ suppression by zinc in Refs.~\cite{Cheng2010,Li2012} are attributed to the technological difficulties in zinc doping at atmospheric pressure and, possibly, that in work~\cite{Cheng2010} zinc concentration was not exceeding 2\% in polycrystalline samples.
Consistent study of zinc effect on the superconductivity in LaFe$_{1-y}$Zn$_y$AsO$_{1-x}$F$_{x}$ revealed the dependence of this effect on $x$ -- $T_c$ slightly increases in underdoped samples ($x=0.05$), stays practically unchanged at optimal doping ($x=0.1$), and becomes rapidly suppressed in overdoped samples ($x=0.15$)~\cite{Li2010}.
In BaFe$_{1.89-2x}$Zn$_{2x}$Co$_{0.11}$As$_2$, zinc suppress $T_c$ with the rate of $3.63$~K per 1\% of Zn~\cite{Li2011}, that is considerably weaker than expected from the Abrikosov-Gor'kov theory.
Chemical substitution in LaFe$_{1-x}M_x$PO$_{0.95}$F$_{0.05}$ results in the $T_c$ suppression rate of $-2.2$~K per 1\% for $M$ = Co and $-9.3$~K per 1\% for $M$ = Mn~\cite{Suzuki2010}. According to the magnetoresistance measurements, the authors of Ref.~\cite{Suzuki2010} claim that cobalt (manganese) is a nonmagnetic (magnetic) impurity.
For K$_{0.8}$Fe$_{2-y-x}M_x$Se$_2$ with $M$ = Cr, Co, and Zn, the rapid suppression of $T_c$ is observed that is absent for $M$ = Mn~\cite{Tan2011}. At the same time, based on electronic paramagnetic resonance (EPR) measurements, the authors of Ref.~\cite{Tan2011} claim that an introduction of Cr, Co, and Zn cause the formation of the large local moments, in contrast to the Mn case.
Replacement of iron in Fe$_{1-y}M_y$Te$_{0.65}$Se$_{0.35}$ ($M$ = Co, Ni, Cu) results in the following $T_c$ suppression rates: $5.8$, $2.6$, and $1.3$~K per 1\% for Cu, Ni, and Co, respectively~\cite{Bezusyy2012}.
Strong $T_c$ suppression was observed in LaFe$_{1-x}$Zn$_x$AsO$_{0.85}$ with the rate of $9$~K per 1\% of Zn~\cite{Guo2010imp}.
Isovalent replacement of potassium with sodium in K$_{1-x}$Na$_x$Fe$_2$As$_2$ cause the drop of $T_c$ from $3.5$~K at $x = 0$ to $2.8$~K at $x = 0.07$~\cite{Kim2014}. With  isovalent ruthenium doping in NdFe$_{1-y}$Ru$_y$AsO$_{0.89}$F$_{0.11}$~\cite{Lee2010} and LaFe$_{1-y}$Ru$_y$AsO$_{0.89}$F$_{0.11}$~\cite{Satomi2010}, $T_c$ decreases much weaker than at the iron replacement with cobalt in NdFe$_{1-y}$Co$_y$AsO$_{0.89}$F$_{0.11}$ and than expected from the Abrikosov-Gor'kov theory. In SmFe$_{1-x}$Ru$_x$AsO$_{0.85}$F$_{0.15}$, the isovalent substitution of iron with ruthenium results in the rapid (slow) $T_c$ suppression for $x < 0.5$ ($x > 0.5$)~\cite{Tropeano2010}. The authors of Ref.~\cite{Tropeano2010} connected such a change in the behavior to the change of the role played by ruthenium -- initially it plays a role of the nonmagnetic impurity and than, for $x > 0.5$, the metallic behavior is restored due to the large ruthenium concentration and its contribution to the band structure.

In the K(Fe$_{1-x}$Co$_x$)$_2$As$_2$ system, cobalt doping cause the same rapid $T_c$ suppression as in cuprates YBa$_2$(Cu$_{1-x}$Zn$_x$)$_3$O$_{6.93}$ and La$_{1.85}$Sr$_{0.15}$Cu$_{1-x}$Ni$_x$O$_4$, and at $x \approx 0.4$, superconductivity vanishes~\cite{Wang2014}. Perhaps, analogy with cuprates arise here due to the presence of line nodes in both cuprates and KFe$_2$As$_2$~\cite{MaitiKorshunov2012}.

There are also an unusual situations, for example, LaO$_{0.9}$F$_{0.1}$FeAs$_{1-\delta}$, where the arsenic disorder with $\delta \approx 0.06$ cause not the decrease, but the slight increase of $T_c$~\cite{Grinenko2011}.

Let us switch to the irradiation studies. Here the situation in general is less diversified than at the chemical substitution of ions. In particular, the suppression of $T_c$ is observed, though it is much weaker than expected from the Abrikosov-Gor'kov expression. This is true for the irradiation by neutrons of LaFeAsO$_{0.9}$F$_{0.1}$~\cite{Karkin2009}, by protons of Ba(Fe$_{1-x}$Co$_x$)$_2$As$_2$ ($x=0.045$, 0.075, 0.113)~\cite{Nakajima2010} and Ba(Fe$_{0.9}$Co$_{0.1}$)$_2$As$_2$~\cite{Schilling2016}, by electrons of Ba$_{1-x}$K$_x$Fe$_2$As$_2$ ($x=0.19$, 0.26, 0.32, 0.34)~\cite{Cho2014} and Ba(Fe$_{1-x}$Ru$_x$)$_2$As$_2$ ($x=0.24$)~\cite{Prozorov2014}, by alpha-particles of NdFeAsO$_{0.7}$F$_{0.3}$~\cite{Tarantini2010}, and by heavy ions of Ba(Fe$_{1-x}M_x$)$_2$As$_2$ ($M$ = Co, Ni)~\cite{Kim2010} and Ba(Fe$_{1-x}$Co$_x$)$_2$As$_2$~\cite{Murphy2013}. In the latter case, optimally doped Ba$_{0.6}$K$_{0.4}$Fe$_2$As$_2$ stays apart because the effect on $T_c$ was not observed in it~\cite{Salovich2013}.

On the separate note, there are works on an electron irradiation of BaFe$_2$(As$_{1-x}$P$_x$)$_2$~\cite{Mizukami2014} and SrFe$_2$(As$_{1-x}$P$_x$)$_2$~\cite{Strehlow2014}, where apparently ``accidental'' nodes in the nodal $s_\pm$ state were removed with increasing disorder, as it was predicted earlier theoretically in Ref.~\cite{v_mishra_09}. However, this effect was not observed with the proton irradiation of BaFe$_2$(As$_{1-x}$P$_x$)$_2$~\cite{Smylie2016}.

Summarizing, there is a suppression of superconductivity in most cases. At the same time, the $T_c$ decrease rate is much lower than expected from the Abrikosov-Gor'kov expression.

\section{Dynamical properties of dirty superconductors \label{sec:experiments}}

One of the important features of the discussed $s_{\pm} \to s_{++}$ and $s_{++} \to s_{\pm}$ transitions is the gapless superconductivity, which has a direct relation to experiments on iron-based materials. In particular, since one of the gaps change sign it necessarily goes through zero that is corresponds to the gapless state. Therefore, the transition should manifest itself in various dynamical properties of the superconducting state. Those properties are, first, the density of states~(\ref{eq.N}) that can probed in tunneling experiments and ARPES, second, temperature dependence of the London penetration depth $\lambda_{L}$~(\ref{eq.lambda}), and, third, a frequency dependence of the optical conductivity $\sigma(\omega)$~(\ref{eq:sigma}). More subtle effect impurity scattering have on a dynamical spin susceptibility and, thus on $1/T_1T$ -- the NMR spin-lattice relaxation rate $1/T_1$ normalized by the temperature $T$. We discuss these points in details in following Sections~\ref{subsec:dos}--\ref{subsec:1T1}. We choose coupling constants to be the same as in Section~\ref{subsec:magnimpresults}.

\subsection{Density of states and penetration depth \label{subsec:dos}}

\begin{figure}
\centering
\includegraphics[width=0.6\textwidth]{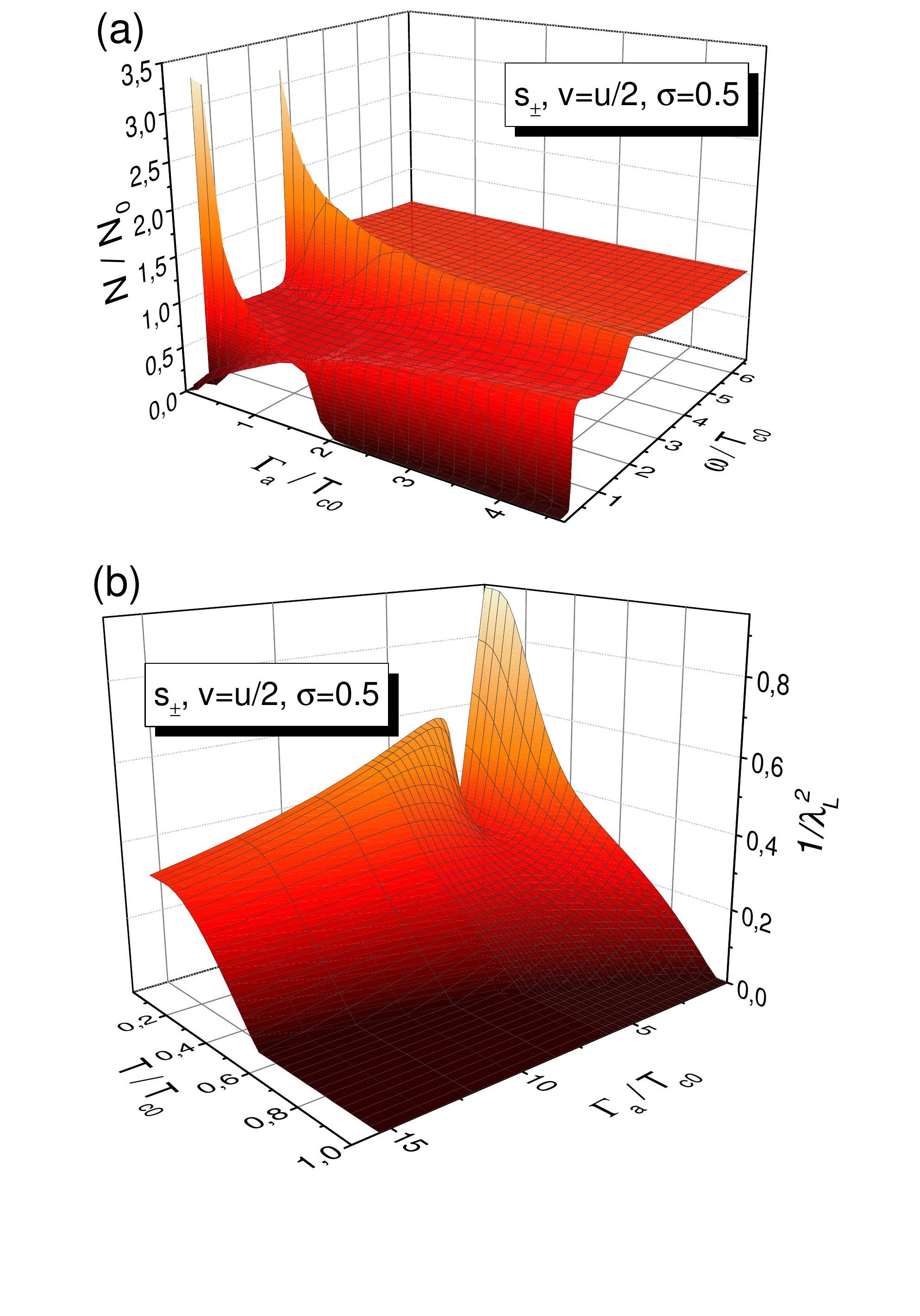}
\caption{(a) Density of states $N(\omega)$ normalized by the density of states in the normal state $N_0$ as a function of frequency $\omega$ and nonmagnetic impurity scattering rate $\GammaN_a$ in the $s_\pm$ superconductors with $\la \lambda \ra > 0$, $\sigmaN = 0.5$, $\etaN = 0.5$,
and $N = N_a + N_b$. (b) $1 / \lambda_L^2$ normalized by the total plasma frequency as a function of $\GammaN_a$ and temperature $T$.}
\label{fig:DOSlambda}
\end{figure}

At first, we discuss the transition from the $s_{\pm}$ to $s_{++}$ state induced by the nonmagnetic impurities. A total density of states $N(\omega)$ calculated using the expression~(\ref{eq.N}) for systems with $\la \lambda \ra > 0$ is shown in Figure~\ref{fig:DOSlambda}a. With the increasing impurity scattering rate, the smaller gap closes resulting in the finite residual density of state at zero frequency, $N(\omega = 0)$, and then reopens. Such behavior is reflected in the temperature dependence of the London penetration depth~(\ref{eq.lambda}), shown in Figure~\ref{fig:DOSlambda}b. here we present results correspondingly normalized by the plasma frequency $\omega_{P\alpha}$. Evidently, $1 / \lambda_L^2$ in a clean limit has an activation temperature dependence determined by the smaller gap, and then it transforms to the $T^2$ behavior in the gapless state, and finally shows a new activation regime in the $s_{++}$ state. In other words, at $\GammaN_a = 0$ we have a typical two-gap dependence~\cite{Golubov2002}. For larger values of the scattering rate when two gaps are almost equal, the temperature dependence of the penetration depth becomes as in a single-band superconductor.

\begin{figure}[ht]
\centering
\includegraphics[width=0.49\textwidth]{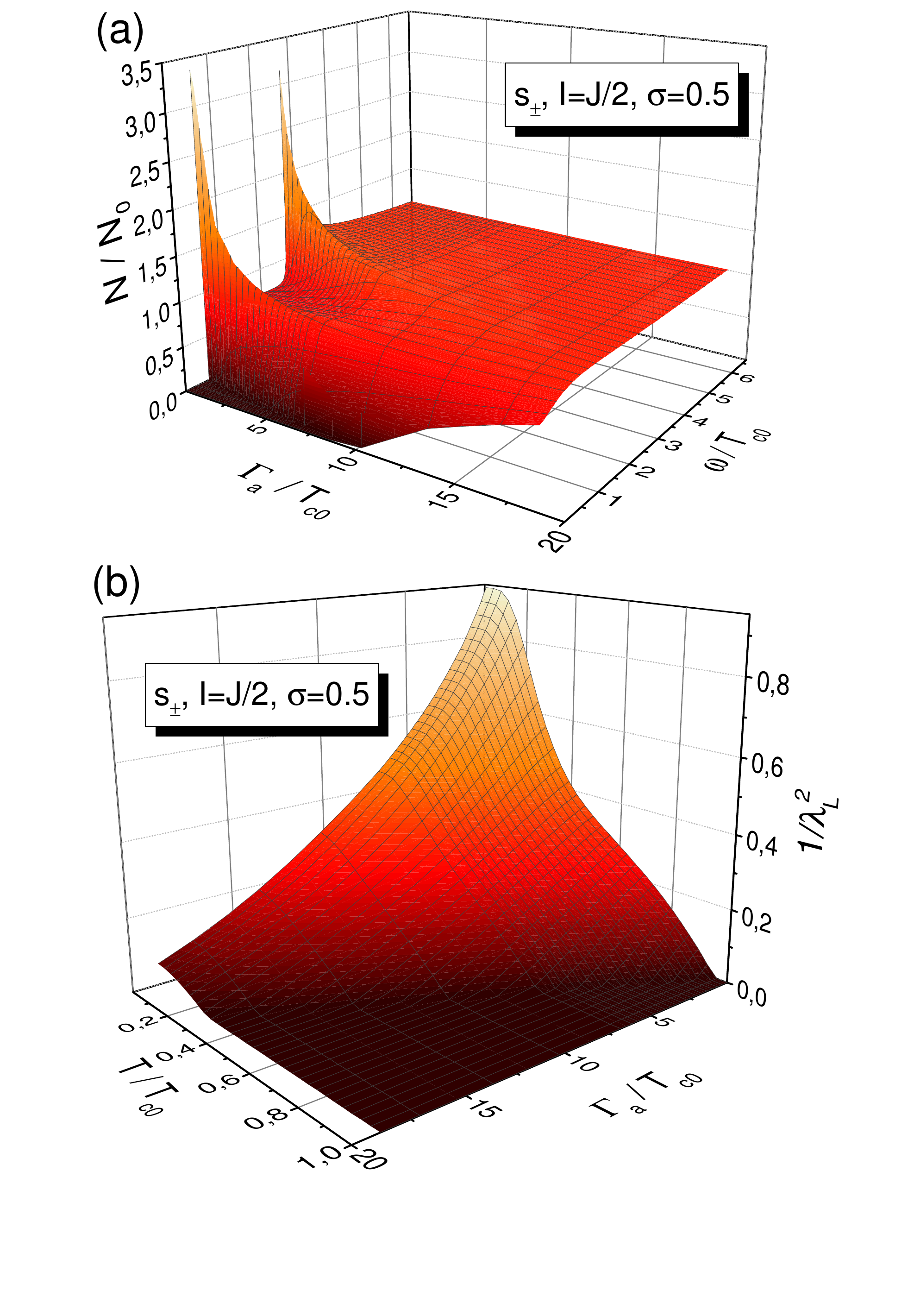}
\includegraphics[width=0.49\textwidth]{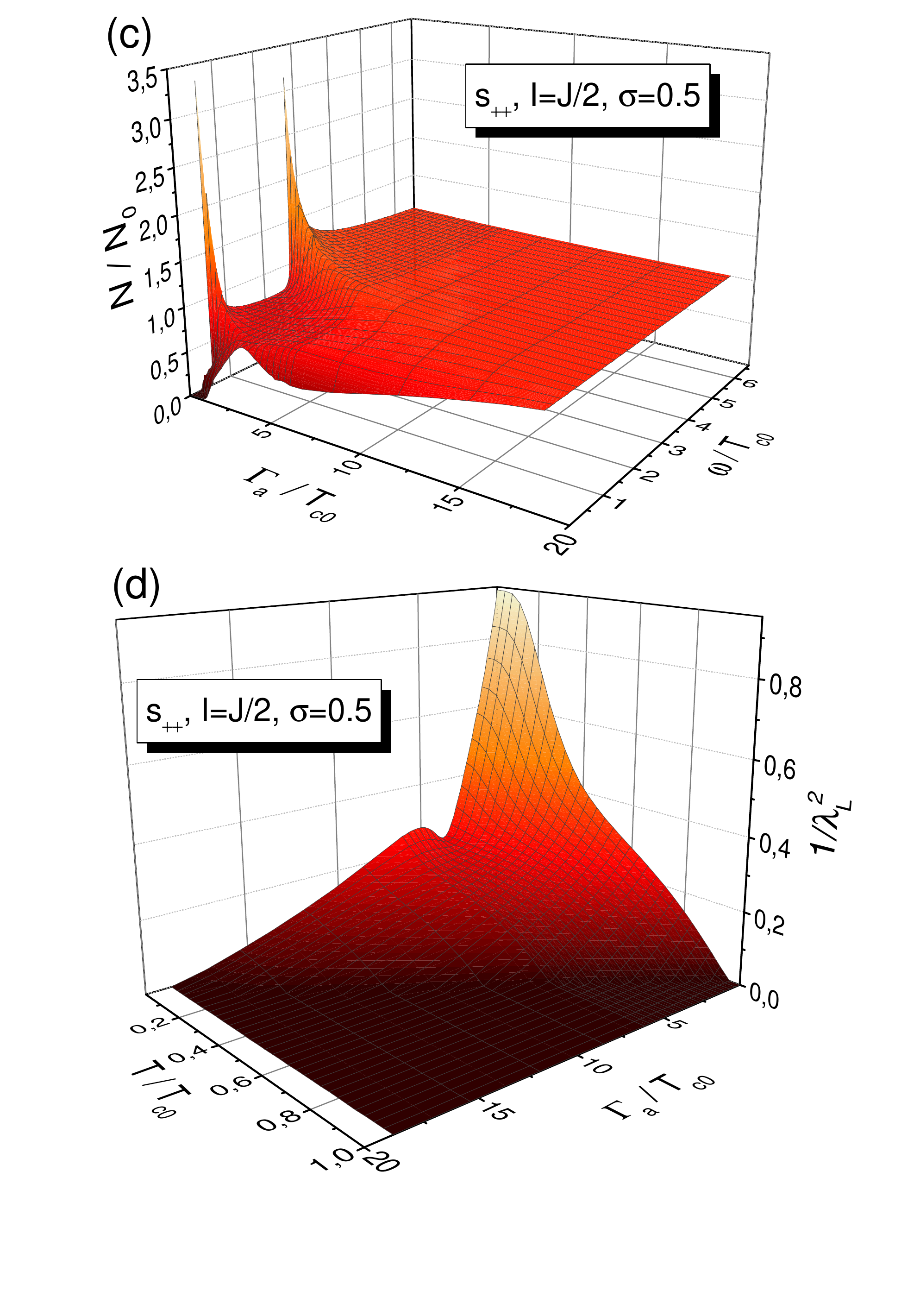}
\caption{(a,c) Density of states $N(\omega)$ normalized by the density of states in the normal state $N_0$ as a function of frequency $\omega$ and magnetic impurity scattering rate $\Gamma_a$. (b,d) Dependence of $1 / \lambda_{L}^2$ normalized by the total plasma frequency on $\Gamma_{a}$ and temperature $T$ for the $s_{\pm}$ superconductor (panels~a and b) and the $s_{++}$ superconductor (panels~c and d) with $\iM=\jM/2$ and $\sigma=0.5$. Note the transition from the $s_{++}$ state to the $s_\pm$ state at $\Gamma_a \sim 4 T_{c0}$
and a gapless region right after it.}
\label{fig:dosspmspp}
\end{figure}
The density of states $N(\omega)$ and the inverse square of the penetration depth $1 / \lambda_L^2$ in the case of magnetic impurities with $\iM = \jM/2$ and $\sigma = 0.5$ are shown in Figure~\ref{fig:dosspmspp} for the $s_\pm$ and $s_{++}$ superconductors. In the former case, we see the expected behavior with the gradually decreasing gaps. Gapless superconductivity with the residual $N(\omega=0)$ occurs for $\Gamma_a > 10 T_{c0}$
when $\mathrm{Re}\Delta_{\alpha}(\omega=0)$ vanishes, that is seen in Figure~\ref{fig:spmsppDeltaPade}b. As for the $s_{++}$ state, with increasing of the impurity scattering rate $\Gamma_a$, the smaller gap vanishes leading to the finite residual density of states $N(\omega=0)$. Then the gap reopens and $\Delta_{b n} \neq 0$ until $T_c$ reaches zero at $\Gamma_a \sim 20 T_{c0}$.
Still, the superconductivity stays gapless with the finite $N(0)$ because $\mathrm{Re}\Delta_{\alpha}(\omega=0) \to 0$ as seen in Figure~\ref{fig:spmsppDeltaPade}d. Penetration depth in the clean limit shows the activation behavior determined by the smaller gap. In the case of the $s_{++}$ state, penetration depth becomes proportional to $T^{2}$ in the gapless regime causing the significant reduction of it near $\Gamma_a \sim 4 T_{c0}$
(Figure~\ref{fig:dosspmspp}d), and then penetration depth shows activation temperature dependence in the $s_{\pm}$ state after the transition.

\subsection{ARPES \label{subsec:ARPES}}

Presence of the gapless state should definitely manifest itself in ARPES spectra. Total measured photoemission current intensity $I(\k,\omega)$ in the sudden approximation is equal to
\beq
 I(\k,\omega) = \sum_{\alpha} |M_{\alpha}(\k, \omega)|^2 f(\omega) A_{\alpha}(\k,\omega),
\eeq
where $M(\k,\omega)$ is the matrix element of one-electron dipole interaction depending on the initial and final states of the photoelectron, photon energy and its polarization, $f(\omega)$ is the Fermi function, and $A_{\alpha}(\k,\omega)$ is the spectral function. Latter can be expressed through the analytical continuation of Green's function~(\ref{eq.Gfull}) to the real frequencies as
\beq
 A_{\alpha}(\k, \omega) = - \frac{1}{2\pi} \mathrm{Tr} \left[ \mathrm{Im}\hat{G}^{\alpha \alpha}(\k,\omega) \hat{\tau}_{0} \right] = -\frac{1}{\pi} \mathrm{Im}\frac{\tilde{\omega}_{\alpha}(\omega)}{\tilde{\omega}_{\alpha}^2(\omega) - \xi_{\k \alpha}^2 - \tilde{\phi}_{\alpha}^2(\omega)}.
\label{eq:Akomega}
\eeq
Note, here we have a ``bare'' dispersion $\xi_{\k \alpha}$ because the self-energy in our approximation does not depends on momentum and makes corresponding contributions neither to the dispersion nor to the chemical potential shift.

Contribution of the electron-boson interaction to the self-energy $\Sigma_{0 \alpha}(\k,\omega)$ vanishes in the weak coupling approximation~\cite{Efremov2013}. Therefore, in the model with the isotropic self-energy we have $\Sigma_{0 \alpha}(\omega) \to 0$ and $\Sigma_{2 \alpha}(\omega) \to \Delta_{\alpha}(\omega)$. Then the spectral function takes the following form~\cite{Efremov2013},
\beq
 A_{\alpha}(\k,\omega) = \frac{1}{\pi} \mathrm{Im}\frac{\omega}{D_\alpha} \left[ 1 + \ii \sum\limits_{\beta} \frac{\GammaN_{\alpha \beta}}{\sqrt{\omega^2 - \Delta_{\beta}^2(\omega)}} \right],
\eeq
where
\beq
 D_\alpha = \xi_{\k \alpha}^2 + \left[ \Delta_\alpha^2(\omega) - \omega^2 \right] \left[ 1 + \ii \sum\limits_{\beta} \frac{\GammaN_{\alpha \beta}}{\sqrt{\omega^2 - \Delta_{\beta}^2(\omega)}} \right]^2.
\eeq

To be more specific, let $\Delta_b$ be the smaller gap. Two cases should be distinguished -- one with the full total gap and the gapless one. In the first case, $A_\alpha(\k,\omega)$ vanishes at energies below $\Delta_{\alpha}$. On the other hand, the spectral function of the same band in the gapless regime with $\Delta_b \to 0$ behaves in the same way as it would do in the normal state,
\beq
 A_{b}(\k,\omega) = \frac{1}{\pi} \mathrm{Im} \frac{\omega \left[1 + \ii \sum\limits_{\beta} \GammaN_{b \beta} / |\omega| \right]}{\xi_{\k b}^2 - \omega^2 \left( 1 + \ii \sum\limits_{\beta} \GammaN_{b \beta}/|\omega| \right)^2}.
\eeq

\begin{figure}[ht]
\centering
(a)\includegraphics[height=2.1in]{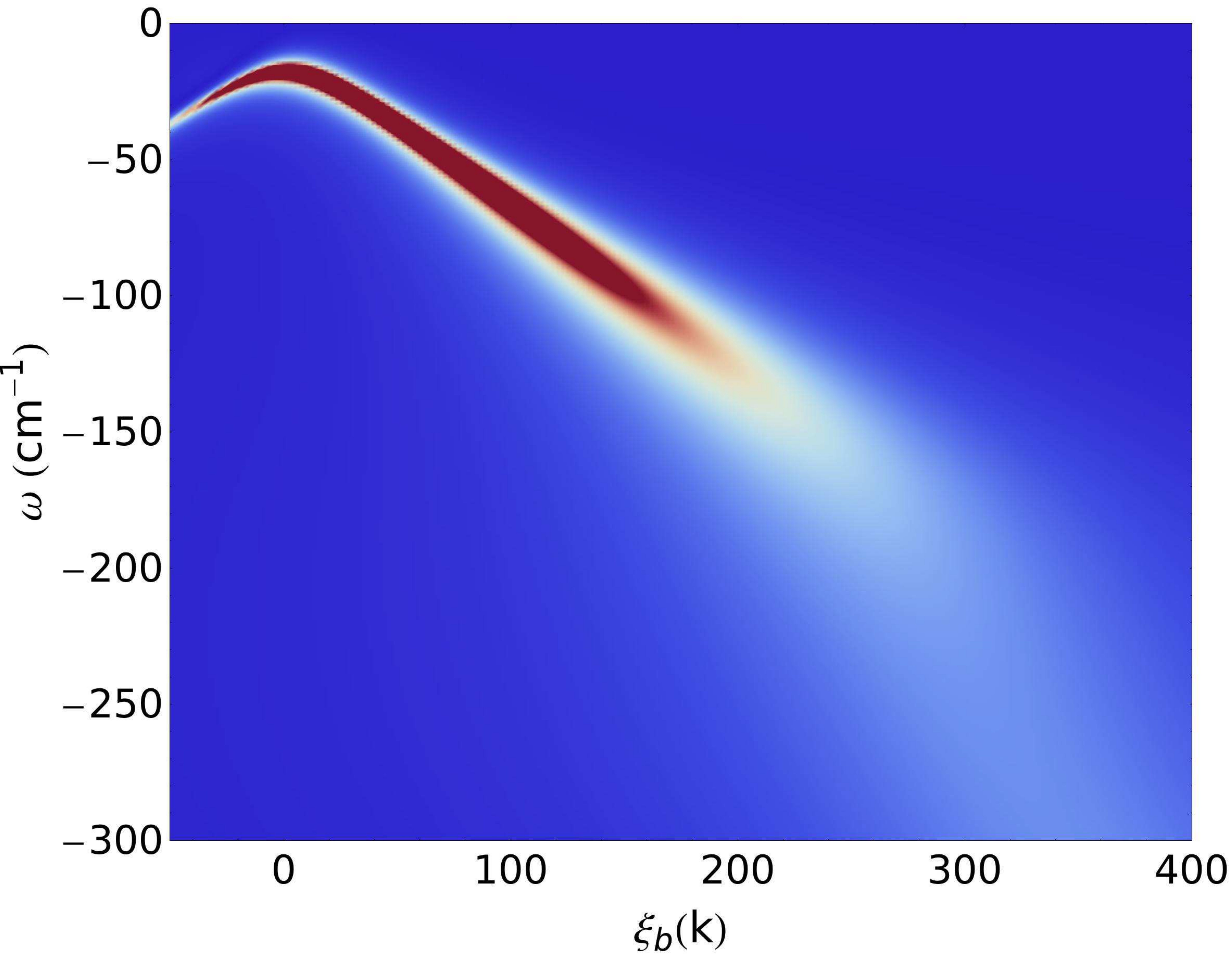}
(b)\includegraphics[height=2.1in]{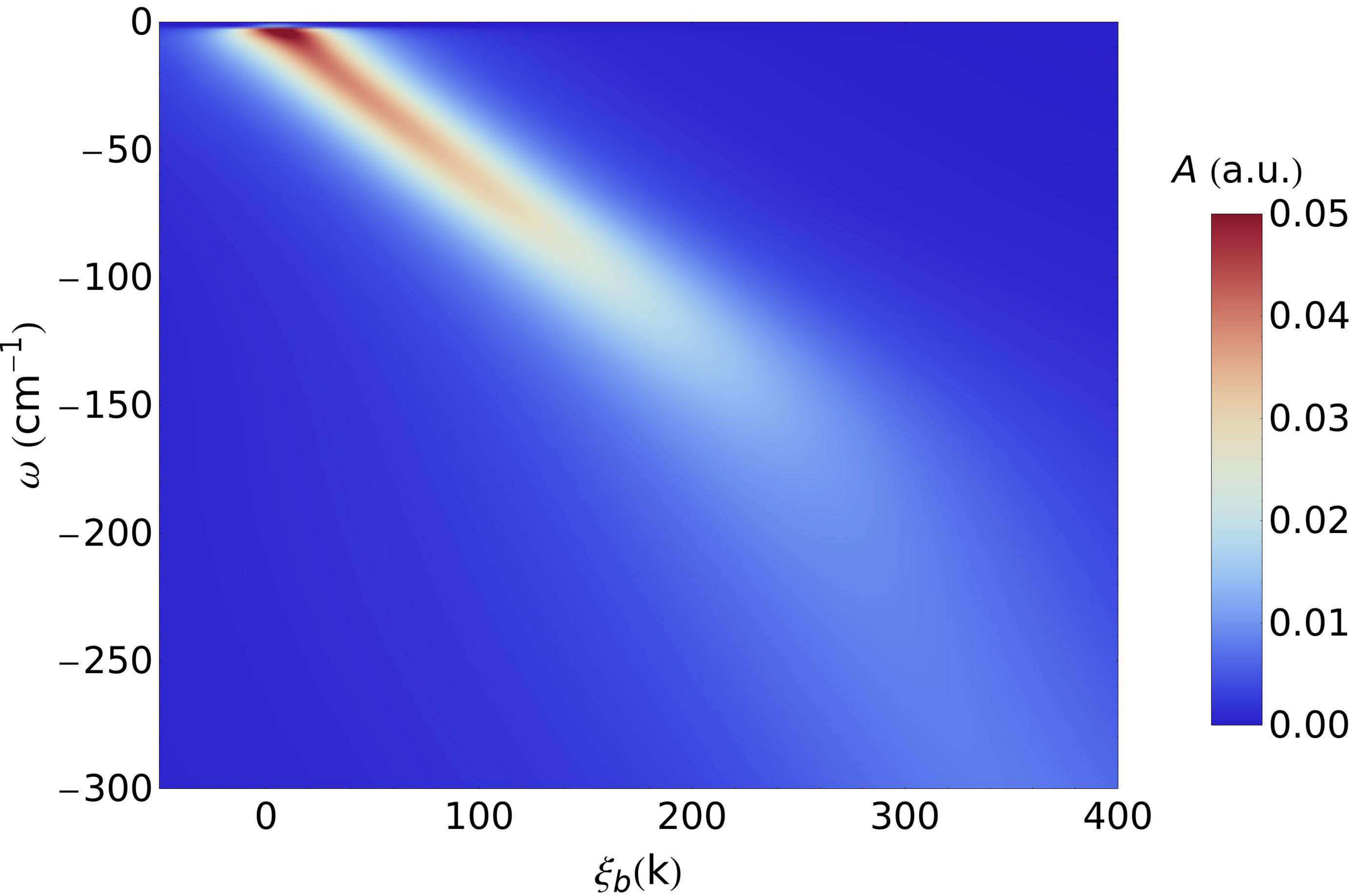}
\caption{(Color online.) Spectral function $A_{b}(\k,\omega)$ of the band $b$ with the smaller gap in the clean limit (a) and in the gapless regime (b) with the finite nonmagnetic impurity scattering rate $\GammaN_a = 1.33 T_{c0}$~\cite{Efremov2013}.
}
\label{fig:Akomega}
\end{figure}

The fermionic spectral function $A_{b}(\k,\omega)$ for the band $b$ calculated via expression~(\ref{eq:Akomega}) is shown in Figure~\ref{fig:Akomega}. For the sake of argument, we show calculation for $\abs{\Delta_b} < \abs{\Delta_a}$ and the scattering on nonmagnetic impurities with $\etaN = 0.5$ and $\sigmaN = 0.5$, although present results retain for the case of magnetic impurities. In the clean limit (Figure~\ref{fig:Akomega}a), behavior of $A_{b}(\k,\omega)$ at small $\omega$ and $\xi_{\k b}$ determined by the presence of the superconducting gap in the spectrum of excitations. On the other hand, at the $s_{\pm}$ to $s_{++}$ disorder-induced transition, $A_{b}(\k,\omega)$ shows the absence of the gap (Figure~\ref{fig:Akomega}b). With further increase of the scattering rate $\GammaN_a$, when the transition already happened, gap in the spectrum of $b$-band reappears. Therefore, ARPES measurements in the superconducting state at different impurity concentrations would help to detect the disorder-induced transition.

\subsection{Optical conductivity \label{subsec:optics}}

\begin{figure}[ht]
\centering
\includegraphics[width=0.7\textwidth]{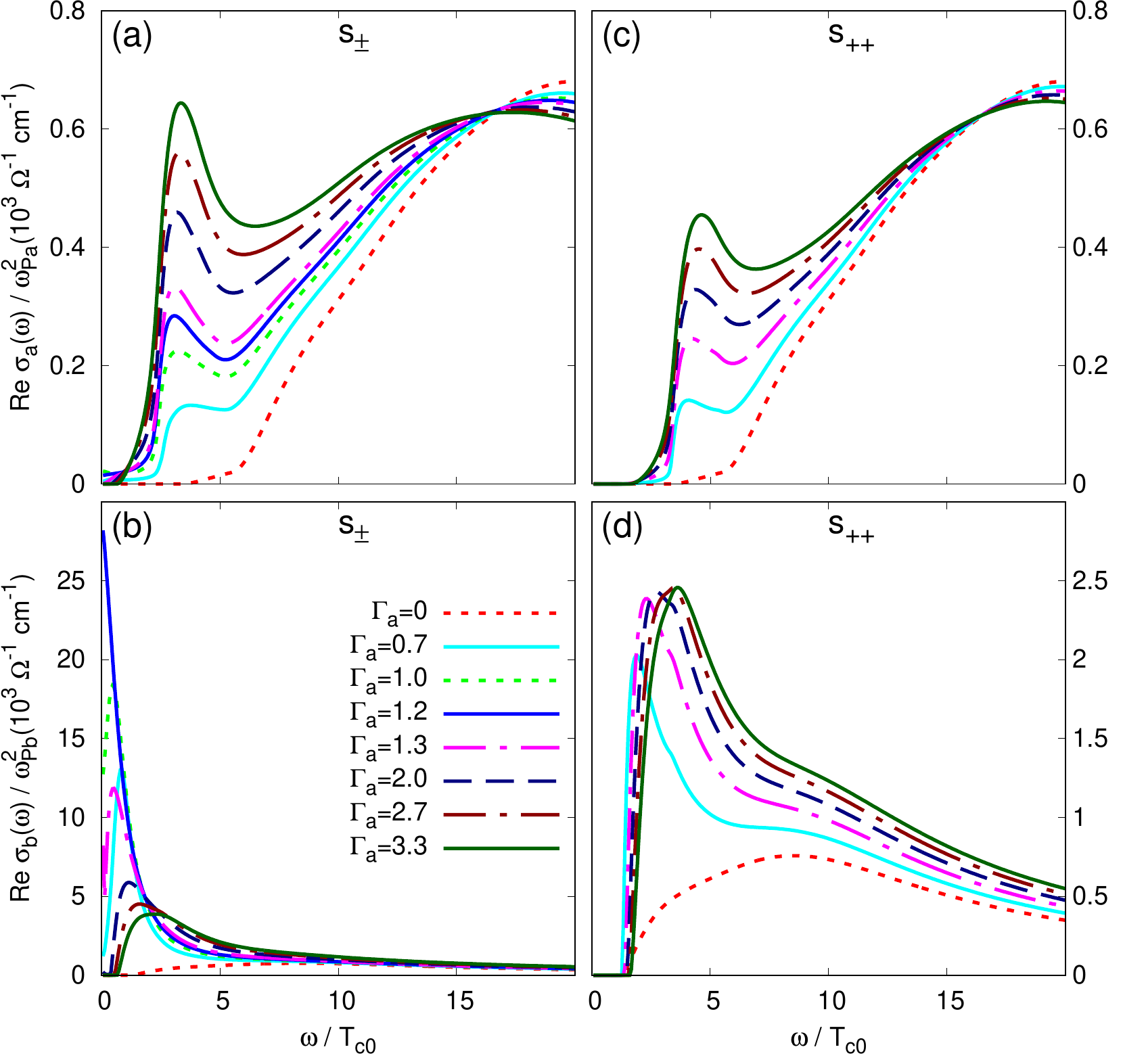}
\caption{(Color online.) Frequency dependence of real part of the optical conductivity, $\mathrm{Re}\sigma_\alpha(\omega)$, of bands $\alpha = a$ (a,c) and $\alpha = b$ (b,d) for the $s_\pm$ (panels~a and b) and $s_{++}$ (panels~c and d) superconductors at different nonmagnetic impurity scattering rates $\GammaN_a$ (in units of $T_{c0}$). Temperature is $T = 0.03 T_{c0}$,
$\sigmaN = 0.5$, and $\etaN = 0.5$~\cite{Efremov2013}.}
\label{fig:Re_sigma}
\end{figure}

Considering nonmagnetic impurities as an example, here we show how the optical conductivity changes its behavior with increasing impurity scattering rate and, in particular, near the transition between the $s_\pm$ and $s_{++}$ states. Figure~\ref{fig:Re_sigma} shows the optical conductivity $\mathrm{Re}\sigma(\omega) = \sum_{\alpha} \mathrm{Re}\sigma_\alpha(\omega)$ calculated as the solution of equations~(\ref{eq:sigma}) and~(\ref{eq:Pi_omega}) at different rates of disorder with $\etaN = 0.5$ and $\sigmaN = 0.5$. Due to the presence of the superconducting gap, in the clean limit we have $\mathrm{Re}\sigma_\alpha(\omega) = 0$ at small frequencies $\omega < 2\Delta_\alpha$. With the increase of the impurity scattering rate in the $s_\pm$ state, as opposed to the $s_{++}$ superconductor, the range of zero value of $\mathrm{Re}\sigma_b(\omega)$ for the band $b$ diminishes and the peak above $2\Delta_b$ becomes narrower. This is surely due to the decrease of the gap $\Delta_b$ with approaching the $s_\pm \to s_{++}$ transition (Figure~\ref{fig:spmsppDeltaPadeNonmag}b). It is clearly seen in Figure~\ref{fig:Re_sigma}b, that near the $s_\pm \to s_{++}$ transition at $\GammaN_a \sim 1.2 T_{c0}$
the Drude peak appears. This peak is typical for the normal metal. The reason is vanishing of the gap in the $b$-band, i.e., the gapless superconductivity regime at the transition. With further increase of $\GammaN_a$, the optical conductivity regains the form of the full gap superconductor though with a smaller value of gap than it was initially.

Described behavior differs significantly from the behavior of the $s_{++}$ superconductor shown in Figures~\ref{fig:Re_sigma}c and~d, where gaps converge in the limit of the infinite impurity scattering rate.

\begin{figure}[ht]
\centering
\includegraphics[width=0.6\textwidth]{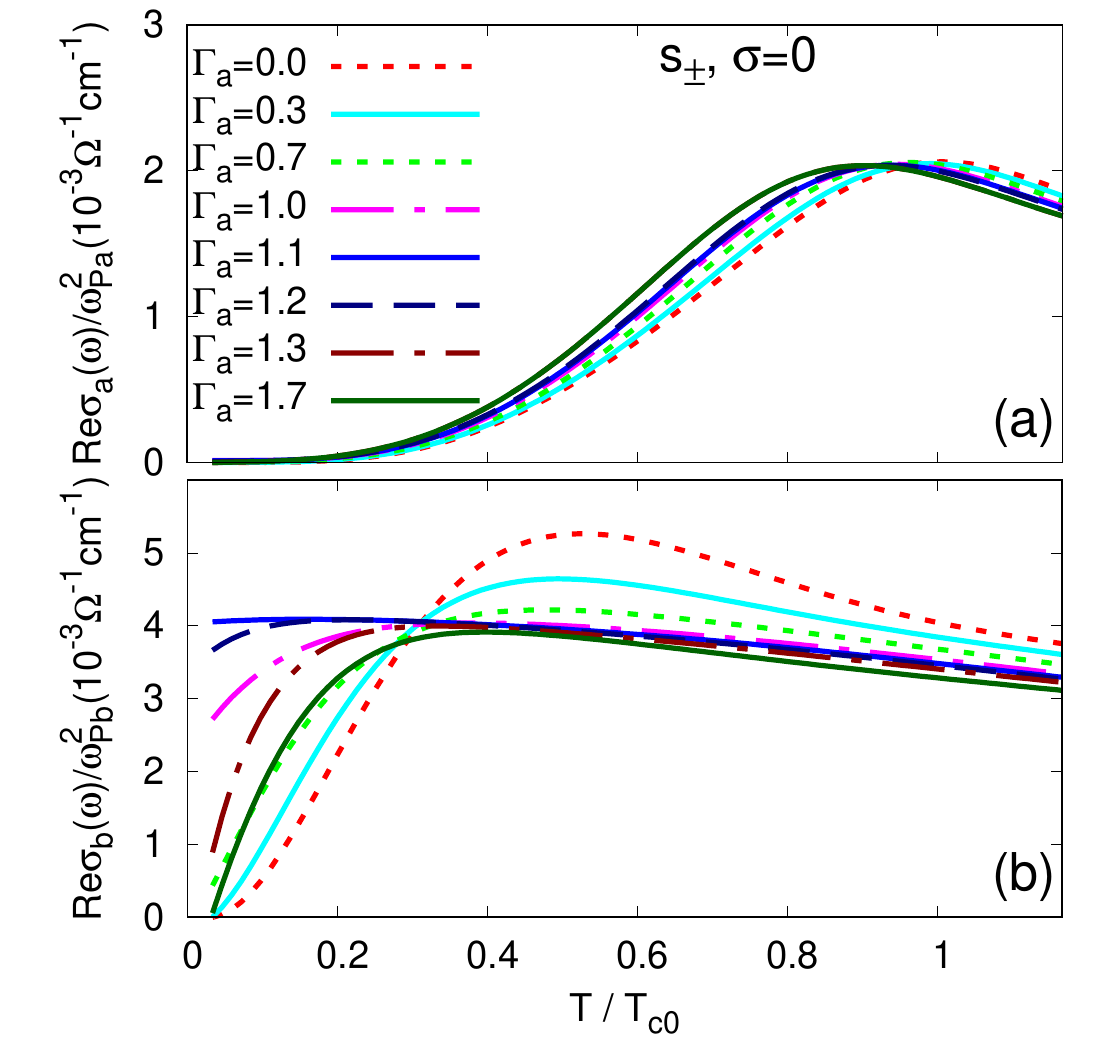}
\caption{(Color online.) Temperature dependence of the real part of the polarization operator at a fixed frequency, $\mathrm{Re}\sigma_\alpha(\omega=5\mathrm{cm}^{-1})$, for the two band $\alpha = a$ (a) and $\alpha = b$ (b) in the $s_{\pm}$ state at different nonmagnetic impurity scattering rates $\GammaN_a$ (in units of $T_{c0}$). Interband scattering was chosen to be equal to $\Gamma_{intra}=6.7 T_{c0}$
for the sake of presentation~\cite{Efremov2013}.}
\label{fig:Re_sigma_T}
\end{figure}

Temperature dependence of the optical conductivity $\mathrm{Re}\sigma(\omega)$ at the small frequency for the $s_\pm$ superconductor is shown in Figure~\ref{fig:Re_sigma_T}. Evidently, the low-temperature contribution to the optical conductivity of the band $b$ increases with increasing scattering rate $\GammaN_a$ before the transition to the $s_{++}$ state at $\GammaN_a \sim 1.1 T_{c0}$,
and then the contribution decreases.

\begin{figure}[ht]
\centering
\includegraphics[width=0.7\textwidth]{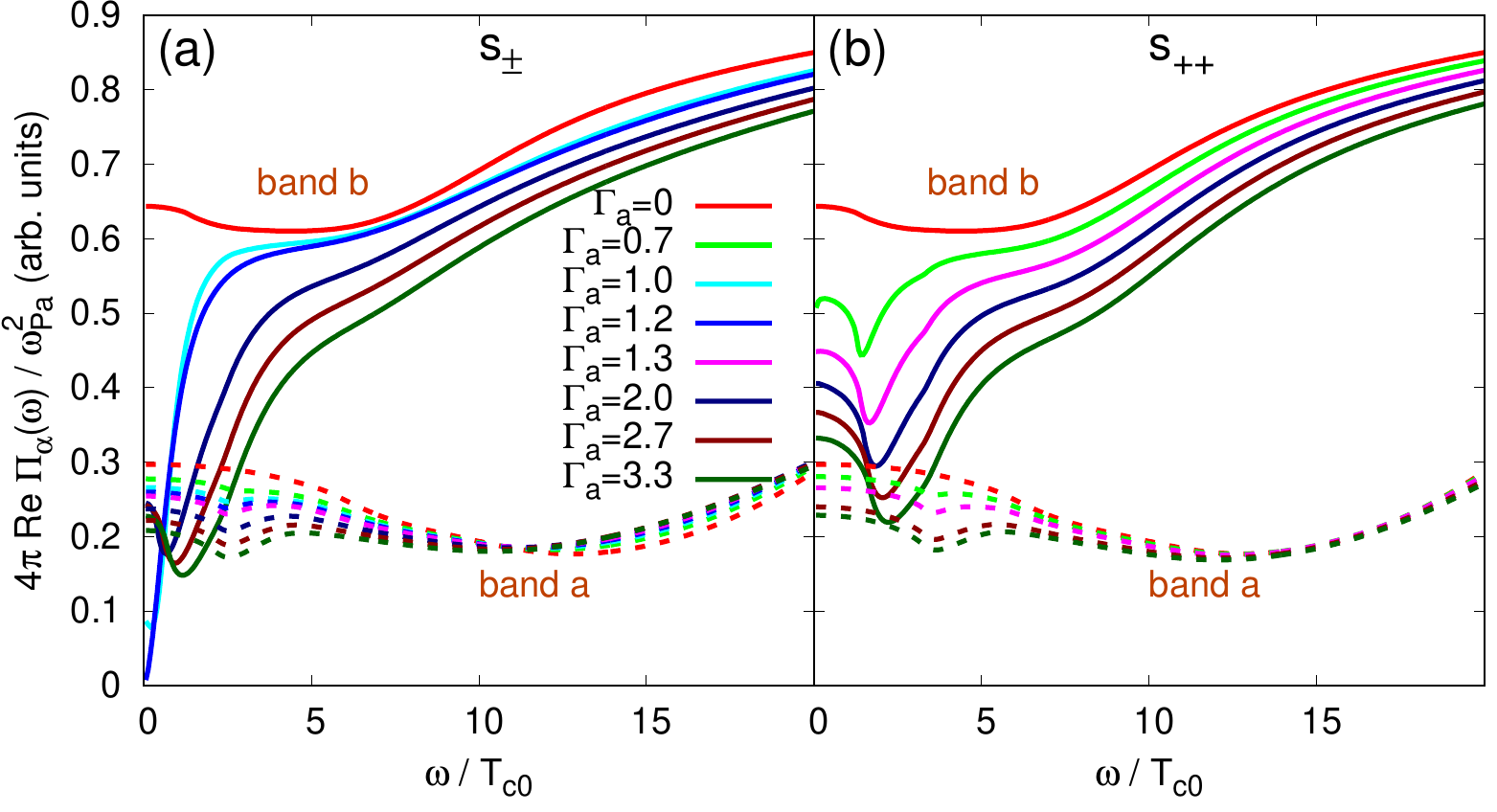}
\caption{(Color online.) Real part of the polarization operator, $\mathrm{Re}\Pi_\alpha(\omega)$, for the $s_{\pm}$ (a) and $s_{++}$ (b) superconductors at different nonmagnetic impurity scattering rates $\GammaN_a$ (in units of $T_{c0}$). Temperature is $T = 0.03 T_{c0}$,
$\sigmaN = 0.5$, and $\etaN = 0.5$~\cite{Efremov2013}.}
\label{fig:Re_Pi}
\end{figure}

Imaginary part of the optical conductivity, $\mathrm{Im}\sigma(\omega)$, is proportional to the real part of the polarization operator $\Pi(\omega)$, as seen from its definition~(\ref{eq:sigma}). Frequency dependence of $\Pi(\omega)$ for the $s_{\pm}$ and $s_{++}$ states in the presence of the impurity scattering is shown in Figure~\ref{fig:Re_Pi}. There is a dip at frequency $\omega = 2 \Delta_{\alpha}(\omega)$ in the case of the $s_{++}$ superconductor. This agrees with results for single-band superconductors~\cite{Marsiglio1996}. In the $s_{\pm}$ state, interesting features are observed for the band $b$: first, the location of the dip is a nonmonotonic function of the scattering rate, and, second, the dip disappears in the gapless regime near the $s_{\pm} \to s_{++}$ transition.

\begin{figure}[ht]
\centering
\includegraphics[width=\textwidth]{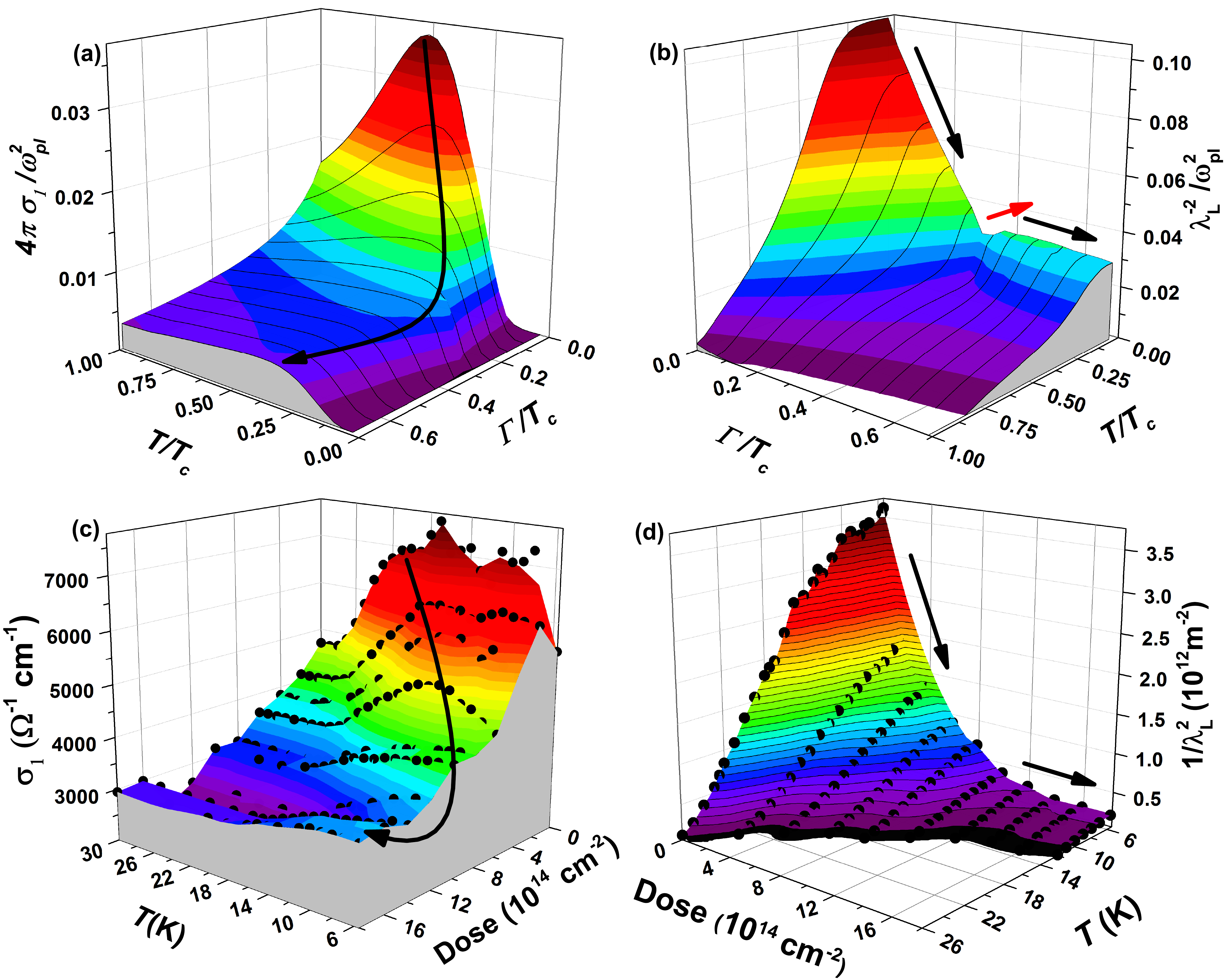}
\caption{(Color online.) Calculated (a,b) and experimental (c,d) dependencies of the real part of the optical conductivity $\sigma_1$ (panels~a and c) and $\lambda_L^{-2}$ (panels~b and d) on temperature $T$ and level of disorder. Latter is related to the nonmagnetic impurity scattering rate $\GammaN_a$ in the theoretical calculations and in the experiment on superconducting Ba(Fe$_{0.9}$Co$_{0.1}$)$_2$As$_2$, it is the proton irradiation dose (number of protons per 1~cm$^2$)~\cite{Schilling2016}.}
\label{fig:sigmaexp}
\end{figure}

Comparison of theoretical and experimental dependencies of the optical conductivity at THz frequencies and the London penetration depth on the dose of proton irradiation~\cite{Schilling2016} is shown in Figure~\ref{fig:sigmaexp}. It is interesting to follow the behavior of the coherence peak in the real part of the optical conductivity $\sigma_1(T, \omega \to 0)$. The peak is analogous to the Hebel-Slichter peak discussed in Section~\ref{subsec:1T1}. With the increase of the irradiation dose, the peak disappears and then reappears again (Figure~\ref{fig:sigmaexp}c). Such a behavior is a signature of the gradual close of the smaller gap and its later reopening. It is exactly the process taking place at the $s_{\pm} \to s_{++}$ transition. General trend of the penetration depth behavior is the same in the theory and in the experiment, as evident from the comparison of Figures~\ref{fig:sigmaexp}b and d. In the experiment, however, it was not possible to ``catch'' the region of the $s_{\pm} \to s_{++}$ transition itself. Latter is marked in Figure~\ref{fig:sigmaexp}b by the red arrow.

\subsection{NMR spin-lattice relaxation rate $1/T_1T$ \label{subsec:1T1}}

In addition to the Knight shift, which allows one to distinguish between singlet and triplet pairing, NMR can probe the spin-lattice relaxation rate $1/T_1$. Since we are going to discuss Fe-based materials, later we imply NMR at the iron nuclei. Effect of the nuclei formfactors is not very important here compared to e.g. cuprates. It is confirmed by a good agreement between $1/T_1T$ data on different nucleus ($^{57}$Fe, $^{75}$As, $^{59}$Co, and $^{139}$La) in 122 and 1111 systems~\cite{m_yashima_09,Nakai2008,Ning2008,Nakai2010}. It is also experimentally claimed~\cite{MatanoBKFA} that the hyperfine coupling $A_{hf}(\q)$ is most probably does not depend on the wave vector $\q$.

The spin-lattice relaxation rate determined by the spin susceptibility integrated over the Brillouin zone,
\begin{equation}
 \frac{1}{T_1 T} \propto \lim_{\omega \to 0} \sum_{\q} \frac{\mathrm{Im}\chi(\q,\omega)}{\omega}.
\end{equation}
As in the case with the spin resonance~\cite{KorshunovEreminResonance2008,t_maier_08b,HirschfeldKorshunov2011,Inosov2016}, $1/T_1$ carries information about the underlying gap symmetry and structure. For example, an isotropic $s$-wave state is characterized by a Hebel-Slichter peak just below $T_c$ and an exponential low-$T$ temperature dependence. It is well-known that $d$-wave superconductors exhibit weak or absent peak and demonstrate $T_1^{-1} \sim T^3$ behavior for $T \ll T_c$.

\begin{figure}
\centering
\includegraphics[width=0.85\textwidth]{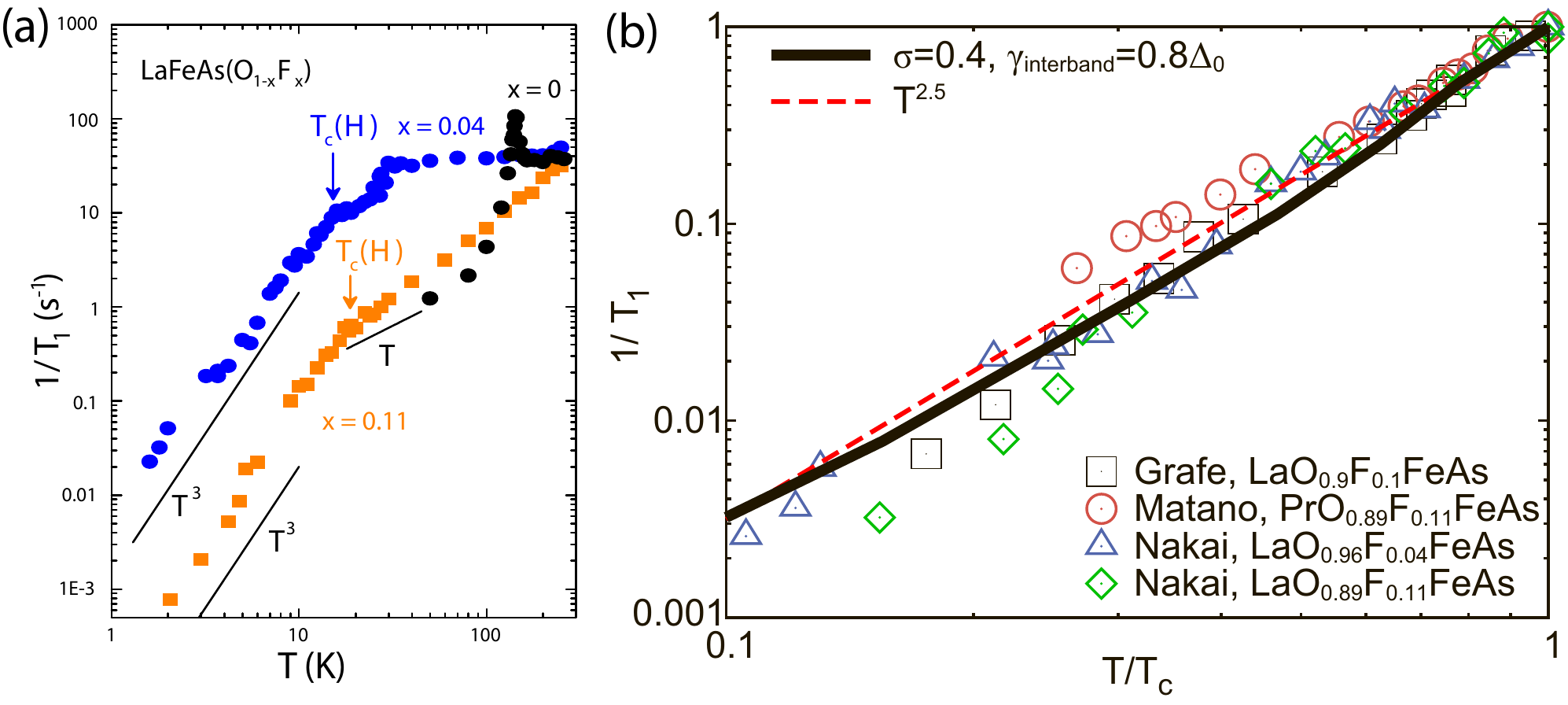}
\caption{Temperature dependence of $1/T_1$ in iron-based materials.
(a) Experimental results for the 1111 system from Ref.~\cite{Nakai2008}.
(b) Log-log plot summarizing experimental data from several groups~\cite{k_matano_08,h_grafe_08,Nakai2008}, theoretical curve for the $s_\pm$ superconductor with intermediate strength of impurity scattering ($0 \leq \sigma \leq 1$) and pairbreaking parameter $\gamma_\mathrm{interband} = 0.4 \Delta_0$, and $T^{2.5}$ curve to demonstrate a power-law dependence~\cite{ParkerKorshunov2008}.}
\label{fig:nmr}
\end{figure}

In the case of iron-based materials, the situation is somewhat more complicated. Typical data are shown in Figure~\ref{fig:nmr}a. Apparently, there is no peak below $T_c$ and the temperature dependence does not follow the same simple power or exponential law. However, simple arguments can enable us to understand the main features found in experiments.

In case of a weakly coupled clean two-band superconductor below $T_c$, assuming that the main contribution to $\mathrm{Im}\chi(\q,\omega)$ comes from interband interactions, we have the following expression for the inverse NMR spin-lattice relaxation rate:
\begin{equation}
 \frac{1}{T_1 T} \propto \sum_{\k \k'} \left[1 + \frac{\Delta_{\k} \Delta_{\k'}}{E_{\k} E_{\k'}} \right] \left( -\frac{\partial f(E_{\k})}{\partial E_{\k}} \right) \delta\left( E_{\k} - E_{\k'} \right),
 \label{eq:1T1}
\end{equation}
where $E_{\k}$ is the quasiparticle energy in the superconducting state, $\k$ and $\k' = \k + \q$ lie on hole and electron Fermi sheets, respectively. Thus $\q$ is the vector connecting hole and electron sheets. The equation above follows from the expression for the ``bare'' susceptibility $\chi_0(\q,\omega)$ at zero temperature and for a vanishing frequency. It is written in the special way to emphasize the role of coherence factors for the dominating interband processes. The coherence factor in square brackets in~(\ref{eq:1T1}) gives rise to an important distinction between different symmetries of the gap. They play a similar role in the formation of the spin resonance peak in inelastic neutron scattering~\cite{KorshunovEreminResonance2008}. In the NMR $1/T_1$ coherence factors, the internal sign is different from that in coherence factors entering the spin susceptibility related to neutron scattering. For the isotropic $s_{++}$-state with $\Delta_{\k} = \Delta_{\k'} = \Delta$ we have
\begin{equation}
 \frac{1}{T_1} \propto \int\limits_{\Delta(T)}^{\infty} dE \frac{E^2 + \Delta^2}{E^2 - \Delta^2} ~ \mathrm{sech}^2\left(\frac{E}{2T}\right).
\end{equation}
The denominator gives rise to a peak for temperatures $T \lesssim T_c$ near $T_c$, which is the Hebel-Slichter peak. As pointed out earlier~\cite{mazin_08}, it is suppressed for the $s_{\pm}$=state. Indeed, if $\Delta_{\k} = -\Delta_{\k'} = \Delta$,
\begin{equation}
 \frac{1}{T_1} \propto \int\limits_{\Delta(T)}^{\infty} dE \frac{E^2 - \Delta^2}{E^2 - \Delta^2} ~ \mathrm{sech}^2\left(\frac{E}{2T}\right) = \int\limits_{\Delta(T)}^{\infty} dE ~ \mathrm{sech}^2\left(\frac{E}{2T}\right),
\label{eq:1T1spm}
\end{equation}
which is a monotonically decreasing function with decreasing temperature for $T < T_c$.  The same can be shown for a more general $s_\pm$-state with $|\Delta_{\k}| \neq |\Delta_{\k'}|$~\cite{ParkerKorshunov2008}.

It is well known that pair-breaking impurity scattering dramatically increases the subgap density of states just below $T_c$, and even a weak magnetic scattering can broaden and eliminate the Hebel-Slichter peak in conventional superconductors. In the case of the sign-changing gap, the same effect is present due to nonmagnetic interband scattering~\cite{Golubov1997}. Since the Hebel-Slichter peak is not present in iron-based materials even in a clean case, see equation~(\ref{eq:1T1spm}), the pair-breaking effect is more subtle: it changes an exponential behavior for $T < T_c$ to a more power-law like one. If the impurity-induced bound state lies at the Fermi level, the relaxation rate acquires a low-temperature linear in temperature Korringa-like term over a range of temperatures corresponding to the impurity bandwidth~\cite{PJHconsequences}.

Qualitative arguments suggest that neither pure Born nor pure unitary limits with a simple isotropic $s_\pm$-state are well suited for explaining the observed $1/T_1$ behavior: the former leads to an exponential behavior at low temperatures in a relatively clean system, the latter to Korringa behavior. Various data on the 1111 systems appeared to be between these two limits~\cite{k_matano_08,h_grafe_08,Nakai2008}, see Figure~\ref{fig:nmr}b. Results of the $1/T_1$ calculation for the simple $s_{\pm}$ state is also shown there~\cite{ParkerKorshunov2008}. We observe that the $s_{\pm}$ state result exhibits no coherence peak and as opposed to the Born and unitary limits, intermediate case with $\sigma$ not equal to 0 or 1 is capable of reproducing the experimental behavior of $1/T_1$~\cite{ParkerKorshunov2008,Chubukov2008,Senga2008,Senga2009,Bang2009}. These results, taken alone, should not be taken as an evidence for an isotropic $s_\pm$ state, since the strong gap anisotropy is probably present in some of these systems, and will also lead to a higher density of quasiparticles contributing at intermediate temperatures.

Regarding other systems, data obtained on BaFe$_2$(As$_{1-x}$P$_x$)$_2$ shows a linear in temperature term in $1/T_1$ for an optimally doped sample, crossing over to something roughly approximating $\sim T^3$ above $\sim 0.1 T_c$~\cite{Nakai2010,NakaiPdoped2010}, consistent with reports of nodes in this material from other probes. In Ba$_{0.68}$K$_{0.32}$Fe$_2$As$_2$, $1/T_1$ shows an exponential decrease below $T \approx 0.45 T_c$ consistent with a full $s_\pm$ gap~\cite{LiBKFA}. Finally, consistent with other measurements, NMR in the LiFeAs system also shows a full gap~\cite{Li111}.

\section{Conclusions \label{sec:conclusion}}

Summarizing, the disorder in multiband systems may have an unexpected impact on the superconductivity. It is especially important in cases of MgB$_2$, iron pnictides and iron chalcogenides as well as for the approximate treatment of the $d$-wave superconductors like cuprates where parts of the Fermi surface with different signs of the gap to some extent can be considered as different bands. As an example here we considered the problem of scattering on nonmagnetic and magnetic impurities in two-band superconductors with $s_{++}$ and $s_\pm$ order parameter types.

For the nonmagnetic disorder, $T_c$ is more stable against impurity scattering compared to the trivial generalization of the Abrikosov-Gor'kov theory~(\ref{eq.AG}). The exact rate of the $T_c$ suppression depends on the relation between intra- and interband coupling constants. Depending on the sign of the averaged coupling constant, $\la \lambda \ra$, originating from interelectron interactions, $s_\pm$ superconductors can be divided into two types. First type belongs to the largely discussed in the literature case with $\la \lambda \ra < 0$ where the superconductivity primarily determined by interband scattering. In such systems, $T_c$ suppressed with increasing disorder and vanishes the critical value of the scattering rate. Second type of the $s_\pm$ state has $\la \lambda \ra > 0$ and is characterized by the finite value of $T_c$ for the increasing disorder while signs of order parameters for different bands become equal. Latter imply the transition from the $s_\pm$ to the $s_{++}$ state. The case of $\la \lambda \ra > 0$ corresponds to the sizeable intraband attraction. In spite of this attraction, even a weak interband repulsion leads to the opposite phases of order parameters in two different bands, i.e., $s_\pm$ state. Note, the strong intraband attraction in the two-band model considered here may be a consequence of a large intraband pairing amplitude, as well as a result of the downfolding procedure of the realistic multiband model onto the two-band model. Large intraband pairing amplitude could be a result of the electron-phonon interaction and/or orbital fluctuations. Downfolding procedure of the multiband model with the small intraband attractive pairing potential and the large band-asymmetric interband repulsion also may result in the \textit{effective} string intraband attraction in the two-band model. Such a case was considered in Ref.~\cite{Charnukha2011} for the initial four-band model.

Regarding the magnetic disorder, generally, the superconducting state is destroyed with the increase of scattering on magnetic impurities. There are, however, few special cases with the absence of a complete $T_c$ suppression, in which it saturates for the large impurity scattering rate. Such situation occurs in the unitary limit and in the $s_{++}$ and $s_\pm$ states with the interband-only impurity potential. Remarkably, in this case, the $s_{++}$ superconductor is robust against magnetic disorder not by itself, but due to the transition to the $s_\pm$ state insensitive to impurity scattering. Latter is in line with the qualitative arguments on the analogy between the magnetic impurities in the $s_\pm$ state and nonmagnetic impurities in the isotropic $s_{++}$ state: that is, equations have the same form and since the $s_{++}$ state is robust against nonmagnetic impurities, then $s_\pm$ state is robust against the magnetic ones. Note, the finite intraband component of the scattering potential leads to the complete $T_c$ suppression, though with a slower rate compared to the Abrikosov-Gor'kov theory result for the single-band superconductors. In this case, even the $s_{++} \to s_\pm$ transition can't save the superconductivity from a collapse.

\begin{figure}
\centering
\includegraphics[width=0.7\textwidth]{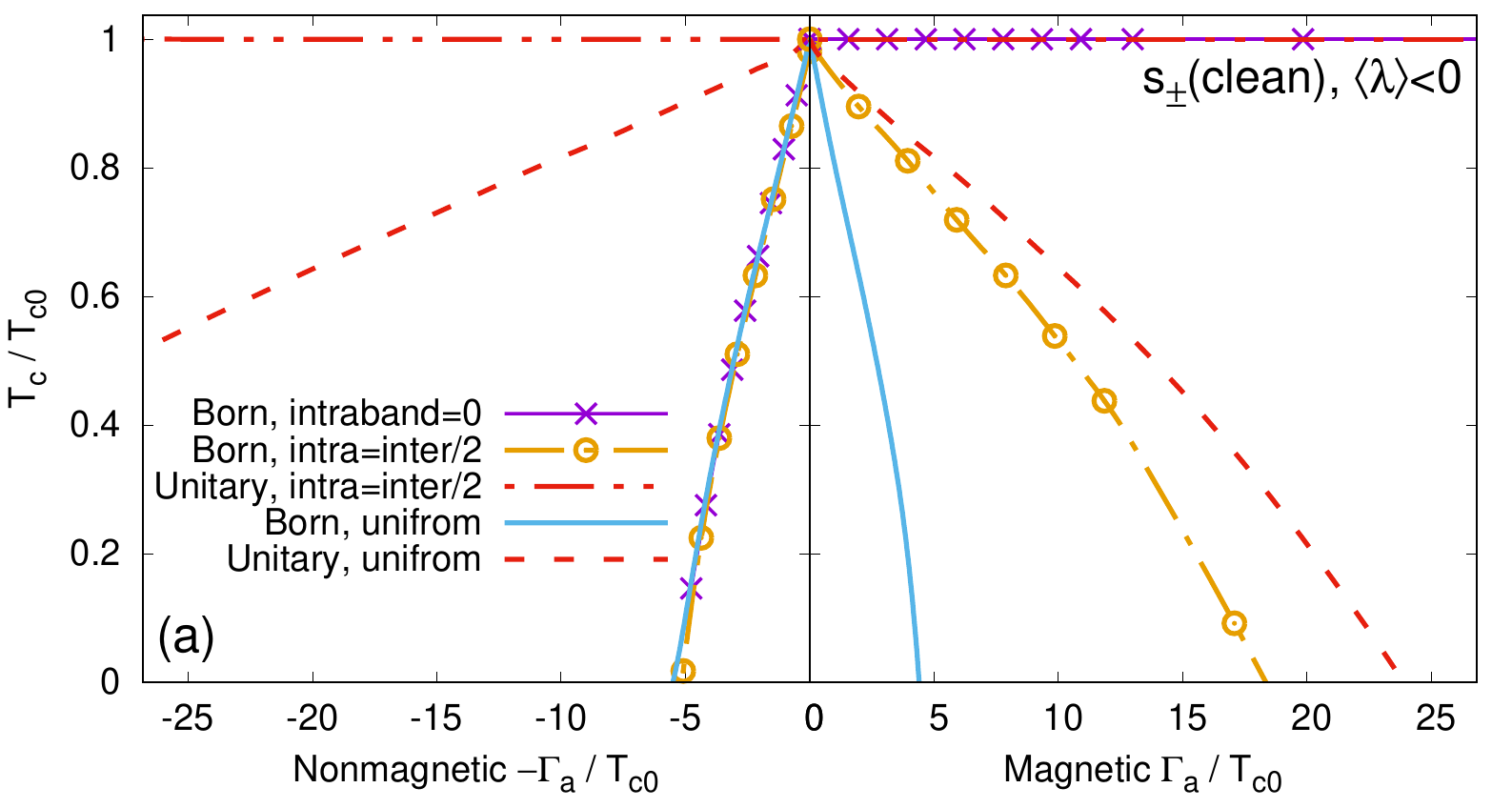}
\includegraphics[width=0.7\textwidth]{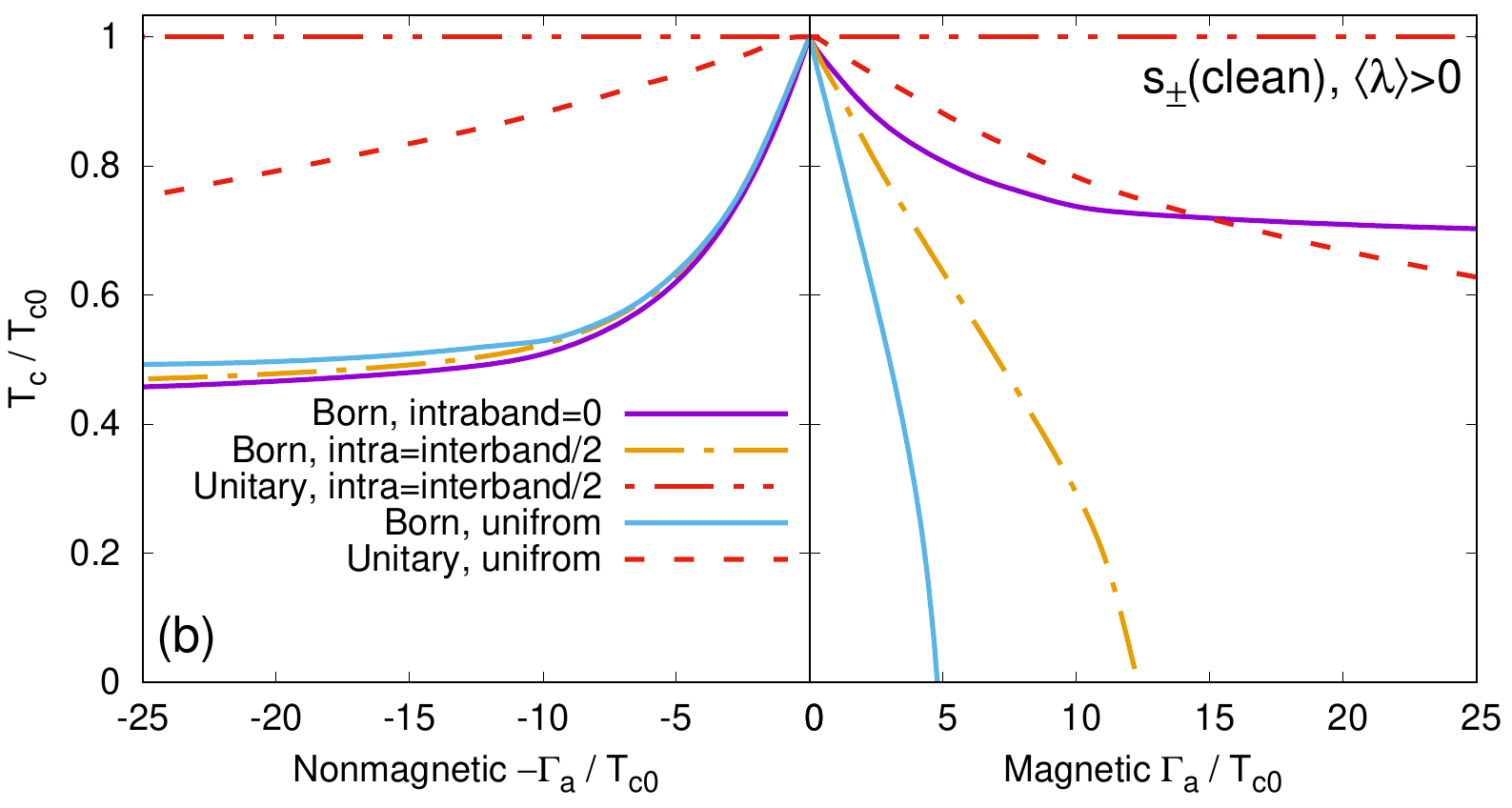}
\includegraphics[width=0.7\textwidth]{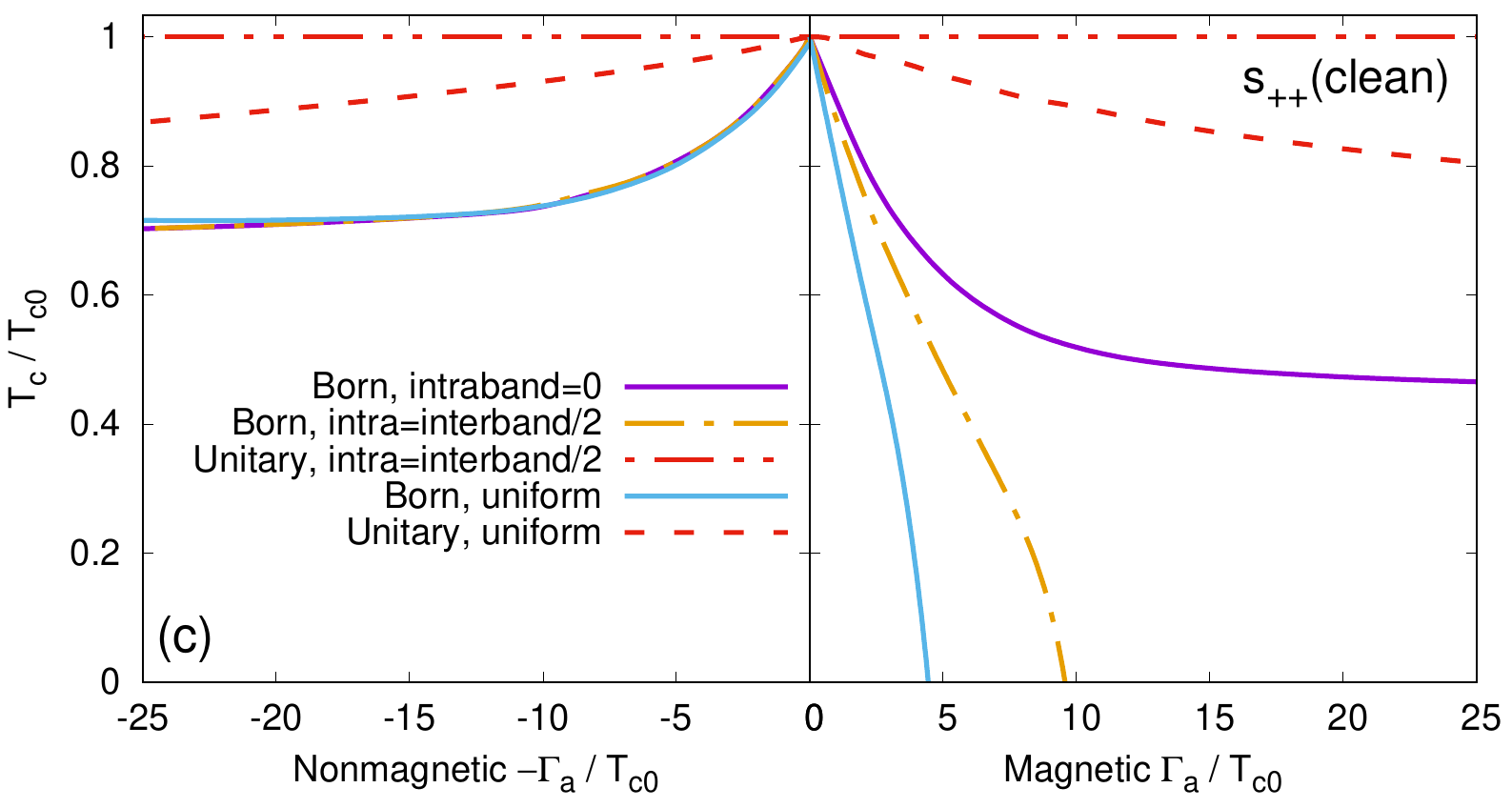}
\caption{Dependence of $T_c$ on nonmagnetic (on the left) and magnetic (on the right) impurity scattering rates $\Gamma_a$.
(a) $s_\pm$ superconductor with the negative averaged coupling constant.
(b) Superconductor with the positive averaged coupling constant that has the $s_\pm$ gap symmetry in the clean case ($\Gamma_a=0$): for the nonmagnetic impurities, a transition from the $s_\pm$ to the $s_{++}$ state occurs at particular values of $\Gamma_a$.
(c) Superconductor with the $s_{++}$ gap symmetry in the clean case: for the magnetic impurities, a transition to the $s_\pm$ state occurs at particular values of $\Gamma_a$. Various curves represent Born ($\sigma = 0$) and unitary ($\sigma = 1$) limits obtained for the different relation between intra- and interband impurity potentials, that is, absence of the intraband potential ($\eta = 0$), interband potential is twice the intraband one ($\eta = 0.5$), and the uniform impurity potential ($\eta = 1$).
}
\label{fig:spmsppTcMagnNonmag}
\end{figure}

Summary plots of the $T_c$ dependence on the impurity scattering rate $\Gamma_a$ for the $s_\pm$ and $s_{++}$ superconductors is shown in Figure~\ref{fig:spmsppTcMagnNonmag}. For both the $s_\pm$ state with the positive averaged coupling constant and the $s_{++}$ state, the nonmagnetic disorder does not completely destroys superconductivity. The reason for this in the case of $s_\pm$ superconductor is, however, different from that for the $s_{++}$ state: there is a transition to the $s_{++}$ state in the former case. Similarly, the reason for the absence of complete $T_c$ suppression by the interband ($\eta = 0$) magnetic disorder for the $s_{++}$ state is the transition to the $s_\pm$ state which is robust against the scattering on magnetic impurities due to the Anderson's theorem analog. In the unitary limit, results for all cases are the same except for the uniform impurity potential ($\eta = 1$) with the fall of $T_c$. Note, while the exact form of the impurity potential is not known, it is hard to imagine that its intra- and interband parts would be equal in different cases, e.g., adding Zn or proton irradiation. This means that practically the case with $\eta = 1$ is highly improbable.

The general conclusion on the transition between states with different gap structures is the following: if the system has two interactions in the clean limit, dominating (1) and subdominating (2), and the interaction (2) may induce superconducting state that is robust against impurities then the system will transform to this state as soon as the order due to the interaction (1) is destroyed by a disorder. That is, $s_\pm$ state occurs due to the interband interaction while $s_{++}$ state originates mainly from the intraband one. And if initially there was an $s_\pm$ state without the intraband component then the interband nonmagnetic impurities would completely destroy this state and suppress $T_c$ to zero. If there is an $s_\pm$ state with the intraband component (even the small one) of the interaction then the same impurities would suppress $s_\pm$ state but due to the residual intraband interaction the $s_{++}$ state that can't be destroyed by the nonmagnetic impurities would stabilize. It is the $s_\pm \to s_{++}$ transition. For the magnetic impurities, the situation is reversed. If initially there was an $s_{++}$ state without the interband component of the superconducting interaction then the interband magnetic impurities would destroy it. However, the presence of even a small interband interaction would result in the $s_\pm$ state after the $s_{++}$ state is suppressed by the magnetic disorder. It would be the $s_{++} \to s_\pm$ transition.

Since these transitions go through the gapless regime, they should manifest itself in thermodynamical and transport properties. For example, they can be observed in optical and tunneling experiments, as well as in a photoemission spectroscopy and tunneling conductivity in iron-based superconductors and other multiband systems. That is, since the smaller gap vanishes near the transition, ARPES should demonstrate the gapless spectra and optical conductivity would reveal the ``recovery'' of the Drude-like frequency dependence of $\mathrm{Re}\sigma(\omega)$.

\subsection{Acknowledgements}

We would like to thank A. Bianconi, A.A. Golubov, B.P. Gorshunov, I.M. Eremin, M.V. Eremin, D.V. Efremov, B. Keimer, I.I. Mazin, R. Prozorov, M.V. Sadovskii, D.J. Scalapino, M. Tanatar, P.J. Hirschfeld, and A.V. Chubukov for useful discussions. M.M. Korshunov is grateful to Max-Planck-Institut f\"{u}r Festk\"{o}rperforschung and B. Keimer for the hospitality during his visit. We acknowledge partial support by RFBR (grant 16-02-00098) and Government Support of the Leading Scientific Schools of the Russian Federation (NSh-7559.2016.2).

\bibliography{mmkbibl1}

\begin{thebibliography}{179}
\expandafter\ifx\csname natexlab\endcsname\relax\def\natexlab#1{#1}\fi
\expandafter\ifx\csname bibnamefont\endcsname\relax
  \def\bibnamefont#1{#1}\fi
\expandafter\ifx\csname bibfnamefont\endcsname\relax
  \def\bibfnamefont#1{#1}\fi
\expandafter\ifx\csname citenamefont\endcsname\relax
  \def\citenamefont#1{#1}\fi
\expandafter\ifx\csname url\endcsname\relax
  \def\url#1{\texttt{#1}}\fi
\expandafter\ifx\csname urlprefix\endcsname\relax\def\urlprefix{URL }\fi
\providecommand{\bibinfo}[2]{#2}
\providecommand{\eprint}[2][]{\url{#2}}

\bibitem[{\citenamefont{Bardeen et~al.}(1957)\citenamefont{Bardeen, Cooper, and
  Schrieffer}}]{bcs}
\bibinfo{author}{\bibfnamefont{J.}~\bibnamefont{Bardeen}},
  \bibinfo{author}{\bibfnamefont{L.~N.} \bibnamefont{Cooper}},
  \bibnamefont{and} \bibinfo{author}{\bibfnamefont{J.~R.}
  \bibnamefont{Schrieffer}}, \bibinfo{journal}{Phys. Rev.}
  \textbf{\bibinfo{volume}{108}}, \bibinfo{pages}{1175} (\bibinfo{year}{1957}),
  \urlprefix\url{http://link.aps.org/doi/10.1103/PhysRev.108.1175}.

\bibitem[{\citenamefont{Abrikosov et~al.}(1963)\citenamefont{Abrikosov, Gorkov,
  and Dzyaloshinsky}}]{AGD1962eng}
\bibinfo{author}{\bibfnamefont{A.~A.} \bibnamefont{Abrikosov}},
  \bibinfo{author}{\bibfnamefont{L.~P.} \bibnamefont{Gorkov}},
  \bibnamefont{and} \bibinfo{author}{\bibfnamefont{I.~E.}
  \bibnamefont{Dzyaloshinsky}}, \emph{\bibinfo{title}{Methods of Quantum Field
  Theory in Stattistical Physics}} (\bibinfo{publisher}{Prentice-Hall},
  \bibinfo{address}{Englewood Clifs, N.J.}, \bibinfo{year}{1963}).

\bibitem[{\citenamefont{Volovik and Gor'kov}(1984)}]{Volovik1984eng}
\bibinfo{author}{\bibfnamefont{G.E.} \bibnamefont{Volovik}} \bibnamefont{and}
  \bibinfo{author}{\bibfnamefont{L.P.} \bibnamefont{Gor'kov}},
  \bibinfo{journal}{JETP Lett.} \textbf{\bibinfo{volume}{39}},
  \bibinfo{pages}{674} (\bibinfo{year}{1984}),
  \urlprefix\url{http://www.jetpletters.ac.ru/ps/1304/article_19706.shtml}.

\bibitem[{\citenamefont{Volovik and Gor'kov}(1985)}]{Volovik1985}
\bibinfo{author}{\bibfnamefont{G.E.} \bibnamefont{Volovik}} \bibnamefont{and}
  \bibinfo{author}{\bibfnamefont{L.P.} \bibnamefont{Gor'kov}},
  \bibinfo{journal}{JETP} \textbf{\bibinfo{volume}{61}}, \bibinfo{pages}{843}
  (\bibinfo{year}{1985}),
  \urlprefix\url{http://www.jetp.ac.ru/cgi-bin/r/index/e/61/4/p843?a=list}.

\bibitem[{\citenamefont{Mineev and Samokhin}(1999)}]{MineevSamokhin1998eng}
\bibinfo{author}{\bibfnamefont{V.~P.} \bibnamefont{Mineev}} \bibnamefont{and}
  \bibinfo{author}{\bibfnamefont{K.~V.} \bibnamefont{Samokhin}},
  \emph{\bibinfo{title}{Introduction to Unconventional Superconductivity}}
  (\bibinfo{publisher}{Gordon and Breach Scie. Publ.},
  \bibinfo{address}{Amsterdam}, \bibinfo{year}{1999}).

\bibitem[{\citenamefont{Riseborough et~al.}(2008)\citenamefont{Riseborough,
  Schmiedeshoff, and Smith}}]{Riseborough1998}
\bibinfo{author}{\bibfnamefont{P.~S.} \bibnamefont{Riseborough}},
  \bibinfo{author}{\bibfnamefont{G.~M.} \bibnamefont{Schmiedeshoff}},
  \bibnamefont{and} \bibinfo{author}{\bibfnamefont{J.~L.} \bibnamefont{Smith}},
  in \emph{\bibinfo{booktitle}{Superconductivity Volume 2: Novel
  Superconductors}}, edited by \bibinfo{editor}{\bibfnamefont{K.H.}
  \bibnamefont{Bennemann}} \bibnamefont{and}
  \bibinfo{editor}{\bibfnamefont{J.B.} \bibnamefont{Ketterson}}
  (\bibinfo{publisher}{Springer-Verlag}, \bibinfo{address}{Berlin Heidelberg},
  \bibinfo{year}{2008}), pp. \bibinfo{pages}{1031--1154},
  \urlprefix\url{http://dx.doi.org/10.1007/978-3-540-73253-2}.

\bibitem[{\citenamefont{Vonsovsky et~al.}(1982)\citenamefont{Vonsovsky,
  Izyumov, and Kurmaev}}]{VonsovskiiIzyumovKurmaev1982}
\bibinfo{author}{\bibfnamefont{S.~V.} \bibnamefont{Vonsovsky}},
  \bibinfo{author}{\bibfnamefont{Yu.~A.} \bibnamefont{Izyumov}},
  \bibnamefont{and} \bibinfo{author}{\bibfnamefont{E.~Z.}
  \bibnamefont{Kurmaev}}, \emph{\bibinfo{title}{Superconductivity of Transition
  Metals, Their Alloys and Compounds}} (\bibinfo{publisher}{Springer-Verlag},
  \bibinfo{address}{Berlin}, \bibinfo{year}{1982}).

\bibitem[{\citenamefont{Bednorz and M{\"u}ller}(1986)}]{bednorz-muller}
\bibinfo{author}{\bibfnamefont{J.~G.} \bibnamefont{Bednorz}} \bibnamefont{and}
  \bibinfo{author}{\bibfnamefont{K.~A.} \bibnamefont{M{\"u}ller}},
  \bibinfo{journal}{Zeitschrift f{\"u}r Physik B Condensed Matter}
  \textbf{\bibinfo{volume}{64}}, \bibinfo{pages}{189} (\bibinfo{year}{1986}),
  ISSN \bibinfo{issn}{1431-584X},
  \urlprefix\url{http://dx.doi.org/10.1007/BF01303701}.

\bibitem[{\citenamefont{Sigrist and Ueda}(1991)}]{SigristUeda}
\bibinfo{author}{\bibfnamefont{M.}~\bibnamefont{Sigrist}} \bibnamefont{and}
  \bibinfo{author}{\bibfnamefont{K.}~\bibnamefont{Ueda}},
  \bibinfo{journal}{Rev. Mod. Phys.} \textbf{\bibinfo{volume}{63}},
  \bibinfo{pages}{239} (\bibinfo{year}{1991}),
  \urlprefix\url{http://link.aps.org/doi/10.1103/RevModPhys.63.239}.

\bibitem[{\citenamefont{Hebard et~al.}(1991)\citenamefont{Hebard, Rosseinsky,
  Haddon, Murphy, Glarum, Palstra, Ramirez, and Kortan}}]{Hebard1991}
\bibinfo{author}{\bibfnamefont{A.~F.} \bibnamefont{Hebard}},
  \bibinfo{author}{\bibfnamefont{M.~J.} \bibnamefont{Rosseinsky}},
  \bibinfo{author}{\bibfnamefont{R.~C.} \bibnamefont{Haddon}},
  \bibinfo{author}{\bibfnamefont{D.~W.} \bibnamefont{Murphy}},
  \bibinfo{author}{\bibfnamefont{S.~H.} \bibnamefont{Glarum}},
  \bibinfo{author}{\bibfnamefont{T.~T.~M.} \bibnamefont{Palstra}},
  \bibinfo{author}{\bibfnamefont{A.~P.} \bibnamefont{Ramirez}},
  \bibnamefont{and} \bibinfo{author}{\bibfnamefont{A.~R.}
  \bibnamefont{Kortan}}, \bibinfo{journal}{Nature}
  \textbf{\bibinfo{volume}{350}}, \bibinfo{pages}{600} (\bibinfo{year}{1991}),
  \urlprefix\url{http://dx.doi.org/10.1038/350600a0}.

\bibitem[{\citenamefont{Nagamatsu et~al.}(2001)\citenamefont{Nagamatsu,
  Nakagawa, Muranaka, Zenitani, and Akimitsu}}]{Nagamatsu2001}
\bibinfo{author}{\bibfnamefont{J.}~\bibnamefont{Nagamatsu}},
  \bibinfo{author}{\bibfnamefont{N.}~\bibnamefont{Nakagawa}},
  \bibinfo{author}{\bibfnamefont{T.}~\bibnamefont{Muranaka}},
  \bibinfo{author}{\bibfnamefont{Y.}~\bibnamefont{Zenitani}}, \bibnamefont{and}
  \bibinfo{author}{\bibfnamefont{J.}~\bibnamefont{Akimitsu}},
  \bibinfo{journal}{Nature} \textbf{\bibinfo{volume}{410}}, \bibinfo{pages}{63}
  (\bibinfo{year}{2001}), ISSN \bibinfo{issn}{0028-0836},
  \urlprefix\url{http://dx.doi.org/10.1038/35065039}.

\bibitem[{\citenamefont{Kamihara et~al.}(2008)\citenamefont{Kamihara, Watanabe,
  Hirano, and Hosono}}]{y_kamihara_08}
\bibinfo{author}{\bibfnamefont{Y.}~\bibnamefont{Kamihara}},
  \bibinfo{author}{\bibfnamefont{T.}~\bibnamefont{Watanabe}},
  \bibinfo{author}{\bibfnamefont{M.}~\bibnamefont{Hirano}}, \bibnamefont{and}
  \bibinfo{author}{\bibfnamefont{H.}~\bibnamefont{Hosono}},
  \bibinfo{journal}{Journal of the American Chemical Society}
  \textbf{\bibinfo{volume}{130}}, \bibinfo{pages}{3296} (\bibinfo{year}{2008}),
  \bibinfo{note}{pMID: 18293989},
  \urlprefix\url{http://dx.doi.org/10.1021/ja800073m}.

\bibitem[{\citenamefont{Drozdov et~al.}(2014)\citenamefont{Drozdov, Eremets,
  and Troyan}}]{Drozdov2014}
\bibinfo{author}{\bibfnamefont{A.~P.} \bibnamefont{Drozdov}},
  \bibinfo{author}{\bibfnamefont{M.~I.} \bibnamefont{Eremets}},
  \bibnamefont{and} \bibinfo{author}{\bibfnamefont{I.~A.}
  \bibnamefont{Troyan}}, \bibinfo{journal}{ArXiv e-prints}
  (\bibinfo{year}{2014}), \eprint{http://arxiv.org/abs/1412.0460},
  \urlprefix\url{http://arxiv.org/abs/1412.0460}.

\bibitem[{\citenamefont{Drozdov et~al.}(2015)\citenamefont{Drozdov, Eremets,
  Troyan, Ksenofontov, and Shylin}}]{Drozdov2015}
\bibinfo{author}{\bibfnamefont{A.~P.} \bibnamefont{Drozdov}},
  \bibinfo{author}{\bibfnamefont{M.~I.} \bibnamefont{Eremets}},
  \bibinfo{author}{\bibfnamefont{I.~A.} \bibnamefont{Troyan}},
  \bibinfo{author}{\bibfnamefont{V.}~\bibnamefont{Ksenofontov}},
  \bibnamefont{and} \bibinfo{author}{\bibfnamefont{S.~I.}
  \bibnamefont{Shylin}}, \bibinfo{journal}{Nature}
  \textbf{\bibinfo{volume}{525}}, \bibinfo{pages}{73} (\bibinfo{year}{2015}),
  \urlprefix\url{http://dx.doi.org/10.1038/nature14964}.

\bibitem[{\citenamefont{Hirschfeld et~al.}(2011)\citenamefont{Hirschfeld,
  Korshunov, and Mazin}}]{HirschfeldKorshunov2011}
\bibinfo{author}{\bibfnamefont{P.~J.} \bibnamefont{Hirschfeld}},
  \bibinfo{author}{\bibfnamefont{M.~M.} \bibnamefont{Korshunov}},
  \bibnamefont{and} \bibinfo{author}{\bibfnamefont{I.~I.} \bibnamefont{Mazin}},
  \bibinfo{journal}{Reports on Progress in Physics}
  \textbf{\bibinfo{volume}{74}}, \bibinfo{pages}{124508}
  (\bibinfo{year}{2011}),
  \urlprefix\url{http://stacks.iop.org/0034-4885/74/i=12/a=124508}.

\bibitem[{\citenamefont{Mazin et~al.}(2008)\citenamefont{Mazin, Singh,
  Johannes, and Du}}]{mazin_08}
\bibinfo{author}{\bibfnamefont{I.~I.} \bibnamefont{Mazin}},
  \bibinfo{author}{\bibfnamefont{D.~J.} \bibnamefont{Singh}},
  \bibinfo{author}{\bibfnamefont{M.~D.} \bibnamefont{Johannes}},
  \bibnamefont{and} \bibinfo{author}{\bibfnamefont{M.~H.} \bibnamefont{Du}},
  \bibinfo{journal}{Phys. Rev. Lett.} \textbf{\bibinfo{volume}{101}},
  \bibinfo{pages}{057003} (\bibinfo{year}{2008}),
  \urlprefix\url{http://link.aps.org/doi/10.1103/PhysRevLett.101.057003}.

\bibitem[{\citenamefont{Graser et~al.}(2009)\citenamefont{Graser, Maier,
  Hirschfeld, and Scalapino}}]{Graser2009}
\bibinfo{author}{\bibfnamefont{S.}~\bibnamefont{Graser}},
  \bibinfo{author}{\bibfnamefont{T.A.} \bibnamefont{Maier}},
  \bibinfo{author}{\bibfnamefont{P.J.} \bibnamefont{Hirschfeld}},
  \bibnamefont{and} \bibinfo{author}{\bibfnamefont{D.J.}
  \bibnamefont{Scalapino}}, \bibinfo{journal}{New Journal of Physics}
  \textbf{\bibinfo{volume}{11}}, \bibinfo{pages}{025016}
  (\bibinfo{year}{2009}),
  \urlprefix\url{http://stacks.iop.org/1367-2630/11/i=2/a=025016}.

\bibitem[{\citenamefont{Kuroki et~al.}(2008)\citenamefont{Kuroki, Onari, Arita,
  Usui, Tanaka, Kontani, and Aoki}}]{k_kuroki_08}
\bibinfo{author}{\bibfnamefont{Kazuhiko} \bibnamefont{Kuroki}},
  \bibinfo{author}{\bibfnamefont{Seiichiro} \bibnamefont{Onari}},
  \bibinfo{author}{\bibfnamefont{Ryotaro} \bibnamefont{Arita}},
  \bibinfo{author}{\bibfnamefont{Hidetomo} \bibnamefont{Usui}},
  \bibinfo{author}{\bibfnamefont{Yukio} \bibnamefont{Tanaka}},
  \bibinfo{author}{\bibfnamefont{Hiroshi} \bibnamefont{Kontani}},
  \bibnamefont{and} \bibinfo{author}{\bibfnamefont{Hideo} \bibnamefont{Aoki}},
  \bibinfo{journal}{Phys. Rev. Lett.} \textbf{\bibinfo{volume}{101}},
  \bibinfo{pages}{087004} (\bibinfo{year}{2008}),
  \urlprefix\url{http://link.aps.org/doi/10.1103/PhysRevLett.101.087004}.

\bibitem[{\citenamefont{Maiti et~al.}(2011{\natexlab{a}})\citenamefont{Maiti,
  Korshunov, Maier, Hirschfeld, and Chubukov}}]{MaitiKorshunovPRB2011}
\bibinfo{author}{\bibfnamefont{S.}~\bibnamefont{Maiti}},
  \bibinfo{author}{\bibfnamefont{M.~M.} \bibnamefont{Korshunov}},
  \bibinfo{author}{\bibfnamefont{T.~A.} \bibnamefont{Maier}},
  \bibinfo{author}{\bibfnamefont{P.~J.} \bibnamefont{Hirschfeld}},
  \bibnamefont{and} \bibinfo{author}{\bibfnamefont{A.~V.}
  \bibnamefont{Chubukov}}, \bibinfo{journal}{Phys. Rev. B}
  \textbf{\bibinfo{volume}{84}}, \bibinfo{pages}{224505}
  (\bibinfo{year}{2011}{\natexlab{a}}),
  \urlprefix\url{http://link.aps.org/doi/10.1103/PhysRevB.84.224505}.

\bibitem[{\citenamefont{Korshunov}(2014)}]{Korshunov2014eng}
\bibinfo{author}{\bibfnamefont{M~M} \bibnamefont{Korshunov}},
  \bibinfo{journal}{Physics-Uspekhi} \textbf{\bibinfo{volume}{57}},
  \bibinfo{pages}{813} (\bibinfo{year}{2014}),
  \urlprefix\url{http://stacks.iop.org/1063-7869/57/i=8/a=813}.

\bibitem[{\citenamefont{Kontani and Onari}(2010)}]{Kontani}
\bibinfo{author}{\bibfnamefont{H.}~\bibnamefont{Kontani}} \bibnamefont{and}
  \bibinfo{author}{\bibfnamefont{S.}~\bibnamefont{Onari}},
  \bibinfo{journal}{Phys. Rev. Lett.} \textbf{\bibinfo{volume}{104}},
  \bibinfo{pages}{157001} (\bibinfo{year}{2010}),
  \urlprefix\url{http://link.aps.org/doi/10.1103/PhysRevLett.104.157001}.

\bibitem[{\citenamefont{Boeri et~al.}(2008)\citenamefont{Boeri, Dolgov, and
  Golubov}}]{Boeri_08}
\bibinfo{author}{\bibfnamefont{L.}~\bibnamefont{Boeri}},
  \bibinfo{author}{\bibfnamefont{O.~V.} \bibnamefont{Dolgov}},
  \bibnamefont{and} \bibinfo{author}{\bibfnamefont{A.~A.}
  \bibnamefont{Golubov}}, \bibinfo{journal}{Phys. Rev. Lett.}
  \textbf{\bibinfo{volume}{101}}, \bibinfo{pages}{026403}
  (\bibinfo{year}{2008}),
  \urlprefix\url{http://link.aps.org/doi/10.1103/PhysRevLett.101.026403}.

\bibitem[{\citenamefont{Kuli\'{c} et~al.}(2009)\citenamefont{Kuli\'{c},
  Drechsler, and Dolgov}}]{Kulic2009}
\bibinfo{author}{\bibfnamefont{M.~L.} \bibnamefont{Kuli\'{c}}},
  \bibinfo{author}{\bibfnamefont{S.-L.} \bibnamefont{Drechsler}},
  \bibnamefont{and} \bibinfo{author}{\bibfnamefont{O.~V.}
  \bibnamefont{Dolgov}}, \bibinfo{journal}{EPL (Europhysics Letters)}
  \textbf{\bibinfo{volume}{85}}, \bibinfo{pages}{47008} (\bibinfo{year}{2009}),
  \urlprefix\url{http://stacks.iop.org/0295-5075/85/i=4/a=47008}.

\bibitem[{\citenamefont{Fujioka et~al.}(2014)\citenamefont{Fujioka, Denholme,
  Tanaka, Takeya, Yamaguchi, and Takano}}]{SmFeAsOTc}
\bibinfo{author}{\bibfnamefont{M.}~\bibnamefont{Fujioka}},
  \bibinfo{author}{\bibfnamefont{S.~J.} \bibnamefont{Denholme}},
  \bibinfo{author}{\bibfnamefont{M.}~\bibnamefont{Tanaka}},
  \bibinfo{author}{\bibfnamefont{H.}~\bibnamefont{Takeya}},
  \bibinfo{author}{\bibfnamefont{T.}~\bibnamefont{Yamaguchi}},
  \bibnamefont{and} \bibinfo{author}{\bibfnamefont{Y.}~\bibnamefont{Takano}},
  \bibinfo{journal}{Applied Physics Letters} \textbf{\bibinfo{volume}{105}},
  \bibinfo{eid}{102602} (\bibinfo{year}{2014}),
  \urlprefix\url{http://scitation.aip.org/content/aip/journal/apl/105/10/10.1063/1.4895574}.

\bibitem[{\citenamefont{Qing-Yan et~al.}(2012)\citenamefont{Qing-Yan, Zhi,
  Wen-Hao, Zuo-Cheng, Jin-Song, Wei, Hao, Yun-Bo, Peng, Kai et~al.}}]{FeSeTc}
\bibinfo{author}{\bibfnamefont{Wang} \bibnamefont{Qing-Yan}},
  \bibinfo{author}{\bibfnamefont{Li}~\bibnamefont{Zhi}},
  \bibinfo{author}{\bibfnamefont{Zhang} \bibnamefont{Wen-Hao}},
  \bibinfo{author}{\bibfnamefont{Zhang} \bibnamefont{Zuo-Cheng}},
  \bibinfo{author}{\bibfnamefont{Zhang} \bibnamefont{Jin-Song}},
  \bibinfo{author}{\bibfnamefont{Li}~\bibnamefont{Wei}},
  \bibinfo{author}{\bibfnamefont{Ding} \bibnamefont{Hao}},
  \bibinfo{author}{\bibfnamefont{Ou}~\bibnamefont{Yun-Bo}},
  \bibinfo{author}{\bibfnamefont{Deng} \bibnamefont{Peng}},
  \bibinfo{author}{\bibfnamefont{Chang} \bibnamefont{Kai}},
  \bibnamefont{et~al.}, \bibinfo{journal}{Chinese Physics Letters}
  \textbf{\bibinfo{volume}{29}}, \bibinfo{pages}{037402}
  (\bibinfo{year}{2012}),
  \urlprefix\url{http://stacks.iop.org/0256-307X/29/i=3/a=037402}.

\bibitem[{\citenamefont{Liu et~al.}(2012)\citenamefont{Liu, Zhang, Mou, He, Ou,
  Wang, Li, Wang, Zhao, He et~al.}}]{FeSeARPES}
\bibinfo{author}{\bibfnamefont{D.}~\bibnamefont{Liu}},
  \bibinfo{author}{\bibfnamefont{W.}~\bibnamefont{Zhang}},
  \bibinfo{author}{\bibfnamefont{D.}~\bibnamefont{Mou}},
  \bibinfo{author}{\bibfnamefont{J.}~\bibnamefont{He}},
  \bibinfo{author}{\bibfnamefont{Y.-B.} \bibnamefont{Ou}},
  \bibinfo{author}{\bibfnamefont{Q.-Y.} \bibnamefont{Wang}},
  \bibinfo{author}{\bibfnamefont{Z.}~\bibnamefont{Li}},
  \bibinfo{author}{\bibfnamefont{L.}~\bibnamefont{Wang}},
  \bibinfo{author}{\bibfnamefont{L.}~\bibnamefont{Zhao}},
  \bibinfo{author}{\bibfnamefont{S.}~\bibnamefont{He}}, \bibnamefont{et~al.},
  \bibinfo{journal}{Nat. Commun.} \textbf{\bibinfo{volume}{3}},
  \bibinfo{pages}{931} (\bibinfo{year}{2012}),
  \urlprefix\url{http://dx.doi.org/10.1038/ncomms1946}.

\bibitem[{\citenamefont{He et~al.}(2013)\citenamefont{He, He, Zhang, Zhao, Liu,
  Liu, Mou, Ou, Wang, Li et~al.}}]{HeFeSeAnneal}
\bibinfo{author}{\bibfnamefont{Shaolong} \bibnamefont{He}},
  \bibinfo{author}{\bibfnamefont{Junfeng} \bibnamefont{He}},
  \bibinfo{author}{\bibfnamefont{Wenhao} \bibnamefont{Zhang}},
  \bibinfo{author}{\bibfnamefont{Lin} \bibnamefont{Zhao}},
  \bibinfo{author}{\bibfnamefont{Defa} \bibnamefont{Liu}},
  \bibinfo{author}{\bibfnamefont{Xu}~\bibnamefont{Liu}},
  \bibinfo{author}{\bibfnamefont{Daixiang} \bibnamefont{Mou}},
  \bibinfo{author}{\bibfnamefont{Yun-Bo} \bibnamefont{Ou}},
  \bibinfo{author}{\bibfnamefont{Qing-Yan} \bibnamefont{Wang}},
  \bibinfo{author}{\bibfnamefont{Zhi} \bibnamefont{Li}}, \bibnamefont{et~al.},
  \bibinfo{journal}{Nat Mater} \textbf{\bibinfo{volume}{12}},
  \bibinfo{pages}{605} (\bibinfo{year}{2013}), ISSN \bibinfo{issn}{1476-1122},
  \urlprefix\url{http://dx.doi.org/10.1038/nmat3648}.

\bibitem[{\citenamefont{Tan et~al.}(2013)\citenamefont{Tan, Zhang, Xia, Ye,
  Chen, Xie, Peng, Xu, Fan, Xu et~al.}}]{TanFeSeARPES}
\bibinfo{author}{\bibfnamefont{Shiyong} \bibnamefont{Tan}},
  \bibinfo{author}{\bibfnamefont{Yan} \bibnamefont{Zhang}},
  \bibinfo{author}{\bibfnamefont{Miao} \bibnamefont{Xia}},
  \bibinfo{author}{\bibfnamefont{Zirong} \bibnamefont{Ye}},
  \bibinfo{author}{\bibfnamefont{Fei} \bibnamefont{Chen}},
  \bibinfo{author}{\bibfnamefont{Xin} \bibnamefont{Xie}},
  \bibinfo{author}{\bibfnamefont{Rui} \bibnamefont{Peng}},
  \bibinfo{author}{\bibfnamefont{Difei} \bibnamefont{Xu}},
  \bibinfo{author}{\bibfnamefont{Qin} \bibnamefont{Fan}},
  \bibinfo{author}{\bibfnamefont{Haichao} \bibnamefont{Xu}},
  \bibnamefont{et~al.}, \bibinfo{journal}{Nat Mater}
  \textbf{\bibinfo{volume}{12}}, \bibinfo{pages}{634} (\bibinfo{year}{2013}),
  ISSN \bibinfo{issn}{1476-1122},
  \urlprefix\url{http://dx.doi.org/10.1038/nmat3654}.

\bibitem[{\citenamefont{Ge et~al.}(2015)\citenamefont{Ge, Liu, Liu, Gao, Qian,
  Xue, Liu, and Jia}}]{GeFeSe100K}
\bibinfo{author}{\bibfnamefont{Jian-Feng} \bibnamefont{Ge}},
  \bibinfo{author}{\bibfnamefont{Zhi-Long} \bibnamefont{Liu}},
  \bibinfo{author}{\bibfnamefont{Canhua} \bibnamefont{Liu}},
  \bibinfo{author}{\bibfnamefont{Chun-Lei} \bibnamefont{Gao}},
  \bibinfo{author}{\bibfnamefont{Dong} \bibnamefont{Qian}},
  \bibinfo{author}{\bibfnamefont{Qi-Kun} \bibnamefont{Xue}},
  \bibinfo{author}{\bibfnamefont{Ying} \bibnamefont{Liu}}, \bibnamefont{and}
  \bibinfo{author}{\bibfnamefont{Jin-Feng} \bibnamefont{Jia}},
  \bibinfo{journal}{Nat Mater} \textbf{\bibinfo{volume}{14}},
  \bibinfo{pages}{285} (\bibinfo{year}{2015}), ISSN \bibinfo{issn}{1476-1122},
  \urlprefix\url{http://dx.doi.org/10.1038/nmat4153}.

\bibitem[{\citenamefont{Golubov et~al.}(2011)\citenamefont{Golubov, Dolgov,
  Boris, Charnukha, Sun, Lin, Shevchun, Korobenko, Trunin, and
  Zverev}}]{Golubov_10eng}
\bibinfo{author}{\bibfnamefont{A.~A.} \bibnamefont{Golubov}},
  \bibinfo{author}{\bibfnamefont{O.~V.} \bibnamefont{Dolgov}},
  \bibinfo{author}{\bibfnamefont{A.~V.} \bibnamefont{Boris}},
  \bibinfo{author}{\bibfnamefont{A.}~\bibnamefont{Charnukha}},
  \bibinfo{author}{\bibfnamefont{D.~L.} \bibnamefont{Sun}},
  \bibinfo{author}{\bibfnamefont{C.~T.} \bibnamefont{Lin}},
  \bibinfo{author}{\bibfnamefont{A.~F.} \bibnamefont{Shevchun}},
  \bibinfo{author}{\bibfnamefont{A.~V.} \bibnamefont{Korobenko}},
  \bibinfo{author}{\bibfnamefont{M.~R.} \bibnamefont{Trunin}},
  \bibnamefont{and} \bibinfo{author}{\bibfnamefont{V.~N.}
  \bibnamefont{Zverev}}, \bibinfo{journal}{JETP Letters}
  \textbf{\bibinfo{volume}{94}}, \bibinfo{pages}{333} (\bibinfo{year}{2011}),
  ISSN \bibinfo{issn}{1090-6487},
  \urlprefix\url{http://dx.doi.org/10.1134/S0021364011160041}.

\bibitem[{\citenamefont{Kordyuk}(2015)}]{KordyukPseudogapReview}
\bibinfo{author}{\bibfnamefont{A.~A.} \bibnamefont{Kordyuk}},
  \bibinfo{journal}{Low Temperature Physics} \textbf{\bibinfo{volume}{41}},
  \bibinfo{pages}{319} (\bibinfo{year}{2015}),
  \urlprefix\url{http://scitation.aip.org/content/aip/journal/ltp/41/5/10.1063/1.4919371}.

\bibitem[{\citenamefont{Kuchinskii and Sadovskii}(2008)}]{Kuchinskii2008eng}
\bibinfo{author}{\bibfnamefont{E.~Z.} \bibnamefont{Kuchinskii}}
  \bibnamefont{and} \bibinfo{author}{\bibfnamefont{M.~V.}
  \bibnamefont{Sadovskii}}, \bibinfo{journal}{JETP Letters}
  \textbf{\bibinfo{volume}{88}}, \bibinfo{pages}{192} (\bibinfo{year}{2008}),
  ISSN \bibinfo{issn}{1090-6487},
  \urlprefix\url{http://dx.doi.org/10.1134/S0021364008150101}.

\bibitem[{\citenamefont{Sawatzky et~al.}(2009)\citenamefont{Sawatzky, Elfimov,
  van~den Brink, and Zaanen}}]{Sawatzky2009}
\bibinfo{author}{\bibfnamefont{G.~A.} \bibnamefont{Sawatzky}},
  \bibinfo{author}{\bibfnamefont{I.~S.} \bibnamefont{Elfimov}},
  \bibinfo{author}{\bibfnamefont{J.}~\bibnamefont{van~den Brink}},
  \bibnamefont{and} \bibinfo{author}{\bibfnamefont{J.}~\bibnamefont{Zaanen}},
  \bibinfo{journal}{EPL (Europhysics Letters)} \textbf{\bibinfo{volume}{86}},
  \bibinfo{pages}{17006} (\bibinfo{year}{2009}),
  \urlprefix\url{http://stacks.iop.org/0295-5075/86/i=1/a=17006}.

\bibitem[{\citenamefont{Nakamura et~al.}(2011)\citenamefont{Nakamura, Arita,
  and Ikeda}}]{Nakamura2011}
\bibinfo{author}{\bibfnamefont{K.}~\bibnamefont{Nakamura}},
  \bibinfo{author}{\bibfnamefont{R.}~\bibnamefont{Arita}}, \bibnamefont{and}
  \bibinfo{author}{\bibfnamefont{H.}~\bibnamefont{Ikeda}},
  \bibinfo{journal}{Phys. Rev. B} \textbf{\bibinfo{volume}{83}},
  \bibinfo{pages}{144512} (\bibinfo{year}{2011}),
  \urlprefix\url{http://link.aps.org/doi/10.1103/PhysRevB.83.144512}.

\bibitem[{\citenamefont{Brouet et~al.}(2009)\citenamefont{Brouet, Marsi,
  Mansart, Nicolaou, Taleb-Ibrahimi, Le~F\`evre, Bertran, Rullier-Albenque,
  Forget, and Colson}}]{Brouet2009}
\bibinfo{author}{\bibfnamefont{V.}~\bibnamefont{Brouet}},
  \bibinfo{author}{\bibfnamefont{M.}~\bibnamefont{Marsi}},
  \bibinfo{author}{\bibfnamefont{B.}~\bibnamefont{Mansart}},
  \bibinfo{author}{\bibfnamefont{A.}~\bibnamefont{Nicolaou}},
  \bibinfo{author}{\bibfnamefont{A.}~\bibnamefont{Taleb-Ibrahimi}},
  \bibinfo{author}{\bibfnamefont{P.}~\bibnamefont{Le~F\`evre}},
  \bibinfo{author}{\bibfnamefont{F.}~\bibnamefont{Bertran}},
  \bibinfo{author}{\bibfnamefont{F.}~\bibnamefont{Rullier-Albenque}},
  \bibinfo{author}{\bibfnamefont{A.}~\bibnamefont{Forget}}, \bibnamefont{and}
  \bibinfo{author}{\bibfnamefont{D.}~\bibnamefont{Colson}},
  \bibinfo{journal}{Phys. Rev. B} \textbf{\bibinfo{volume}{80}},
  \bibinfo{pages}{165115} (\bibinfo{year}{2009}),
  \urlprefix\url{http://link.aps.org/doi/10.1103/PhysRevB.80.165115}.

\bibitem[{\citenamefont{Kuroki et~al.}(2009)\citenamefont{Kuroki, Usui, Onari,
  Arita, and Aoki}}]{k_kuroki_09}
\bibinfo{author}{\bibfnamefont{Kazuhiko} \bibnamefont{Kuroki}},
  \bibinfo{author}{\bibfnamefont{Hidetomo} \bibnamefont{Usui}},
  \bibinfo{author}{\bibfnamefont{Seiichiro} \bibnamefont{Onari}},
  \bibinfo{author}{\bibfnamefont{Ryotaro} \bibnamefont{Arita}},
  \bibnamefont{and} \bibinfo{author}{\bibfnamefont{Hideo} \bibnamefont{Aoki}},
  \bibinfo{journal}{Phys. Rev. B} \textbf{\bibinfo{volume}{79}},
  \bibinfo{pages}{224511} (\bibinfo{year}{2009}),
  \urlprefix\url{http://link.aps.org/doi/10.1103/PhysRevB.79.224511}.

\bibitem[{\citenamefont{Mizuguchi et~al.}(2010)\citenamefont{Mizuguchi, Hara,
  Deguchi, Tsuda, Yamaguchi, Takeda, Kotegawa, Tou, and
  Takano}}]{Mizuguchi2010}
\bibinfo{author}{\bibfnamefont{Y.}~\bibnamefont{Mizuguchi}},
  \bibinfo{author}{\bibfnamefont{Y.}~\bibnamefont{Hara}},
  \bibinfo{author}{\bibfnamefont{K.}~\bibnamefont{Deguchi}},
  \bibinfo{author}{\bibfnamefont{S.}~\bibnamefont{Tsuda}},
  \bibinfo{author}{\bibfnamefont{T.}~\bibnamefont{Yamaguchi}},
  \bibinfo{author}{\bibfnamefont{K.}~\bibnamefont{Takeda}},
  \bibinfo{author}{\bibfnamefont{H.}~\bibnamefont{Kotegawa}},
  \bibinfo{author}{\bibfnamefont{H.}~\bibnamefont{Tou}}, \bibnamefont{and}
  \bibinfo{author}{\bibfnamefont{Y.}~\bibnamefont{Takano}},
  \bibinfo{journal}{Superconductor Science and Technology}
  \textbf{\bibinfo{volume}{23}}, \bibinfo{pages}{054013}
  (\bibinfo{year}{2010}),
  \urlprefix\url{http://stacks.iop.org/0953-2048/23/i=5/a=054013}.

\bibitem[{\citenamefont{Kuchinskii et~al.}(2010)\citenamefont{Kuchinskii,
  Nekrasov, and Sadovskii}}]{Kuchinskii2010eng}
\bibinfo{author}{\bibfnamefont{E.~Z.} \bibnamefont{Kuchinskii}},
  \bibinfo{author}{\bibfnamefont{I.~A.} \bibnamefont{Nekrasov}},
  \bibnamefont{and} \bibinfo{author}{\bibfnamefont{M.~V.}
  \bibnamefont{Sadovskii}}, \bibinfo{journal}{JETP Letters}
  \textbf{\bibinfo{volume}{91}}, \bibinfo{pages}{518} (\bibinfo{year}{2010}),
  ISSN \bibinfo{issn}{1090-6487},
  \urlprefix\url{http://dx.doi.org/10.1134/S0021364010100061}.

\bibitem[{\citenamefont{Hirschfeld}(2016)}]{Hirschfeld2016}
\bibinfo{author}{\bibfnamefont{Peter~J.} \bibnamefont{Hirschfeld}},
  \bibinfo{journal}{Comptes Rendus Physique} \textbf{\bibinfo{volume}{17}},
  \bibinfo{pages}{197 } (\bibinfo{year}{2016}), ISSN \bibinfo{issn}{1631-0705},
  \urlprefix\url{http://www.sciencedirect.com/science/article/pii/S1631070515001693}.

\bibitem[{\citenamefont{Anderson}(1959)}]{Anderson1959}
\bibinfo{author}{\bibfnamefont{P.W.} \bibnamefont{Anderson}},
  \bibinfo{journal}{Journal of Physics and Chemistry of Solids}
  \textbf{\bibinfo{volume}{11}}, \bibinfo{pages}{26} (\bibinfo{year}{1959}),
  ISSN \bibinfo{issn}{0022-3697},
  \urlprefix\url{http://www.sciencedirect.com/science/article/pii/0022369759900368}.

\bibitem[{\citenamefont{Abrikosov and Gor'kov}(1961)}]{AGeng}
\bibinfo{author}{\bibfnamefont{A.~A.} \bibnamefont{Abrikosov}}
  \bibnamefont{and} \bibinfo{author}{\bibfnamefont{L.~P.}
  \bibnamefont{Gor'kov}}, \bibinfo{journal}{Sov. Phys. JETP}
  \textbf{\bibinfo{volume}{12}}, \bibinfo{pages}{1243} (\bibinfo{year}{1961}).

\bibitem[{\citenamefont{Tarascon et~al.}(1990)\citenamefont{Tarascon, Wang,
  Kivelson, Bagley, Hull, and Ramesh}}]{Tarascon1990}
\bibinfo{author}{\bibfnamefont{J.~M.} \bibnamefont{Tarascon}},
  \bibinfo{author}{\bibfnamefont{E.}~\bibnamefont{Wang}},
  \bibinfo{author}{\bibfnamefont{S.}~\bibnamefont{Kivelson}},
  \bibinfo{author}{\bibfnamefont{B.~G.} \bibnamefont{Bagley}},
  \bibinfo{author}{\bibfnamefont{G.~W.} \bibnamefont{Hull}}, \bibnamefont{and}
  \bibinfo{author}{\bibfnamefont{R.}~\bibnamefont{Ramesh}},
  \bibinfo{journal}{Phys. Rev. B} \textbf{\bibinfo{volume}{42}},
  \bibinfo{pages}{218} (\bibinfo{year}{1990}),
  \urlprefix\url{http://link.aps.org/doi/10.1103/PhysRevB.42.218}.

\bibitem[{\citenamefont{Jayaram et~al.}(1995)\citenamefont{Jayaram, Chen, and
  Callaway}}]{Jayaram1995}
\bibinfo{author}{\bibfnamefont{B.}~\bibnamefont{Jayaram}},
  \bibinfo{author}{\bibfnamefont{H.}~\bibnamefont{Chen}}, \bibnamefont{and}
  \bibinfo{author}{\bibfnamefont{J.}~\bibnamefont{Callaway}},
  \bibinfo{journal}{Phys. Rev. B} \textbf{\bibinfo{volume}{52}},
  \bibinfo{pages}{3742} (\bibinfo{year}{1995}),
  \urlprefix\url{http://link.aps.org/doi/10.1103/PhysRevB.52.3742}.

\bibitem[{\citenamefont{Brinkmann et~al.}(1996)\citenamefont{Brinkmann, Bach,
  and Westerholt}}]{Brinkmann1996}
\bibinfo{author}{\bibfnamefont{Matthias} \bibnamefont{Brinkmann}},
  \bibinfo{author}{\bibfnamefont{Heinrich} \bibnamefont{Bach}},
  \bibnamefont{and} \bibinfo{author}{\bibfnamefont{Kurt}
  \bibnamefont{Westerholt}}, \bibinfo{journal}{Phys. Rev. B}
  \textbf{\bibinfo{volume}{54}}, \bibinfo{pages}{6680} (\bibinfo{year}{1996}),
  \urlprefix\url{http://link.aps.org/doi/10.1103/PhysRevB.54.6680}.

\bibitem[{\citenamefont{Markert et~al.}(1989)\citenamefont{Markert,
  Dalichaouch, and Marle}}]{Markert1989}
\bibinfo{author}{\bibfnamefont{J.~T.} \bibnamefont{Markert}},
  \bibinfo{author}{\bibfnamefont{Y.}~\bibnamefont{Dalichaouch}},
  \bibnamefont{and} \bibinfo{author}{\bibfnamefont{M.~B.} \bibnamefont{Marle}},
  in \emph{\bibinfo{booktitle}{Physical Properties of High Temperature
  Superconductors I}}, edited by \bibinfo{editor}{\bibfnamefont{D.~N.}
  \bibnamefont{Ginsberg}} (\bibinfo{publisher}{World Scientific},
  \bibinfo{address}{Singapore}, \bibinfo{year}{1989}),
  chap.~\bibinfo{chapter}{6}, pp. \bibinfo{pages}{265--337},
  \urlprefix\url{http://www.worldscientific.com/doi/abs/10.1142/9789814434249_0006}.

\bibitem[{\citenamefont{Xiao et~al.}(1990)\citenamefont{Xiao, Cieplak, Xiao,
  and Chien}}]{Xiao1990}
\bibinfo{author}{\bibfnamefont{Gang} \bibnamefont{Xiao}},
  \bibinfo{author}{\bibfnamefont{Marta~Z.} \bibnamefont{Cieplak}},
  \bibinfo{author}{\bibfnamefont{J.~Q.} \bibnamefont{Xiao}}, \bibnamefont{and}
  \bibinfo{author}{\bibfnamefont{C.~L.} \bibnamefont{Chien}},
  \bibinfo{journal}{Phys. Rev. B} \textbf{\bibinfo{volume}{42}},
  \bibinfo{pages}{8752} (\bibinfo{year}{1990}),
  \urlprefix\url{http://link.aps.org/doi/10.1103/PhysRevB.42.8752}.

\bibitem[{\citenamefont{Ting et~al.}(1992)\citenamefont{Ting, Pernambuco-Wise,
  Crow, Manousakis, and Weaver}}]{Ting1992}
\bibinfo{author}{\bibfnamefont{S.~T.} \bibnamefont{Ting}},
  \bibinfo{author}{\bibfnamefont{P.}~\bibnamefont{Pernambuco-Wise}},
  \bibinfo{author}{\bibfnamefont{J.~E.} \bibnamefont{Crow}},
  \bibinfo{author}{\bibfnamefont{E.}~\bibnamefont{Manousakis}},
  \bibnamefont{and} \bibinfo{author}{\bibfnamefont{J.}~\bibnamefont{Weaver}},
  \bibinfo{journal}{Phys. Rev. B} \textbf{\bibinfo{volume}{46}},
  \bibinfo{pages}{11772} (\bibinfo{year}{1992}),
  \urlprefix\url{http://link.aps.org/doi/10.1103/PhysRevB.46.11772}.

\bibitem[{\citenamefont{Ovchinnikov}(1995)}]{Ovchinnikov1995eng}
\bibinfo{author}{\bibfnamefont{S.~G.} \bibnamefont{Ovchinnikov}},
  \bibinfo{journal}{Phys. Solid State} \textbf{\bibinfo{volume}{37}},
  \bibinfo{pages}{2007} (\bibinfo{year}{1995}),
  \urlprefix\url{http://journals.ioffe.ru/articles/17239}.

\bibitem[{\citenamefont{Gaididei and Loktev}(1988)}]{Gaididei1988}
\bibinfo{author}{\bibfnamefont{Yu.~B.} \bibnamefont{Gaididei}}
  \bibnamefont{and} \bibinfo{author}{\bibfnamefont{V.~M.}
  \bibnamefont{Loktev}}, \bibinfo{journal}{physica status solidi (b)}
  \textbf{\bibinfo{volume}{147}}, \bibinfo{pages}{307} (\bibinfo{year}{1988}),
  ISSN \bibinfo{issn}{1521-3951},
  \urlprefix\url{http://dx.doi.org/10.1002/pssb.2221470135}.

\bibitem[{\citenamefont{Kluge et~al.}(1995)\citenamefont{Kluge, Koike,
  Fujiwara, Kato, Noji, and Saito}}]{Kluge1995}
\bibinfo{author}{\bibfnamefont{T.}~\bibnamefont{Kluge}},
  \bibinfo{author}{\bibfnamefont{Y.}~\bibnamefont{Koike}},
  \bibinfo{author}{\bibfnamefont{A.}~\bibnamefont{Fujiwara}},
  \bibinfo{author}{\bibfnamefont{M.}~\bibnamefont{Kato}},
  \bibinfo{author}{\bibfnamefont{T.}~\bibnamefont{Noji}}, \bibnamefont{and}
  \bibinfo{author}{\bibfnamefont{Y.}~\bibnamefont{Saito}},
  \bibinfo{journal}{Phys. Rev. B} \textbf{\bibinfo{volume}{52}},
  \bibinfo{pages}{R727} (\bibinfo{year}{1995}),
  \urlprefix\url{http://link.aps.org/doi/10.1103/PhysRevB.52.R727}.

\bibitem[{\citenamefont{Davydov et~al.}(1988)\citenamefont{Davydov, Kar'kin,
  Mirmel'shtein, Berger, Voronin, Parkhomenko, Kozhevnikov, Cheshnitskii, and
  Goshchitski}}]{Davydov1988eng}
\bibinfo{author}{\bibfnamefont{S.~A.} \bibnamefont{Davydov}},
  \bibinfo{author}{\bibfnamefont{A.~E.} \bibnamefont{Kar'kin}},
  \bibinfo{author}{\bibfnamefont{A.~V.} \bibnamefont{Mirmel'shtein}},
  \bibinfo{author}{\bibfnamefont{I.~F.} \bibnamefont{Berger}},
  \bibinfo{author}{\bibfnamefont{V.~I.} \bibnamefont{Voronin}},
  \bibinfo{author}{\bibfnamefont{V.~D.} \bibnamefont{Parkhomenko}},
  \bibinfo{author}{\bibfnamefont{V.~L.} \bibnamefont{Kozhevnikov}},
  \bibinfo{author}{\bibfnamefont{S.~M.} \bibnamefont{Cheshnitskii}},
  \bibnamefont{and} \bibinfo{author}{\bibfnamefont{B.~N.}
  \bibnamefont{Goshchitski}}, \bibinfo{journal}{JETP Lett.}
  \textbf{\bibinfo{volume}{47}}, \bibinfo{pages}{234} (\bibinfo{year}{1988}),
  \urlprefix\url{http://www.jetpletters.ac.ru/ps/1092/article_16492.shtml}.

\bibitem[{\citenamefont{Aleksashin et~al.}(1989)\citenamefont{Aleksashin,
  Voronin, Verkhovskii, Goshchitskii, Davydov, Zhdanov, Kar'kin, Kozhevnikov,
  Mirmel'shtein, Mikhalev et~al.}}]{Aleksashin1989}
\bibinfo{author}{\bibfnamefont{B.~A.} \bibnamefont{Aleksashin}},
  \bibinfo{author}{\bibfnamefont{V.~P.} \bibnamefont{Voronin}},
  \bibinfo{author}{\bibfnamefont{S.~V.} \bibnamefont{Verkhovskii}},
  \bibinfo{author}{\bibfnamefont{B.~N.} \bibnamefont{Goshchitskii}},
  \bibinfo{author}{\bibfnamefont{S.~A.} \bibnamefont{Davydov}},
  \bibinfo{author}{\bibfnamefont{Yu.~I.} \bibnamefont{Zhdanov}},
  \bibinfo{author}{\bibfnamefont{A.~E.} \bibnamefont{Kar'kin}},
  \bibinfo{author}{\bibfnamefont{V.~L.} \bibnamefont{Kozhevnikov}},
  \bibinfo{author}{\bibfnamefont{A.~V.} \bibnamefont{Mirmel'shtein}},
  \bibinfo{author}{\bibfnamefont{K.~N.} \bibnamefont{Mikhalev}},
  \bibnamefont{et~al.}, \bibinfo{journal}{JETP} \textbf{\bibinfo{volume}{68}},
  \bibinfo{pages}{382} (\bibinfo{year}{1989}), \bibinfo{note}{russian original
  - ZhETF, Vol. 95, No. 2, p. 678, February 1989},
  \urlprefix\url{http://www.jetp.ac.ru/cgi-bin/e/index/e/68/2/p382?a=list}.

\bibitem[{\citenamefont{Anan'ev et~al.}(1998)\citenamefont{Anan'ev, Zhdanov,
  Gerashchenko, Mikhalev, Verkhovskii, Medvedev, Okulova, Chebotaev, and
  Goshchitskii}}]{Ananyev1998eng}
\bibinfo{author}{\bibfnamefont{A.~V.} \bibnamefont{Anan'ev}},
  \bibinfo{author}{\bibfnamefont{Yu.~I.} \bibnamefont{Zhdanov}},
  \bibinfo{author}{\bibfnamefont{A.~P.} \bibnamefont{Gerashchenko}},
  \bibinfo{author}{\bibfnamefont{K.~N.} \bibnamefont{Mikhalev}},
  \bibinfo{author}{\bibfnamefont{S.~V.} \bibnamefont{Verkhovskii}},
  \bibinfo{author}{\bibfnamefont{E.~Yu.} \bibnamefont{Medvedev}},
  \bibinfo{author}{\bibfnamefont{K.~A.} \bibnamefont{Okulova}},
  \bibinfo{author}{\bibfnamefont{N.~I.} \bibnamefont{Chebotaev}},
  \bibnamefont{and} \bibinfo{author}{\bibfnamefont{V.~N.}
  \bibnamefont{Goshchitskii}}, \bibinfo{journal}{Journal of Experimental and
  Theoretical Physics Letters} \textbf{\bibinfo{volume}{67}},
  \bibinfo{pages}{182} (\bibinfo{year}{1998}), ISSN \bibinfo{issn}{1090-6487},
  \urlprefix\url{http://dx.doi.org/10.1134/1.567648}.

\bibitem[{\citenamefont{Ovchinnikov}(1997)}]{Ovchinnikov1997eng}
\bibinfo{author}{\bibfnamefont{S.~G.} \bibnamefont{Ovchinnikov}},
  \bibinfo{journal}{Physics-Uspekhi} \textbf{\bibinfo{volume}{40}},
  \bibinfo{pages}{993} (\bibinfo{year}{1997}),
  \urlprefix\url{http://stacks.iop.org/1063-7869/40/i=10/a=R02}.

\bibitem[{\citenamefont{Kuli\'{c}}(2000)}]{Kulic2000}
\bibinfo{author}{\bibfnamefont{Miodrag~L.} \bibnamefont{Kuli\'{c}}},
  \bibinfo{journal}{Physics Reports} \textbf{\bibinfo{volume}{338}},
  \bibinfo{pages}{1} (\bibinfo{year}{2000}), ISSN \bibinfo{issn}{0370-1573},
  \urlprefix\url{http://www.sciencedirect.com/science/article/pii/S0370157300000089}.

\bibitem[{\citenamefont{Hussey}(2002)}]{Hussey2002}
\bibinfo{author}{\bibfnamefont{N.E.} \bibnamefont{Hussey}},
  \bibinfo{journal}{Advances in Physics} \textbf{\bibinfo{volume}{51}},
  \bibinfo{pages}{1685} (\bibinfo{year}{2002}),
  \urlprefix\url{http://dx.doi.org/10.1080/00018730210164638}.

\bibitem[{\citenamefont{Hirschfeld and Atkinson}(2002)}]{Hirschfeld2002}
\bibinfo{author}{\bibfnamefont{P.~J.} \bibnamefont{Hirschfeld}}
  \bibnamefont{and} \bibinfo{author}{\bibfnamefont{W.~A.}
  \bibnamefont{Atkinson}}, \bibinfo{journal}{Journal of Low Temperature
  Physics} \textbf{\bibinfo{volume}{126}}, \bibinfo{pages}{881}
  (\bibinfo{year}{2002}), ISSN \bibinfo{issn}{1573-7357},
  \urlprefix\url{http://dx.doi.org/10.1023/A:1013838523587}.

\bibitem[{\citenamefont{Balatsky et~al.}(2006)\citenamefont{Balatsky, Vekhter,
  and Zhu}}]{Balatsky2006}
\bibinfo{author}{\bibfnamefont{A.~V.} \bibnamefont{Balatsky}},
  \bibinfo{author}{\bibfnamefont{I.}~\bibnamefont{Vekhter}}, \bibnamefont{and}
  \bibinfo{author}{\bibfnamefont{Jian-Xin} \bibnamefont{Zhu}},
  \bibinfo{journal}{Rev. Mod. Phys.} \textbf{\bibinfo{volume}{78}},
  \bibinfo{pages}{373} (\bibinfo{year}{2006}),
  \urlprefix\url{http://link.aps.org/doi/10.1103/RevModPhys.78.373}.

\bibitem[{\citenamefont{Alloul et~al.}(2009)\citenamefont{Alloul, Bobroff,
  Gabay, and Hirschfeld}}]{Alloul2009}
\bibinfo{author}{\bibfnamefont{H.}~\bibnamefont{Alloul}},
  \bibinfo{author}{\bibfnamefont{J.}~\bibnamefont{Bobroff}},
  \bibinfo{author}{\bibfnamefont{M.}~\bibnamefont{Gabay}}, \bibnamefont{and}
  \bibinfo{author}{\bibfnamefont{P.~J.} \bibnamefont{Hirschfeld}},
  \bibinfo{journal}{Rev. Mod. Phys.} \textbf{\bibinfo{volume}{81}},
  \bibinfo{pages}{45} (\bibinfo{year}{2009}),
  \urlprefix\url{http://link.aps.org/doi/10.1103/RevModPhys.81.45}.

\bibitem[{\citenamefont{Pogorelov et~al.}(2011)\citenamefont{Pogorelov, Santos,
  and Loktev}}]{Pogorelov2011}
\bibinfo{author}{\bibfnamefont{Yu.~G.} \bibnamefont{Pogorelov}},
  \bibinfo{author}{\bibfnamefont{M.~C.} \bibnamefont{Santos}},
  \bibnamefont{and} \bibinfo{author}{\bibfnamefont{V.~M.}
  \bibnamefont{Loktev}}, \bibinfo{journal}{Low Temperature Physics}
  \textbf{\bibinfo{volume}{37}}, \bibinfo{pages}{633} (\bibinfo{year}{2011}),
  \urlprefix\url{http://scitation.aip.org/content/aip/journal/ltp/37/8/10.1063/1.3651472}.

\bibitem[{\citenamefont{Radtke et~al.}(1993)\citenamefont{Radtke, Levin,
  Sch\"uttler, and Norman}}]{Radtke1993}
\bibinfo{author}{\bibfnamefont{R.~J.} \bibnamefont{Radtke}},
  \bibinfo{author}{\bibfnamefont{K.}~\bibnamefont{Levin}},
  \bibinfo{author}{\bibfnamefont{H.-B.} \bibnamefont{Sch\"uttler}},
  \bibnamefont{and} \bibinfo{author}{\bibfnamefont{M.~R.}
  \bibnamefont{Norman}}, \bibinfo{journal}{Phys. Rev. B}
  \textbf{\bibinfo{volume}{48}}, \bibinfo{pages}{653} (\bibinfo{year}{1993}),
  \urlprefix\url{http://link.aps.org/doi/10.1103/PhysRevB.48.653}.

\bibitem[{\citenamefont{Preosti et~al.}(1994)\citenamefont{Preosti, Kim, and
  Muzikar}}]{Preosti1994}
\bibinfo{author}{\bibfnamefont{Gianfranco} \bibnamefont{Preosti}},
  \bibinfo{author}{\bibfnamefont{Heesang} \bibnamefont{Kim}}, \bibnamefont{and}
  \bibinfo{author}{\bibfnamefont{Paul} \bibnamefont{Muzikar}},
  \bibinfo{journal}{Phys. Rev. B} \textbf{\bibinfo{volume}{50}},
  \bibinfo{pages}{1259} (\bibinfo{year}{1994}),
  \urlprefix\url{http://link.aps.org/doi/10.1103/PhysRevB.50.1259}.

\bibitem[{\citenamefont{Fehrenbacher and Norman}(1994)}]{Fehrenbacher1994}
\bibinfo{author}{\bibfnamefont{R.}~\bibnamefont{Fehrenbacher}}
  \bibnamefont{and} \bibinfo{author}{\bibfnamefont{M.~R.}
  \bibnamefont{Norman}}, \bibinfo{journal}{Phys. Rev. B}
  \textbf{\bibinfo{volume}{50}}, \bibinfo{pages}{3495} (\bibinfo{year}{1994}),
  \urlprefix\url{http://link.aps.org/doi/10.1103/PhysRevB.50.3495}.

\bibitem[{\citenamefont{Balatsky et~al.}(1995)\citenamefont{Balatsky, Salkola,
  and Rosengren}}]{Balatsky1995}
\bibinfo{author}{\bibfnamefont{A.~V.} \bibnamefont{Balatsky}},
  \bibinfo{author}{\bibfnamefont{M.~I.} \bibnamefont{Salkola}},
  \bibnamefont{and}
  \bibinfo{author}{\bibfnamefont{A.}~\bibnamefont{Rosengren}},
  \bibinfo{journal}{Phys. Rev. B} \textbf{\bibinfo{volume}{51}},
  \bibinfo{pages}{15547} (\bibinfo{year}{1995}),
  \urlprefix\url{http://link.aps.org/doi/10.1103/PhysRevB.51.15547}.

\bibitem[{\citenamefont{Hara\ifmmode~\acute{n}\else \'{n}\fi{} and
  Nagi}(1996)}]{Haran1996}
\bibinfo{author}{\bibfnamefont{Grzegorz}
  \bibnamefont{Hara\ifmmode~\acute{n}\else \'{n}\fi{}}} \bibnamefont{and}
  \bibinfo{author}{\bibfnamefont{A.~D.~S.} \bibnamefont{Nagi}},
  \bibinfo{journal}{Phys. Rev. B} \textbf{\bibinfo{volume}{54}},
  \bibinfo{pages}{15463} (\bibinfo{year}{1996}),
  \urlprefix\url{http://link.aps.org/doi/10.1103/PhysRevB.54.15463}.

\bibitem[{\citenamefont{Franz et~al.}(1997)\citenamefont{Franz, Kallin,
  Berlinsky, and Salkola}}]{Franz1997}
\bibinfo{author}{\bibfnamefont{M.}~\bibnamefont{Franz}},
  \bibinfo{author}{\bibfnamefont{C.}~\bibnamefont{Kallin}},
  \bibinfo{author}{\bibfnamefont{A.~J.} \bibnamefont{Berlinsky}},
  \bibnamefont{and} \bibinfo{author}{\bibfnamefont{M.~I.}
  \bibnamefont{Salkola}}, \bibinfo{journal}{Phys. Rev. B}
  \textbf{\bibinfo{volume}{56}}, \bibinfo{pages}{7882} (\bibinfo{year}{1997}),
  \urlprefix\url{http://link.aps.org/doi/10.1103/PhysRevB.56.7882}.

\bibitem[{\citenamefont{Kuli\'{c} and Oudovenko}(1997)}]{Kulic1997}
\bibinfo{author}{\bibfnamefont{Miodrag~L.} \bibnamefont{Kuli\'{c}}}
  \bibnamefont{and} \bibinfo{author}{\bibfnamefont{Viktor}
  \bibnamefont{Oudovenko}}, \bibinfo{journal}{Solid State Communications}
  \textbf{\bibinfo{volume}{104}}, \bibinfo{pages}{375 } (\bibinfo{year}{1997}),
  ISSN \bibinfo{issn}{0038-1098},
  \urlprefix\url{http://www.sciencedirect.com/science/article/pii/S0038109897003955}.

\bibitem[{\citenamefont{Kuli\ifmmode\acute{c}\else\'{c}\fi{} and
  Dolgov}(1999)}]{Kulic1999}
\bibinfo{author}{\bibfnamefont{Miodrag~L.}
  \bibnamefont{Kuli\ifmmode\acute{c}\else\'{c}\fi{}}} \bibnamefont{and}
  \bibinfo{author}{\bibfnamefont{Oleg~V.} \bibnamefont{Dolgov}},
  \bibinfo{journal}{Phys. Rev. B} \textbf{\bibinfo{volume}{60}},
  \bibinfo{pages}{13062} (\bibinfo{year}{1999}),
  \urlprefix\url{http://link.aps.org/doi/10.1103/PhysRevB.60.13062}.

\bibitem[{\citenamefont{Chen and Schrieffer}(2002)}]{Chen2002}
\bibinfo{author}{\bibfnamefont{Qijin} \bibnamefont{Chen}} \bibnamefont{and}
  \bibinfo{author}{\bibfnamefont{J.~R.} \bibnamefont{Schrieffer}},
  \bibinfo{journal}{Phys. Rev. B} \textbf{\bibinfo{volume}{66}},
  \bibinfo{pages}{014512} (\bibinfo{year}{2002}),
  \urlprefix\url{http://link.aps.org/doi/10.1103/PhysRevB.66.014512}.

\bibitem[{\citenamefont{Fujita et~al.}(2005)\citenamefont{Fujita, Noda, Kojima,
  Eisaki, and Uchida}}]{Fujita2005}
\bibinfo{author}{\bibfnamefont{K.}~\bibnamefont{Fujita}},
  \bibinfo{author}{\bibfnamefont{T.}~\bibnamefont{Noda}},
  \bibinfo{author}{\bibfnamefont{K.~M.} \bibnamefont{Kojima}},
  \bibinfo{author}{\bibfnamefont{H.}~\bibnamefont{Eisaki}}, \bibnamefont{and}
  \bibinfo{author}{\bibfnamefont{S.}~\bibnamefont{Uchida}},
  \bibinfo{journal}{Phys. Rev. Lett.} \textbf{\bibinfo{volume}{95}},
  \bibinfo{pages}{097006} (\bibinfo{year}{2005}),
  \urlprefix\url{http://link.aps.org/doi/10.1103/PhysRevLett.95.097006}.

\bibitem[{\citenamefont{Hirschfeld et~al.}(1989)\citenamefont{Hirschfeld,
  W\"olfle, Sauls, Einzel, and Putikka}}]{Hirschfeld1989}
\bibinfo{author}{\bibfnamefont{P.~J.} \bibnamefont{Hirschfeld}},
  \bibinfo{author}{\bibfnamefont{P.}~\bibnamefont{W\"olfle}},
  \bibinfo{author}{\bibfnamefont{J.~A.} \bibnamefont{Sauls}},
  \bibinfo{author}{\bibfnamefont{D.}~\bibnamefont{Einzel}}, \bibnamefont{and}
  \bibinfo{author}{\bibfnamefont{W.~O.} \bibnamefont{Putikka}},
  \bibinfo{journal}{Phys. Rev. B} \textbf{\bibinfo{volume}{40}},
  \bibinfo{pages}{6695} (\bibinfo{year}{1989}),
  \urlprefix\url{http://link.aps.org/doi/10.1103/PhysRevB.40.6695}.

\bibitem[{\citenamefont{Hirschfeld and Goldenfeld}(1993)}]{Hirschfeld1993}
\bibinfo{author}{\bibfnamefont{Peter~J.} \bibnamefont{Hirschfeld}}
  \bibnamefont{and} \bibinfo{author}{\bibfnamefont{Nigel}
  \bibnamefont{Goldenfeld}}, \bibinfo{journal}{Phys. Rev. B}
  \textbf{\bibinfo{volume}{48}}, \bibinfo{pages}{4219} (\bibinfo{year}{1993}),
  \urlprefix\url{http://link.aps.org/doi/10.1103/PhysRevB.48.4219}.

\bibitem[{\citenamefont{Hirschfeld et~al.}(1994)\citenamefont{Hirschfeld,
  Putikka, and Scalapino}}]{Hirschfeld1994}
\bibinfo{author}{\bibfnamefont{P.~J.} \bibnamefont{Hirschfeld}},
  \bibinfo{author}{\bibfnamefont{W.~O.} \bibnamefont{Putikka}},
  \bibnamefont{and} \bibinfo{author}{\bibfnamefont{D.~J.}
  \bibnamefont{Scalapino}}, \bibinfo{journal}{Phys. Rev. B}
  \textbf{\bibinfo{volume}{50}}, \bibinfo{pages}{10250} (\bibinfo{year}{1994}),
  \urlprefix\url{http://link.aps.org/doi/10.1103/PhysRevB.50.10250}.

\bibitem[{\citenamefont{Quinlan et~al.}(1996)\citenamefont{Quinlan, Hirschfeld,
  and Scalapino}}]{Quinlan1996}
\bibinfo{author}{\bibfnamefont{S.~M.} \bibnamefont{Quinlan}},
  \bibinfo{author}{\bibfnamefont{P.~J.} \bibnamefont{Hirschfeld}},
  \bibnamefont{and} \bibinfo{author}{\bibfnamefont{D.~J.}
  \bibnamefont{Scalapino}}, \bibinfo{journal}{Phys. Rev. B}
  \textbf{\bibinfo{volume}{53}}, \bibinfo{pages}{8575} (\bibinfo{year}{1996}),
  \urlprefix\url{http://link.aps.org/doi/10.1103/PhysRevB.53.8575}.

\bibitem[{\citenamefont{Hirschfeld et~al.}(1997)\citenamefont{Hirschfeld,
  Quinlan, and Scalapino}}]{Hirschfeld1997}
\bibinfo{author}{\bibfnamefont{P.~J.} \bibnamefont{Hirschfeld}},
  \bibinfo{author}{\bibfnamefont{S.~M.} \bibnamefont{Quinlan}},
  \bibnamefont{and} \bibinfo{author}{\bibfnamefont{D.~J.}
  \bibnamefont{Scalapino}}, \bibinfo{journal}{Phys. Rev. B}
  \textbf{\bibinfo{volume}{55}}, \bibinfo{pages}{12742} (\bibinfo{year}{1997}),
  \urlprefix\url{http://link.aps.org/doi/10.1103/PhysRevB.55.12742}.

\bibitem[{\citenamefont{Duffy et~al.}(2001)\citenamefont{Duffy, Hirschfeld, and
  Scalapino}}]{Duffy2001}
\bibinfo{author}{\bibfnamefont{Daniel} \bibnamefont{Duffy}},
  \bibinfo{author}{\bibfnamefont{P.~J.} \bibnamefont{Hirschfeld}},
  \bibnamefont{and} \bibinfo{author}{\bibfnamefont{Douglas~J.}
  \bibnamefont{Scalapino}}, \bibinfo{journal}{Phys. Rev. B}
  \textbf{\bibinfo{volume}{64}}, \bibinfo{pages}{224522}
  (\bibinfo{year}{2001}),
  \urlprefix\url{http://link.aps.org/doi/10.1103/PhysRevB.64.224522}.

\bibitem[{\citenamefont{Atkinson and Hirschfeld}(2002)}]{Atkinson2002}
\bibinfo{author}{\bibfnamefont{W.~A.} \bibnamefont{Atkinson}} \bibnamefont{and}
  \bibinfo{author}{\bibfnamefont{P.~J.} \bibnamefont{Hirschfeld}},
  \bibinfo{journal}{Phys. Rev. Lett.} \textbf{\bibinfo{volume}{88}},
  \bibinfo{pages}{187003} (\bibinfo{year}{2002}),
  \urlprefix\url{http://link.aps.org/doi/10.1103/PhysRevLett.88.187003}.

\bibitem[{\citenamefont{Tsai and Hirschfeld}(2002)}]{Tsai2002}
\bibinfo{author}{\bibfnamefont{Shan-Wen} \bibnamefont{Tsai}} \bibnamefont{and}
  \bibinfo{author}{\bibfnamefont{P.~J.} \bibnamefont{Hirschfeld}},
  \bibinfo{journal}{Phys. Rev. Lett.} \textbf{\bibinfo{volume}{89}},
  \bibinfo{pages}{147004} (\bibinfo{year}{2002}),
  \urlprefix\url{http://link.aps.org/doi/10.1103/PhysRevLett.89.147004}.

\bibitem[{\citenamefont{Zhu et~al.}(2003)\citenamefont{Zhu, Atkinson, and
  Hirschfeld}}]{Zhu2003}
\bibinfo{author}{\bibfnamefont{Lingyin} \bibnamefont{Zhu}},
  \bibinfo{author}{\bibfnamefont{W.~A.} \bibnamefont{Atkinson}},
  \bibnamefont{and} \bibinfo{author}{\bibfnamefont{P.~J.}
  \bibnamefont{Hirschfeld}}, \bibinfo{journal}{Phys. Rev. B}
  \textbf{\bibinfo{volume}{67}}, \bibinfo{pages}{094508}
  (\bibinfo{year}{2003}),
  \urlprefix\url{http://link.aps.org/doi/10.1103/PhysRevB.67.094508}.

\bibitem[{\citenamefont{Preosti and Muzikar}(1996)}]{Muzikar1996}
\bibinfo{author}{\bibfnamefont{G.}~\bibnamefont{Preosti}} \bibnamefont{and}
  \bibinfo{author}{\bibfnamefont{P.}~\bibnamefont{Muzikar}},
  \bibinfo{journal}{Phys. Rev. B} \textbf{\bibinfo{volume}{54}},
  \bibinfo{pages}{3489} (\bibinfo{year}{1996}),
  \urlprefix\url{http://link.aps.org/doi/10.1103/PhysRevB.54.3489}.

\bibitem[{\citenamefont{Shimizu et~al.}(2001)\citenamefont{Shimizu, Kimura,
  Furomoto, Takeda, Kontani, Onuki, and Amaya}}]{Shimizu2001}
\bibinfo{author}{\bibfnamefont{Katsuya} \bibnamefont{Shimizu}},
  \bibinfo{author}{\bibfnamefont{Tomohiro} \bibnamefont{Kimura}},
  \bibinfo{author}{\bibfnamefont{Shigeyuki} \bibnamefont{Furomoto}},
  \bibinfo{author}{\bibfnamefont{Keiki} \bibnamefont{Takeda}},
  \bibinfo{author}{\bibfnamefont{Kazuyoshi} \bibnamefont{Kontani}},
  \bibinfo{author}{\bibfnamefont{Yoshichika} \bibnamefont{Onuki}},
  \bibnamefont{and} \bibinfo{author}{\bibfnamefont{Kiichi}
  \bibnamefont{Amaya}}, \bibinfo{journal}{Nature}
  \textbf{\bibinfo{volume}{412}}, \bibinfo{pages}{316} (\bibinfo{year}{2001}),
  ISSN \bibinfo{issn}{0028-0836},
  \urlprefix\url{http://dx.doi.org/10.1038/35085536}.

\bibitem[{\citenamefont{Bose et~al.}(2003)\citenamefont{Bose, Dolgov, Kortus,
  Jepsen, and Andersen}}]{Bose2003}
\bibinfo{author}{\bibfnamefont{S.~K.} \bibnamefont{Bose}},
  \bibinfo{author}{\bibfnamefont{O.~V.} \bibnamefont{Dolgov}},
  \bibinfo{author}{\bibfnamefont{J.}~\bibnamefont{Kortus}},
  \bibinfo{author}{\bibfnamefont{O.}~\bibnamefont{Jepsen}}, \bibnamefont{and}
  \bibinfo{author}{\bibfnamefont{O.~K.} \bibnamefont{Andersen}},
  \bibinfo{journal}{Phys. Rev. B} \textbf{\bibinfo{volume}{67}},
  \bibinfo{pages}{214518} (\bibinfo{year}{2003}),
  \urlprefix\url{http://link.aps.org/doi/10.1103/PhysRevB.67.214518}.

\bibitem[{\citenamefont{Leb\`egue}(2007)}]{s_lebegue_07}
\bibinfo{author}{\bibfnamefont{S.}~\bibnamefont{Leb\`egue}},
  \bibinfo{journal}{Phys. Rev. B} \textbf{\bibinfo{volume}{75}},
  \bibinfo{pages}{035110} (\bibinfo{year}{2007}),
  \urlprefix\url{http://link.aps.org/doi/10.1103/PhysRevB.75.035110}.

\bibitem[{\citenamefont{Singh and Du}(2008)}]{d_singh_08}
\bibinfo{author}{\bibfnamefont{D.~J.} \bibnamefont{Singh}} \bibnamefont{and}
  \bibinfo{author}{\bibfnamefont{M.-H.} \bibnamefont{Du}},
  \bibinfo{journal}{Phys. Rev. Lett.} \textbf{\bibinfo{volume}{100}},
  \bibinfo{pages}{237003} (\bibinfo{year}{2008}),
  \urlprefix\url{http://link.aps.org/doi/10.1103/PhysRevLett.100.237003}.

\bibitem[{\citenamefont{Cao et~al.}(2008)\citenamefont{Cao, Hirschfeld, and
  Cheng}}]{c_cao_08}
\bibinfo{author}{\bibfnamefont{Chao} \bibnamefont{Cao}},
  \bibinfo{author}{\bibfnamefont{P.~J.} \bibnamefont{Hirschfeld}},
  \bibnamefont{and} \bibinfo{author}{\bibfnamefont{Hai-Ping}
  \bibnamefont{Cheng}}, \bibinfo{journal}{Phys. Rev. B}
  \textbf{\bibinfo{volume}{77}}, \bibinfo{pages}{220506}
  (\bibinfo{year}{2008}),
  \urlprefix\url{http://link.aps.org/doi/10.1103/PhysRevB.77.220506}.

\bibitem[{\citenamefont{Kordyuk}(2012)}]{Kordyuk}
\bibinfo{author}{\bibfnamefont{A.~A.} \bibnamefont{Kordyuk}},
  \bibinfo{journal}{Low Temperature Physics} \textbf{\bibinfo{volume}{38}},
  \bibinfo{pages}{888} (\bibinfo{year}{2012}),
  \urlprefix\url{http://scitation.aip.org/content/aip/journal/ltp/38/9/10.1063/1.4752092}.

\bibitem[{\citenamefont{Brouet et~al.}(2012)\citenamefont{Brouet, Jensen, Lin,
  Taleb-Ibrahimi, Le~F\`evre, Bertran, Lin, Ku, Forget, and Colson}}]{Brouet}
\bibinfo{author}{\bibfnamefont{V.}~\bibnamefont{Brouet}},
  \bibinfo{author}{\bibfnamefont{M.~Fuglsang} \bibnamefont{Jensen}},
  \bibinfo{author}{\bibfnamefont{Ping-Hui} \bibnamefont{Lin}},
  \bibinfo{author}{\bibfnamefont{A.}~\bibnamefont{Taleb-Ibrahimi}},
  \bibinfo{author}{\bibfnamefont{P.}~\bibnamefont{Le~F\`evre}},
  \bibinfo{author}{\bibfnamefont{F.}~\bibnamefont{Bertran}},
  \bibinfo{author}{\bibfnamefont{Chia-Hui} \bibnamefont{Lin}},
  \bibinfo{author}{\bibfnamefont{Wei} \bibnamefont{Ku}},
  \bibinfo{author}{\bibfnamefont{A.}~\bibnamefont{Forget}}, \bibnamefont{and}
  \bibinfo{author}{\bibfnamefont{D.}~\bibnamefont{Colson}},
  \bibinfo{journal}{Phys. Rev. B} \textbf{\bibinfo{volume}{86}},
  \bibinfo{pages}{075123} (\bibinfo{year}{2012}),
  \urlprefix\url{http://link.aps.org/doi/10.1103/PhysRevB.86.075123}.

\bibitem[{\citenamefont{Ning et~al.}(2009)\citenamefont{Ning, Ahilan, Imai,
  Sefat, Jin, McGuire, Sales, and Mandrus}}]{f_ning_08}
\bibinfo{author}{\bibfnamefont{F.}~\bibnamefont{Ning}},
  \bibinfo{author}{\bibfnamefont{K.}~\bibnamefont{Ahilan}},
  \bibinfo{author}{\bibfnamefont{T.}~\bibnamefont{Imai}},
  \bibinfo{author}{\bibfnamefont{A.~S.} \bibnamefont{Sefat}},
  \bibinfo{author}{\bibfnamefont{R.}~\bibnamefont{Jin}},
  \bibinfo{author}{\bibfnamefont{M.~A.} \bibnamefont{McGuire}},
  \bibinfo{author}{\bibfnamefont{B.~C.} \bibnamefont{Sales}}, \bibnamefont{and}
  \bibinfo{author}{\bibfnamefont{D.}~\bibnamefont{Mandrus}},
  \bibinfo{journal}{Journal of the Physical Society of Japan}
  \textbf{\bibinfo{volume}{78}}, \bibinfo{pages}{013711}
  (\bibinfo{year}{2009}),
  \urlprefix\url{http://dx.doi.org/10.1143/JPSJ.78.013711}.

\bibitem[{\citenamefont{Grafe et~al.}(2008)\citenamefont{Grafe, Paar, Lang,
  Curro, Behr, Werner, Hamann-Borrero, Hess, Leps, Klingeler
  et~al.}}]{h_grafe_08}
\bibinfo{author}{\bibfnamefont{H.-J.} \bibnamefont{Grafe}},
  \bibinfo{author}{\bibfnamefont{D.}~\bibnamefont{Paar}},
  \bibinfo{author}{\bibfnamefont{G.}~\bibnamefont{Lang}},
  \bibinfo{author}{\bibfnamefont{N.~J.} \bibnamefont{Curro}},
  \bibinfo{author}{\bibfnamefont{G.}~\bibnamefont{Behr}},
  \bibinfo{author}{\bibfnamefont{J.}~\bibnamefont{Werner}},
  \bibinfo{author}{\bibfnamefont{J.}~\bibnamefont{Hamann-Borrero}},
  \bibinfo{author}{\bibfnamefont{C.}~\bibnamefont{Hess}},
  \bibinfo{author}{\bibfnamefont{N.}~\bibnamefont{Leps}},
  \bibinfo{author}{\bibfnamefont{R.}~\bibnamefont{Klingeler}},
  \bibnamefont{et~al.}, \bibinfo{journal}{Phys. Rev. Lett.}
  \textbf{\bibinfo{volume}{101}}, \bibinfo{pages}{047003}
  (\bibinfo{year}{2008}),
  \urlprefix\url{http://link.aps.org/doi/10.1103/PhysRevLett.101.047003}.

\bibitem[{\citenamefont{Matano et~al.}(2008)\citenamefont{Matano, Ren, Dong,
  Sun, Zhao, and qing Zheng}}]{k_matano_08}
\bibinfo{author}{\bibfnamefont{K.}~\bibnamefont{Matano}},
  \bibinfo{author}{\bibfnamefont{Z.~A.} \bibnamefont{Ren}},
  \bibinfo{author}{\bibfnamefont{X.~L.} \bibnamefont{Dong}},
  \bibinfo{author}{\bibfnamefont{L.~L.} \bibnamefont{Sun}},
  \bibinfo{author}{\bibfnamefont{Z.~X.} \bibnamefont{Zhao}}, \bibnamefont{and}
  \bibinfo{author}{\bibfnamefont{Guo} \bibnamefont{qing Zheng}},
  \bibinfo{journal}{EPL (Europhysics Letters)} \textbf{\bibinfo{volume}{83}},
  \bibinfo{pages}{57001} (\bibinfo{year}{2008}),
  \urlprefix\url{http://stacks.iop.org/0295-5075/83/i=5/a=57001}.

\bibitem[{\citenamefont{Matano et~al.}(2009)\citenamefont{Matano, Li, Sun, Sun,
  Lin, Ichioka, and qing Zheng}}]{MatanoBKFA}
\bibinfo{author}{\bibfnamefont{K.}~\bibnamefont{Matano}},
  \bibinfo{author}{\bibfnamefont{Z.}~\bibnamefont{Li}},
  \bibinfo{author}{\bibfnamefont{G.~L.} \bibnamefont{Sun}},
  \bibinfo{author}{\bibfnamefont{D.~L.} \bibnamefont{Sun}},
  \bibinfo{author}{\bibfnamefont{C.~T.} \bibnamefont{Lin}},
  \bibinfo{author}{\bibfnamefont{M.}~\bibnamefont{Ichioka}}, \bibnamefont{and}
  \bibinfo{author}{\bibfnamefont{Guo} \bibnamefont{qing Zheng}},
  \bibinfo{journal}{EPL (Europhysics Letters)} \textbf{\bibinfo{volume}{87}},
  \bibinfo{pages}{27012} (\bibinfo{year}{2009}),
  \urlprefix\url{http://stacks.iop.org/0295-5075/87/i=2/a=27012}.

\bibitem[{\citenamefont{Yashima et~al.}(2009)\citenamefont{Yashima, Nishimura,
  Mukuda, Kitaoka, Miyazawa, Shirage, Kihou, Kito, Eisaki, and
  Iyo}}]{m_yashima_09}
\bibinfo{author}{\bibfnamefont{M.}~\bibnamefont{Yashima}},
  \bibinfo{author}{\bibfnamefont{H.}~\bibnamefont{Nishimura}},
  \bibinfo{author}{\bibfnamefont{H.}~\bibnamefont{Mukuda}},
  \bibinfo{author}{\bibfnamefont{Y.}~\bibnamefont{Kitaoka}},
  \bibinfo{author}{\bibfnamefont{K.}~\bibnamefont{Miyazawa}},
  \bibinfo{author}{\bibfnamefont{P.~M.} \bibnamefont{Shirage}},
  \bibinfo{author}{\bibfnamefont{K.}~\bibnamefont{Kihou}},
  \bibinfo{author}{\bibfnamefont{H.}~\bibnamefont{Kito}},
  \bibinfo{author}{\bibfnamefont{H.}~\bibnamefont{Eisaki}}, \bibnamefont{and}
  \bibinfo{author}{\bibfnamefont{A.}~\bibnamefont{Iyo}},
  \bibinfo{journal}{Journal of the Physical Society of Japan}
  \textbf{\bibinfo{volume}{78}}, \bibinfo{pages}{103702}
  (\bibinfo{year}{2009}),
  \urlprefix\url{http://dx.doi.org/10.1143/JPSJ.78.103702}.

\bibitem[{\citenamefont{Jegli\v{c} et~al.}(2010)\citenamefont{Jegli\v{c},
  Poto\v{c}nik, Klanj\v{s}ek, Bobnar, Jagodi\v{c}, Koch, Rosner, Margadonna,
  Lv, Guloy et~al.}}]{Jeglic111}
\bibinfo{author}{\bibfnamefont{P.}~\bibnamefont{Jegli\v{c}}},
  \bibinfo{author}{\bibfnamefont{A.}~\bibnamefont{Poto\v{c}nik}},
  \bibinfo{author}{\bibfnamefont{M.}~\bibnamefont{Klanj\v{s}ek}},
  \bibinfo{author}{\bibfnamefont{M.}~\bibnamefont{Bobnar}},
  \bibinfo{author}{\bibfnamefont{M.}~\bibnamefont{Jagodi\v{c}}},
  \bibinfo{author}{\bibfnamefont{K.}~\bibnamefont{Koch}},
  \bibinfo{author}{\bibfnamefont{H.}~\bibnamefont{Rosner}},
  \bibinfo{author}{\bibfnamefont{S.}~\bibnamefont{Margadonna}},
  \bibinfo{author}{\bibfnamefont{B.}~\bibnamefont{Lv}},
  \bibinfo{author}{\bibfnamefont{A.~M.} \bibnamefont{Guloy}},
  \bibnamefont{et~al.}, \bibinfo{journal}{Phys. Rev. B}
  \textbf{\bibinfo{volume}{81}}, \bibinfo{pages}{140511}
  (\bibinfo{year}{2010}),
  \urlprefix\url{http://link.aps.org/doi/10.1103/PhysRevB.81.140511}.

\bibitem[{\citenamefont{Li et~al.}(2010)\citenamefont{Li, Ooe, Wang, Liu, Jin,
  Ichioka, and Zheng}}]{Li111}
\bibinfo{author}{\bibfnamefont{Z.}~\bibnamefont{Li}},
  \bibinfo{author}{\bibfnamefont{Y.}~\bibnamefont{Ooe}},
  \bibinfo{author}{\bibfnamefont{X.-C.} \bibnamefont{Wang}},
  \bibinfo{author}{\bibfnamefont{Q.-Q.} \bibnamefont{Liu}},
  \bibinfo{author}{\bibfnamefont{C.-Q.} \bibnamefont{Jin}},
  \bibinfo{author}{\bibfnamefont{M.}~\bibnamefont{Ichioka}}, \bibnamefont{and}
  \bibinfo{author}{\bibfnamefont{G.-q.} \bibnamefont{Zheng}},
  \bibinfo{journal}{Journal of the Physical Society of Japan}
  \textbf{\bibinfo{volume}{79}}, \bibinfo{pages}{083702}
  (\bibinfo{year}{2010}),
  \urlprefix\url{http://dx.doi.org/10.1143/JPSJ.79.083702}.

\bibitem[{\citenamefont{Nakai et~al.}(2010{\natexlab{a}})\citenamefont{Nakai,
  Iye, Kitagawa, Ishida, Kasahara, Shibauchi, Matsuda, and
  Terashima}}]{Nakai2010}
\bibinfo{author}{\bibfnamefont{Y.}~\bibnamefont{Nakai}},
  \bibinfo{author}{\bibfnamefont{T.}~\bibnamefont{Iye}},
  \bibinfo{author}{\bibfnamefont{S.}~\bibnamefont{Kitagawa}},
  \bibinfo{author}{\bibfnamefont{K.}~\bibnamefont{Ishida}},
  \bibinfo{author}{\bibfnamefont{S.}~\bibnamefont{Kasahara}},
  \bibinfo{author}{\bibfnamefont{T.}~\bibnamefont{Shibauchi}},
  \bibinfo{author}{\bibfnamefont{Y.}~\bibnamefont{Matsuda}}, \bibnamefont{and}
  \bibinfo{author}{\bibfnamefont{T.}~\bibnamefont{Terashima}},
  \bibinfo{journal}{Phys. Rev. B} \textbf{\bibinfo{volume}{81}},
  \bibinfo{pages}{020503} (\bibinfo{year}{2010}{\natexlab{a}}),
  \urlprefix\url{http://link.aps.org/doi/10.1103/PhysRevB.81.020503}.

\bibitem[{\citenamefont{Zhang et~al.}(2009)\citenamefont{Zhang, Oh, Liu, Yan,
  Kim, Greene, and Takeuchi}}]{Greene_tunneling}
\bibinfo{author}{\bibfnamefont{X.}~\bibnamefont{Zhang}},
  \bibinfo{author}{\bibfnamefont{Y.~S.} \bibnamefont{Oh}},
  \bibinfo{author}{\bibfnamefont{Y.}~\bibnamefont{Liu}},
  \bibinfo{author}{\bibfnamefont{L.}~\bibnamefont{Yan}},
  \bibinfo{author}{\bibfnamefont{K.~H.} \bibnamefont{Kim}},
  \bibinfo{author}{\bibfnamefont{R.~L.} \bibnamefont{Greene}},
  \bibnamefont{and} \bibinfo{author}{\bibfnamefont{I.}~\bibnamefont{Takeuchi}},
  \bibinfo{journal}{Phys. Rev. Lett.} \textbf{\bibinfo{volume}{102}},
  \bibinfo{pages}{147002} (\bibinfo{year}{2009}),
  \urlprefix\url{http://link.aps.org/doi/10.1103/PhysRevLett.102.147002}.

\bibitem[{\citenamefont{Hicks et~al.}(2009)\citenamefont{Hicks, Lippman, Huber,
  Ren, Yang, Zhao, and Moler}}]{KAM}
\bibinfo{author}{\bibfnamefont{C.~W.} \bibnamefont{Hicks}},
  \bibinfo{author}{\bibfnamefont{T.~M.} \bibnamefont{Lippman}},
  \bibinfo{author}{\bibfnamefont{M.~E.} \bibnamefont{Huber}},
  \bibinfo{author}{\bibfnamefont{Z.-A.} \bibnamefont{Ren}},
  \bibinfo{author}{\bibfnamefont{J.}~\bibnamefont{Yang}},
  \bibinfo{author}{\bibfnamefont{Z.-X.} \bibnamefont{Zhao}}, \bibnamefont{and}
  \bibinfo{author}{\bibfnamefont{K.~A.} \bibnamefont{Moler}},
  \bibinfo{journal}{Journal of the Physical Society of Japan}
  \textbf{\bibinfo{volume}{78}}, \bibinfo{pages}{013708}
  (\bibinfo{year}{2009}),
  \urlprefix\url{http://dx.doi.org/10.1143/JPSJ.78.013708}.

\bibitem[{\citenamefont{Geshkenbein and Larkin}(1986)}]{Geshkenbein1986eng}
\bibinfo{author}{\bibfnamefont{V.~B.} \bibnamefont{Geshkenbein}}
  \bibnamefont{and} \bibinfo{author}{\bibfnamefont{A.~I.}
  \bibnamefont{Larkin}}, \bibinfo{journal}{JETP Lett.}
  \textbf{\bibinfo{volume}{43}}, \bibinfo{pages}{395} (\bibinfo{year}{1986}),
  \urlprefix\url{http://www.jetpletters.ac.ru/ps/1404/article_21329.shtml}.

\bibitem[{\citenamefont{Maier et~al.}(2011)\citenamefont{Maier, Graser,
  Hirschfeld, and Scalapino}}]{Maier2011}
\bibinfo{author}{\bibfnamefont{T.~A.} \bibnamefont{Maier}},
  \bibinfo{author}{\bibfnamefont{S.}~\bibnamefont{Graser}},
  \bibinfo{author}{\bibfnamefont{P.~J.} \bibnamefont{Hirschfeld}},
  \bibnamefont{and} \bibinfo{author}{\bibfnamefont{D.~J.}
  \bibnamefont{Scalapino}}, \bibinfo{journal}{Phys. Rev. B}
  \textbf{\bibinfo{volume}{83}}, \bibinfo{pages}{100515}
  (\bibinfo{year}{2011}),
  \urlprefix\url{http://link.aps.org/doi/10.1103/PhysRevB.83.100515}.

\bibitem[{\citenamefont{Wang et~al.}(2011)\citenamefont{Wang, Yang, Gao, Lu,
  Xiang, and Lee}}]{Wang2011}
\bibinfo{author}{\bibfnamefont{F.}~\bibnamefont{Wang}},
  \bibinfo{author}{\bibfnamefont{F.}~\bibnamefont{Yang}},
  \bibinfo{author}{\bibfnamefont{M.}~\bibnamefont{Gao}},
  \bibinfo{author}{\bibfnamefont{Z.-Y.} \bibnamefont{Lu}},
  \bibinfo{author}{\bibfnamefont{T.}~\bibnamefont{Xiang}}, \bibnamefont{and}
  \bibinfo{author}{\bibfnamefont{D.-H.} \bibnamefont{Lee}},
  \bibinfo{journal}{EPL (Europhysics Letters)} \textbf{\bibinfo{volume}{93}},
  \bibinfo{pages}{57003} (\bibinfo{year}{2011}),
  \urlprefix\url{http://stacks.iop.org/0295-5075/93/i=5/a=57003}.

\bibitem[{\citenamefont{Das and Balatsky}(2011)}]{Das2011}
\bibinfo{author}{\bibfnamefont{T.}~\bibnamefont{Das}} \bibnamefont{and}
  \bibinfo{author}{\bibfnamefont{A.~V.} \bibnamefont{Balatsky}},
  \bibinfo{journal}{Phys. Rev. B} \textbf{\bibinfo{volume}{84}},
  \bibinfo{pages}{014521} (\bibinfo{year}{2011}),
  \urlprefix\url{http://link.aps.org/doi/10.1103/PhysRevB.84.014521}.

\bibitem[{\citenamefont{Maiti et~al.}(2011{\natexlab{b}})\citenamefont{Maiti,
  Korshunov, Maier, Hirschfeld, and Chubukov}}]{MaitiKorshunovPRL2011}
\bibinfo{author}{\bibfnamefont{S.}~\bibnamefont{Maiti}},
  \bibinfo{author}{\bibfnamefont{M.~M.} \bibnamefont{Korshunov}},
  \bibinfo{author}{\bibfnamefont{T.~A.} \bibnamefont{Maier}},
  \bibinfo{author}{\bibfnamefont{P.~J.} \bibnamefont{Hirschfeld}},
  \bibnamefont{and} \bibinfo{author}{\bibfnamefont{A.~V.}
  \bibnamefont{Chubukov}}, \bibinfo{journal}{Phys. Rev. Lett.}
  \textbf{\bibinfo{volume}{107}}, \bibinfo{pages}{147002}
  (\bibinfo{year}{2011}{\natexlab{b}}),
  \urlprefix\url{http://link.aps.org/doi/10.1103/PhysRevLett.107.147002}.

\bibitem[{\citenamefont{Mazin}(2011)}]{Mazin2011}
\bibinfo{author}{\bibfnamefont{I.~I.} \bibnamefont{Mazin}},
  \bibinfo{journal}{Phys. Rev. B} \textbf{\bibinfo{volume}{84}},
  \bibinfo{pages}{024529} (\bibinfo{year}{2011}),
  \urlprefix\url{http://link.aps.org/doi/10.1103/PhysRevB.84.024529}.

\bibitem[{\citenamefont{Mishra et~al.}(2009)\citenamefont{Mishra, Boyd, Graser,
  Maier, Hirschfeld, and Scalapino}}]{v_mishra_09}
\bibinfo{author}{\bibfnamefont{V.}~\bibnamefont{Mishra}},
  \bibinfo{author}{\bibfnamefont{G.}~\bibnamefont{Boyd}},
  \bibinfo{author}{\bibfnamefont{S.}~\bibnamefont{Graser}},
  \bibinfo{author}{\bibfnamefont{T.}~\bibnamefont{Maier}},
  \bibinfo{author}{\bibfnamefont{P.~J.} \bibnamefont{Hirschfeld}},
  \bibnamefont{and} \bibinfo{author}{\bibfnamefont{D.~J.}
  \bibnamefont{Scalapino}}, \bibinfo{journal}{Phys. Rev. B}
  \textbf{\bibinfo{volume}{79}}, \bibinfo{pages}{094512}
  (\bibinfo{year}{2009}).

\bibitem[{\citenamefont{Mizukami et~al.}(2014)\citenamefont{Mizukami,
  Konczykowski, Kawamoto, Kurata, Kasahara, Hashimoto, Mishra, Kreisel, Wang,
  Hirschfeld et~al.}}]{Mizukami2014}
\bibinfo{author}{\bibfnamefont{Y.}~\bibnamefont{Mizukami}},
  \bibinfo{author}{\bibfnamefont{M.}~\bibnamefont{Konczykowski}},
  \bibinfo{author}{\bibfnamefont{Y.}~\bibnamefont{Kawamoto}},
  \bibinfo{author}{\bibfnamefont{S.}~\bibnamefont{Kurata}},
  \bibinfo{author}{\bibfnamefont{S.}~\bibnamefont{Kasahara}},
  \bibinfo{author}{\bibfnamefont{K.}~\bibnamefont{Hashimoto}},
  \bibinfo{author}{\bibfnamefont{V.}~\bibnamefont{Mishra}},
  \bibinfo{author}{\bibfnamefont{A.}~\bibnamefont{Kreisel}},
  \bibinfo{author}{\bibfnamefont{Y.}~\bibnamefont{Wang}},
  \bibinfo{author}{\bibfnamefont{P.~J.} \bibnamefont{Hirschfeld}},
  \bibnamefont{et~al.}, \bibinfo{journal}{Nat. Commun.}
  \textbf{\bibinfo{volume}{5}}, \bibinfo{pages}{5657} (\bibinfo{year}{2014}),
  \urlprefix\url{http://dx.doi.org/10.1038/ncomms6657}.

\bibitem[{\citenamefont{Karkin et~al.}(2009)\citenamefont{Karkin, Werner, Behr,
  and Goshchitskii}}]{Karkin2009}
\bibinfo{author}{\bibfnamefont{A.~E.} \bibnamefont{Karkin}},
  \bibinfo{author}{\bibfnamefont{J.}~\bibnamefont{Werner}},
  \bibinfo{author}{\bibfnamefont{G.}~\bibnamefont{Behr}}, \bibnamefont{and}
  \bibinfo{author}{\bibfnamefont{B.~N.} \bibnamefont{Goshchitskii}},
  \bibinfo{journal}{Phys. Rev. B} \textbf{\bibinfo{volume}{80}},
  \bibinfo{pages}{174512} (\bibinfo{year}{2009}),
  \urlprefix\url{http://link.aps.org/doi/10.1103/PhysRevB.80.174512}.

\bibitem[{\citenamefont{Cheng et~al.}(2010)\citenamefont{Cheng, Shen, Hu, and
  Wen}}]{Cheng2010}
\bibinfo{author}{\bibfnamefont{P.}~\bibnamefont{Cheng}},
  \bibinfo{author}{\bibfnamefont{B.}~\bibnamefont{Shen}},
  \bibinfo{author}{\bibfnamefont{J.}~\bibnamefont{Hu}}, \bibnamefont{and}
  \bibinfo{author}{\bibfnamefont{H.-H.} \bibnamefont{Wen}},
  \bibinfo{journal}{Phys. Rev. B} \textbf{\bibinfo{volume}{81}},
  \bibinfo{pages}{174529} (\bibinfo{year}{2010}),
  \urlprefix\url{http://link.aps.org/doi/10.1103/PhysRevB.81.174529}.

\bibitem[{\citenamefont{Li et~al.}(2012{\natexlab{a}})\citenamefont{Li, Tong,
  Tao, Feng, Cao, Chen, chun Zhang, and an~Xu}}]{Li2010}
\bibinfo{author}{\bibfnamefont{Yuke} \bibnamefont{Li}},
  \bibinfo{author}{\bibfnamefont{Jun} \bibnamefont{Tong}},
  \bibinfo{author}{\bibfnamefont{Qian} \bibnamefont{Tao}},
  \bibinfo{author}{\bibfnamefont{Chunmu} \bibnamefont{Feng}},
  \bibinfo{author}{\bibfnamefont{Guanghan} \bibnamefont{Cao}},
  \bibinfo{author}{\bibfnamefont{Weiqiang} \bibnamefont{Chen}},
  \bibinfo{author}{\bibfnamefont{Fu}~\bibnamefont{chun Zhang}},
  \bibnamefont{and} \bibinfo{author}{\bibfnamefont{Zhu} \bibnamefont{an~Xu}},
  \bibinfo{journal}{New Journal of Physics} \textbf{\bibinfo{volume}{12}},
  \bibinfo{pages}{083008} (\bibinfo{year}{2012}{\natexlab{a}}),
  \urlprefix\url{http://stacks.iop.org/1367-2630/12/i=8/a=083008}.

\bibitem[{\citenamefont{Nakajima et~al.}(2010)\citenamefont{Nakajima, Taen,
  Tsuchiya, Tamegai, Kitamura, and Murakami}}]{Nakajima2010}
\bibinfo{author}{\bibfnamefont{Y.}~\bibnamefont{Nakajima}},
  \bibinfo{author}{\bibfnamefont{T.}~\bibnamefont{Taen}},
  \bibinfo{author}{\bibfnamefont{Y.}~\bibnamefont{Tsuchiya}},
  \bibinfo{author}{\bibfnamefont{T.}~\bibnamefont{Tamegai}},
  \bibinfo{author}{\bibfnamefont{H.}~\bibnamefont{Kitamura}}, \bibnamefont{and}
  \bibinfo{author}{\bibfnamefont{T.}~\bibnamefont{Murakami}},
  \bibinfo{journal}{Phys. Rev. B} \textbf{\bibinfo{volume}{82}},
  \bibinfo{pages}{220504} (\bibinfo{year}{2010}),
  \urlprefix\url{http://link.aps.org/doi/10.1103/PhysRevB.82.220504}.

\bibitem[{\citenamefont{Tropeano et~al.}(2010)\citenamefont{Tropeano, Cimberle,
  Ferdeghini, Lamura, Martinelli, Palenzona, Pallecchi, Sala, Sheikin,
  Bernardini et~al.}}]{Tropeano2010}
\bibinfo{author}{\bibfnamefont{M.}~\bibnamefont{Tropeano}},
  \bibinfo{author}{\bibfnamefont{M.~R.} \bibnamefont{Cimberle}},
  \bibinfo{author}{\bibfnamefont{C.}~\bibnamefont{Ferdeghini}},
  \bibinfo{author}{\bibfnamefont{G.}~\bibnamefont{Lamura}},
  \bibinfo{author}{\bibfnamefont{A.}~\bibnamefont{Martinelli}},
  \bibinfo{author}{\bibfnamefont{A.}~\bibnamefont{Palenzona}},
  \bibinfo{author}{\bibfnamefont{I.}~\bibnamefont{Pallecchi}},
  \bibinfo{author}{\bibfnamefont{A.}~\bibnamefont{Sala}},
  \bibinfo{author}{\bibfnamefont{I.}~\bibnamefont{Sheikin}},
  \bibinfo{author}{\bibfnamefont{F.}~\bibnamefont{Bernardini}},
  \bibnamefont{et~al.}, \bibinfo{journal}{Phys. Rev. B}
  \textbf{\bibinfo{volume}{81}}, \bibinfo{pages}{184504}
  (\bibinfo{year}{2010}),
  \urlprefix\url{http://link.aps.org/doi/10.1103/PhysRevB.81.184504}.

\bibitem[{\citenamefont{Kim et~al.}(2014)\citenamefont{Kim, Tanatar, Liu, Sims,
  Zhang, Dai, Lograsso, and Prozorov}}]{Kim2014}
\bibinfo{author}{\bibfnamefont{H.}~\bibnamefont{Kim}},
  \bibinfo{author}{\bibfnamefont{M.~A.} \bibnamefont{Tanatar}},
  \bibinfo{author}{\bibfnamefont{Yong} \bibnamefont{Liu}},
  \bibinfo{author}{\bibfnamefont{Zachary~Cole} \bibnamefont{Sims}},
  \bibinfo{author}{\bibfnamefont{Chenglin} \bibnamefont{Zhang}},
  \bibinfo{author}{\bibfnamefont{Pengcheng} \bibnamefont{Dai}},
  \bibinfo{author}{\bibfnamefont{T.~A.} \bibnamefont{Lograsso}},
  \bibnamefont{and} \bibinfo{author}{\bibfnamefont{R.}~\bibnamefont{Prozorov}},
  \bibinfo{journal}{Phys. Rev. B} \textbf{\bibinfo{volume}{89}},
  \bibinfo{pages}{174519} (\bibinfo{year}{2014}),
  \urlprefix\url{http://link.aps.org/doi/10.1103/PhysRevB.89.174519}.

\bibitem[{\citenamefont{Prozorov et~al.}(2014)\citenamefont{Prozorov,
  Ko\ifmmode~\acute{n}\else \'{n}\fi{}czykowski, Tanatar, Thaler, Bud'ko,
  Canfield, Mishra, and Hirschfeld}}]{Prozorov2014}
\bibinfo{author}{\bibfnamefont{R.}~\bibnamefont{Prozorov}},
  \bibinfo{author}{\bibfnamefont{M.}~\bibnamefont{Ko\ifmmode~\acute{n}\else
  \'{n}\fi{}czykowski}}, \bibinfo{author}{\bibfnamefont{M.~A.}
  \bibnamefont{Tanatar}},
  \bibinfo{author}{\bibfnamefont{A.}~\bibnamefont{Thaler}},
  \bibinfo{author}{\bibfnamefont{S.~L.} \bibnamefont{Bud'ko}},
  \bibinfo{author}{\bibfnamefont{P.~C.} \bibnamefont{Canfield}},
  \bibinfo{author}{\bibfnamefont{V.}~\bibnamefont{Mishra}}, \bibnamefont{and}
  \bibinfo{author}{\bibfnamefont{P.~J.} \bibnamefont{Hirschfeld}},
  \bibinfo{journal}{Phys. Rev. X} \textbf{\bibinfo{volume}{4}},
  \bibinfo{pages}{041032} (\bibinfo{year}{2014}),
  \urlprefix\url{http://link.aps.org/doi/10.1103/PhysRevX.4.041032}.

\bibitem[{\citenamefont{Tarantini et~al.}(2010)\citenamefont{Tarantini, Putti,
  Gurevich, Shen, Singh, Rowell, Newman, Larbalestier, Cheng, Jia
  et~al.}}]{Tarantini2010}
\bibinfo{author}{\bibfnamefont{C.}~\bibnamefont{Tarantini}},
  \bibinfo{author}{\bibfnamefont{M.}~\bibnamefont{Putti}},
  \bibinfo{author}{\bibfnamefont{A.}~\bibnamefont{Gurevich}},
  \bibinfo{author}{\bibfnamefont{Y.}~\bibnamefont{Shen}},
  \bibinfo{author}{\bibfnamefont{R.~K.} \bibnamefont{Singh}},
  \bibinfo{author}{\bibfnamefont{J.~M.} \bibnamefont{Rowell}},
  \bibinfo{author}{\bibfnamefont{N.}~\bibnamefont{Newman}},
  \bibinfo{author}{\bibfnamefont{D.~C.} \bibnamefont{Larbalestier}},
  \bibinfo{author}{\bibfnamefont{Peng} \bibnamefont{Cheng}},
  \bibinfo{author}{\bibfnamefont{Ying} \bibnamefont{Jia}},
  \bibnamefont{et~al.}, \bibinfo{journal}{Phys. Rev. Lett.}
  \textbf{\bibinfo{volume}{104}}, \bibinfo{pages}{087002}
  (\bibinfo{year}{2010}),
  \urlprefix\url{http://link.aps.org/doi/10.1103/PhysRevLett.104.087002}.

\bibitem[{\citenamefont{Tan et~al.}(2011)\citenamefont{Tan, Zhang, Xi, Ling,
  Zhang, Tong, Yu, Feng, Yu, Pi et~al.}}]{Tan2011}
\bibinfo{author}{\bibfnamefont{D.}~\bibnamefont{Tan}},
  \bibinfo{author}{\bibfnamefont{C.}~\bibnamefont{Zhang}},
  \bibinfo{author}{\bibfnamefont{C.}~\bibnamefont{Xi}},
  \bibinfo{author}{\bibfnamefont{L.}~\bibnamefont{Ling}},
  \bibinfo{author}{\bibfnamefont{L.}~\bibnamefont{Zhang}},
  \bibinfo{author}{\bibfnamefont{W.}~\bibnamefont{Tong}},
  \bibinfo{author}{\bibfnamefont{Y.}~\bibnamefont{Yu}},
  \bibinfo{author}{\bibfnamefont{G.}~\bibnamefont{Feng}},
  \bibinfo{author}{\bibfnamefont{H.}~\bibnamefont{Yu}},
  \bibinfo{author}{\bibfnamefont{L.}~\bibnamefont{Pi}}, \bibnamefont{et~al.},
  \bibinfo{journal}{Phys. Rev. B} \textbf{\bibinfo{volume}{84}},
  \bibinfo{pages}{014502} (\bibinfo{year}{2011}),
  \urlprefix\url{http://link.aps.org/doi/10.1103/PhysRevB.84.014502}.

\bibitem[{\citenamefont{Grinenko et~al.}(2011)\citenamefont{Grinenko, Kikoin,
  Drechsler, Fuchs, Nenkov, Wurmehl, Hammerath, Lang, Grafe, Holzapfel
  et~al.}}]{Grinenko2011}
\bibinfo{author}{\bibfnamefont{V.}~\bibnamefont{Grinenko}},
  \bibinfo{author}{\bibfnamefont{K.}~\bibnamefont{Kikoin}},
  \bibinfo{author}{\bibfnamefont{S.-L.} \bibnamefont{Drechsler}},
  \bibinfo{author}{\bibfnamefont{G.}~\bibnamefont{Fuchs}},
  \bibinfo{author}{\bibfnamefont{K.}~\bibnamefont{Nenkov}},
  \bibinfo{author}{\bibfnamefont{S.}~\bibnamefont{Wurmehl}},
  \bibinfo{author}{\bibfnamefont{F.}~\bibnamefont{Hammerath}},
  \bibinfo{author}{\bibfnamefont{G.}~\bibnamefont{Lang}},
  \bibinfo{author}{\bibfnamefont{H.-J.} \bibnamefont{Grafe}},
  \bibinfo{author}{\bibfnamefont{B.}~\bibnamefont{Holzapfel}},
  \bibnamefont{et~al.}, \bibinfo{journal}{Phys. Rev. B}
  \textbf{\bibinfo{volume}{84}}, \bibinfo{pages}{134516}
  (\bibinfo{year}{2011}),
  \urlprefix\url{http://link.aps.org/doi/10.1103/PhysRevB.84.134516}.

\bibitem[{\citenamefont{Li et~al.}(2012{\natexlab{b}})\citenamefont{Li, Guo,
  Zhang, Yuan, Tsujimoto, Wang, Sathish, Sun, Yu, Yi et~al.}}]{Li2012}
\bibinfo{author}{\bibfnamefont{J.}~\bibnamefont{Li}},
  \bibinfo{author}{\bibfnamefont{Y.~F.} \bibnamefont{Guo}},
  \bibinfo{author}{\bibfnamefont{S.~B.} \bibnamefont{Zhang}},
  \bibinfo{author}{\bibfnamefont{J.}~\bibnamefont{Yuan}},
  \bibinfo{author}{\bibfnamefont{Y.}~\bibnamefont{Tsujimoto}},
  \bibinfo{author}{\bibfnamefont{X.}~\bibnamefont{Wang}},
  \bibinfo{author}{\bibfnamefont{C.~I.} \bibnamefont{Sathish}},
  \bibinfo{author}{\bibfnamefont{Y.}~\bibnamefont{Sun}},
  \bibinfo{author}{\bibfnamefont{S.}~\bibnamefont{Yu}},
  \bibinfo{author}{\bibfnamefont{W.}~\bibnamefont{Yi}}, \bibnamefont{et~al.},
  \bibinfo{journal}{Phys. Rev. B} \textbf{\bibinfo{volume}{85}},
  \bibinfo{pages}{214509} (\bibinfo{year}{2012}{\natexlab{b}}),
  \urlprefix\url{http://link.aps.org/doi/10.1103/PhysRevB.85.214509}.

\bibitem[{\citenamefont{Schilling et~al.}(2016)\citenamefont{Schilling,
  Baumgartner, Gorshunov, Zhukova, Dravin, Mitsen, Efremov, Dolgov, Iida,
  Dressel et~al.}}]{Schilling2016}
\bibinfo{author}{\bibfnamefont{M.~B.} \bibnamefont{Schilling}},
  \bibinfo{author}{\bibfnamefont{A.}~\bibnamefont{Baumgartner}},
  \bibinfo{author}{\bibfnamefont{B.}~\bibnamefont{Gorshunov}},
  \bibinfo{author}{\bibfnamefont{E.~S.} \bibnamefont{Zhukova}},
  \bibinfo{author}{\bibfnamefont{V.~A.} \bibnamefont{Dravin}},
  \bibinfo{author}{\bibfnamefont{K.~V.} \bibnamefont{Mitsen}},
  \bibinfo{author}{\bibfnamefont{D.~V.} \bibnamefont{Efremov}},
  \bibinfo{author}{\bibfnamefont{O.~V.} \bibnamefont{Dolgov}},
  \bibinfo{author}{\bibfnamefont{K.}~\bibnamefont{Iida}},
  \bibinfo{author}{\bibfnamefont{M.}~\bibnamefont{Dressel}},
  \bibnamefont{et~al.}, \bibinfo{journal}{Phys. Rev. B}
  \textbf{\bibinfo{volume}{93}}, \bibinfo{pages}{174515}
  (\bibinfo{year}{2016}),
  \urlprefix\url{http://link.aps.org/doi/10.1103/PhysRevB.93.174515}.

\bibitem[{\citenamefont{Smylie et~al.}(2016)\citenamefont{Smylie, Leroux,
  Mishra, Fang, Taddei, Chmaissem, Claus, Kayani, Snezhko, Welp
  et~al.}}]{Smylie2016}
\bibinfo{author}{\bibfnamefont{M.~P.} \bibnamefont{Smylie}},
  \bibinfo{author}{\bibfnamefont{M.}~\bibnamefont{Leroux}},
  \bibinfo{author}{\bibfnamefont{V.}~\bibnamefont{Mishra}},
  \bibinfo{author}{\bibfnamefont{L.}~\bibnamefont{Fang}},
  \bibinfo{author}{\bibfnamefont{K.~M.} \bibnamefont{Taddei}},
  \bibinfo{author}{\bibfnamefont{O.}~\bibnamefont{Chmaissem}},
  \bibinfo{author}{\bibfnamefont{H.}~\bibnamefont{Claus}},
  \bibinfo{author}{\bibfnamefont{A.}~\bibnamefont{Kayani}},
  \bibinfo{author}{\bibfnamefont{A.}~\bibnamefont{Snezhko}},
  \bibinfo{author}{\bibfnamefont{U.}~\bibnamefont{Welp}}, \bibnamefont{et~al.},
  \bibinfo{journal}{Phys. Rev. B} \textbf{\bibinfo{volume}{93}},
  \bibinfo{pages}{115119} (\bibinfo{year}{2016}),
  \urlprefix\url{http://link.aps.org/doi/10.1103/PhysRevB.93.115119}.

\bibitem[{\citenamefont{Strehlow et~al.}(2014)\citenamefont{Strehlow,
  Ko\ifmmode~\acute{n}\else \'{n}\fi{}czykowski, Murphy, Teknowijoyo, Cho,
  Tanatar, Kobayashi, Miyasaka, Tajima, and Prozorov}}]{Strehlow2014}
\bibinfo{author}{\bibfnamefont{C.~P.} \bibnamefont{Strehlow}},
  \bibinfo{author}{\bibfnamefont{M.}~\bibnamefont{Ko\ifmmode~\acute{n}\else
  \'{n}\fi{}czykowski}}, \bibinfo{author}{\bibfnamefont{J.~A.}
  \bibnamefont{Murphy}},
  \bibinfo{author}{\bibfnamefont{S.}~\bibnamefont{Teknowijoyo}},
  \bibinfo{author}{\bibfnamefont{K.}~\bibnamefont{Cho}},
  \bibinfo{author}{\bibfnamefont{M.~A.} \bibnamefont{Tanatar}},
  \bibinfo{author}{\bibfnamefont{T.}~\bibnamefont{Kobayashi}},
  \bibinfo{author}{\bibfnamefont{S.}~\bibnamefont{Miyasaka}},
  \bibinfo{author}{\bibfnamefont{S.}~\bibnamefont{Tajima}}, \bibnamefont{and}
  \bibinfo{author}{\bibfnamefont{R.}~\bibnamefont{Prozorov}},
  \bibinfo{journal}{Phys. Rev. B} \textbf{\bibinfo{volume}{90}},
  \bibinfo{pages}{020508} (\bibinfo{year}{2014}),
  \urlprefix\url{http://link.aps.org/doi/10.1103/PhysRevB.90.020508}.

\bibitem[{\citenamefont{Cho et~al.}(2014)\citenamefont{Cho,
  Ko\ifmmode~\acute{n}\else \'{n}\fi{}czykowski, Murphy, Kim, Tanatar,
  Straszheim, Shen, Wen, and Prozorov}}]{Cho2014}
\bibinfo{author}{\bibfnamefont{K.}~\bibnamefont{Cho}},
  \bibinfo{author}{\bibfnamefont{M.}~\bibnamefont{Ko\ifmmode~\acute{n}\else
  \'{n}\fi{}czykowski}},
  \bibinfo{author}{\bibfnamefont{J.}~\bibnamefont{Murphy}},
  \bibinfo{author}{\bibfnamefont{H.}~\bibnamefont{Kim}},
  \bibinfo{author}{\bibfnamefont{M.~A.} \bibnamefont{Tanatar}},
  \bibinfo{author}{\bibfnamefont{W.~E.} \bibnamefont{Straszheim}},
  \bibinfo{author}{\bibfnamefont{B.}~\bibnamefont{Shen}},
  \bibinfo{author}{\bibfnamefont{H.~H.} \bibnamefont{Wen}}, \bibnamefont{and}
  \bibinfo{author}{\bibfnamefont{R.}~\bibnamefont{Prozorov}},
  \bibinfo{journal}{Phys. Rev. B} \textbf{\bibinfo{volume}{90}},
  \bibinfo{pages}{104514} (\bibinfo{year}{2014}),
  \urlprefix\url{http://link.aps.org/doi/10.1103/PhysRevB.90.104514}.

\bibitem[{\citenamefont{Kim et~al.}(2010)\citenamefont{Kim, Gordon, Tanatar,
  Hua, Welp, Kwok, Ni, Bud'ko, Canfield, Vorontsov et~al.}}]{Kim2010}
\bibinfo{author}{\bibfnamefont{H.}~\bibnamefont{Kim}},
  \bibinfo{author}{\bibfnamefont{R.~T.} \bibnamefont{Gordon}},
  \bibinfo{author}{\bibfnamefont{M.~A.} \bibnamefont{Tanatar}},
  \bibinfo{author}{\bibfnamefont{J.}~\bibnamefont{Hua}},
  \bibinfo{author}{\bibfnamefont{U.}~\bibnamefont{Welp}},
  \bibinfo{author}{\bibfnamefont{W.~K.} \bibnamefont{Kwok}},
  \bibinfo{author}{\bibfnamefont{N.}~\bibnamefont{Ni}},
  \bibinfo{author}{\bibfnamefont{S.~L.} \bibnamefont{Bud'ko}},
  \bibinfo{author}{\bibfnamefont{P.~C.} \bibnamefont{Canfield}},
  \bibinfo{author}{\bibfnamefont{A.~B.} \bibnamefont{Vorontsov}},
  \bibnamefont{et~al.}, \bibinfo{journal}{Phys. Rev. B}
  \textbf{\bibinfo{volume}{82}}, \bibinfo{pages}{060518}
  (\bibinfo{year}{2010}),
  \urlprefix\url{http://link.aps.org/doi/10.1103/PhysRevB.82.060518}.

\bibitem[{\citenamefont{Murphy et~al.}(2013)\citenamefont{Murphy, Tanatar, Kim,
  Kwok, Welp, Graf, Brooks, Bud'ko, Canfield, and Prozorov}}]{Murphy2013}
\bibinfo{author}{\bibfnamefont{J.}~\bibnamefont{Murphy}},
  \bibinfo{author}{\bibfnamefont{M.~A.} \bibnamefont{Tanatar}},
  \bibinfo{author}{\bibfnamefont{Hyunsoo} \bibnamefont{Kim}},
  \bibinfo{author}{\bibfnamefont{W.}~\bibnamefont{Kwok}},
  \bibinfo{author}{\bibfnamefont{U.}~\bibnamefont{Welp}},
  \bibinfo{author}{\bibfnamefont{D.}~\bibnamefont{Graf}},
  \bibinfo{author}{\bibfnamefont{J.~S.} \bibnamefont{Brooks}},
  \bibinfo{author}{\bibfnamefont{S.~L.} \bibnamefont{Bud'ko}},
  \bibinfo{author}{\bibfnamefont{P.~C.} \bibnamefont{Canfield}},
  \bibnamefont{and} \bibinfo{author}{\bibfnamefont{R.}~\bibnamefont{Prozorov}},
  \bibinfo{journal}{Phys. Rev. B} \textbf{\bibinfo{volume}{88}},
  \bibinfo{pages}{054514} (\bibinfo{year}{2013}),
  \urlprefix\url{http://link.aps.org/doi/10.1103/PhysRevB.88.054514}.

\bibitem[{\citenamefont{Salovich et~al.}(2013)\citenamefont{Salovich, Kim,
  Ghosh, Giannetta, Kwok, Welp, Shen, Zhu, Wen, Tanatar et~al.}}]{Salovich2013}
\bibinfo{author}{\bibfnamefont{N.~W.} \bibnamefont{Salovich}},
  \bibinfo{author}{\bibfnamefont{Hyunsoo} \bibnamefont{Kim}},
  \bibinfo{author}{\bibfnamefont{Ajay~K.} \bibnamefont{Ghosh}},
  \bibinfo{author}{\bibfnamefont{R.~W.} \bibnamefont{Giannetta}},
  \bibinfo{author}{\bibfnamefont{W.}~\bibnamefont{Kwok}},
  \bibinfo{author}{\bibfnamefont{U.}~\bibnamefont{Welp}},
  \bibinfo{author}{\bibfnamefont{B.}~\bibnamefont{Shen}},
  \bibinfo{author}{\bibfnamefont{S.}~\bibnamefont{Zhu}},
  \bibinfo{author}{\bibfnamefont{H.-H.} \bibnamefont{Wen}},
  \bibinfo{author}{\bibfnamefont{M.~A.} \bibnamefont{Tanatar}},
  \bibnamefont{et~al.}, \bibinfo{journal}{Phys. Rev. B}
  \textbf{\bibinfo{volume}{87}}, \bibinfo{pages}{180502}
  (\bibinfo{year}{2013}),
  \urlprefix\url{http://link.aps.org/doi/10.1103/PhysRevB.87.180502}.

\bibitem[{\citenamefont{{Gerashenko} et~al.}(2009)\citenamefont{{Gerashenko},
  {Verkhovskii}, {Karkin}, {Voronin}, {Kazantsev}, {Goshchitskii}, {Werner},
  and {Behr}}}]{Gerashenko2009}
\bibinfo{author}{\bibfnamefont{A.}~\bibnamefont{{Gerashenko}}},
  \bibinfo{author}{\bibfnamefont{S.}~\bibnamefont{{Verkhovskii}}},
  \bibinfo{author}{\bibfnamefont{A.}~\bibnamefont{{Karkin}}},
  \bibinfo{author}{\bibfnamefont{V.}~\bibnamefont{{Voronin}}},
  \bibinfo{author}{\bibfnamefont{A.}~\bibnamefont{{Kazantsev}}},
  \bibinfo{author}{\bibfnamefont{B.}~\bibnamefont{{Goshchitskii}}},
  \bibinfo{author}{\bibfnamefont{J.}~\bibnamefont{{Werner}}}, \bibnamefont{and}
  \bibinfo{author}{\bibfnamefont{G.}~\bibnamefont{{Behr}}},
  \bibinfo{journal}{ArXiv e-prints}  (\bibinfo{year}{2009}),
  \urlprefix\url{http://arxiv.org/abs/0911.2127}.

\bibitem[{\citenamefont{Zhao et~al.}(2008)\citenamefont{Zhao, Huang, de~la
  Cruz, Li, Lynn, Chen, Green, Chen, Li, Li et~al.}}]{Zhao2008}
\bibinfo{author}{\bibfnamefont{Jun} \bibnamefont{Zhao}},
  \bibinfo{author}{\bibfnamefont{Q.}~\bibnamefont{Huang}},
  \bibinfo{author}{\bibfnamefont{Clarina} \bibnamefont{de~la Cruz}},
  \bibinfo{author}{\bibfnamefont{Shiliang} \bibnamefont{Li}},
  \bibinfo{author}{\bibfnamefont{J.~W.} \bibnamefont{Lynn}},
  \bibinfo{author}{\bibfnamefont{Y.}~\bibnamefont{Chen}},
  \bibinfo{author}{\bibfnamefont{M.~A.} \bibnamefont{Green}},
  \bibinfo{author}{\bibfnamefont{G.~F.} \bibnamefont{Chen}},
  \bibinfo{author}{\bibfnamefont{G.}~\bibnamefont{Li}},
  \bibinfo{author}{\bibfnamefont{Z.}~\bibnamefont{Li}}, \bibnamefont{et~al.},
  \bibinfo{journal}{Nat. Mater.} \textbf{\bibinfo{volume}{7}},
  \bibinfo{pages}{953} (\bibinfo{year}{2008}), ISSN \bibinfo{issn}{1476-1122},
  \urlprefix\url{http://dx.doi.org/10.1038/nmat2315}.

\bibitem[{\citenamefont{Efremov et~al.}(2011)\citenamefont{Efremov, Korshunov,
  Dolgov, Golubov, and Hirschfeld}}]{EfremovKorshunov2011}
\bibinfo{author}{\bibfnamefont{D.~V.} \bibnamefont{Efremov}},
  \bibinfo{author}{\bibfnamefont{M.~M.} \bibnamefont{Korshunov}},
  \bibinfo{author}{\bibfnamefont{O.~V.} \bibnamefont{Dolgov}},
  \bibinfo{author}{\bibfnamefont{A.~A.} \bibnamefont{Golubov}},
  \bibnamefont{and} \bibinfo{author}{\bibfnamefont{P.~J.}
  \bibnamefont{Hirschfeld}}, \bibinfo{journal}{Phys. Rev. B}
  \textbf{\bibinfo{volume}{84}}, \bibinfo{pages}{180512}
  (\bibinfo{year}{2011}),
  \urlprefix\url{http://link.aps.org/doi/10.1103/PhysRevB.84.180512}.

\bibitem[{\citenamefont{Korshunov et~al.}(2014)\citenamefont{Korshunov,
  Efremov, Golubov, and Dolgov}}]{KorshunovMagn2014}
\bibinfo{author}{\bibfnamefont{M.~M.} \bibnamefont{Korshunov}},
  \bibinfo{author}{\bibfnamefont{D.~V.} \bibnamefont{Efremov}},
  \bibinfo{author}{\bibfnamefont{A.~A.} \bibnamefont{Golubov}},
  \bibnamefont{and} \bibinfo{author}{\bibfnamefont{O.~V.}
  \bibnamefont{Dolgov}}, \bibinfo{journal}{Phys. Rev. B}
  \textbf{\bibinfo{volume}{90}}, \bibinfo{pages}{134517}
  (\bibinfo{year}{2014}),
  \urlprefix\url{http://link.aps.org/doi/10.1103/PhysRevB.90.134517}.

\bibitem[{\citenamefont{Castellani et~al.}(1978)\citenamefont{Castellani,
  Natoli, and Ranninger}}]{Castallani1978}
\bibinfo{author}{\bibfnamefont{C.}~\bibnamefont{Castellani}},
  \bibinfo{author}{\bibfnamefont{C.~R.} \bibnamefont{Natoli}},
  \bibnamefont{and}
  \bibinfo{author}{\bibfnamefont{J.}~\bibnamefont{Ranninger}},
  \bibinfo{journal}{Phys. Rev. B} \textbf{\bibinfo{volume}{18}},
  \bibinfo{pages}{4945} (\bibinfo{year}{1978}),
  \urlprefix\url{http://link.aps.org/doi/10.1103/PhysRevB.18.4945}.

\bibitem[{\citenamefont{Ole\'{s}}(1983)}]{Oles1983}
\bibinfo{author}{\bibfnamefont{A.~M.} \bibnamefont{Ole\'{s}}},
  \bibinfo{journal}{Phys. Rev. B} \textbf{\bibinfo{volume}{28}},
  \bibinfo{pages}{327} (\bibinfo{year}{1983}),
  \urlprefix\url{http://link.aps.org/doi/10.1103/PhysRevB.28.327}.

\bibitem[{\citenamefont{Berk and Schrieffer}(1966)}]{BerkSchrieffer}
\bibinfo{author}{\bibfnamefont{N.~F.} \bibnamefont{Berk}} \bibnamefont{and}
  \bibinfo{author}{\bibfnamefont{J.~R.} \bibnamefont{Schrieffer}},
  \bibinfo{journal}{Phys. Rev. Lett.} \textbf{\bibinfo{volume}{17}},
  \bibinfo{pages}{433} (\bibinfo{year}{1966}),
  \urlprefix\url{http://link.aps.org/doi/10.1103/PhysRevLett.17.433}.

\bibitem[{\citenamefont{Allen and Mitrovic}(1982)}]{allen}
\bibinfo{author}{\bibfnamefont{P.~B.} \bibnamefont{Allen}} \bibnamefont{and}
  \bibinfo{author}{\bibfnamefont{B.}~\bibnamefont{Mitrovic}}, in
  \emph{\bibinfo{booktitle}{Solid State Physics: Advances in Research and
  Applications}}, edited by
  \bibinfo{editor}{\bibfnamefont{H.}~\bibnamefont{Erenreich}},
  \bibinfo{editor}{\bibfnamefont{F.}~\bibnamefont{Zeitz}}, \bibnamefont{and}
  \bibinfo{editor}{\bibfnamefont{D.}~\bibnamefont{Turnbull}}
  (\bibinfo{publisher}{Academic}, \bibinfo{address}{New York},
  \bibinfo{year}{1982}), vol.~\bibinfo{volume}{37}, pp. \bibinfo{pages}{1--92},
  ISBN \bibinfo{isbn}{0126077371}.

\bibitem[{\citenamefont{Parker et~al.}(2008)\citenamefont{Parker, Dolgov,
  Korshunov, Golubov, and Mazin}}]{ParkerKorshunov2008}
\bibinfo{author}{\bibfnamefont{D.}~\bibnamefont{Parker}},
  \bibinfo{author}{\bibfnamefont{O.~V.} \bibnamefont{Dolgov}},
  \bibinfo{author}{\bibfnamefont{M.~M.} \bibnamefont{Korshunov}},
  \bibinfo{author}{\bibfnamefont{A.~A.} \bibnamefont{Golubov}},
  \bibnamefont{and} \bibinfo{author}{\bibfnamefont{I.~I.} \bibnamefont{Mazin}},
  \bibinfo{journal}{Phys. Rev. B} \textbf{\bibinfo{volume}{78}},
  \bibinfo{pages}{134524} (\bibinfo{year}{2008}).

\bibitem[{\citenamefont{Popovich et~al.}(2010)\citenamefont{Popovich, Boris,
  Dolgov, Golubov, Sun, Lin, Kremer, and Keimer}}]{Popovich2010}
\bibinfo{author}{\bibfnamefont{P.}~\bibnamefont{Popovich}},
  \bibinfo{author}{\bibfnamefont{A.~V.} \bibnamefont{Boris}},
  \bibinfo{author}{\bibfnamefont{O.~V.} \bibnamefont{Dolgov}},
  \bibinfo{author}{\bibfnamefont{A.~A.} \bibnamefont{Golubov}},
  \bibinfo{author}{\bibfnamefont{D.~L.} \bibnamefont{Sun}},
  \bibinfo{author}{\bibfnamefont{C.~T.} \bibnamefont{Lin}},
  \bibinfo{author}{\bibfnamefont{R.~K.} \bibnamefont{Kremer}},
  \bibnamefont{and} \bibinfo{author}{\bibfnamefont{B.}~\bibnamefont{Keimer}},
  \bibinfo{journal}{Phys. Rev. Lett.} \textbf{\bibinfo{volume}{105}},
  \bibinfo{pages}{027003} (\bibinfo{year}{2010}),
  \urlprefix\url{http://link.aps.org/doi/10.1103/PhysRevLett.105.027003}.

\bibitem[{\citenamefont{Charnukha et~al.}(2011)\citenamefont{Charnukha, Dolgov,
  Golubov, Matiks, Sun, Lin, Keimer, and Boris}}]{Charnukha2011}
\bibinfo{author}{\bibfnamefont{A.}~\bibnamefont{Charnukha}},
  \bibinfo{author}{\bibfnamefont{O.~V.} \bibnamefont{Dolgov}},
  \bibinfo{author}{\bibfnamefont{A.~A.} \bibnamefont{Golubov}},
  \bibinfo{author}{\bibfnamefont{Y.}~\bibnamefont{Matiks}},
  \bibinfo{author}{\bibfnamefont{D.~L.} \bibnamefont{Sun}},
  \bibinfo{author}{\bibfnamefont{C.~T.} \bibnamefont{Lin}},
  \bibinfo{author}{\bibfnamefont{B.}~\bibnamefont{Keimer}}, \bibnamefont{and}
  \bibinfo{author}{\bibfnamefont{A.~V.} \bibnamefont{Boris}},
  \bibinfo{journal}{Phys. Rev. B} \textbf{\bibinfo{volume}{84}},
  \bibinfo{pages}{174511} (\bibinfo{year}{2011}),
  \urlprefix\url{http://link.aps.org/doi/10.1103/PhysRevB.84.174511}.

\bibitem[{\citenamefont{Ohashi}(2004)}]{Ohashi2004}
\bibinfo{author}{\bibfnamefont{Y.}~\bibnamefont{Ohashi}},
  \bibinfo{journal}{Physica C: Superconductivity}
  \textbf{\bibinfo{volume}{412–414, Part 1}}, \bibinfo{pages}{41 }
  (\bibinfo{year}{2004}), ISSN \bibinfo{issn}{0921-4534},
  \bibinfo{note}{proceedings of the 16th International Symposium on
  Superconductivity (ISS 2003). Advances in Superconductivity XVI. Part I},
  \urlprefix\url{http://www.sciencedirect.com/science/article/pii/S0921453404006185}.

\bibitem[{\citenamefont{Bickers et~al.}(1990)\citenamefont{Bickers, Scalapino,
  Collins, and Schlesinger}}]{Bickers1990}
\bibinfo{author}{\bibfnamefont{N.~E.} \bibnamefont{Bickers}},
  \bibinfo{author}{\bibfnamefont{D.~J.} \bibnamefont{Scalapino}},
  \bibinfo{author}{\bibfnamefont{R.~T.} \bibnamefont{Collins}},
  \bibnamefont{and}
  \bibinfo{author}{\bibfnamefont{Z.}~\bibnamefont{Schlesinger}},
  \bibinfo{journal}{Phys. Rev. B} \textbf{\bibinfo{volume}{42}},
  \bibinfo{pages}{67} (\bibinfo{year}{1990}),
  \urlprefix\url{http://link.aps.org/doi/10.1103/PhysRevB.42.67}.

\bibitem[{\citenamefont{Nam}(1967{\natexlab{a}})}]{Nam1967_I}
\bibinfo{author}{\bibfnamefont{Sang~Boo} \bibnamefont{Nam}},
  \bibinfo{journal}{Phys. Rev.} \textbf{\bibinfo{volume}{156}},
  \bibinfo{pages}{470} (\bibinfo{year}{1967}{\natexlab{a}}),
  \urlprefix\url{http://link.aps.org/doi/10.1103/PhysRev.156.470}.

\bibitem[{\citenamefont{Nam}(1967{\natexlab{b}})}]{Nam1967_II}
\bibinfo{author}{\bibfnamefont{Sang~Boo} \bibnamefont{Nam}},
  \bibinfo{journal}{Phys. Rev.} \textbf{\bibinfo{volume}{156}},
  \bibinfo{pages}{487} (\bibinfo{year}{1967}{\natexlab{b}}),
  \urlprefix\url{http://link.aps.org/doi/10.1103/PhysRev.156.487}.

\bibitem[{\citenamefont{Lee et~al.}(1989)\citenamefont{Lee, Rainer, and
  Zimmermann}}]{Lee1989}
\bibinfo{author}{\bibfnamefont{W.}~\bibnamefont{Lee}},
  \bibinfo{author}{\bibfnamefont{D.}~\bibnamefont{Rainer}}, \bibnamefont{and}
  \bibinfo{author}{\bibfnamefont{W.}~\bibnamefont{Zimmermann}},
  \bibinfo{journal}{Physica C: Superconductivity}
  \textbf{\bibinfo{volume}{159}}, \bibinfo{pages}{535 } (\bibinfo{year}{1989}),
  ISSN \bibinfo{issn}{0921-4534},
  \urlprefix\url{http://www.sciencedirect.com/science/article/pii/0921453489912847}.

\bibitem[{\citenamefont{Dolgov et~al.}(1990)\citenamefont{Dolgov, Golubov, and
  Shulga}}]{Dolgov1990}
\bibinfo{author}{\bibfnamefont{O.V.} \bibnamefont{Dolgov}},
  \bibinfo{author}{\bibfnamefont{A.A.} \bibnamefont{Golubov}},
  \bibnamefont{and} \bibinfo{author}{\bibfnamefont{S.V.} \bibnamefont{Shulga}},
  \bibinfo{journal}{Physics Letters A} \textbf{\bibinfo{volume}{147}},
  \bibinfo{pages}{317 } (\bibinfo{year}{1990}), ISSN \bibinfo{issn}{0375-9601},
  \urlprefix\url{http://www.sciencedirect.com/science/article/pii/037596019090457Y}.

\bibitem[{\citenamefont{Marsiglio}(1991)}]{Marsiglio1991}
\bibinfo{author}{\bibfnamefont{F.}~\bibnamefont{Marsiglio}},
  \bibinfo{journal}{Phys. Rev. B} \textbf{\bibinfo{volume}{44}},
  \bibinfo{pages}{5373} (\bibinfo{year}{1991}),
  \urlprefix\url{http://link.aps.org/doi/10.1103/PhysRevB.44.5373}.

\bibitem[{\citenamefont{Akis and Carbotte}(1991)}]{Akis1991}
\bibinfo{author}{\bibfnamefont{R.}~\bibnamefont{Akis}} \bibnamefont{and}
  \bibinfo{author}{\bibfnamefont{J.P.} \bibnamefont{Carbotte}},
  \bibinfo{journal}{Solid State Communications} \textbf{\bibinfo{volume}{79}},
  \bibinfo{pages}{577 } (\bibinfo{year}{1991}), ISSN \bibinfo{issn}{0038-1098},
  \urlprefix\url{http://www.sciencedirect.com/science/article/pii/003810989190913G}.

\bibitem[{\citenamefont{Sadovskii}(1997)}]{Sadovskii1997}
\bibinfo{author}{\bibfnamefont{M.~V.} \bibnamefont{Sadovskii}},
  \bibinfo{journal}{Physics Reports} \textbf{\bibinfo{volume}{282}},
  \bibinfo{pages}{225} (\bibinfo{year}{1997}), ISSN \bibinfo{issn}{0370-1573},
  \urlprefix\url{http://www.sciencedirect.com/science/article/pii/S0370157396000361}.

\bibitem[{\citenamefont{Sadovskii}(2000)}]{SadovskiiBook2000}
\bibinfo{author}{\bibfnamefont{M.~V.} \bibnamefont{Sadovskii}},
  \emph{\bibinfo{title}{Superconductivity and Localization}}
  (\bibinfo{publisher}{World Scientific Publishing Co. Pte. Inc.},
  \bibinfo{address}{Singapore}, \bibinfo{year}{2000}),
  \urlprefix\url{http://www.worldscientific.com/worldscibooks/10.1142/4321}.

\bibitem[{\citenamefont{Golubov and Mazin}(1997)}]{Golubov1997}
\bibinfo{author}{\bibfnamefont{A.~A.} \bibnamefont{Golubov}} \bibnamefont{and}
  \bibinfo{author}{\bibfnamefont{I.~I.} \bibnamefont{Mazin}},
  \bibinfo{journal}{Phys. Rev. B} \textbf{\bibinfo{volume}{55}},
  \bibinfo{pages}{15146} (\bibinfo{year}{1997}),
  \urlprefix\url{http://link.aps.org/doi/10.1103/PhysRevB.55.15146}.

\bibitem[{\citenamefont{Chubukov et~al.}(2008)\citenamefont{Chubukov, Efremov,
  and Eremin}}]{Chubukov2008}
\bibinfo{author}{\bibfnamefont{A.~V.} \bibnamefont{Chubukov}},
  \bibinfo{author}{\bibfnamefont{D.~V.} \bibnamefont{Efremov}},
  \bibnamefont{and} \bibinfo{author}{\bibfnamefont{I.}~\bibnamefont{Eremin}},
  \bibinfo{journal}{Phys. Rev. B} \textbf{\bibinfo{volume}{78}},
  \bibinfo{pages}{134512} (\bibinfo{year}{2008}),
  \urlprefix\url{http://link.aps.org/doi/10.1103/PhysRevB.78.134512}.

\bibitem[{\citenamefont{Senga and Kontani}(2008)}]{Senga2008}
\bibinfo{author}{\bibfnamefont{Y.}~\bibnamefont{Senga}} \bibnamefont{and}
  \bibinfo{author}{\bibfnamefont{H.}~\bibnamefont{Kontani}},
  \bibinfo{journal}{Journal of the Physical Society of Japan}
  \textbf{\bibinfo{volume}{77}}, \bibinfo{pages}{113710}
  (\bibinfo{year}{2008}),
  \urlprefix\url{http://dx.doi.org/10.1143/JPSJ.77.113710}.

\bibitem[{\citenamefont{Bang et~al.}(2009)\citenamefont{Bang, Choi, and
  Won}}]{Bang2009}
\bibinfo{author}{\bibfnamefont{Y.}~\bibnamefont{Bang}},
  \bibinfo{author}{\bibfnamefont{H.-Y.} \bibnamefont{Choi}}, \bibnamefont{and}
  \bibinfo{author}{\bibfnamefont{H.}~\bibnamefont{Won}},
  \bibinfo{journal}{Phys. Rev. B} \textbf{\bibinfo{volume}{79}},
  \bibinfo{pages}{054529} (\bibinfo{year}{2009}),
  \urlprefix\url{http://link.aps.org/doi/10.1103/PhysRevB.79.054529}.

\bibitem[{\citenamefont{Golubov and Mazin}(1995)}]{Golubov1995}
\bibinfo{author}{\bibfnamefont{A.A.} \bibnamefont{Golubov}} \bibnamefont{and}
  \bibinfo{author}{\bibfnamefont{I.I.} \bibnamefont{Mazin}},
  \bibinfo{journal}{Physica C: Superconductivity}
  \textbf{\bibinfo{volume}{243}}, \bibinfo{pages}{153 } (\bibinfo{year}{1995}),
  ISSN \bibinfo{issn}{0921-4534},
  \urlprefix\url{http://www.sciencedirect.com/science/article/pii/0921453494024456}.

\bibitem[{\citenamefont{Kemper et~al.}(2009)\citenamefont{Kemper, Cao,
  Hirschfeld, and Cheng}}]{a_kemper_09}
\bibinfo{author}{\bibfnamefont{A.~F.} \bibnamefont{Kemper}},
  \bibinfo{author}{\bibfnamefont{C.}~\bibnamefont{Cao}},
  \bibinfo{author}{\bibfnamefont{P.~J.} \bibnamefont{Hirschfeld}},
  \bibnamefont{and} \bibinfo{author}{\bibfnamefont{H.-P.} \bibnamefont{Cheng}},
  \bibinfo{journal}{Phys. Rev. B} \textbf{\bibinfo{volume}{80}},
  \bibinfo{pages}{104511} (\bibinfo{year}{2009}),
  \urlprefix\url{http://link.aps.org/doi/10.1103/PhysRevB.80.104511}.

\bibitem[{\citenamefont{Schopohl and Scharnberg}(1977)}]{Schopohl1977}
\bibinfo{author}{\bibfnamefont{N.}~\bibnamefont{Schopohl}} \bibnamefont{and}
  \bibinfo{author}{\bibfnamefont{K.}~\bibnamefont{Scharnberg}},
  \bibinfo{journal}{Solid State Communications} \textbf{\bibinfo{volume}{22}},
  \bibinfo{pages}{371 } (\bibinfo{year}{1977}), ISSN \bibinfo{issn}{0038-1098},
  \urlprefix\url{http://www.sciencedirect.com/science/article/pii/0038109877910699}.

\bibitem[{\citenamefont{Ummarino}(2007)}]{Ummarino2007}
\bibinfo{author}{\bibfnamefont{G.A.} \bibnamefont{Ummarino}},
  \bibinfo{journal}{Journal of Superconductivity and Novel Magnetism}
  \textbf{\bibinfo{volume}{20}}, \bibinfo{pages}{639} (\bibinfo{year}{2007}),
  ISSN \bibinfo{issn}{1557-1939},
  \urlprefix\url{http://dx.doi.org/10.1007/s10948-007-0259-y}.

\bibitem[{\citenamefont{Yao et~al.}(2012)\citenamefont{Yao, Chen, Li, Cao,
  Jiang, Wang, Xu, and Zhang}}]{Yao2012}
\bibinfo{author}{\bibfnamefont{Z.-J.} \bibnamefont{Yao}},
  \bibinfo{author}{\bibfnamefont{W.-Q.} \bibnamefont{Chen}},
  \bibinfo{author}{\bibfnamefont{Y.-k.} \bibnamefont{Li}},
  \bibinfo{author}{\bibfnamefont{G.-h.} \bibnamefont{Cao}},
  \bibinfo{author}{\bibfnamefont{H.-M.} \bibnamefont{Jiang}},
  \bibinfo{author}{\bibfnamefont{Q.-E.} \bibnamefont{Wang}},
  \bibinfo{author}{\bibfnamefont{Z.-a.} \bibnamefont{Xu}}, \bibnamefont{and}
  \bibinfo{author}{\bibfnamefont{F.-C.} \bibnamefont{Zhang}},
  \bibinfo{journal}{Phys. Rev. B} \textbf{\bibinfo{volume}{86}},
  \bibinfo{pages}{184515} (\bibinfo{year}{2012}),
  \urlprefix\url{http://link.aps.org/doi/10.1103/PhysRevB.86.184515}.

\bibitem[{\citenamefont{Chen et~al.}(2013)\citenamefont{Chen, Tai, Ting, Graf,
  Dai, and Zhu}}]{Chen2013}
\bibinfo{author}{\bibfnamefont{H.}~\bibnamefont{Chen}},
  \bibinfo{author}{\bibfnamefont{Y.-Y.} \bibnamefont{Tai}},
  \bibinfo{author}{\bibfnamefont{C.~S.} \bibnamefont{Ting}},
  \bibinfo{author}{\bibfnamefont{M.~J.} \bibnamefont{Graf}},
  \bibinfo{author}{\bibfnamefont{J.}~\bibnamefont{Dai}}, \bibnamefont{and}
  \bibinfo{author}{\bibfnamefont{J.-X.} \bibnamefont{Zhu}},
  \bibinfo{journal}{Phys. Rev. B} \textbf{\bibinfo{volume}{88}},
  \bibinfo{pages}{184509} (\bibinfo{year}{2013}),
  \urlprefix\url{http://link.aps.org/doi/10.1103/PhysRevB.88.184509}.

\bibitem[{\citenamefont{Ambegaokar and Griffin}(1965)}]{ambeg}
\bibinfo{author}{\bibfnamefont{V.}~\bibnamefont{Ambegaokar}} \bibnamefont{and}
  \bibinfo{author}{\bibfnamefont{A.}~\bibnamefont{Griffin}},
  \bibinfo{journal}{Phys. Rev.} \textbf{\bibinfo{volume}{137}},
  \bibinfo{pages}{A1151} (\bibinfo{year}{1965}),
  \urlprefix\url{http://link.aps.org/doi/10.1103/PhysRev.137.A1151}.

\bibitem[{\citenamefont{Korshunov et~al.}(2016)\citenamefont{Korshunov,
  Togushova, and Dolgov}}]{Korshunov2016}
\bibinfo{author}{\bibfnamefont{M.~M.} \bibnamefont{Korshunov}},
  \bibinfo{author}{\bibfnamefont{Yu.~N.} \bibnamefont{Togushova}},
  \bibnamefont{and} \bibinfo{author}{\bibfnamefont{O.~V.}
  \bibnamefont{Dolgov}}, \bibinfo{journal}{Journal of Superconductivity and
  Novel Magnetism} \textbf{\bibinfo{volume}{29}}, \bibinfo{pages}{1089}
  (\bibinfo{year}{2016}), ISSN \bibinfo{issn}{1557-1947},
  \urlprefix\url{http://dx.doi.org/10.1007/s10948-016-3385-6}.

\bibitem[{\citenamefont{Li and Wang}(2009)}]{Li2009}
\bibinfo{author}{\bibfnamefont{J.}~\bibnamefont{Li}} \bibnamefont{and}
  \bibinfo{author}{\bibfnamefont{Y.}~\bibnamefont{Wang}}, \bibinfo{journal}{EPL
  (Europhysics Letters)} \textbf{\bibinfo{volume}{88}}, \bibinfo{pages}{17009}
  (\bibinfo{year}{2009}),
  \urlprefix\url{http://stacks.iop.org/0295-5075/88/i=1/a=17009}.

\bibitem[{\citenamefont{Stanev and Koshelev}(2012)}]{Stanev2012}
\bibinfo{author}{\bibfnamefont{V.~G.} \bibnamefont{Stanev}} \bibnamefont{and}
  \bibinfo{author}{\bibfnamefont{A.~E.} \bibnamefont{Koshelev}},
  \bibinfo{journal}{Phys. Rev. B} \textbf{\bibinfo{volume}{86}},
  \bibinfo{pages}{174515} (\bibinfo{year}{2012}),
  \urlprefix\url{http://link.aps.org/doi/10.1103/PhysRevB.86.174515}.

\bibitem[{\citenamefont{Cheng et~al.}(2013)\citenamefont{Cheng, Shen, Han, and
  Wen}}]{Cheng2013}
\bibinfo{author}{\bibfnamefont{Peng} \bibnamefont{Cheng}},
  \bibinfo{author}{\bibfnamefont{Bing} \bibnamefont{Shen}},
  \bibinfo{author}{\bibfnamefont{Fei} \bibnamefont{Han}}, \bibnamefont{and}
  \bibinfo{author}{\bibfnamefont{Hai-Hu} \bibnamefont{Wen}},
  \bibinfo{journal}{EPL (Europhysics Letters)} \textbf{\bibinfo{volume}{104}},
  \bibinfo{pages}{37007} (\bibinfo{year}{2013}),
  \urlprefix\url{http://stacks.iop.org/0295-5075/104/i=3/a=37007}.

\bibitem[{\citenamefont{Li et~al.}(2011{\natexlab{a}})\citenamefont{Li, Guo,
  Zhang, Yu, Tsujimoto, Kontani, Yamaura, and Takayama-Muromachi}}]{Li2011}
\bibinfo{author}{\bibfnamefont{Jun} \bibnamefont{Li}},
  \bibinfo{author}{\bibfnamefont{Yanfeng} \bibnamefont{Guo}},
  \bibinfo{author}{\bibfnamefont{Shoubao} \bibnamefont{Zhang}},
  \bibinfo{author}{\bibfnamefont{Shan} \bibnamefont{Yu}},
  \bibinfo{author}{\bibfnamefont{Yoshihiro} \bibnamefont{Tsujimoto}},
  \bibinfo{author}{\bibfnamefont{Hiroshi} \bibnamefont{Kontani}},
  \bibinfo{author}{\bibfnamefont{Kazunari} \bibnamefont{Yamaura}},
  \bibnamefont{and} \bibinfo{author}{\bibfnamefont{Eiji}
  \bibnamefont{Takayama-Muromachi}}, \bibinfo{journal}{Phys. Rev. B}
  \textbf{\bibinfo{volume}{84}}, \bibinfo{pages}{020513}
  (\bibinfo{year}{2011}{\natexlab{a}}),
  \urlprefix\url{http://link.aps.org/doi/10.1103/PhysRevB.84.020513}.

\bibitem[{\citenamefont{Suzuki et~al.}(2010)\citenamefont{Suzuki, Miyasaka, and
  Tajima}}]{Suzuki2010}
\bibinfo{author}{\bibfnamefont{S.}~\bibnamefont{Suzuki}},
  \bibinfo{author}{\bibfnamefont{S.}~\bibnamefont{Miyasaka}}, \bibnamefont{and}
  \bibinfo{author}{\bibfnamefont{S.}~\bibnamefont{Tajima}},
  \bibinfo{journal}{Physica C: Superconductivity} \textbf{\bibinfo{volume}{470,
  Supplement 1}}, \bibinfo{pages}{S330 } (\bibinfo{year}{2010}), ISSN
  \bibinfo{issn}{0921-4534}, \bibinfo{note}{proceedings of the 9th
  International Conference on Materials and Mechanisms of Superconductivity},
  \urlprefix\url{http://www.sciencedirect.com/science/article/pii/S0921453410002157}.

\bibitem[{\citenamefont{Bezusyy et~al.}(2012)\citenamefont{Bezusyy, Gawryluk,
  Berkowski, and Cieplak}}]{Bezusyy2012}
\bibinfo{author}{\bibfnamefont{V.L.} \bibnamefont{Bezusyy}},
  \bibinfo{author}{\bibfnamefont{D.J.} \bibnamefont{Gawryluk}},
  \bibinfo{author}{\bibfnamefont{M.}~\bibnamefont{Berkowski}},
  \bibnamefont{and} \bibinfo{author}{\bibfnamefont{M.Z.}
  \bibnamefont{Cieplak}}, \bibinfo{journal}{Acta Physica Polonica A}
  \textbf{\bibinfo{volume}{121}}, \bibinfo{pages}{816} (\bibinfo{year}{2012}),
  \urlprefix\url{http://dx.doi.org/10.12693/APhysPolA.121.816}.

\bibitem[{\citenamefont{Guo et~al.}(2010)\citenamefont{Guo, Shi, Yu, Belik,
  Matsushita, Tanaka, Katsuya, Kobayashi, Nowik, Felner et~al.}}]{Guo2010imp}
\bibinfo{author}{\bibfnamefont{Y.~F.} \bibnamefont{Guo}},
  \bibinfo{author}{\bibfnamefont{Y.~G.} \bibnamefont{Shi}},
  \bibinfo{author}{\bibfnamefont{S.}~\bibnamefont{Yu}},
  \bibinfo{author}{\bibfnamefont{A.~A.} \bibnamefont{Belik}},
  \bibinfo{author}{\bibfnamefont{Y.}~\bibnamefont{Matsushita}},
  \bibinfo{author}{\bibfnamefont{M.}~\bibnamefont{Tanaka}},
  \bibinfo{author}{\bibfnamefont{Y.}~\bibnamefont{Katsuya}},
  \bibinfo{author}{\bibfnamefont{K.}~\bibnamefont{Kobayashi}},
  \bibinfo{author}{\bibfnamefont{I.}~\bibnamefont{Nowik}},
  \bibinfo{author}{\bibfnamefont{I.}~\bibnamefont{Felner}},
  \bibnamefont{et~al.}, \bibinfo{journal}{Phys. Rev. B}
  \textbf{\bibinfo{volume}{82}}, \bibinfo{pages}{054506}
  (\bibinfo{year}{2010}),
  \urlprefix\url{http://link.aps.org/doi/10.1103/PhysRevB.82.054506}.

\bibitem[{\citenamefont{Lee et~al.}(2010)\citenamefont{Lee, Satomi, Kobayashi,
  and Sato}}]{Lee2010}
\bibinfo{author}{\bibfnamefont{S.~C.} \bibnamefont{Lee}},
  \bibinfo{author}{\bibfnamefont{E.}~\bibnamefont{Satomi}},
  \bibinfo{author}{\bibfnamefont{Y.}~\bibnamefont{Kobayashi}},
  \bibnamefont{and} \bibinfo{author}{\bibfnamefont{M.}~\bibnamefont{Sato}},
  \bibinfo{journal}{Journal of the Physical Society of Japan}
  \textbf{\bibinfo{volume}{79}}, \bibinfo{pages}{023702}
  (\bibinfo{year}{2010}), \eprint{http://dx.doi.org/10.1143/JPSJ.79.023702},
  \urlprefix\url{http://dx.doi.org/10.1143/JPSJ.79.023702}.

\bibitem[{\citenamefont{Satomi et~al.}(2010)\citenamefont{Satomi, Lee,
  Kobayashi, and Sato}}]{Satomi2010}
\bibinfo{author}{\bibfnamefont{Erika} \bibnamefont{Satomi}},
  \bibinfo{author}{\bibfnamefont{Sang~Chul} \bibnamefont{Lee}},
  \bibinfo{author}{\bibfnamefont{Yoshiaki} \bibnamefont{Kobayashi}},
  \bibnamefont{and} \bibinfo{author}{\bibfnamefont{Masatoshi}
  \bibnamefont{Sato}}, \bibinfo{journal}{Journal of the Physical Society of
  Japan} \textbf{\bibinfo{volume}{79}}, \bibinfo{pages}{094702}
  (\bibinfo{year}{2010}), \eprint{http://dx.doi.org/10.1143/JPSJ.79.094702},
  \urlprefix\url{http://dx.doi.org/10.1143/JPSJ.79.094702}.

\bibitem[{\citenamefont{Wang et~al.}(2014)\citenamefont{Wang, Zhou, Luo, Hong,
  Yan, Ying, Cheng, Ye, Xiang, Li et~al.}}]{Wang2014}
\bibinfo{author}{\bibfnamefont{A.~F.} \bibnamefont{Wang}},
  \bibinfo{author}{\bibfnamefont{S.~Y.} \bibnamefont{Zhou}},
  \bibinfo{author}{\bibfnamefont{X.~G.} \bibnamefont{Luo}},
  \bibinfo{author}{\bibfnamefont{X.~C.} \bibnamefont{Hong}},
  \bibinfo{author}{\bibfnamefont{Y.~J.} \bibnamefont{Yan}},
  \bibinfo{author}{\bibfnamefont{J.~J.} \bibnamefont{Ying}},
  \bibinfo{author}{\bibfnamefont{P.}~\bibnamefont{Cheng}},
  \bibinfo{author}{\bibfnamefont{G.~J.} \bibnamefont{Ye}},
  \bibinfo{author}{\bibfnamefont{Z.~J.} \bibnamefont{Xiang}},
  \bibinfo{author}{\bibfnamefont{S.~Y.} \bibnamefont{Li}},
  \bibnamefont{et~al.}, \bibinfo{journal}{Phys. Rev. B}
  \textbf{\bibinfo{volume}{89}}, \bibinfo{pages}{064510}
  (\bibinfo{year}{2014}),
  \urlprefix\url{http://link.aps.org/doi/10.1103/PhysRevB.89.064510}.

\bibitem[{\citenamefont{Maiti et~al.}(2012)\citenamefont{Maiti, Korshunov, and
  Chubukov}}]{MaitiKorshunov2012}
\bibinfo{author}{\bibfnamefont{S.}~\bibnamefont{Maiti}},
  \bibinfo{author}{\bibfnamefont{M.~M.} \bibnamefont{Korshunov}},
  \bibnamefont{and} \bibinfo{author}{\bibfnamefont{A.~V.}
  \bibnamefont{Chubukov}}, \bibinfo{journal}{Phys. Rev. B}
  \textbf{\bibinfo{volume}{85}}, \bibinfo{pages}{014511}
  (\bibinfo{year}{2012}),
  \urlprefix\url{http://link.aps.org/doi/10.1103/PhysRevB.85.014511}.

\bibitem[{\citenamefont{Golubov et~al.}(2002)\citenamefont{Golubov, Brinkman,
  Dolgov, Kortus, and Jepsen}}]{Golubov2002}
\bibinfo{author}{\bibfnamefont{A.~A.} \bibnamefont{Golubov}},
  \bibinfo{author}{\bibfnamefont{A.}~\bibnamefont{Brinkman}},
  \bibinfo{author}{\bibfnamefont{O.~V.} \bibnamefont{Dolgov}},
  \bibinfo{author}{\bibfnamefont{J.}~\bibnamefont{Kortus}}, \bibnamefont{and}
  \bibinfo{author}{\bibfnamefont{O.}~\bibnamefont{Jepsen}},
  \bibinfo{journal}{Phys. Rev. B} \textbf{\bibinfo{volume}{66}},
  \bibinfo{pages}{054524} (\bibinfo{year}{2002}),
  \urlprefix\url{http://link.aps.org/doi/10.1103/PhysRevB.66.054524}.

\bibitem[{\citenamefont{V. Efremov et~al.}(2013)\citenamefont{V. Efremov,
  A. Golubov, and Dolgov}}]{Efremov2013}
\bibinfo{author}{\bibfnamefont{D.}~\bibnamefont{V. Efremov}},
  \bibinfo{author}{\bibfnamefont{A.}~\bibnamefont{A. Golubov}},
  \bibnamefont{and} \bibinfo{author}{\bibfnamefont{O.~V.}
  \bibnamefont{Dolgov}}, \bibinfo{journal}{New Journal of Physics}
  \textbf{\bibinfo{volume}{15}}, \bibinfo{pages}{013002}
  (\bibinfo{year}{2013}),
  \urlprefix\url{http://stacks.iop.org/1367-2630/15/i=1/a=013002}.

\bibitem[{\citenamefont{Marsiglio et~al.}(1996)\citenamefont{Marsiglio,
  Carbotte, Puchkov, and Timusk}}]{Marsiglio1996}
\bibinfo{author}{\bibfnamefont{F.}~\bibnamefont{Marsiglio}},
  \bibinfo{author}{\bibfnamefont{J.~P.} \bibnamefont{Carbotte}},
  \bibinfo{author}{\bibfnamefont{A.}~\bibnamefont{Puchkov}}, \bibnamefont{and}
  \bibinfo{author}{\bibfnamefont{T.}~\bibnamefont{Timusk}},
  \bibinfo{journal}{Phys. Rev. B} \textbf{\bibinfo{volume}{53}},
  \bibinfo{pages}{9433} (\bibinfo{year}{1996}),
  \urlprefix\url{http://link.aps.org/doi/10.1103/PhysRevB.53.9433}.

\bibitem[{\citenamefont{Nakai et~al.}(2008)\citenamefont{Nakai, Ishida,
  Kamihara, Hirano, and Hosono}}]{Nakai2008}
\bibinfo{author}{\bibfnamefont{Y.}~\bibnamefont{Nakai}},
  \bibinfo{author}{\bibfnamefont{K.}~\bibnamefont{Ishida}},
  \bibinfo{author}{\bibfnamefont{Y.}~\bibnamefont{Kamihara}},
  \bibinfo{author}{\bibfnamefont{M.}~\bibnamefont{Hirano}}, \bibnamefont{and}
  \bibinfo{author}{\bibfnamefont{H.}~\bibnamefont{Hosono}},
  \bibinfo{journal}{Journal of the Physical Society of Japan}
  \textbf{\bibinfo{volume}{77}}, \bibinfo{pages}{073701}
  (\bibinfo{year}{2008}),
  \urlprefix\url{http://dx.doi.org/10.1143/JPSJ.77.073701}.

\bibitem[{\citenamefont{Ning et~al.}(2008)\citenamefont{Ning, Ahilan, Imai,
  Sefat, Jin, McGuire, Sales, and Mandrus}}]{Ning2008}
\bibinfo{author}{\bibfnamefont{F.}~\bibnamefont{Ning}},
  \bibinfo{author}{\bibfnamefont{K.}~\bibnamefont{Ahilan}},
  \bibinfo{author}{\bibfnamefont{T.}~\bibnamefont{Imai}},
  \bibinfo{author}{\bibfnamefont{A.~S.} \bibnamefont{Sefat}},
  \bibinfo{author}{\bibfnamefont{R.}~\bibnamefont{Jin}},
  \bibinfo{author}{\bibfnamefont{M.~A.} \bibnamefont{McGuire}},
  \bibinfo{author}{\bibfnamefont{B.~C.} \bibnamefont{Sales}}, \bibnamefont{and}
  \bibinfo{author}{\bibfnamefont{D.}~\bibnamefont{Mandrus}},
  \bibinfo{journal}{Journal of the Physical Society of Japan}
  \textbf{\bibinfo{volume}{77}}, \bibinfo{pages}{103705}
  (\bibinfo{year}{2008}),
  \urlprefix\url{http://dx.doi.org/10.1143/JPSJ.77.103705}.

\bibitem[{\citenamefont{Korshunov and
  Eremin}(2008)}]{KorshunovEreminResonance2008}
\bibinfo{author}{\bibfnamefont{M.~M.} \bibnamefont{Korshunov}}
  \bibnamefont{and} \bibinfo{author}{\bibfnamefont{I.}~\bibnamefont{Eremin}},
  \bibinfo{journal}{Phys. Rev. B} \textbf{\bibinfo{volume}{78}},
  \bibinfo{pages}{140509} (\bibinfo{year}{2008}),
  \urlprefix\url{http://link.aps.org/doi/10.1103/PhysRevB.78.140509}.

\bibitem[{\citenamefont{Maier and Scalapino}(2008)}]{t_maier_08b}
\bibinfo{author}{\bibfnamefont{T.~A.} \bibnamefont{Maier}} \bibnamefont{and}
  \bibinfo{author}{\bibfnamefont{D.~J.} \bibnamefont{Scalapino}},
  \bibinfo{journal}{Phys. Rev. B} \textbf{\bibinfo{volume}{78}},
  \bibinfo{pages}{020514} (\bibinfo{year}{2008}),
  \urlprefix\url{http://link.aps.org/doi/10.1103/PhysRevB.78.020514}.

\bibitem[{\citenamefont{Inosov}(2016)}]{Inosov2016}
\bibinfo{author}{\bibfnamefont{Dmytro~S.} \bibnamefont{Inosov}},
  \bibinfo{journal}{Comptes Rendus Physique} \textbf{\bibinfo{volume}{17}},
  \bibinfo{pages}{60 } (\bibinfo{year}{2016}), ISSN \bibinfo{issn}{1631-0705},
  \urlprefix\url{http://www.sciencedirect.com/science/article/pii/S1631070515000523}.

\bibitem[{\citenamefont{Hirschfeld et~al.}(1988)\citenamefont{Hirschfeld,
  W\"olfle, and Einzel}}]{PJHconsequences}
\bibinfo{author}{\bibfnamefont{P.~J.} \bibnamefont{Hirschfeld}},
  \bibinfo{author}{\bibfnamefont{P.}~\bibnamefont{W\"olfle}}, \bibnamefont{and}
  \bibinfo{author}{\bibfnamefont{D.}~\bibnamefont{Einzel}},
  \bibinfo{journal}{Phys. Rev. B} \textbf{\bibinfo{volume}{37}},
  \bibinfo{pages}{83} (\bibinfo{year}{1988}),
  \urlprefix\url{http://link.aps.org/doi/10.1103/PhysRevB.37.83}.

\bibitem[{\citenamefont{Yuko and Hiroshi}(2009)}]{Senga2009}
\bibinfo{author}{\bibfnamefont{S.}~\bibnamefont{Yuko}} \bibnamefont{and}
  \bibinfo{author}{\bibfnamefont{K.}~\bibnamefont{Hiroshi}},
  \bibinfo{journal}{New Journal of Physics} \textbf{\bibinfo{volume}{11}},
  \bibinfo{pages}{035005} (\bibinfo{year}{2009}),
  \urlprefix\url{http://stacks.iop.org/1367-2630/11/i=3/a=035005}.

\bibitem[{\citenamefont{Nakai et~al.}(2010{\natexlab{b}})\citenamefont{Nakai,
  Iye, Kitagawa, Ishida, Ikeda, Kasahara, Shishido, Shibauchi, Matsuda, and
  Terashima}}]{NakaiPdoped2010}
\bibinfo{author}{\bibfnamefont{Y.}~\bibnamefont{Nakai}},
  \bibinfo{author}{\bibfnamefont{T.}~\bibnamefont{Iye}},
  \bibinfo{author}{\bibfnamefont{S.}~\bibnamefont{Kitagawa}},
  \bibinfo{author}{\bibfnamefont{K.}~\bibnamefont{Ishida}},
  \bibinfo{author}{\bibfnamefont{H.}~\bibnamefont{Ikeda}},
  \bibinfo{author}{\bibfnamefont{S.}~\bibnamefont{Kasahara}},
  \bibinfo{author}{\bibfnamefont{H.}~\bibnamefont{Shishido}},
  \bibinfo{author}{\bibfnamefont{T.}~\bibnamefont{Shibauchi}},
  \bibinfo{author}{\bibfnamefont{Y.}~\bibnamefont{Matsuda}}, \bibnamefont{and}
  \bibinfo{author}{\bibfnamefont{T.}~\bibnamefont{Terashima}},
  \bibinfo{journal}{Phys. Rev. Lett.} \textbf{\bibinfo{volume}{105}},
  \bibinfo{pages}{107003} (\bibinfo{year}{2010}{\natexlab{b}}),
  \urlprefix\url{http://link.aps.org/doi/10.1103/PhysRevLett.105.107003}.

\bibitem[{\citenamefont{Li et~al.}(2011{\natexlab{b}})\citenamefont{Li, Sun,
  Lin, Su, Hu, and Zheng}}]{LiBKFA}
\bibinfo{author}{\bibfnamefont{Z.}~\bibnamefont{Li}},
  \bibinfo{author}{\bibfnamefont{D.~L.} \bibnamefont{Sun}},
  \bibinfo{author}{\bibfnamefont{C.~T.} \bibnamefont{Lin}},
  \bibinfo{author}{\bibfnamefont{Y.~H.} \bibnamefont{Su}},
  \bibinfo{author}{\bibfnamefont{J.~P.} \bibnamefont{Hu}}, \bibnamefont{and}
  \bibinfo{author}{\bibfnamefont{Guo-qing} \bibnamefont{Zheng}},
  \bibinfo{journal}{Phys. Rev. B} \textbf{\bibinfo{volume}{83}},
  \bibinfo{pages}{140506} (\bibinfo{year}{2011}{\natexlab{b}}),
  \urlprefix\url{http://link.aps.org/doi/10.1103/PhysRevB.83.140506}.

\end{thebibliography}

\end{document}